\documentclass{article}
\pdfoutput=1
\usepackage{jcappub}

\title{Dark matter and halo bispectrum in redshift space: theory and applications}

\author[a]{H\'ector Gil-Mar\'in,}
\author[b]{Christian Wagner,}
\author[c]{Jorge Nore\~na,}
\author[d,e,f]{Licia Verde,}
\author[a]{Will Percival}

\affiliation[a]{Institute of Cosmology \& Gravitation, University of Portsmouth, Dennis Sciama Building, Portsmouth PO1 3FX, UK}
\affiliation[b]{Max-Planck-Institut f\"ur Astrophysik, Karl-Schwarzschild Str. 1, 85741 Garching, Germany}
\affiliation[c]{Department of Theoretical Physics and Center for Astroparticle Physics (CAP), 24 quai E. Ansermet, CH-1211 Geneva 4, CH}
\affiliation[d]{ICREA Instituci\'o Catalana de Recerca i Estudis Avan\c{c}ats. Passeig Llu\'is Companys 23, E-08010 Barcelona, Spain}
\affiliation[e]{Institut de Ciencies del Cosmos, Universitat de Barcelona, IEEC-UB, Marti i Franques 1, 08028, Barcelona, Spain}
\affiliation[f]{Institute of Theoretical Astrophysics, University of Oslo, Norway}

\emailAdd{hector.gil@port.ac.uk }

\abstract{We present a phenomenological modification of the standard perturbation theory prediction for the bispectrum in redshift space that allows us to extend the model to mildly non-linear scales over a wide range of redshifts, $z\leq1.5$. Our model require 18 free parameters that are fitted to N-body simulations using the shapes $k_2/k_1=1,\,1.5,\,2.0,\,2.5$. We find that we can describe the bispectrum of dark matter particles with $\sim5\%$ accuracy for $k_i\lesssim0.10\,h/{\rm Mpc}$ at $z=0$, for $k_i\lesssim0.15\,h/{\rm Mpc}$ at $z=0.5$, for $k_i\lesssim0.17\,h/{\rm Mpc}$ at $z=1.0$ and for $k_i\lesssim0.20\,h/{\rm Mpc}$ at $z=1.5$.   For very squeezed triangles with $k_1=k_2\gtrsim0.1\,h{\rm Mpc}^{-1}$ and $k_3\leq0.02\,h{\rm Mpc}^{-1}$, however, neither SPT nor the proposed fitting formula are able to describe the measured dark matter bispectrum with this accuracy. We show that the fitting formula is sufficiently general that can be applied to other intermediate shapes such as $k_2/k_1=1.25,\,1.75,\, {\rm and}\, 2.25$. We also test that the fitting formula is able to describe with similar accuracy the bispectrum of cosmologies with different $\Omega_m$, in the range $0.2\lesssim \Omega_m \lesssim 0.4$, and consequently with different values of the logarithmic grow rate $f$ at $z=0$, $0.4\lesssim f(z=0) \lesssim 0.6$.  We apply this new formula to recover the bias parameters,  $f$ and $\sigma_8$, by combining the redshift space power spectrum  monopole and quadrupole with the bispectrum monopole for  both dark matter particles and haloes. We find that the combination of these three statistics can break the degeneracy between $b_1$, $f$ and $\sigma_8$. For dark matter particles the new model can be used to recover $f$ and $\sigma_8$ with $\sim1\%$ accuracy. For dark matter haloes we find that $f$ and $\sigma_8$ present larger systematic shifts,  $\sim10\%$.  The systematic offsets arise because of limitations in the  modelling of the interplay between bias and redshift space distortions, and represent a limitation  as the statistical errors of forthcoming surveys reach this level. Conveniently, we find that these residual systematics are mitigated for combinations of parameters. In particular, the quantity $f\sigma_8$ is still recovered with  $\sim1\%$ accuracy for the particular halo population and cosmology studied.  The improvement on the modelling of the bispectrum presented in this paper will be  useful for extracting information from current and future galaxy surveys. }
\begin{document}

\maketitle

\section{Introduction}
Over the last decade advances in astronomical spectroscopy and photometry of large samples of galaxies have allowed the galaxy distribution to be measured to unprecedented accuracy. The analysis  of the resulting maps has yielded constraints on  the growth rate of structures, the Universe expansion history as well as  on cosmological parameters.

The successful measurement of cosmological parameters relies on both the accuracy of the theoretical models as well as the precision of the statistics used. In the past, the precision of the measurements was poor and a $\sim10\%$ statistical error on the measurement of the power spectrum and even higher on the bispectrum was the limiting factor for discriminating among models and theories. However, current and forthcoming  surveys (BOSS \citep{BOSS}\footnote{Baryon Oscillator Spectroscopic Survey: \url{http://www.sdss3.org/surveys/boss.php}}, WiggleZ \citep{WiggleZ}\footnote{WiggleZ Dark Energy Survey : \url{http://wigglez.swin.edu.au/}}, DES \citep{DES}\footnote{Dark Energy Survey: \url{http://www.darkenergysurvey.org/}}, EUCLID \citep{EUCLID}\footnote{EUCLID: \url{http://www.euclid-ec.org/}}) are rapidly approaching to the $1\%$ statistical precision for two-point statistics, and are constraining higher-order statistics with similar jump in precision.  This level of precision is comparable to the  accuracy of the theoretical models  that have been developed. Consequently, a large effort has been put into improving the theory, proceeding in different directions. The first step is to upgrade the modelling of the statistics of dark matter in real space \citep{Crocce_Scoccimarro:2006,Okamuraetal:2011,Valageas_Nishimichi:2011,Carlsonetal:2012,Pietroni:2008}. The second is to improve the bias model in order to describe accurately how the galaxies trace dark matter, including non-linear, non-local and non-Gaussian terms \citep{TNS_halo,McDonald_Roy,Saitoetal:2014,Biagetti14}. The third is to accurately model the mapping from real to redshift space statistics \citep{TNS_matter,Vlahetal:2012,Okumuraetal:2012,Reid_White:2011}.  
 Combining these improvements leads to a model to describe the 2-point statistics with accuracy close to 1\%. Although some progress have been made \citep{Scoccimarroetal:1998,Panetal:2007,Sefusatti:2009,Yokoyamaetal:2014,Takushimaetal:2014,Verde_Heavens:2001,Jurgens_Bartelmann,Rampf_Buchert,Rampf_Wong,Angulo14,Baldauf14}, we have not seen a similar improvement  for  higher-order statistics. 
 
Redshift space distortions (RSD), mean that the galaxy  distribution  observed in spectroscopic  surveys is distorted along the line of sight. These distortions depend on the growth  rate of structures, and therefore offer a complementary technique (to those based on the cosmic expansion history) to measure the matter content and to test gravity \citep{Albrechtetal:2006,Amendolaetal:2012,Jenningsetal:2012}. However, the accuracy of these tests relies on the ability of the theoretical models to describe correctly the power spectrum and bispectrum in redshift space. In a previous paper \citep{bispectrum_fitting} we presented an improved phenomenological formula that was able to predict the dark matter bispectrum in real space with a 5\% precision up to scales of $k\leq0.4\,h/{\rm Mpc}$ at $z\leq1.5$.  As a natural continuation,  we now extend that formula to redshift space, making the model closer to observational data and thus useful to test gravity as well as cosmology.

We use a suite of N-body simulations that consists of 60 independent realisations with a total  volume of $\sim829\,[{\rm Gpc}/h]^3$ to constrain the free parameters of the new formula using the measurements of the dark matter redshift space bispectrum monopole. With such a large volume, we ensure that the  statistical errors are much smaller than those of  current or forthcoming surveys. Thus, we can safely quantify the systematic shifts that the fitting formula may introduce. We also use the obtained formulae to model  the bispectrum of dark matter haloes in redshift space, and to recover parameters of cosmological interest such as the logarithmic growth rate  of structure, $f$ and the amplitude of the (linear) power spectrum $\sigma_8$. Comparing the recovered values to the input ones, we can quantify  any possible systematics.

This paper is organised as follows. In \S~\ref{theory_section} we present the state-of-the art theoretical formulae for the 2- and 3-point statistics in Fourier space. In \S~\ref{sims_section} we detail the simulations we use to test these formulae and we also introduce the estimator we use for extracting information from power spectrum and bispectrum measurements. In \S~\ref{ps_section} we present the power spectrum multipole measurements from simulations. In \S~\ref{bis_section} we present the improved fitting formula for the dark matter bispectrum and we compare it to standard perturbation theory predictions. In \S~\ref{haloes_section} we compare how the new formula is able to describe the halo bispectrum. In \S~\ref{applications_section} we show the ability of the formula for recover the biases and growth rate for both dark matter particles and haloes. Finally in \S~\ref{conclusions_section} we discuss the conclusions of this paper.

\section{Theory}\label{theory_section}

The dark  matter density power spectrum, $P_{\delta\delta}$, and bispectrum, $B_{\delta}$, are the two- and three-point correlation functions in Fourier space,
\begin{eqnarray}
\label{P_def}\langle \delta({\bf k})\delta({\bf k}')\rangle &\equiv& (2\pi)^3 \delta^D({\bf k}+{\bf k}')P_{\delta\delta}({\bf k}), \\
\label{B_def}\langle \delta({\bf k}_1)\delta({\bf k}_2)\delta({\bf k}_3)\rangle &\equiv &  (2\pi)^3 \delta^D({\bf k}_1+{\bf k}_2+{\bf k}_3)B_{\delta}({\bf k}_1,{\bf k}_2), 
\end{eqnarray}
where $\delta({\bf k})\equiv\int d^3{\bf x}\, \delta({\bf x})\exp(-i{\bf k}\cdot{\bf x})$ is the Fourier transform of the dark matter overdensity field, $\delta({\bf x})\equiv\rho({\bf x})/\bar{\rho}-1$, $\rho$ is the dark matter density and $\bar\rho$ its mean value;  $\delta^D$ is the Dirac delta distribution and $\langle\ldots\rangle$ the ensemble average (or average over different realisations). Note that the bispectrum function is only defined when the 3 ${\bf k}$-vectors form a closed triangle. Also, we identify   the bispectrum  with the real part of the left hand side of equation~\ref{B_def}. When the distribution of matter is isotropic, the bispectrum is independent of the particular orientation of the triangles. In this case, we can write $B({\bf k}_1,{\bf k}_2)\equiv B(k_1,k_2,k_3)$, where ${\bf k}_3\equiv-{\bf k}_1-{\bf k}_2$.

In order to link the $n$-point statistics for haloes (the same applies to any dark matter tracer) to that of the mass, we need to introduce a  bias model. In this paper, we opt for the halo bias model proposed by \citep{McDonald_Roy}, which is able to account for both non-linearities and non-localities.  This is known as the non-local Eulerian bias model. The gravitational evolution induces non-local terms in the halo distribution, it has been shown that gravitational evolution can induce non-local terms in the halo distribution. These non-local terms appear as a second-order correction for the halo power spectrum. However, for the bispectrum non-localities contribute to leading order. Thus, it is essential to have a  bias model that is able to account for these corrections to the halo bispectrum, even at large scales. According to this  non-local Eulerian model, the halo density field is,
\begin{eqnarray}
\label{delta_galaxy}\delta_{\rm h}({\bf x}) = b_1 \delta({\bf x}) + \frac{1}{2}b_2 [\delta({\bf x})^{2}-\sigma_2]+\frac{1}{2}b_{s^2}[s({\bf x})^2-\langle s^2 \rangle] + \mbox{higher order terms}
\end{eqnarray}
where $\sigma_2$ and $\langle s^2\rangle$  ensure the condition $\langle \delta_{\rm h}\rangle=0$. Non-linearities are included in $\delta({\bf x})^2$, whereas the non-localities are described by the tidal tensor term $s({\bf x})$, 
\begin{eqnarray}
s({\bf x})&\equiv& s_{ij}({\bf x})s_{ij}({\bf x}),\\
s_{ij}({\bf x})&=&\partial_i\partial_j\Phi({\bf x})-\delta_{ij}^{\rm Kr}\delta({\bf x}),
\end{eqnarray}
where $\Phi({\bf x})$ is the gravitational potential, $\nabla^2\Phi({\bf x})=\delta({\bf x})$, and $\delta_{ij}^{\rm Kr}$ is the Kronecker delta. The relation between dark matter  and halo over-densities   is parametrised through the bias parameters: the linear bias term $b_1$, the non-linear bias term $b_2$ , and the non-local bias term $b_{s^2}$. Most higher-order terms only contribute at small scales and we do not consider them here. However, there are some that can be renormalised as large-scale contributions and therefore must be considered for consistency. We will come back to this point in \S{~\ref{sec:ps}. For $b_1=1$, $b_2=0$, $b_{s^2}=0$ and null third order biases,   we recover the unbiased case of dark matter.

In order to have an expression for the halo bias model in $k$-space we Fourier transform Eq.~\ref{delta_galaxy},
\begin{eqnarray}
\label{delta_galaxy_k}
\delta_{\rm h}({\bf k}) &=& b_1 \delta({\bf k}) + \frac{1}{2}b_2 \int \frac{d{\bf q}}{(2\pi)^3}\, \delta({\bf q})\delta({\bf k}-{\bf q})+\\
\nonumber&+&\frac{1}{2}b_{s^2}\int	\frac{d{\bf q}}{(2\pi)^3}\, \delta({\bf q})\delta({\bf k}-{\bf q}) S_2({\bf q},{\bf k}-{\bf q})+ \mbox{higher order terms}
\end{eqnarray}
 where 
\begin{eqnarray}
S_2({\bf q}_1,{\bf q}_2)\equiv\frac{({\bf q}_1\cdot{\bf q}_2)^2}{(q_1q_2)^2}-\frac{1}{3}
\end{eqnarray}
is defined from the Fourier transform of the tidal tensor,
\begin{equation}
s^2({\bf k})\equiv \int d^3{\bf x}\, s^2(x)\exp(-i{\bf k}\cdot{\bf x})=\int\frac{d{\bf k}'}{(2\pi)^3} S_2({\bf k}',{\bf k}-{\bf k}')\delta({\bf k}')\delta({\bf k}-{\bf k}').
\end{equation}
From the bias relation of Eq.~\ref{delta_galaxy_k}, we now derive the expressions for the halo power spectrum and bispectrum.
\subsection{Halo power spectrum}\label{sec:ps}
The halo  power spectrum, $P_{{\rm h},\delta\delta}(k)$, can be written as a function of the statistical moments of dark matter \citep{McDonald_Roy,Florian},
\begin{eqnarray}
\label{galaxyP}P_{{\rm h},\delta\delta}(k)&=&b_1^2 P_{\delta\delta}(k)+2b_2b_1P_{b2,\delta}(k)+2b_{s^2}b_1P_{bs2,\delta}(k)+b_2^2P_{b22}(k)+\\
\nonumber&+& 2b_2b_{s^2}P_{b2s2}(k)+b_{s^2}^2P_{bs22}(k)+2b_1b_{3\rm nl}\sigma_3^2(k)P^{\rm lin}(k),
\end{eqnarray}
where $P^{\rm lin}$ is the linear power spectrum. The definitions of the power spectra quantities, $P_{b2,\delta}$, $P_{bs2,\delta}$, $P_{b22}$, $P_{b2s2}$, $P_{bs22}$ and $\sigma_3^2$ can be found  in \citep{Florian}. Note that the contribution regulated by $b_{3\rm nl}$, does not appear explicitly in Eq.~\ref{delta_galaxy_k}. This contribution arises from various non-local third-order terms after renormalization and is given by the linear power spectrum multiplied with a $k$-dependent factor, which makes its contribution relevant at large scales \citep{McDonald_Roy}.

Analytical and numerical methods have shown that the gravitational evolution of the dark matter density field naturally induces non-local biases terms in Eulerian space in the halo- (and therefore galaxy-) density field, even when the initial conditions are local, i.e. when we have a local Lagrangian bias (see \cite{Catelan98} for initial investigations). Some of these Eulerian non-local bias terms contribute at mildly non-linear scales; hence, they only introduce non-leading order corrections in the shape and amplitude of the power spectrum and bispectrum. On the other hand, other terms contribute at large scales, at the same level as the linear, local bias parameter, $b_1$ \citep{McDonald_Roy, Baldauf, Chan, Saitoetal:2014}. In practice, neglecting the non-local bias terms can produce a mis-estimation of the other bias parameters, even when the analysis is limited to the large, and supposedly linear, scales.  In this paper we assume that the bias is local in Lagrangian space, which is equivalent to say that the Eulerian bias is non-local but with some restrictions. These restrictions are that the initial dark matter field presents a local bias. This assumption relates the values of $b_{s^2}$ and $b_{\rm 3nl}$ to the linear bias: $b_{s^2}=-4/7(b_1-1)$\citep{Baldauf,Chan} and $b_{3\rm nl}=32/315(b_1-1)$ \citep{Florian,Saitoetal:2014}, which have demonstrated to be a good approximation to N-body simulations results \citep{Baldauf, Saitoetal:2014}. Treating $b_{s^2}$ and $b_{3\rm nl}$ as free parameters would account for the general case of a non-local Eulerian bias model. However, in this work we have decided to set the bias model to be local in Lagrangian space because it describes sufficiently accurately the N-body data with less free parameters that the general non-local Eulerian bias model. Although $b_1$ and $b_2$ could also be estimated using the peak background split formalism for a particular halo population, in this paper we treat them as free parameters.
 
 In order to describe  the halo power spectrum and bispectrum in redshift space, we need to incorporate the information of the velocity components into the formalism. $\theta({\bf k})$ is the usual variable that accounts  for the peculiar velocities of dark matter particles, $\theta({\bf k})\equiv[-i{\bf k}\cdot{\bf v}_k]/[af(a)H(a)]$, where $a$ is the scale factor, $H$ the Hubble parameter, $f$ the logarithmic growth rate $d\ln\delta/d\ln a$, and ${\bf v}_k$ the $k$-space components of the velocity.  Two power spectra can be defined using the dark matter velocity field $\theta$,
 \begin{eqnarray}
\label{P2_def}\langle \delta({\bf k})\theta({\bf k}')\rangle &\equiv& (2\pi)^3 \delta^D({\bf k}+{\bf k}')P_{\delta\theta}({\bf k}), \\
\label{P3_def}\langle \theta({\bf k})\theta({\bf k}')\rangle &\equiv& (2\pi)^3 \delta^D({\bf k}+{\bf k}')P_{\theta\theta}({\bf k}), 
 \end{eqnarray}
where $P_{\delta\theta}$ and $P_{\theta\theta}$ are the density-velocity and the velocity-velocity power spectra, respectively, for dark matter.  In this paper we assume that the velocity fields are the same for dark matter and haloes (i.e., no velocity bias). A non-negligible velocity bias would affect the shape of the power spectrum and bispectrum. However, such velocity bias is not expected on large-scales \citep{Desjacques_Sheth,Eliaetal:2011}.
Thus, the cross $\delta\theta$ power spectrum for haloes can be written as, 
  \begin{eqnarray}
P_{{\rm h},\delta\theta}(k)=b_1P_{\delta\theta}(k)+b_2P_{b2,\theta}(k)+b_{s^2}P_{bs2,\theta}(k)+b_{3\rm nl}\sigma_3^2(k)P^{\rm lin}(k).
\end{eqnarray}
As for Eq.~\ref{galaxyP}, the power spectra terms can be found in \cite{Florian}.

 According to \cite{TNS_matter,TNS_halo} (TNS model hereafter), the density halo power spectrum in redshift space can be expressed as a function of density and velocity galaxy statistics in real space,
\begin{eqnarray}
\label{taruya}P^{(s)}_{\rm h}(k,\mu)&=&D^P_{\rm FoG}(k,\mu;\sigma_{\rm FoG}^P)\left[ P_{g,\delta\delta}(k)+2f\mu^2P_{g,\delta\theta}(k)+f^2\mu^4P_{\theta\theta}(k)+\right.\\
\nonumber&+&\left.b_1^3A(k,\mu,f/b_1)+b_1^4B(k,\mu,f/b_1)  \right],
\end{eqnarray}
where $\mu$ denotes the cosine of the angle between the $k$-vector and the line of sight. The definitions of the $A$ and $B$ terms can be found in \cite{TNS_matter} and only incorporate first-order corrections, and  consequently only depend on $b_1$. Higher-order corrections would include $b_2$, but in this paper we ignore their contribution, which is assumed to be subdominant in the studied regime. The factor $D^P_{\rm FoG}(k,\mu,\sigma_{\rm FoG}^P)$ was originally introduced to account for the fully non-linear damping caused by the velocity dispersion of sub-haloes, commonly known as the Fingers of God (FoG) effect. For a haloes-without-structure scenario, no internal velocity dispersion is expected and this term should be set to unity. Previous studies of this model \citep{TNS_halo} have shown that this term is actually needed to describe accurately the halo power spectrum which does not have FoG. The requirement of a FoG term for describing haloes is therefore physically paradoxical. However, we should think of the damping term in the model of Eq.~\ref{taruya} as a general damping required to
correct the model for non-linear effects.
As a practical solution, we therefore allow a $\sigma_{\rm FoG}^P$-term for describing the halo power spectrum. The physical meaning of this term {\it is not} an internal velocity dispersion, but an effective parameter that improves the description of the model. We parametrise the damping term through a one-free-parameter formula of Lorentzian type \citep{Jackson72,DP83},
\begin{eqnarray}
\label{Fog_P}D^P_{\rm FoG}(k,\mu,\sigma_{\rm FoG}^P[z])&=&\left( 1+k^2\mu^2{{\sigma_{\rm FoG}^P}[z]}^2/2  \right)^{-2}
\end{eqnarray}
with $\sigma_{\rm FoG}^P(z)\equiv\sigma_0^P(z)f(z)D(z)$. 

In this work, the dark-matter real-space statistics of Eqs.~\ref{P_def},~\ref{P2_def},~\ref{P3_def}, are given by the 2-loop resummed perturbation theory model described in \citep{ps_model} (2LRPT model hereafter).
This model  has been shown in \cite{ps_model}  to describe   $P_{\delta\delta}$ within $2\%$ accuracy up to $k=0.11\,h/{\rm Mpc}$ for $z=0$; $k=0.15\,h/{\rm Mpc}$ for $z=0.5$; $k=0.22\,h/{\rm Mpc}$ for $z=1$; $k>0.25\,h/{\rm Mpc}$ for $z=1.5$.

It is convenient to express the $P^{(s)}(k,\mu)$ power spectrum as an expansion in the Legendre polynomial basis, $P^{(\ell)}$,  defined as,
\begin{equation}
\label{real_Ps}P^{(\ell)}_{\rm h}(k)=(2\ell+1)\int_0^1 d\mu\, P_{\rm h}^{(s)}(k,\mu)L_{\ell}(\mu),
\end{equation}
where $L_{\ell}$ are the Legendre polynomials of order $\ell$. The first non-vanishing $P^{(\ell)}$ are the monopole ($\ell=0$), quadrupole ($\ell=2$) and hexadecapole ($\ell=4$).
In this paper we only focus on the monopole and quadrupole, as the hexadecapole has low signal-to-noise.

\subsection{Halo bispectrum}\label{sec:halo_bispectrum}
 
The halo bispectrum can be written according to Eq.~\ref{delta_galaxy_k} as, 
\begin{eqnarray}
\label{B_ggg1}B_{\rm h}({\bf k}_1, {\bf k}_2)=b_1^3 B_{\delta}({\bf k}_1, {\bf k}_2) +b_1^2\left[ b_2 P_{\delta\delta}(k_1)P_{\delta\delta}(k_2) + b_{s^2} P_{\delta\delta}(k_1)P_{\delta\delta}(k_2) S_2({\bf k}_1,{\bf k}_2) + {\rm cyc.} \right]
\end{eqnarray}
where we have neglected terms proportional to $b_2^2$ and $b_{s^2}^2$, which are  of higher-order. Applying the tree-level prediction for the dark matter bispectrum we write the halo bispectrum as a function of the matter power spectrum,
\begin{eqnarray}
\label{B_ggg2}B_{\rm h}({\bf k}_1, {\bf k}_2)&=&b_1^3 P_{\delta\delta}(k_1)P_{\delta\delta}(k_2)2F_2({\bf k}_1,{\bf k}_2) + b_1^2b_2 P_{\delta\delta}(k_1)P_{\delta\delta}(k_2) + \\
\nonumber&+&b_1^2b_{s^2} P_{\delta\delta}(k_1)P_{\delta\delta}(k_2) S_2({\bf k}_1,{\bf k}_2) + {\rm cyc.},
\end{eqnarray}
where $F_2$ is given by the second order kernel in standard perturbation theory (SPT) \citep{Jain_Bertschinger},
\begin{equation}
 \label{kernel_spt1} F_2({\bf k}_i,{\bf k}_j)= \frac{5}{7}+\frac{1}{2}\cos(\alpha_{ij})\left(\frac{k_i}{k_j}+\frac{k_j}{k_i}\right)+\frac{2}{7}\cos^2(\alpha_{ij}),
 \end{equation}
 where $\alpha_{ij}$ is the angle between the vectors ${\bf k}_i$ and ${\bf k}_j$.
It was shown that substituting the $F_2$ kernel by an effective kernel, namely $F_2^{\rm eff}$, the dark matter bispectrum description can be extended to mildly non-linear scales with respect to the tree-level prediction  \cite{Scoccimarro_Couchman,bispectrum_fitting},
\begin{eqnarray}
\label{Fkernel_hgm}F_2^{\rm eff}({\bf k}_i,{\bf k}_j)&=&\frac{5}{7}a(n_i,k_i,{\bf a}^F)a(n_j,k_j,{\bf a}^F) +\frac{1}{2}\cos(\alpha_{ij})  \left(\frac{k_i}{k_j}+\frac{k_j}{k_i}\right)b(n_i,k_i,{\bf a}^F)b(n_j,k_j,{\bf a}^F) \nonumber \\
&+&\frac{2}{7} \cos^2(\alpha_{ij})c(n_i,k_i,{\bf a}^F)c(n_j,k_j,{\bf a}^F),
\end{eqnarray}
where the definition of the $a$, $b$ and $c$ functions can be found in Appendix \ref{fit_kernels}.
The set of ${\bf a}^F\equiv\{ a^F_1,a^F_2,\ldots,a_9^F \}$ parameters is an empirical fit to N-body data,
\begin{eqnarray}
\nonumber a^F_1 &=& 0.484\quad\,\,\,\, a^F_4 = 0.392\quad\,\,\,\, a^F_7 = 0.128\\
 a^F_2 &=& 3.740\quad\,\,\,\, a^F_5 = 1.013\quad\,\,\,\, a^F_8 = -0.722\\
\nonumber a^F_3 &=& -0.849\quad a^F_6 = -0.575\quad a^F_9 = -0.926\,.
\label{aF}
\end{eqnarray}
This set of parameters does not depend on redshift and is assumed to be only weakly dependent on cosmology. 
Since the dependence of the $F_2$ kernel with cosmology is assumed weak \citep{Catelan95}, we expect the dependence of ${\bf a}^F$ to be weak as well.  We have shown the validity of this assumption in  Appendix B of \cite{bispectrum_fitting} for different cosmologies with different values of $\sigma_8$ such as, 0.791, 0.834, 0.878, and 0.944 and different spectral indices $n_s$: 0.95, 1.061, 1.087 and 0.985.  This assumption is in line with \cite{Scoccimarro_Couchman,Scoccimarro_2000} who showed that the reduced bispectrum $Q$ in real space has a lack of sensitivity to cosmological parameter such as $\Omega_m$ and $\Omega_\Lambda$ for a fixed $\sigma_8$ even in the non-linear regime.

To describe the halo bispectrum in redshift space on large scales, one can apply the standard perturbation theory formalism.
In this case, the form of the bispectrum formula is the same  as that in Eq.~\ref{B_ggg2}, but substituting the kernels by new redshift space ones  \citep{Scoccimarroetal:1999,Verdeetal:1998,Bernardeauetal:2002,Hivonetal:1995},
\begin{equation}
\label{Bspt}B_{\rm h}^{(s)}({\bf k}_1,{\bf k}_2)=D^B_{\rm FoG}(k_1,k_2,k_3,\sigma^B_{\rm FoG})\left[2P_{\delta\delta}(k_1)\,Z_1({\bf k}_1)\,P_{\delta\delta}(k_2)\,Z_1({\bf k}_2)\,Z_2({\bf k}_1,{\bf k}_2)+\mbox{cyc.}\right]
\end{equation}
where $Z_i$ are the redshift space kernels 
\begin{eqnarray}
\label{Zkernel1}Z_1({\bf k}_i)&\equiv& (b_1+f\mu_i^2) \\
\label{Zkernel2}Z_2({\bf k}_1, {\bf k}_2)&\equiv& b_1\left[ F_2({\bf k}_1, {\bf k}_2)+\frac{f\mu k}{2}\left(\frac{\mu_1}{k_1}+\frac{\mu_2}{k_2}\right)\right]+f\mu^2G_2({\bf k}_1, {\bf k}_2)+\\
\nonumber&+& \frac{f^2\mu k}{2}\mu_1\mu_2\left( \frac{\mu_2}{k_1}+\frac{\mu_1}{k_2} \right)+\frac{b_2}{2}+\frac{b_{s^2}}{2}S_2({\bf k}_1,{\bf k}_2),
\end{eqnarray}
and $\mu\equiv(\mu_1 k_1+\mu_2k_2)/k$, $k^2=({\bf k}_1+{\bf k}_2)^2$. $F_2$ and $G_2$ are the second-order real-space kernels of the densities and velocities respectively \citep{Jain_Bertschinger},
\begin{equation}
  \label{kernel_spt2} G_2({\bf k}_i,{\bf k}_j)= \frac{3}{7}+\frac{1}{2}\cos(\alpha_{ij})\left(\frac{k_i}{k_j}+\frac{k_j}{k_i}\right)+\frac{4}{7}\cos^2(\alpha_{ij}).
 \end{equation}
 Below we will modify this expression to push further into the non-linear regime. $D_{\rm FoG}^B$ stands for the Fingers-of-God (FoG) damping term due to the intra-halo velocity dispersion, analogously to  Eq~\ref{Fog_P}. In this paper we assume that it is described by \citep{Scoccimarroetal:1999},
\begin{eqnarray}
\label{D_fog_sc}D^B_{\rm FoG}(k_1,k_2,k_3,\sigma^B_{\rm FoG}[z])&=&\left( 1+[k_1^2\mu_1^2+k_2^2\mu_2^2+k_3^2\mu_3^2]^2\sigma_{\rm FoG}^B[z]^2/2  \right)^{-2},
\end{eqnarray}
where $\sigma_{\rm FoG}^B(z)\equiv\sigma_0^B(z)f(z)D(z)$. 
In this paper we consider $\sigma_0^P$ and $\sigma_0^B$ to be independent parameters having different roles, with $\sigma_0^P$ acting as a general non-linear damping term while $\sigma_0^B$  only corrects for the FoG.
Note that when $f\rightarrow0$, then $Z_1\rightarrow b_1$ and $Z_2\rightarrow b_1 F_2+ b_2/2 + b_{s^2}/2S_2$, and the real space prediction is recovered. This phenomenological extension has been demonstrated to describe N-body data before \cite{Scoccimarroetal:1999}.

As for the power spectrum,  we can decompose  the redshift space bispectrum in  Legendre basis. In particular the monopole is,
\begin{eqnarray}
B_{\rm h}^{(0)}(k_1,k_2,k_3)=\frac{1}{4\pi}\int_{-1}^{+1}d\mu_1\int_{0}^{2\pi}d\varphi\,B_{\rm h}^{(s)}({\bf k}_1,{\bf k}_2)
\end{eqnarray}
where $\varphi$ has been defined to satisfy, $\mu_{2}\equiv\mu_1 x_{12}-\sqrt{1-\mu_1^2}\sqrt{1-x_{12}^2}\cos\varphi$, with $x_{12}\equiv({\bf k}_1\cdot{\bf k}_2)/(k_1k_2)$. 

The expression for the halo bispectrum monopole can be analytically written in the absence of FoG term ($D_{\rm FoG}^B=1$) as,
\begin{eqnarray}
\label{Bhhh0_real}B_{\rm h}^{(0)}({\bf k}_1, {\bf k}_2)&=&P_{\delta\delta}(k_1)P_{\delta\delta}(k_2)b_1^4\left\{ \frac{1}{b_1}F_2({\bf k}_1,{\bf k}_2){ \mathcal D}_{\rm SQ1}^{(0)}+\frac{1}{b_1}G_2({\bf k}_1,{\bf k}_2){\mathcal D}_{\rm SQ2}^{(0)} \right.\\ 
\nonumber&+&\left[\frac{b_2}{b_1^2}+\frac{b_{s^2}}{b_1^2}S_2({\bf k}_1,{\bf k}_2)\right] \left.{\mathcal D}_{\rm NLB}^{(0)}+{\mathcal D}_{\rm FoG}^{(0)}  \right\}+{\mbox{cyc.}}
\end{eqnarray}
where the $\mathcal D$-terms are defined in \cite{Scoccimarroetal:1999}.

\subsection{Halo shot noise}\label{sec:shot_noise}

We use a simple model for the halo shot noise parametrised by 1 free parameter, $A_{\rm noise}$ that extends the standard Poisson noise model by varying the amplitude but not the scale dependence of the Poisson prediction,
\begin{eqnarray}
\label{A_definitionP}P_{\rm noise}&=& (1-A_{\rm noise})P_{\rm Poisson},  \\
B_{\rm noise}(k_1,k_2,k_3)&=&(1-A_{\rm noise})B_{\rm Poisson}(k_1, k_2, k_3), 
\end{eqnarray}
where \cite{Peebles},
\begin{eqnarray}
P_{\rm Poisson}&=&\frac{1}{\bar{n}},\\
B_{\rm Poisson}(k_1,k_2,k_3)&=&\frac{1}{\bar{n}}\left(P(k_1)+P(k_2)+P(k_3)-\frac{3}{\bar{n}}\right)+\frac{1}{\bar{n}^2}.
\end{eqnarray}
Here $P(k)$ is the measured halo-halo power spectrum without any shot noise subtraction\footnote{In this paper, the Poisson prediction for each N-body realisation will be computed with the measured quantities $\bar{n}$ and $P(k)$ in each corresponding realisation.}.
We fit $A_{\rm noise}$ assuming a uniform prior between -1 and +1. Note that $A_{\rm noise}=0$ corresponds to the pure Poisson shot noise, whereas $A_{\rm noise}<0$ produces a super-Poisson shot noise, and $A_{\rm noise}>0$ a sub-Poisson shot noise. For a halo population in real space we expect that $A_{\rm noise}$ to be slightly sub-Poisson, which is generally associated with halo-exclusion \cite{Casas_Miranda, Manera_Gazta}. In this case, the specific values for $A_{\rm noise}$ would depend on the mass of the considered halo population. Assuming that $A_{\rm noise}$ deviates from the Poisson value uniquely due to exclusion factors (so no cosmology or redshift dependence), given a realistic model that relates the volume of a halo with its mass, $m_h$, a model for $A_{\rm noise}(m_h)$ could be proposed following only geometrical considerations. On the other hand, we expect that an extra clustering along the line of sight is induced by redshift space distortions. Note that since this clustering is not real, it is insensitive to halo-exclusion effects. As shown in \cite{Gil-Marinetal:14}, the effect of redshift-space distortions on $A_{\rm noise}$ is very significant when this parameter is freely varied, so a realistic analytical model for $A_{\rm noise}$ should account for these effects. Since including the redshift space distortion effects is far from being a simple task (especially for the bispectrum) we treat $A_{\rm noise}$ as a free parameter in this paper.

\subsection{Parameter estimation}\label{sec:estimation}

We are interested in estimating a set of parameters, $\bf \Psi$, from the power spectrum and bispectrum. $\bf \Psi$ includes cosmologically interesting parameters such as the bias parameters, the amplitude of the matter power spectrum $\sigma_8$ and the logarithmic growth rate parameter $f$, as well as nuisance parameters, such as shot noise parameters and the damping factors of the FoG terms. 
 We do not allow the spectral index, $n_s$, the Hubble parameter $h$ and the matter and baryon densities, $\Omega_m$  and $\Omega_b$ to vary from their fiducial values, thus assuming a fixed shape of the dark matter linear power spectrum. We refer to these parameters as $\bf \Omega$. 
 
 In principle the optimal way to analyse a statistical quantity is to model its probability density function and proceed to parameter fitting from there. However, the probability distribution for the bispectrum especially in the mildly non-linear regime is not known (although some progress are being made see e.g., \cite{Matsubara:2007}); even if one invokes the central limit theorem and models the distribution of bispectrum modes as a multi-variate Gaussian, the evaluation of its covariance would be challenging (see e.g., eq. 38-42 of \cite{MVH97}, appendix A of \cite{Verdeetal:1998} and discussion above). In addition we want to analyse jointly power spectrum and bispectrum whose joint distribution is not known. Another approach is therefore needed. In order to estimate the set of ${\bf\Psi}$ parameters, we opt for the approach proposed in \cite{Verdeetal:2002}, which consists of introducing a suboptimal but unbiased estimator.  
Thus, following this formalism, we define the $\chi_{\rm diag.}^2$-function as,
\begin{eqnarray}
\nonumber\chi_{\rm diag.}^2({\bf\Psi})&=&\sum_{k-{\rm bins}} \frac{\left[ P_{(i)}^{\rm meas.}(k)-P^{\rm model}(k,{\bf \Psi}; {\bf \Omega}) \right]^2}{\sigma_P(k)^{2}}+\\
\label{chi2_formula}&+&\sum_{{\rm triangles}} \frac{\left[ B^{\rm meas.}_{(i)}(k_1,k_2,k_3)-B^{\rm model}(k_1,k_2,k_3,{\bf \Psi}; {\bf \Omega}) \right]^2}{\sigma_B(k_1,k_2,k_3)^{2}},
\end{eqnarray}
where $\sigma_P$ and $\sigma_B$ are the diagonal terms of the covariance matrix for the power spectrum and bispectrum respectively. Therefore,  Eq.~\ref{chi2_formula} ignores the contribution from off-diagonal terms, and takes into account only the diagonal terms. In this paper, the terms $\sigma_P$ and $\sigma_B$ are computed from the dispersion of the different realisations of dark matter or halo population. By ignoring the off diagonal terms of the covariance matrix in the $\chi_{\rm diag.}^2$ definition we  do not have a maximum likelihood estimator which is  minimum variance and  unbiased.  This estimator is sub-optimal (the variance is not minimal), but is unbiased as we will show explicitly in \S~\ref{applications_dm}. 
 Furthermore,  {\it a)}} the particular  value of the $\chi_{\rm diag.}^2$ at its minimum is meaningless and should not be used to estimate  a goodness of fit and {\it b)} the errors on the parameters cannot be estimated by standard $\chi_{\rm diag.}^2$ differences.

We use a  Nelder-Mead based-method of minimization \citep{NR} to  a set of best-fitting parameters that minimise $\chi_{\rm diag.}^2$ for each  realisation $i$, namely ${\bf \Psi}_{(i)}$. 

Therefore, the main point of this method is that  $\langle{\bf\Psi}_{(i)}\rangle$ is an unbiased estimator of the true set ${\bf\Psi}_{{\rm true}}$ and that the dispersion of ${\bf\Psi}_{(i)}$ is a suitable estimator of the error: ${\bf\Psi}_{{\rm true}}\simeq\langle {\bf\Psi}_{{\rm i}}\rangle \pm \sqrt{\langle {\bf\Psi}_{{\rm i }}^2\rangle-{\langle {\bf\Psi}_{{\rm i}}\rangle}^2}$. This procedure will be applied in \S~\ref{haloes_section} and \S~\ref{applications_section}, using the total amount of realisations (60 for dark matter and 20 for haloes) in order to obtain ${\bf\Psi}$ and their errors.  

As we have mentioned above, an alternative approach is to include also the contribution of the off-diagonal terms in the covariance matrix and proceed similarly. However, given the large number of $k$-bins and triangular configurations considered in this paper ($\sim5000$), the number of different realisations necessary to estimate the full covariance from simulations would be of order of $\gtrsim10.000$ \cite{Hartlap07}. Alternatively, an analytical estimate for the covariance could be used \cite{Sefusatti06}. However, this kind of approach, requires the prediction of higher order point statistics, such the the 4- and 6-point correlation functions in $k$-space. Although these could be estimated at large scales using perturbation theory, they would fail at mildly non-linear scales which we are interested in. Because of this, we have decided not to include the off-diagonal terms in the parameters estimation procedure. 

The effect of ignoring the off-diagonal terms in the $\chi^2$ can be easily estimated when extracting parameters from the power spectrum monopole and quadrupole. In this case, since the number of total $k$-bins is of order of $\sim100$, with ``only" $\sim600$ realisations we are able to estimate the full covariance from simulations and therefore can compare the error-bars of the parameters estimated using $\chi^2_{\rm diag}$ and the full $\chi^2$. We have checked \citep{GMinprep} that the difference in the error-bars is less than 20\% between the two alternatives.

\section{Simulations}\label{sims_section}

The simulations used in this paper consist of two different sets, one providing dark matter density fields (with different cosmological parameters) and the other one halo catalogues. These simulations have been used in previous works (see \citep{bispectrum_fitting} for dark matter and \citep{Whiteetal:2011} for the halo catalogue).

\subsection{Dark matter particles simulations}
\label{dark_matter}
The dark matter particle set of simulations consists of N-body  simulations with a flat  LCDM cosmology listed in Table \ref{table_sims} as ``Sim DM". The box size is $L_b=2.4\,{\rm Gpc}/h$ with periodic boundary conditions and the number of particles is $N_p=768^3$ with 60 independent runs.  The initial conditions were generated at $z=49$ by displacing the particles according to the second-order Lagrangian PT from their initial grid points. The initial  power spectrum of the density fluctuations was computed by CAMB \cite{CAMB}. Taking only the gravitational interaction into account, the simulations were performed with the GADGET-2 code \citep{springel05}. There are four snapshots at redshifts $z=0$, $z=0.5$, $z=1.0$ and $z=1.5$. In order to obtain the dark matter density field from particles we apply the Cloud-in-Cell (CiC) prescription using $512^3$ grid cells. Thus the size of the grid cells is 4.68 Mpc/$h$.  Two additional dark matter particle simulations are listed in Table \ref{table_sims} as ``low-$\Omega_m$" and ``high-$\Omega_m$". The cosmologies of these   simulations have a lower/higher $\Omega_m$ value relative to ``Sim DM" but share the value of $\Omega_b h^2$ and $\Omega_m h^2$ with the ``Sim DM" set. Hence,  the shape of the initial power spectrum is identical to that of ``Sim DM". Additionally, the value of  $\sigma_8$ is chosen such that the amplitude of the linear real-space power spectrum at $z=0$ is the same as that in the ``Sim DM" set when the distances are expressed in Mpc instead of Mpc/$h$. These two additional sets are only used in Appendix \ref{cosmo_appendix} to test explicitly the dependence of the model proposed in this work on the growth rate $f$ since $f\approx \Omega_m^{0.55}$.

 In order to correct for the effects of the grid left by the CiC scheme we have corrected appropriately by \cite{Hockney_Eastwood81,Jing05},
\begin{eqnarray}
\delta^{\rm true}({\bf k})\simeq\delta^{\rm meas.}({\bf k})/W({\bf k}),
\end{eqnarray}
where $\delta^{\rm meas.}$ is the measured overdensity, $\delta^{\rm true}$ is the true overdensity which is traced by the particles and $W$ is the window function that depends on how the particles have been assigned to the grid cells. For CiC $W$ is,
\begin{eqnarray}
W({\bf k})=\left( { j}_0\left[\frac{\pi k_1}{2k_N}\right] { j}_0\left[\frac{\pi k_2}{2k_N}\right] { j_0}\left[\frac{\pi k_3}{2k_N}\right] \right)^{-2},
\end{eqnarray}
 where ${j_0}(x)\equiv\sin(x)/x$ is the spherical Bessel function of 0{\it th} order, ${\bf k}\equiv(k_1,k_2,k_3)$ and $k_N$ is the Nyquist frequency. 
 
 We have not corrected for any aliasing effects. However, we have checked (by increasing the number of grid cells) that for $k\leq0.25\,h{\rm Mpc}^{-1}$ the aliasing effects are  negligible.

\subsection{Dark matter halo catalogue}

The dark matter halo catalogue set  is based on 20 N-body dark matter simulations with an order of magnitude higher mass resolution. The cosmology is listed in Table \ref{table_sims} as ``Sim HC". The box size is $L_b=1.5\,{\rm Gpc}/h$. The mass of the dark matter particles is $m_p=7.6\times10^{10}\,M_\odot/h$, and the minimum halo mass has been selected to be $7.8\times10^{12}\,M_\odot/h$. The halo catalogues are generated by the Friends-of-Friends algorithm \citep{FoF} with a linking length of 0.168 times the mean inter-particle spacing. There is one snapshot at $z=0.55$. In order to extract the halo field, a CiC prescription is  also  used with $512^3$ grid cells, whose size is $2.93\,{\rm Mpc}/h$. The same correction scheme as in \S\ref{dark_matter} has been applied. 

\begin{table}[htdp]
\begin{center}
\begin{tabular}{|c|c|c|c|c|c|c|c|c|c|c|}
\hline
 & $\Omega_\Lambda$ & $\Omega_m$ & $h$ & $\Omega_b h^2$ & $n_s$ & $\sigma_8$ & $z$ & $L_b[{\rm Gpc}]$ & $V\,[{\rm Gpc}^3]$  \\
 \hline
 \hline
 Sim DM & 0.73 & 0.27 & 0.7 & 0.023 & 0.95 & 0.79 & 0 - 1.5  & 3.428 & $40.30\times60$ \\
 \hline
 low-$\Omega_m$ & 0.8 & 0.2 & 0.8133 & 0.023  & 0.95 & 0.872 & 0 & 3.428 & $40.30$  \\
 \hline
 high-$\Omega_m$ & 0.6 & 0.4 & 0.5751 & 0.023 & 0.95 & 0.693 & 0 & 3.428 & $40.30$  \\
 \hline
 Sim HC  & 0.726 & 0.274 & 0.7 & 0.0224 & 0.95 & 0.8 & 0.55 & 2.143 & $9.840\times20$ \\
 \hline
\end{tabular}
\end{center}
\caption{Cosmological parameter of the sets of simulations used in this paper: dark matter only with a fiducial cosmology parameters (DM), two additional dark matter only simulations with a lowered and increased value of $\Omega_m$ (low-$\Omega_m$ and high-$\Omega_m$, respectively) and halo catalogs (HC). Several cosmological parameters are listed: the dark energy density, $\Omega_\Lambda$, matter density $\Omega_m$, Hubble parameter $h$, physical baryon density  $\Omega_bh^2$, primordial power-law power spectrum spectral index $n_s$ and  amplitude of the primordial power spectrum linearly extrapolated at $z=0$, $\sigma_8$. Also the box size of the simulation is provided, $L_b$, and the  volume per realisation times the total number of realisations, $V$.}
\label{table_sims}
\end{table}

\section{Dark matter power spectrum multipoles}\label{ps_section}
We start by illustrating  the ability of the theoretical modelling to describe the real-space dark matter power spectrum as well as its redshift space multipoles. We use the TNS model in combination with the 2LRPT model (hereafter TNS-2LRPT) to describe the power spectrum multipoles.

\begin{figure}
\centering
\includegraphics[scale=0.4]{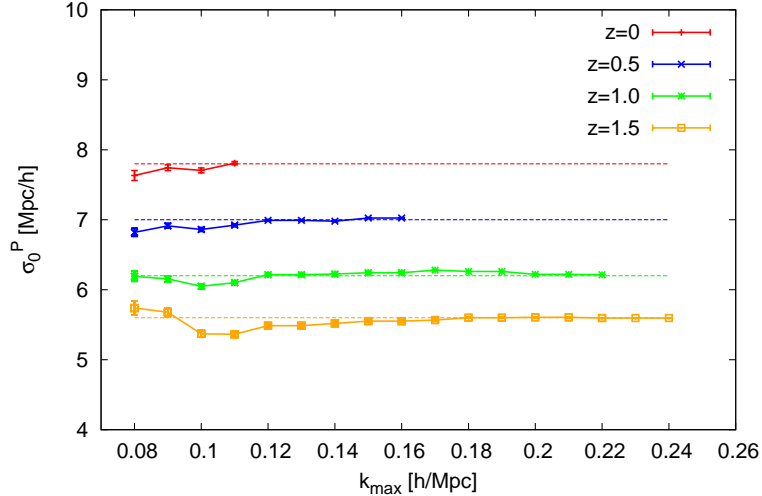}
\caption{Best-fitting power spectrum FoG parameter, $\sigma_0$ as a function of $k_{\rm max}$ (solid lines). The colour notation correspond to the redshift snapshot:  $z=0$ (red), $z=0.5$ (blue), $z=1.0$ (green) and $z=1.5$ (orange) as indicated in the key. Dashed lines show the value at $k_c$, which is given in Table~\ref{table_sigma}. Note that we only plot the best-fitting value up to $k_c(z)$ (scale at which the real space  modelling fails at more than 2\% level).}
\label{plot_sigma}
\end{figure}

The description of the dark matter power spectrum in redshift space requires  the parameter $\sigma_0^P$, as  in Eq.~\ref{Fog_P}. Although there are some prescriptions to approximate this parameter analytically \citep{Scoccimarro04} (if it were only correcting for the FoG), in this paper we treat it as a nuisance parameter to be fitted from N-body  simulations measurements of the monopole and quadrupole.
 We assume that this parameter can in principle change freely with redshift but not with the scale.  In Fig.~\ref{plot_sigma} we show its best-fitting  values as a function of the maximum scale used, for different  redshift snapshots: $z=0$ (red solid line), $z=0.5$ (blue solid line), $z=1.0$ (green solid line) and $z=1.5$ (orange solid line). The dashed lines show the $\sigma_0^P$ value for the smallest scale (largest $k$) we  trust the model, namely $k_c$. This scale is defined as the largest that satisfies, $\left|P_{\rm sim}/P_{\rm model}\right|<1.02$ for the real space power spectrum. The value for $\sigma_0^P(k_{\rm max}=k_c)$  as well as the $k_c$ scale, are shown in Table~\ref{table_sigma} for the four redshift snapshots studied here. Note that in Fig.~\ref{plot_sigma}, the best-fitting $\sigma_0^P$ value is only plotted for $k_{\rm max}\leq k_c$.

\begin{table}[htdp]
\begin{center}
\begin{tabular}{|c|c|c|c|c|}
\hline
$z$ & 0 & 0.5 & 1.0 & 1.5 \\
\hline
\hline
$\sigma_{\rm 0}^P(z)$ [Mpc/$h$] & 7.8 & 7.0 & 6.2 & 5.6 \\
\hline
$D(z)$ & 1.000 & 0.782 & 0.623 & 0.511 \\
\hline
$f(z)$ & 0.483 & 0.723 & 0.852 & 0.915 \\
\hline
$k_{\rm c}(z)$ [$h$/Mpc] & 0.11 & 0.16 & 0.22 & $>0.25$\\
\hline
\end{tabular}
\end{center}
\caption{Best fitting values of $\sigma_{\rm 0}^P$ from the dark matter power spectrum monopole and quadrupole measurements at different $z$. The growth  parameters $D$ and $f$, as well as the $k_{\rm c}$ values  are also  reported.}
\label{table_sigma}
\end{table}

\begin{figure}
\centering
\includegraphics[clip=false, trim= 15mm 0mm 15mm 10mm,scale=0.30]{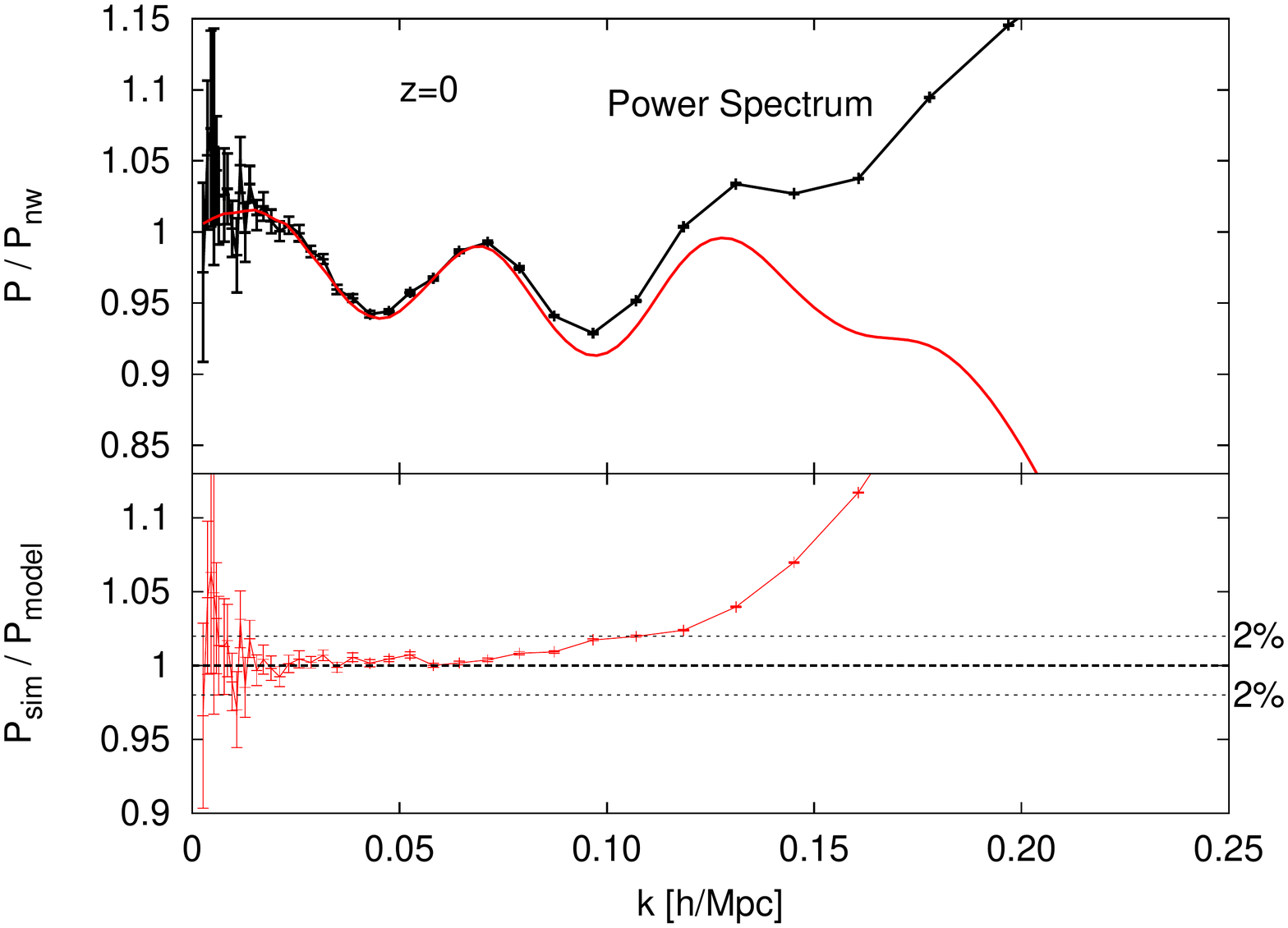}
\includegraphics[clip=false, trim= 15mm 0mm 15mm 10mm,scale=0.30]{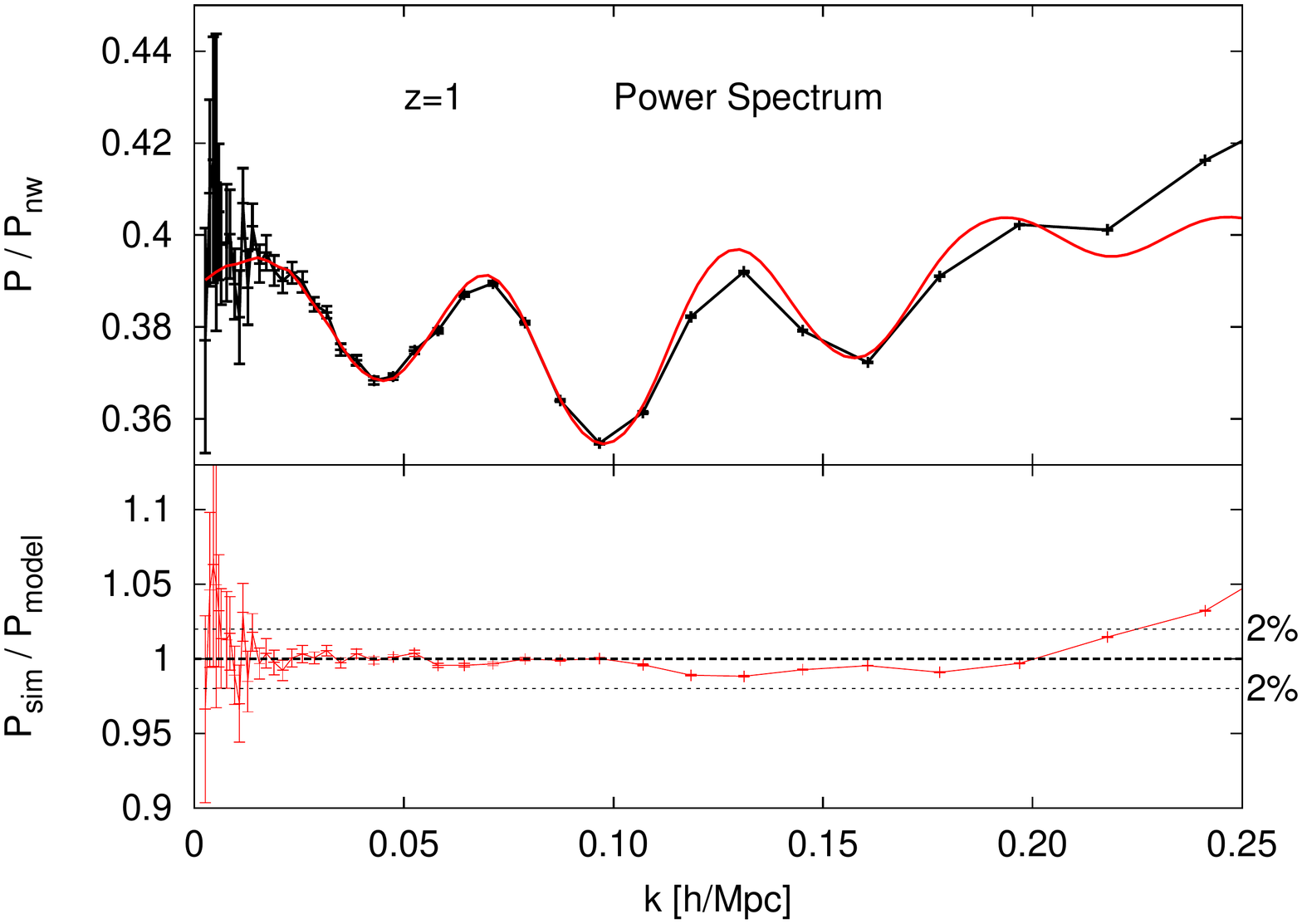}

\includegraphics[clip=false, trim= 15mm 0mm 15mm 10mm,scale=0.30]{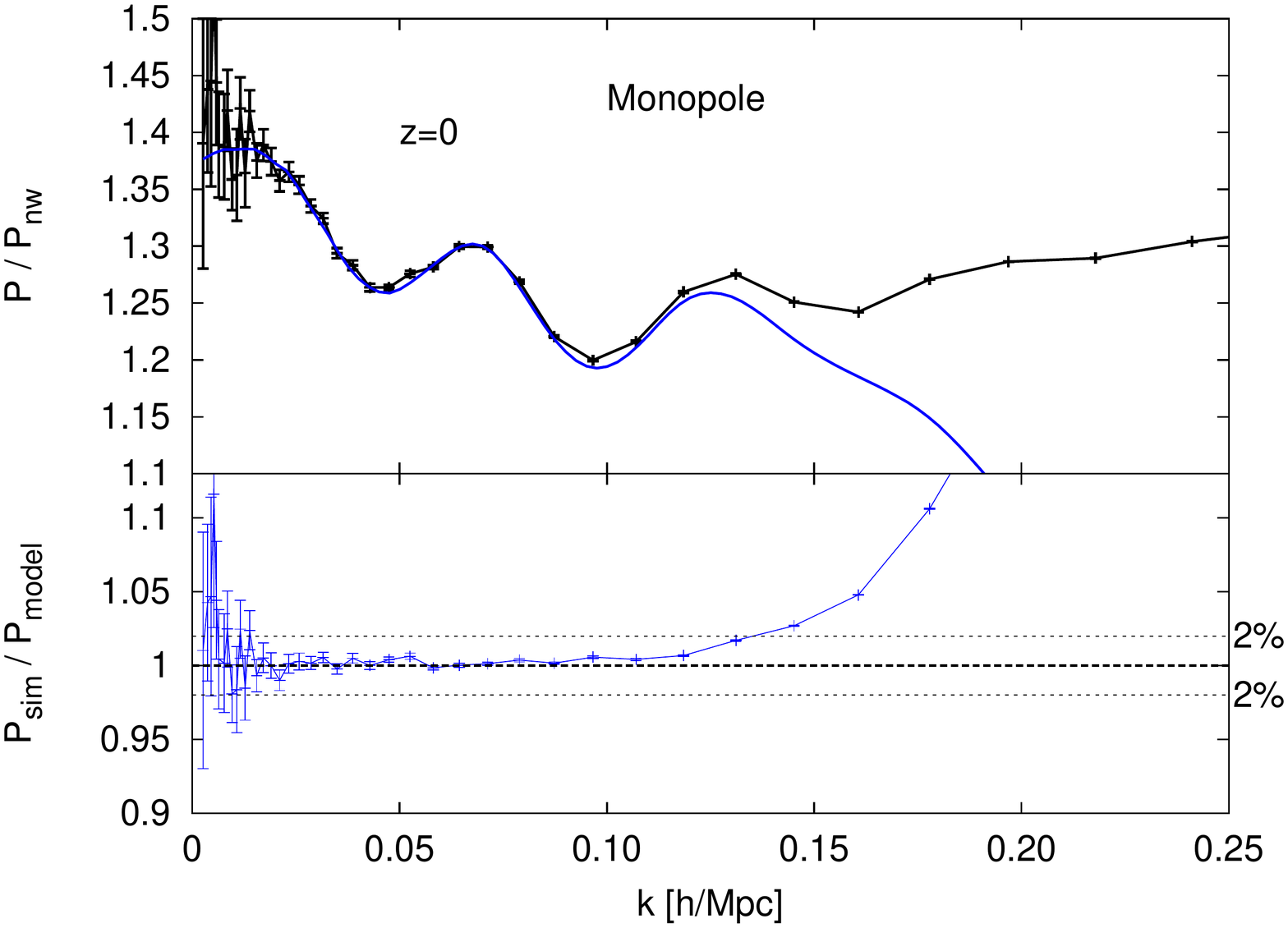}
\includegraphics[clip=false, trim= 15mm 0mm 15mm 10mm,scale=0.30]{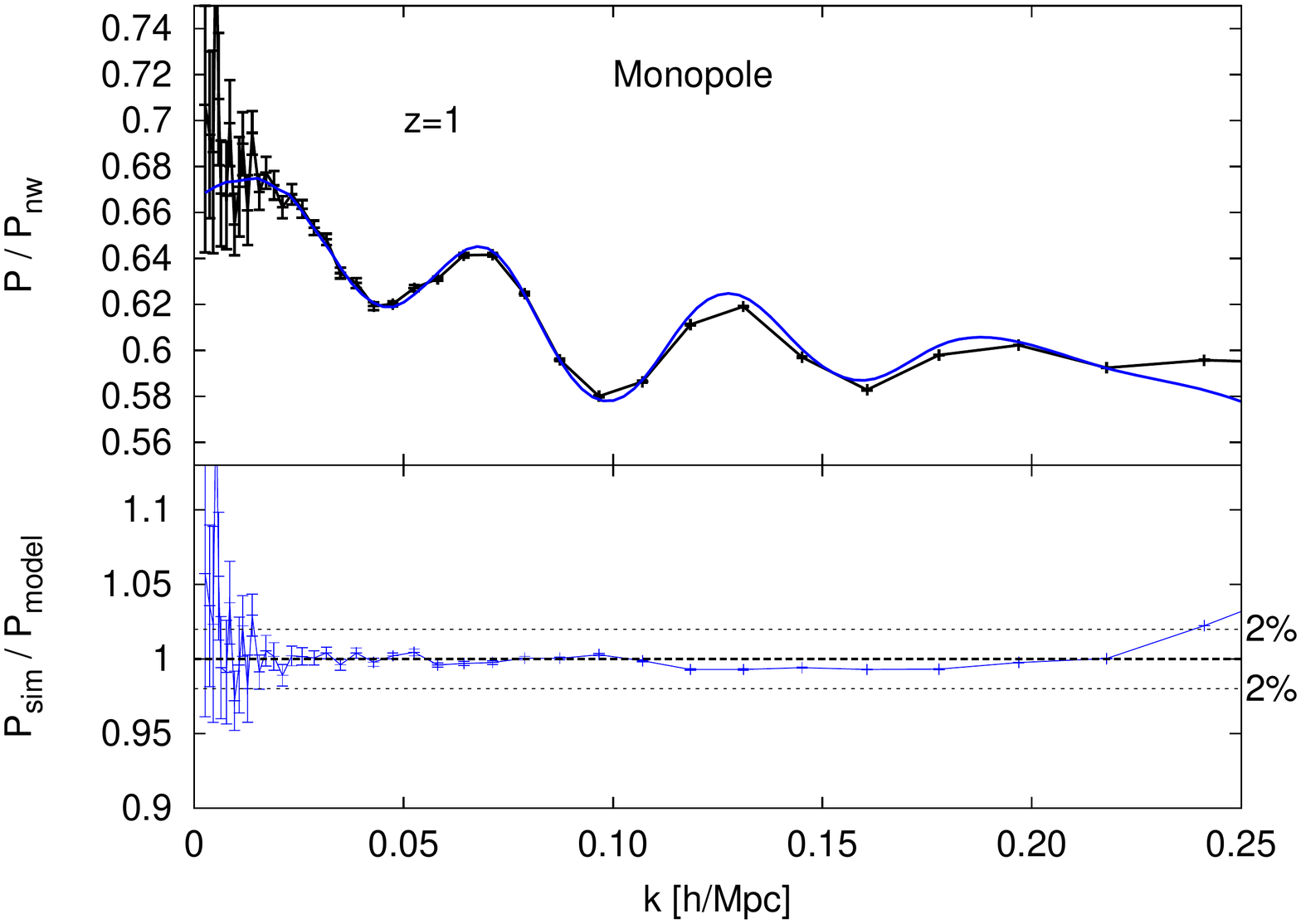}

\includegraphics[clip=false, trim= 15mm 0mm 15mm 10mm,scale=0.30]{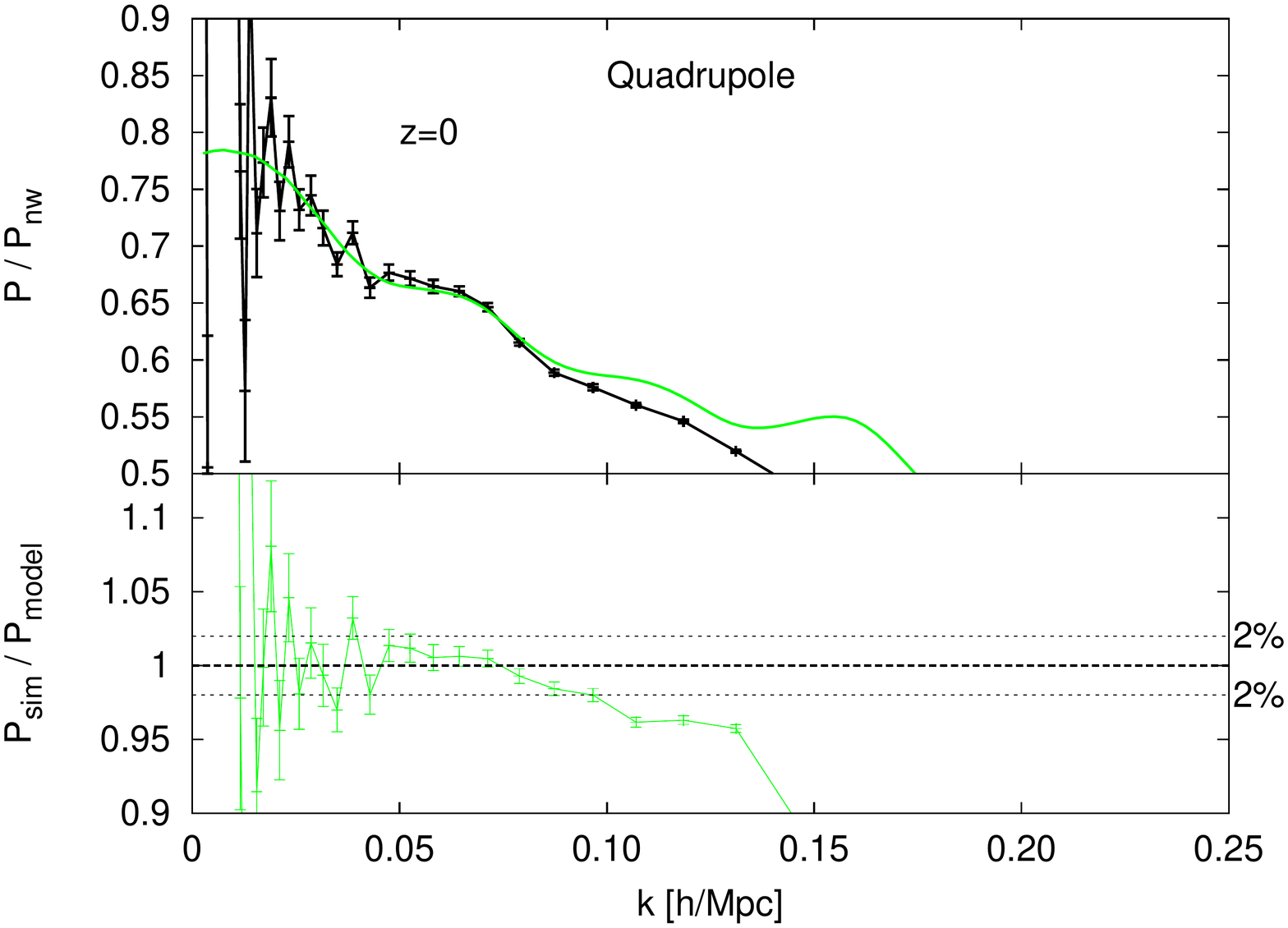}
\includegraphics[clip=false, trim= 15mm 0mm 15mm 10mm,scale=0.30]{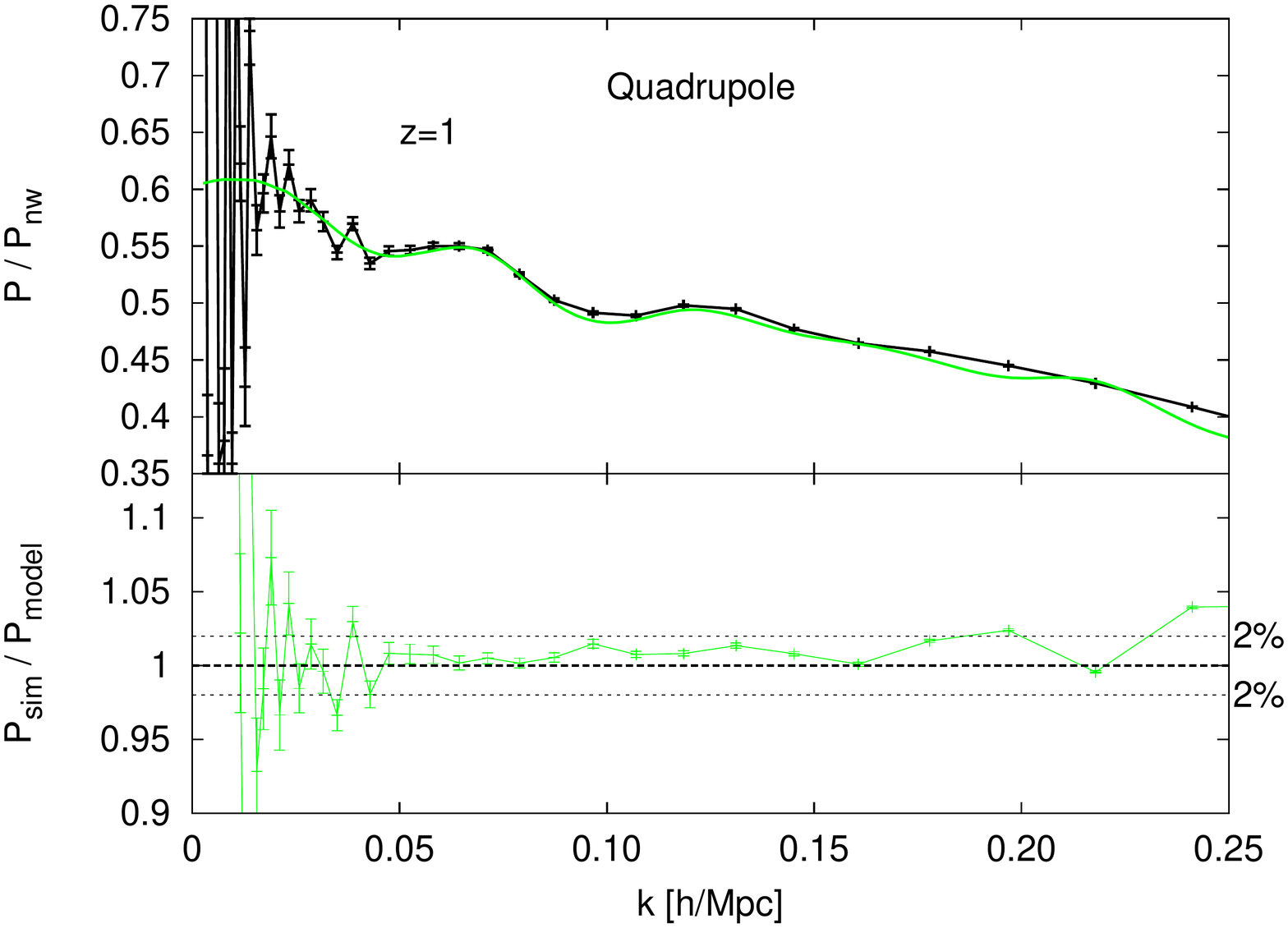}
\caption{Power spectra in real space (top panels), monopoles (middle panels) and quadrupoles (bottom panels), for $z=0$ (left panels) and $z=1$ (right panels). Black symbols are measurements from 60 N-body dark mater simulations, whereas colour lines are predictions for dark matter using the TNS model in combination with 2LRPT model with $\sigma_{\rm FoG}^P$ as a free parameter. The top-subpanels show the power spectrum normalised by a non-wiggle linear model and the bottom sub-panels show the relative deviation of the N-body dark matter measurements to the corresponding models. The values for $\sigma_{\rm FoG}^P$ used in the models are listed in Table~\ref{table_sigma}.}
\label{plot_multipoles}
\end{figure}

From the values of Table~\ref{table_sigma}, we observe that the best-fitting $\sigma_{\rm 0}^P(z)$ depends on  redshift: lower redshift snapshots present a more severe damping at a given scale and therefore $\sigma_{\rm 0}^P$ decreases with $z$. We observe a similar behaviour for the quantity $\sigma_0^P(z)D(z)f(z)$.  However, the quantity $\sigma_{\rm 0}^P(z)f (z)$ presents a weak redshift-dependence for $z=0.5$, $1.0$ and $1.5$, although at $z=0$  has a significantly different value. 

Fig.~\ref{plot_multipoles} presents the dark matter power spectrum in real space (top panels), the dark matter redshift space monopole (middle panel) and quadrupole (bottom panels) for $z=0$ (left panels) and $z=1$ (right panels). In the top sub-panels the power spectrum (normalised to a non-wiggle model) is displayed for N-body measurements (black symbols with error-bars). Solid lines correspond to  the theoretical predictions of the TNS model (with $b_1=1$ and $b_2=b_{s^2}=b_{3\rm nl}=0$) implemented with  2LRPT: red for the real space power spectrum, blue for the redshift space monopole and green for the quadrupole. For the monopole and quadrupole the values of $\sigma_0^P(k_c)$ listed in Table~\ref{table_sigma} have been used. The growth parameters $f$ and $D$ (listed also in Table~\ref{table_sigma}) have been fixed at their true values. The error-bars correspond to the error on the mean of 60 realisations, for a total   volume of $V\simeq829\,[{\rm Gpc}/h]^3$.

From Fig.~\ref{plot_multipoles} we see that the TNS model in combination with 2LRPT model is able to describe the two-point N-body statistics with an accuracy of $\sim1\%$ for the real space power spectrum and redshift space monopole up to $k_c$. The quadrupole is described typically with $\sim2\%$ accuracy. 

In conclusion, the TNS model in combination with the 2LRPT model is able to describe with per-cent accuracy the two-points real space statistics for dark matter particles without any free parameters. The main two-point redshift space statistics, the monopole and the quadrupole, are described with  per-cent precision as well and require one free parameter per redshift snapshot. The function of this parameter is to reduce an excess of power at small scales, typically produced by non-linear processes such as intra-halo  velocity dispersion. 

\section{Dark matter bispectrum modelling in redshift space}\label{bis_section}
The main goal of this paper is to provide a modification of the SPT model prediction for the bispectrum in redshift space given by Eq.~\ref{Bspt} for dark matter and by Eq~\ref{Bhhh0_real} for haloes. Typically the SPT approach works well at large scales and at high redshifts, but breaks down in the mildly non-linear regime. 

 We follow a similar procedure to that in \cite{bispectrum_fitting}. It consist of modifying the SPT kernels into effective kernels with free parameters to be fitted from N-body simulations.  In this case we propose changing the velocity kernel $G_2$ of Eq.~\ref{Zkernel2} by an effective kernel of the same form as $F_2^{\rm eff}$,
\begin{eqnarray}
\label{Gkernel_hgm}G_2^{\rm eff}({\bf k}_i,{\bf k}_j)&=&\frac{3}{7}a(n_i,k_i,{\bf a}^G)a(n_j,k_j,{\bf a}^G) +\frac{1}{2}\cos(\alpha_{ij})  \left(\frac{k_i}{k_j}+\frac{k_j}{k_i}\right)b(n_i,k_i,{\bf a}^G)b(n_j,k_j,{\bf a}^G) \\
&+&\frac{4}{7} \cos^2(\alpha_{ij})c(n_i,k_i,{\bf a}^G)c(n_j,k_j,{\bf a}^G),\nonumber
\end{eqnarray}
where the functions $a$, $b$ and $c$ are the same as used in $F^{\rm eff}_2$ and can be found in Appendix \ref{fit_kernels}. We assume that the set of parameters ${\bf a}^G\equiv\{ a^G_1,a^G_2,\ldots,a_9^G \}$, is redshift-, scale- and shape-independent, and needs to be fitted from the measurement of the redshift space bispectrum monopole in N-body simulations. We consider the damping terms of Eq.~\ref{D_fog_sc} to describe the FoG features of the bispectrum.
We allow this parameter to depend on the redshift and to be independent of the $\sigma_{\rm 0}^P$ of Table~\ref{table_sigma}. 
Assuming $G_2^{\rm eff}$ of this form has no other motivation than being of the same type of modification which was shown to work for $F_2^{\rm eff}$. We will see in the next sections how this {\it ad-hoc} assumption works for mildly non-linear scales.

We consider different approaches to describe the redshift space bispectrum monopole. All of them are based on SPT leading order correction (Eq.~\ref{Bspt}) with different  changes in the definition of the redshift space kernels,
\begin{enumerate}
\item\label{case1} {\bf SPT approach}. We use the SPT prediction of Eq.~\ref{Bspt} with  the SPT kernels $F_2$ and $G_2$ of Eqs.~\ref{kernel_spt1} and~\ref{kernel_spt2}. We include the FoG effect through  the damping functions of Eq.~\ref{D_fog_sc}. This function contains one free parameter, $\sigma_0^B$. We allow this parameter to freely vary with redshift, but we consider it scale- and shape-independent. Hence, this approach has four-free parameters for the whole redshift range.  In order to avoid systematic effects in the power spectra modelling, we use the $P_{\delta\delta}$ measured from simulations (the averaged value among realizations of ``SimDM") instead of the theoretical prediction of the 2LRPT model.  Hereafter we refer the bispectrum prediction of this model as $B^{\rm spt}$.
\item\label{case2} {\bf Hybrid approach}.  We use the prediction of Eq.~\ref{Bspt} taking the effective kernel $F_2^{\rm eff}$ from Eq.~\ref{Fkernel_hgm} instead of the SPT form of $F_2$. The ${\bf a}^F$ values given in Eq.~\ref{aF} from \citep{bispectrum_fitting} are used. We consider the SPT kernel for $G_2$. As for the SPT approach, we use Eq.~\ref{D_fog_sc} to describe the FoG effect.  As before, we use the actual measurement of $P_{\delta\delta}$ in the bispectrum formula.  We refer to the bispectrum prediction from this model as $B^{F}$.
\item\label{case4} {\bf Effective approach}. We use the Eq.~\ref{Bspt} structure with the effective kernel $F_2^{\rm eff}$ of Eq.~\ref{Fkernel_hgm}, and the effective $G_2^{\rm eff}$ kernel from Eq.~\ref{Gkernel_hgm}  with a set of nine free parameters, ${\bf a}^G$, to be fitted. We add  the FoG-term of Eq.~\ref{D_fog_sc} with one extra free parameter per redshift snapshot.  As before, we use the actual measurement of $P_{\delta\delta}$ in the bispectrum formula. We will refer to this model as $B^{\rm FG}$.
\end{enumerate}

We use the method described in \S~\ref{sec:estimation} for estimating ${\bf a}^G$,  fixing the bias parameters, $f$ and $\sigma_8$ to their true values. There are a large number of possible triangular shapes to consider and it is not practical to consider them all. However, it is not necessary to compute all possible triplets as their bispectra are highly correlated. Therefore, 
 here we consider only  triangles  with $k_2/k_1=1.0,\, 1.5,\, 2.0$ and 2.5.  In Appendix \ref{appendixb} we show that the fitting formula is also able to describe other triangular shapes.

 We estimate that $\sim70\%$ of the full information of the bispectrum is contained by these shapes at $k\leq0.1\,h/{\rm Mpc}$.  This estimation is done by measuring the real space dark matter bispectrum for all the triangular configurations. Comparing the variance of  the bias parameters, $b_1$ and $b_2$, estimated from the whole set of triangles, or just a sub-set, we obtain an estimation of the information lost by only selecting few shapes (see Appendix \ref{variance_appendix} for details). 
 Although this estimation is done for dark matter in real space, we do not expect to be very different in redshift space.

  Since we expect that the theory breaks down at different scales at different $z$ we set the fitting range to: $k_i\leq0.15\,h/{\rm Mpc}$ for $z=0$, $k_i\leq0.18\,h/{\rm Mpc}$ for $z=0.5$, $k_i\leq0.21\,h/{\rm Mpc}$ for $z=1.0$ and $k_i\leq0.25\,h/{\rm Mpc}$ for $z=1.5$.  We have checked iteratively that these are the maximum scales for which the { Effective approach} can describe the simulated data with $\lesssim5\%$ accuracy. The set of best-fitting ${\bf a}^G$ parameters are,
\begin{eqnarray}
\nonumber a^G_1 &=& 3.599\quad\,\,\,\, a^G_4 = -3.588\quad a^G_7 = 5.022\\
 a^G_2 &=&-3.879\quad a^G_5 = 0.336\quad\,\,\,\, a^G_8 = -3.104\\
\nonumber a^G_3 &=& 0.518\quad\,\,\,\, a^G_6 = 7.431\quad\,\,\,\,\, a^G_9 = -0.484\,.
\label{aG}
\end{eqnarray}
The fitting process also provides  best-fitting values for $\sigma_{\rm 0}^B$. These values are listed in Table~\ref{sigmab_table} for the different models used. The $\sigma_0^P$ parameters found in \S~\ref{ps_section} are also shown  for reference. Note that we expect that the $\sigma_{\rm 0}^P$ and $\sigma_{\rm 0}^B$ parameters change as a function of the selected tracer as well as a  function of redshift. However, the ${\bf a}^G$ parameter set (as well as ${\bf a}^F$) is assumed to be universal. 

We are aware that this result may depend on the cosmology. However, as we have mentioned for the ${\bf a}^F$ fit, the dependence of the $F_2$ and $G_2$ kernels on cosmology is very weak, so this is expected to hold also for the  ${\bf a}^F$ and ${\bf a}^G$ parameters. Furthermore, since we have performed the fit for a wide range of redshifts, any cosmology dependence that is equivalent to a redshift re-scaling can be described by the model.

\begin{figure}
\centering
\includegraphics[clip=false, trim= 80mm 10mm 22mm 35mm,scale=0.25]{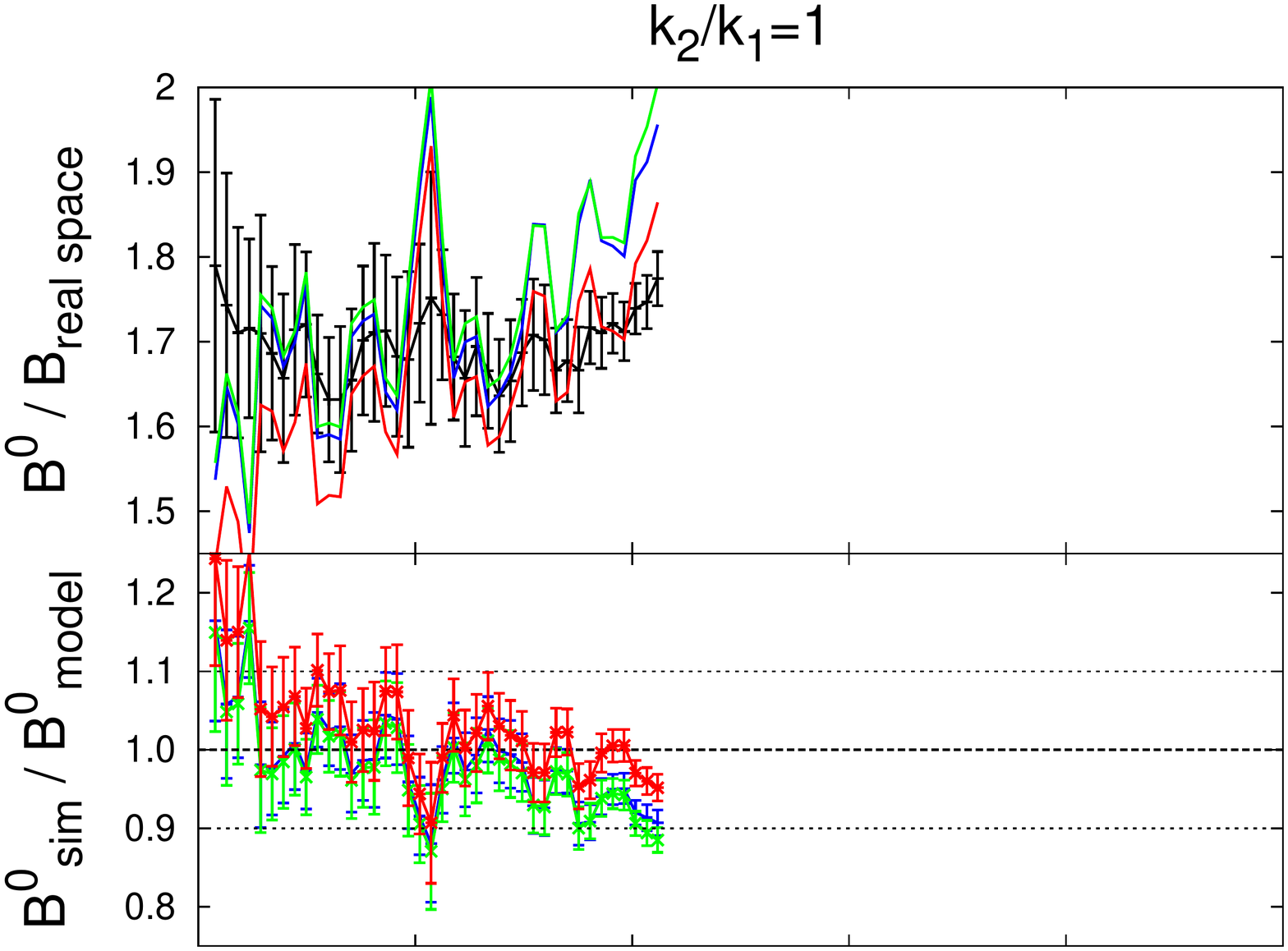}
\includegraphics[clip=false,trim= 25mm 10mm 22mm 35mm, scale=0.25]{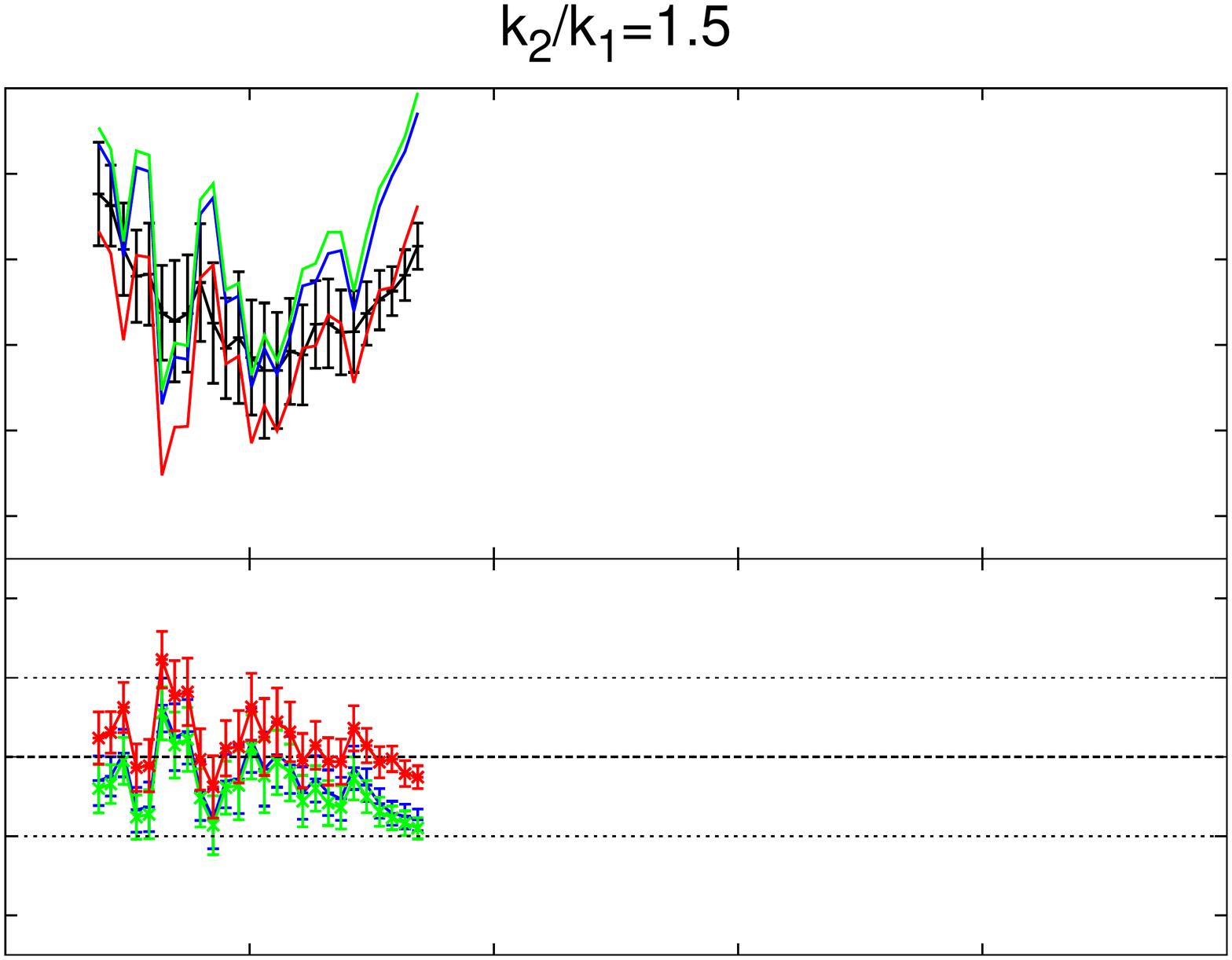}
\includegraphics[clip=false,trim= 25mm 10mm 80mm 35mm, scale=0.25]{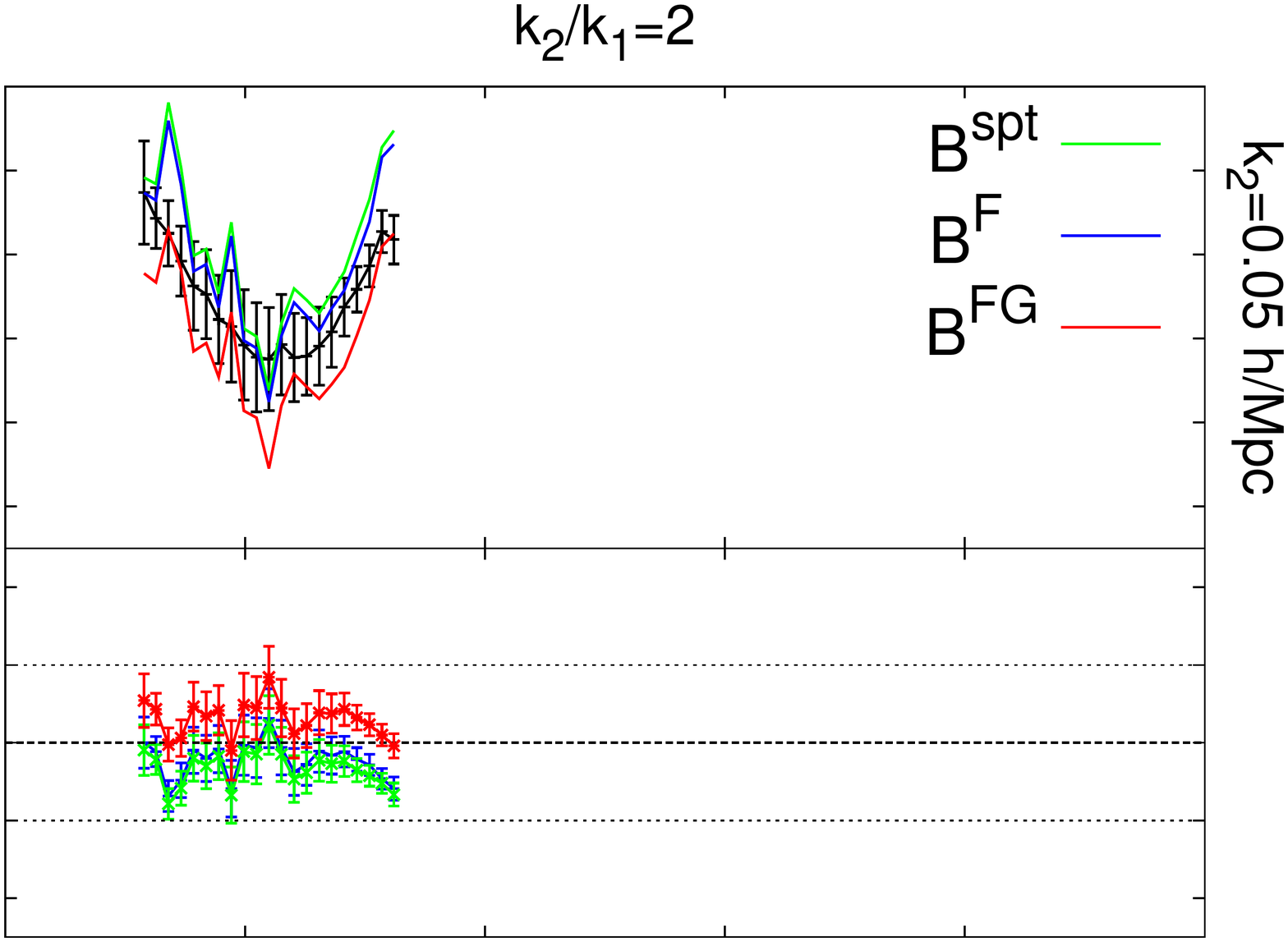}

\includegraphics[clip=false, trim= 80mm 10mm 22mm 35mm,scale=0.25]{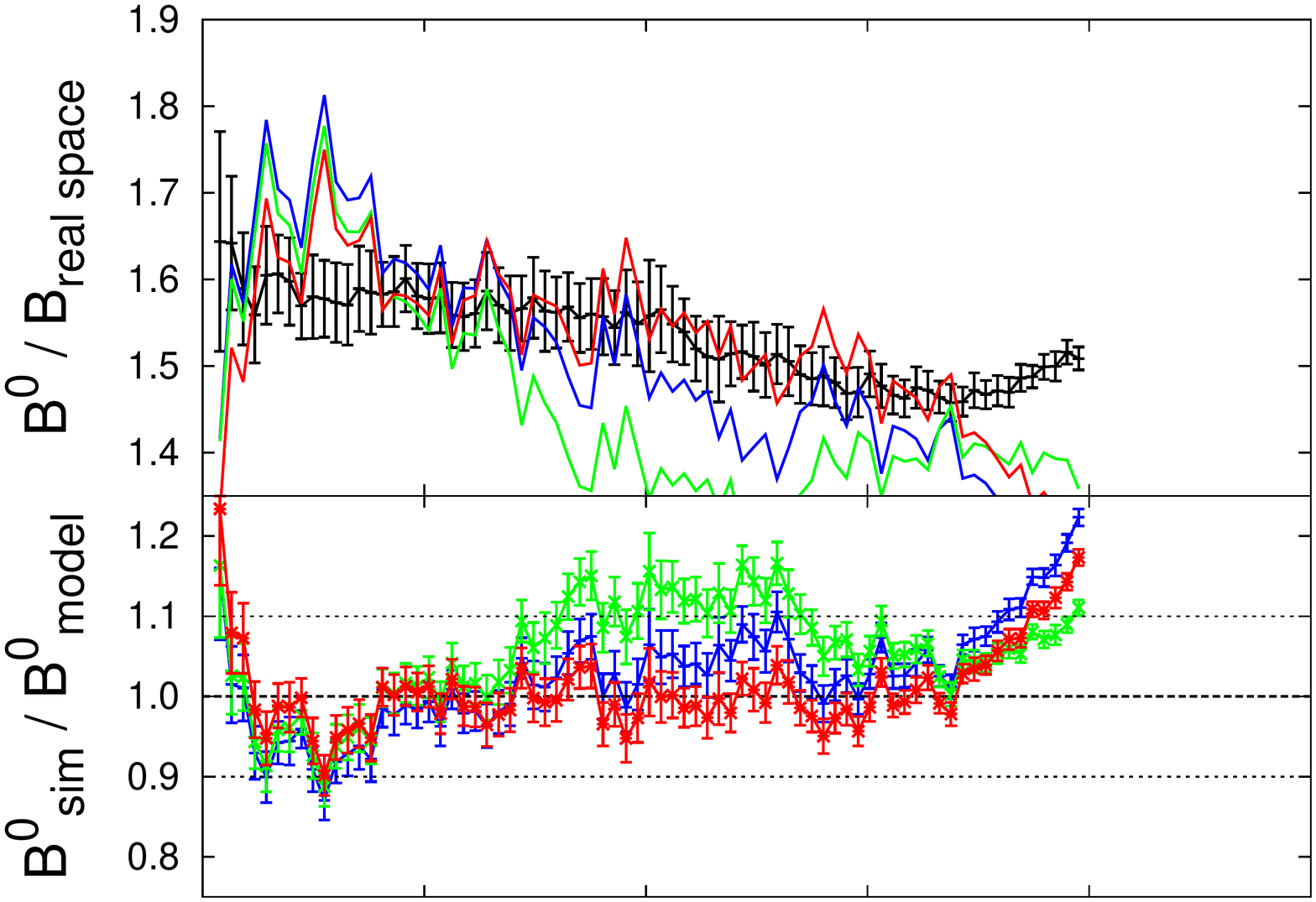}
\includegraphics[clip=false,trim= 25mm 10mm 22mm 35mm, scale=0.25]{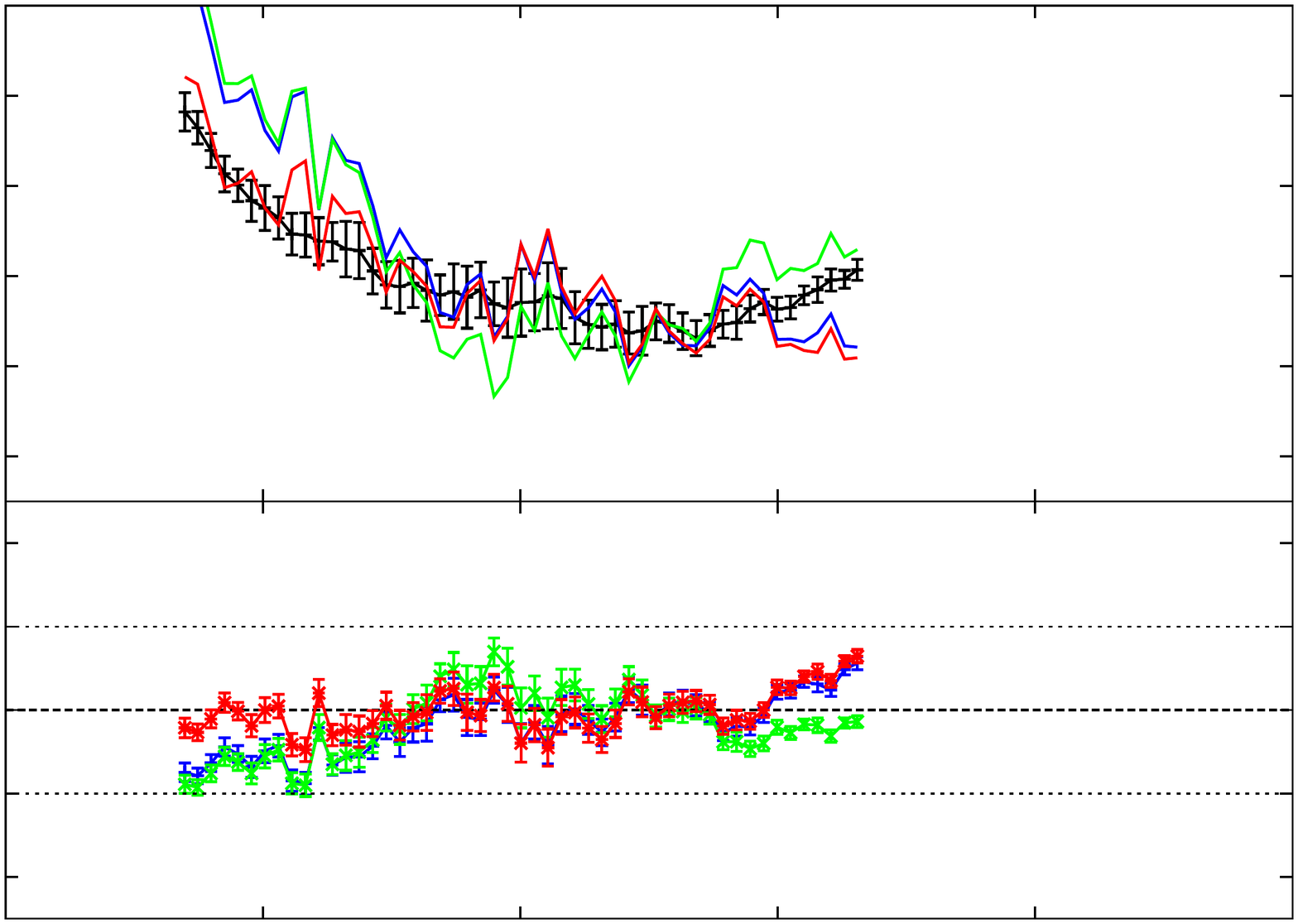}
\includegraphics[clip=false,trim= 25mm 10mm 80mm 35mm, scale=0.25]{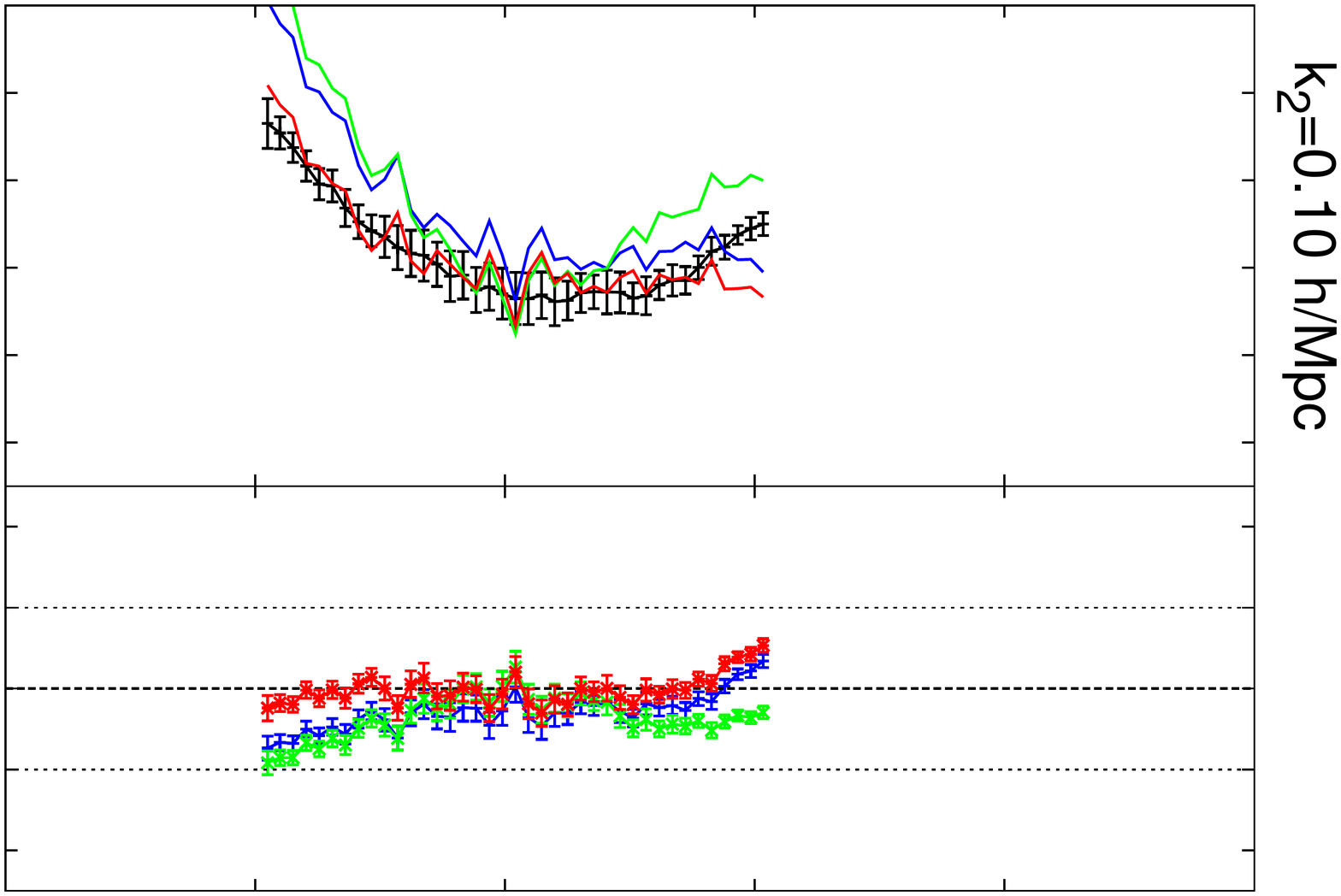}

\includegraphics[clip=false, trim= 80mm 10mm 22mm 35mm,scale=0.25]{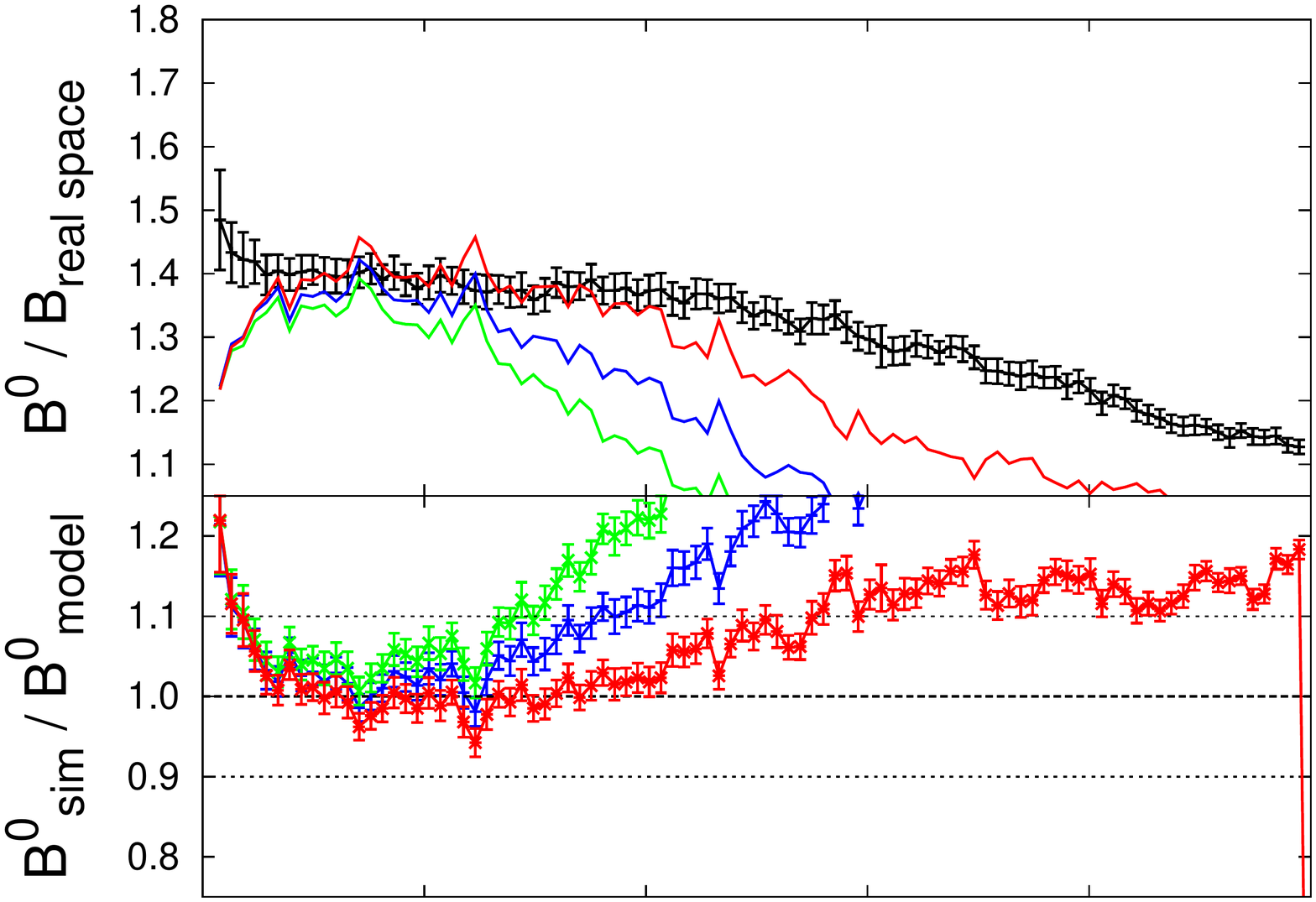}
\includegraphics[clip=false,trim= 25mm 10mm 22mm 35mm, scale=0.25]{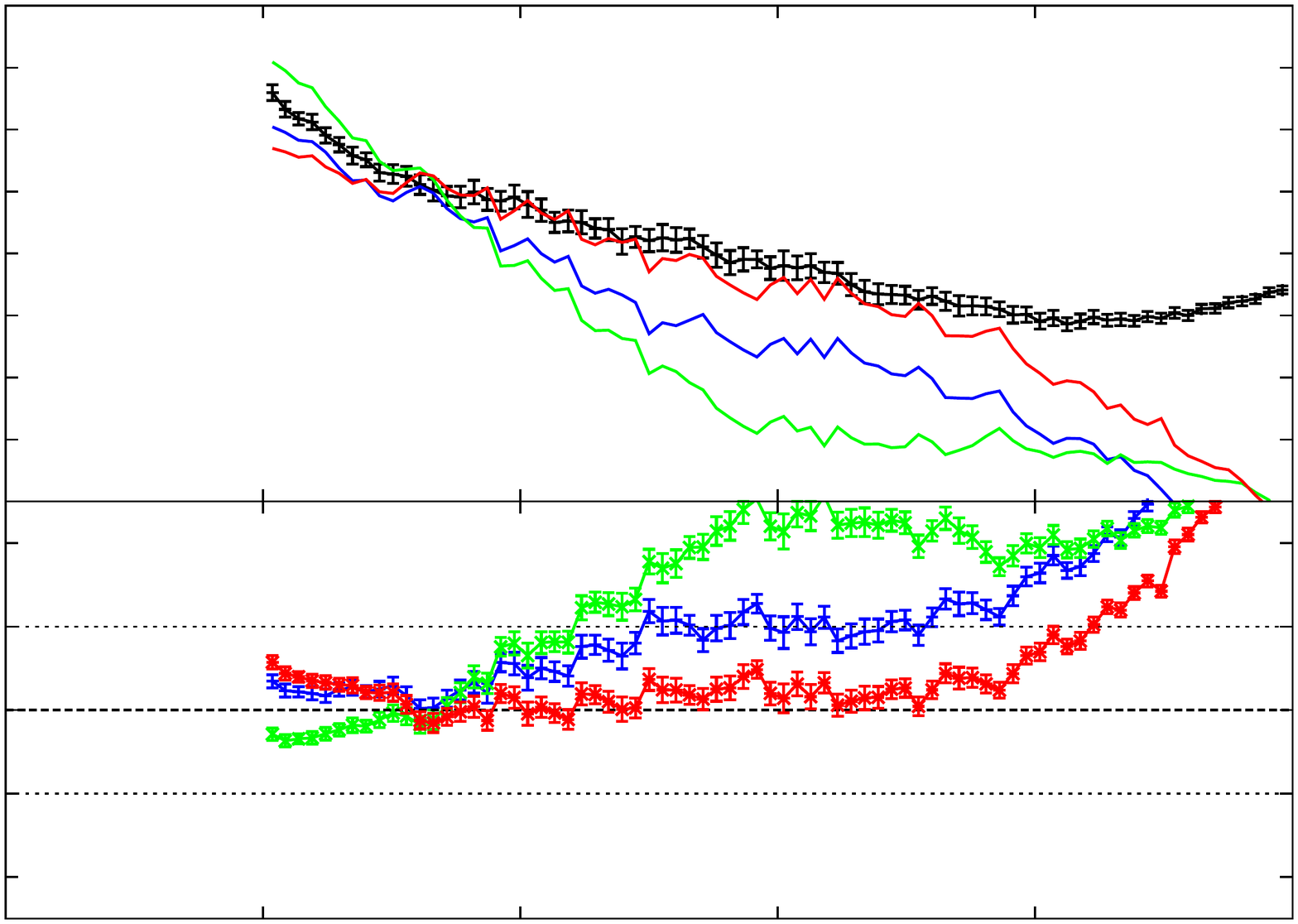}
\includegraphics[clip=false,trim= 25mm 10mm 80mm 35mm, scale=0.25]{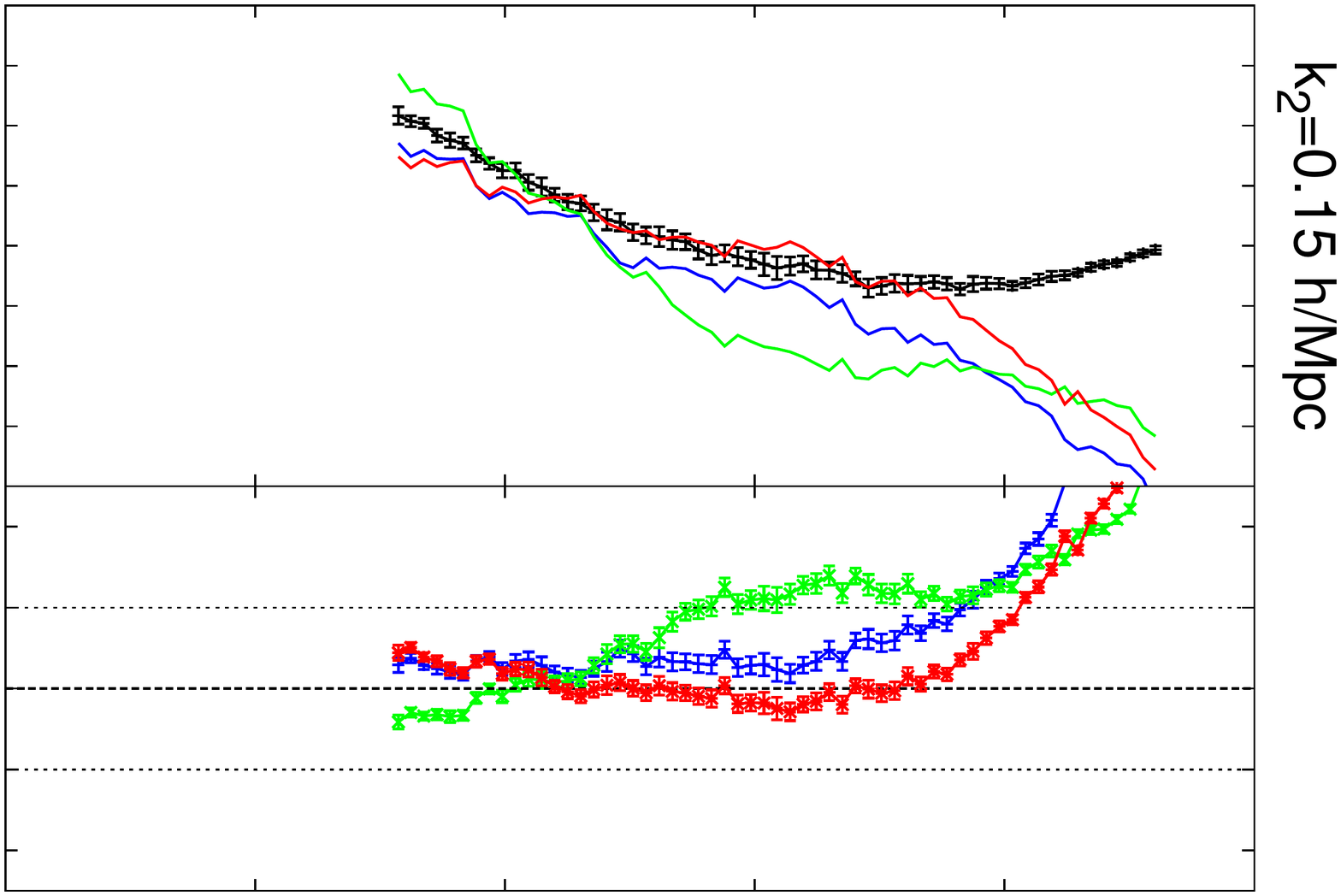}

\includegraphics[clip=false, trim= 80mm 10mm 22mm 35mm,scale=0.25]{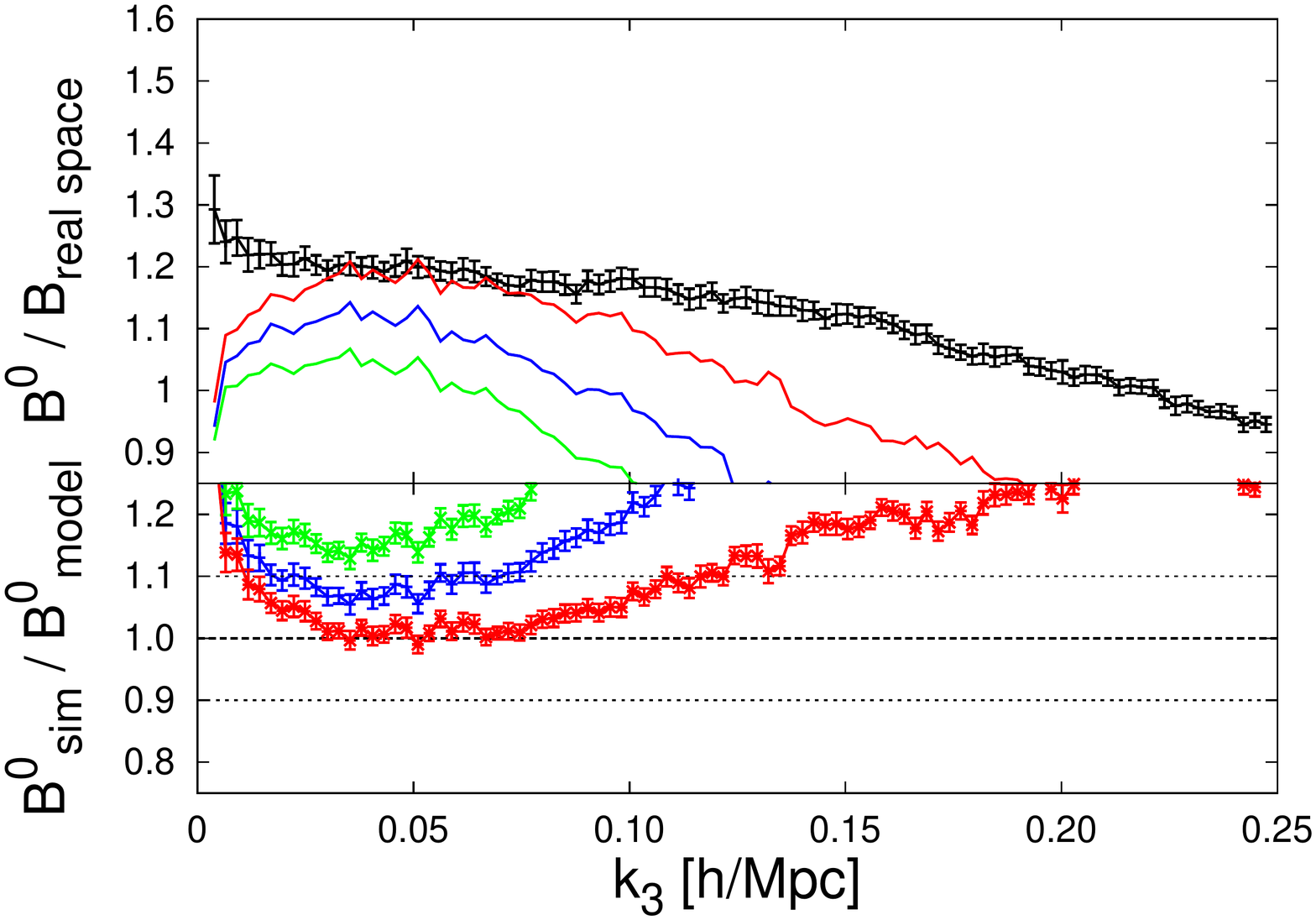}
\includegraphics[clip=false,trim= 25mm 10mm 22mm 35mm, scale=0.25]{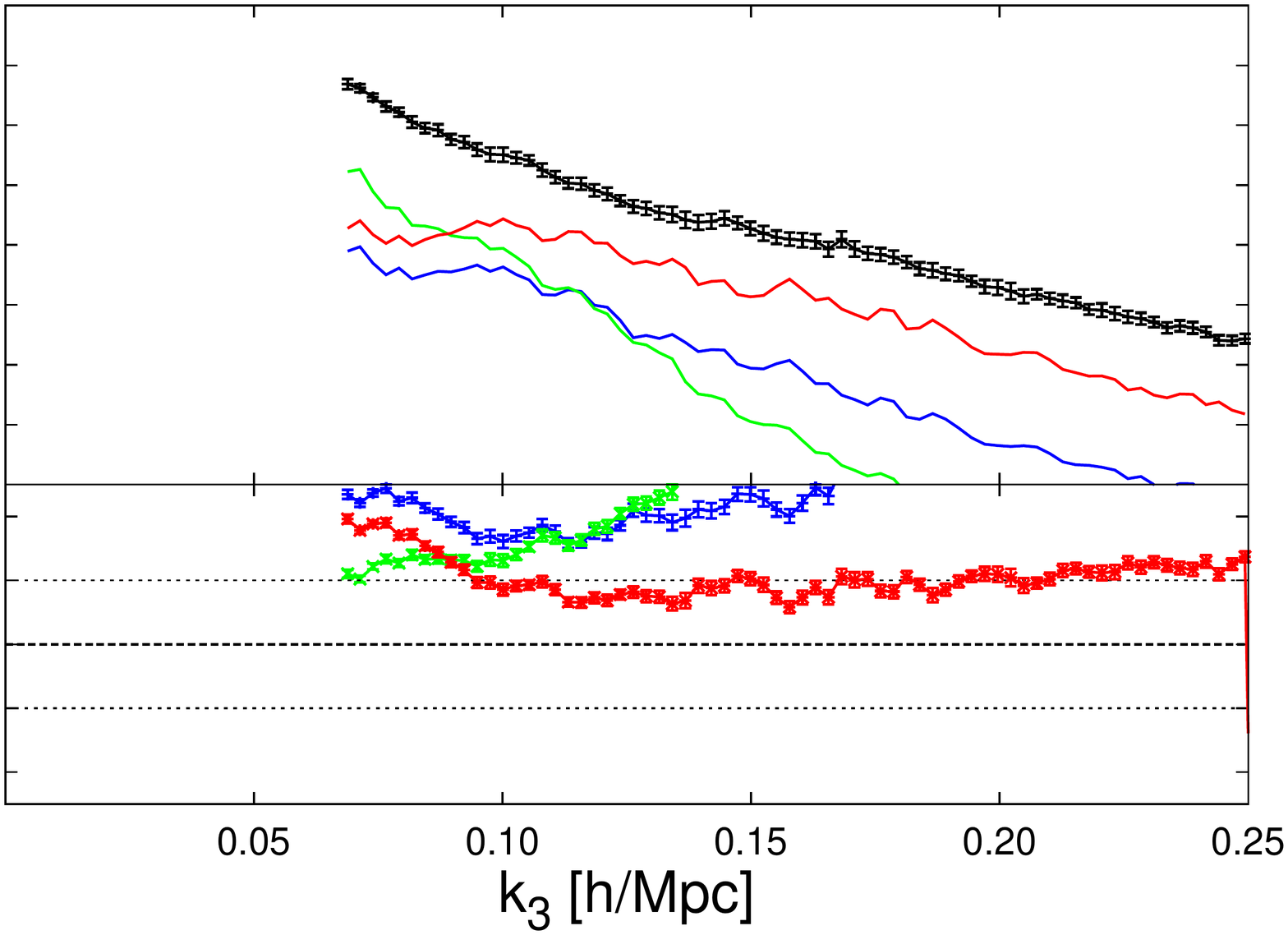}
\includegraphics[clip=false,trim= 25mm 10mm 80mm 35mm, scale=0.25]{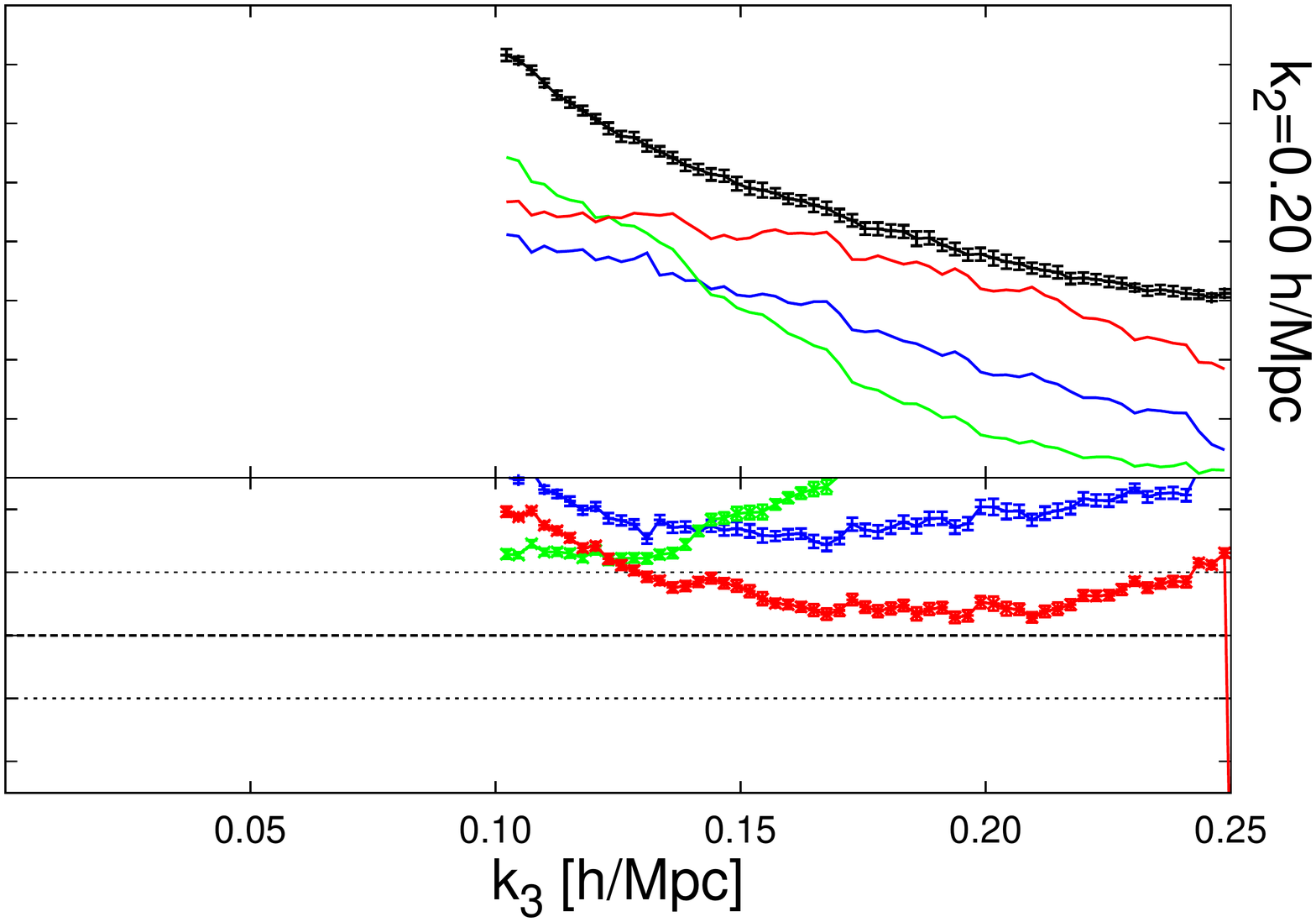}

\caption{Top sub-panels: dark matter monopole bispectrum normalised to the real space matter bispectrum for different triangle configurations. First column, second and third column panels are triangles with $k_2/k_1=1.0$, 1.5 and 2 respectively. Different rows show different scales: first, second, third and forth rows correspond to $k_2=0.05$, 0.10, 0.15 and 0.20\, $h$/Mpc as indicated. Black symbols correspond to N-body simulations whereas colour lines to the different models based on Eq.~\ref{Bspt}: $B^{\rm spt}$ (green lines), $B^{ \rm F}$ (blue line) and $B^{\rm FG}$ (red) (see text for description).  Bottom sub-panels: ratio between the dark matter measurement and the prediction of each model. All panels  are at $z=0$.}
\label{bis1}
\end{figure}

\begin{figure}
\centering
\includegraphics[clip=false, trim= 80mm 10mm 22mm 35mm,scale=0.25]{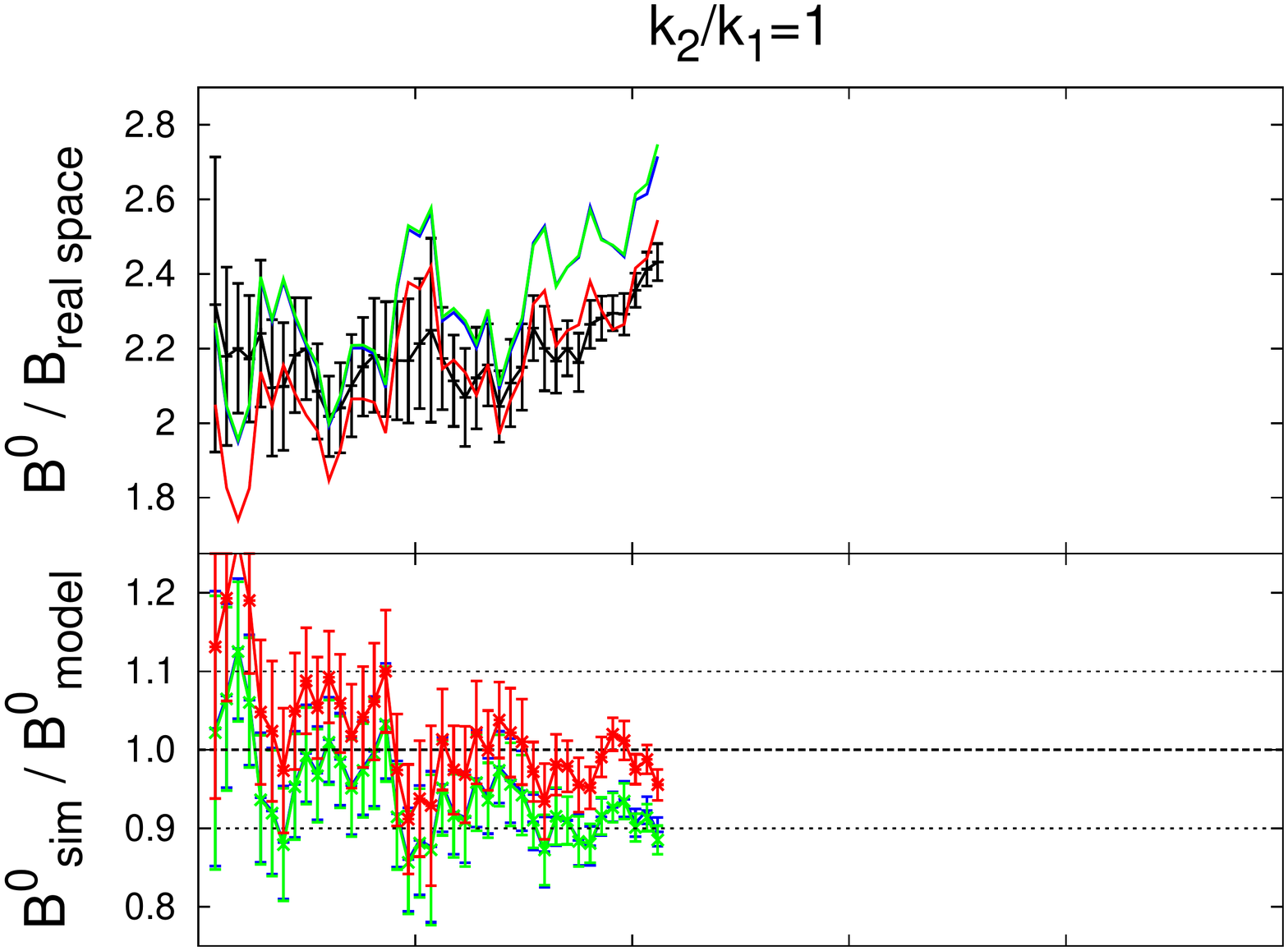}
\includegraphics[clip=false,trim= 25mm 10mm 22mm 35mm, scale=0.25]{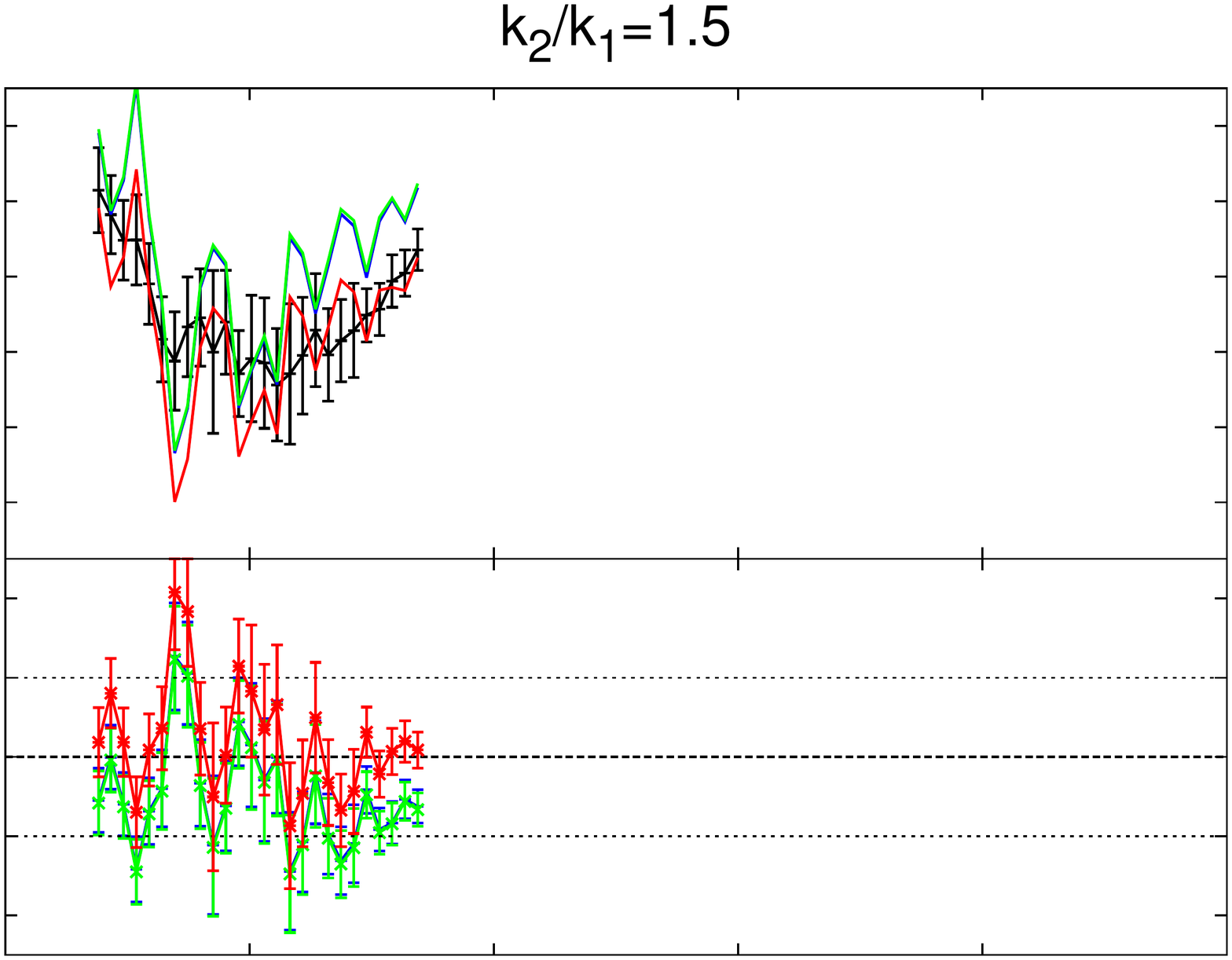}
\includegraphics[clip=false,trim= 25mm 10mm 80mm 35mm, scale=0.25]{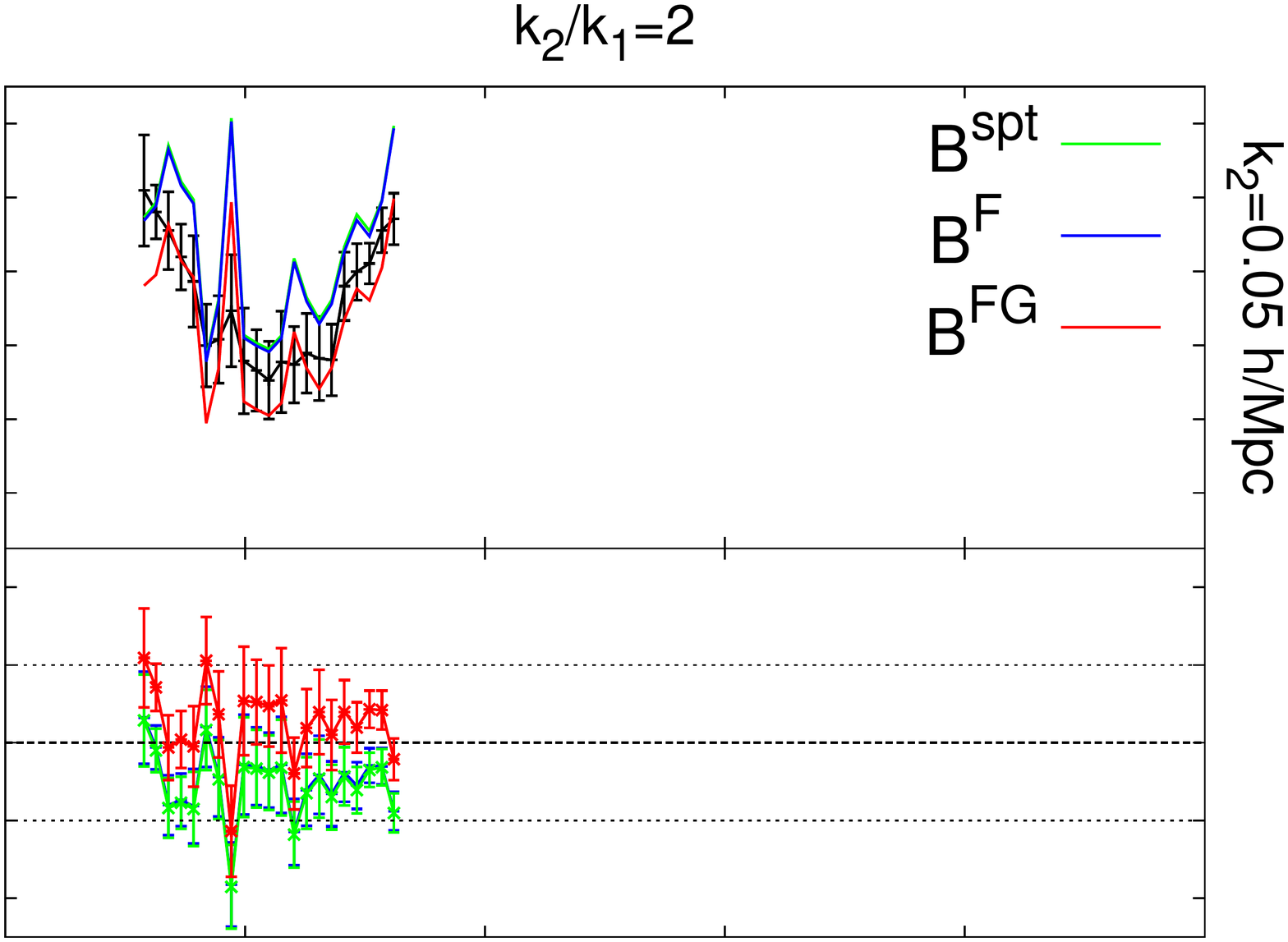}

\includegraphics[clip=false, trim= 80mm 10mm 22mm 35mm,scale=0.25]{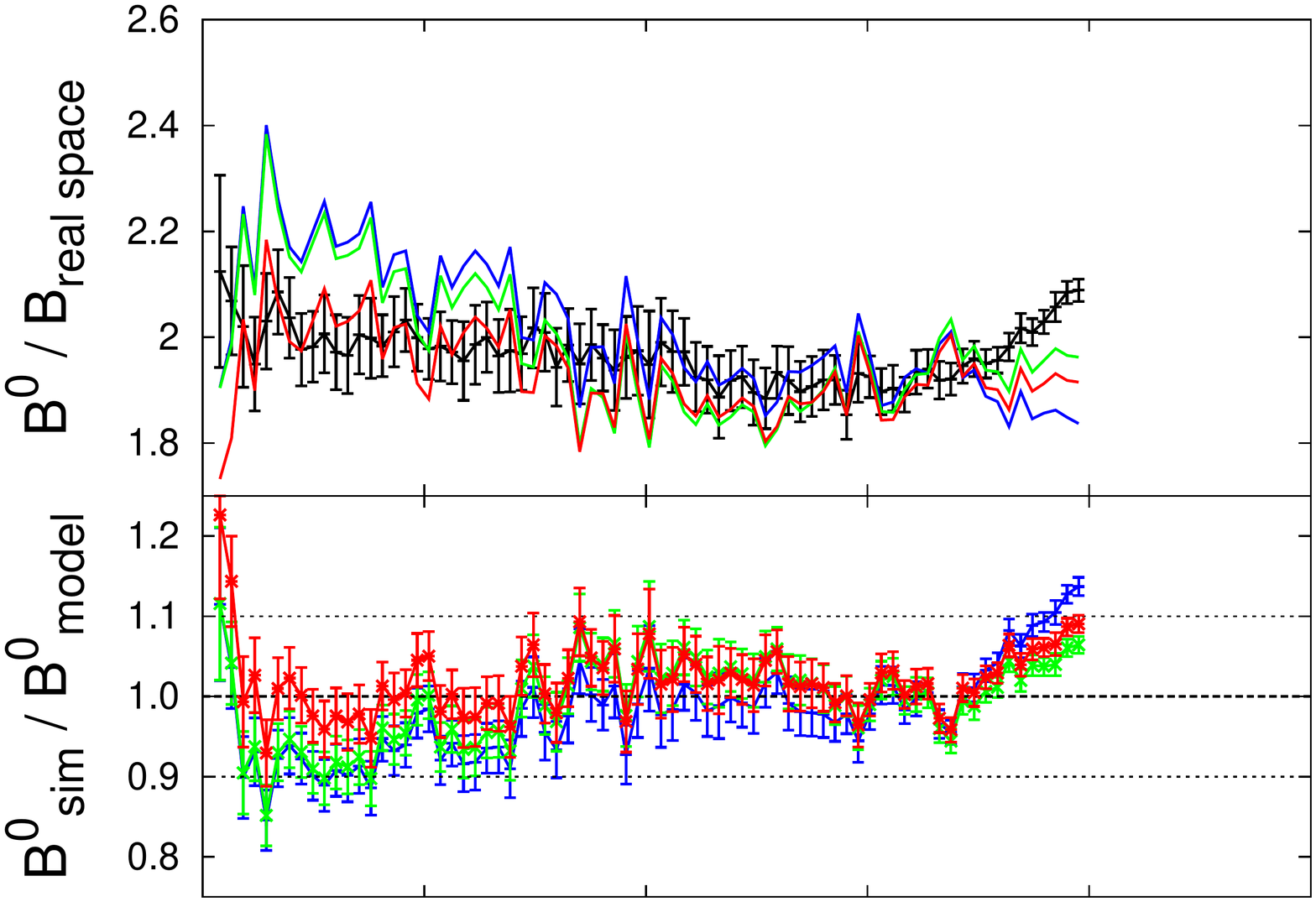}
\includegraphics[clip=false,trim= 25mm 10mm 22mm 35mm, scale=0.25]{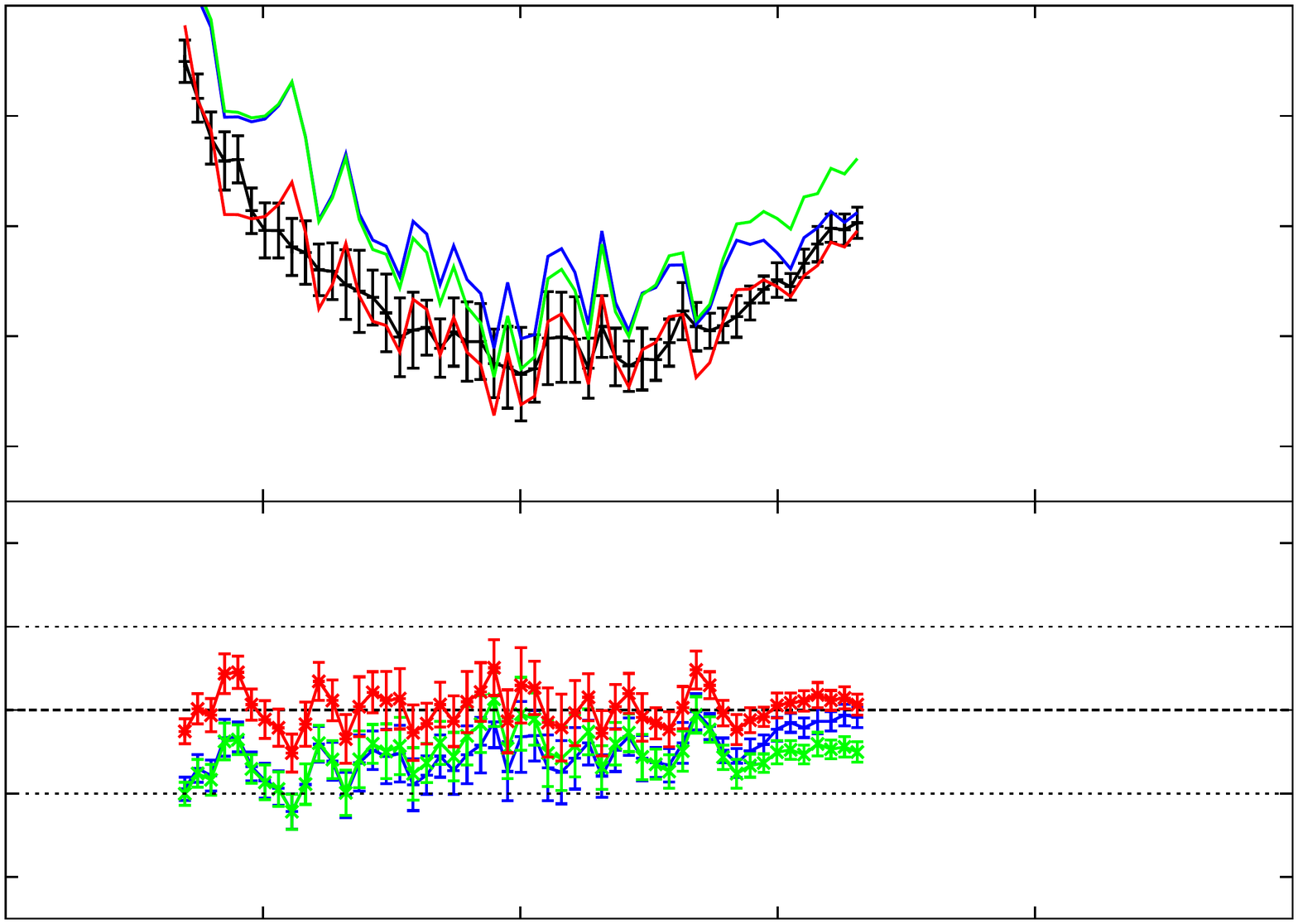}
\includegraphics[clip=false,trim= 25mm 10mm 80mm 35mm, scale=0.25]{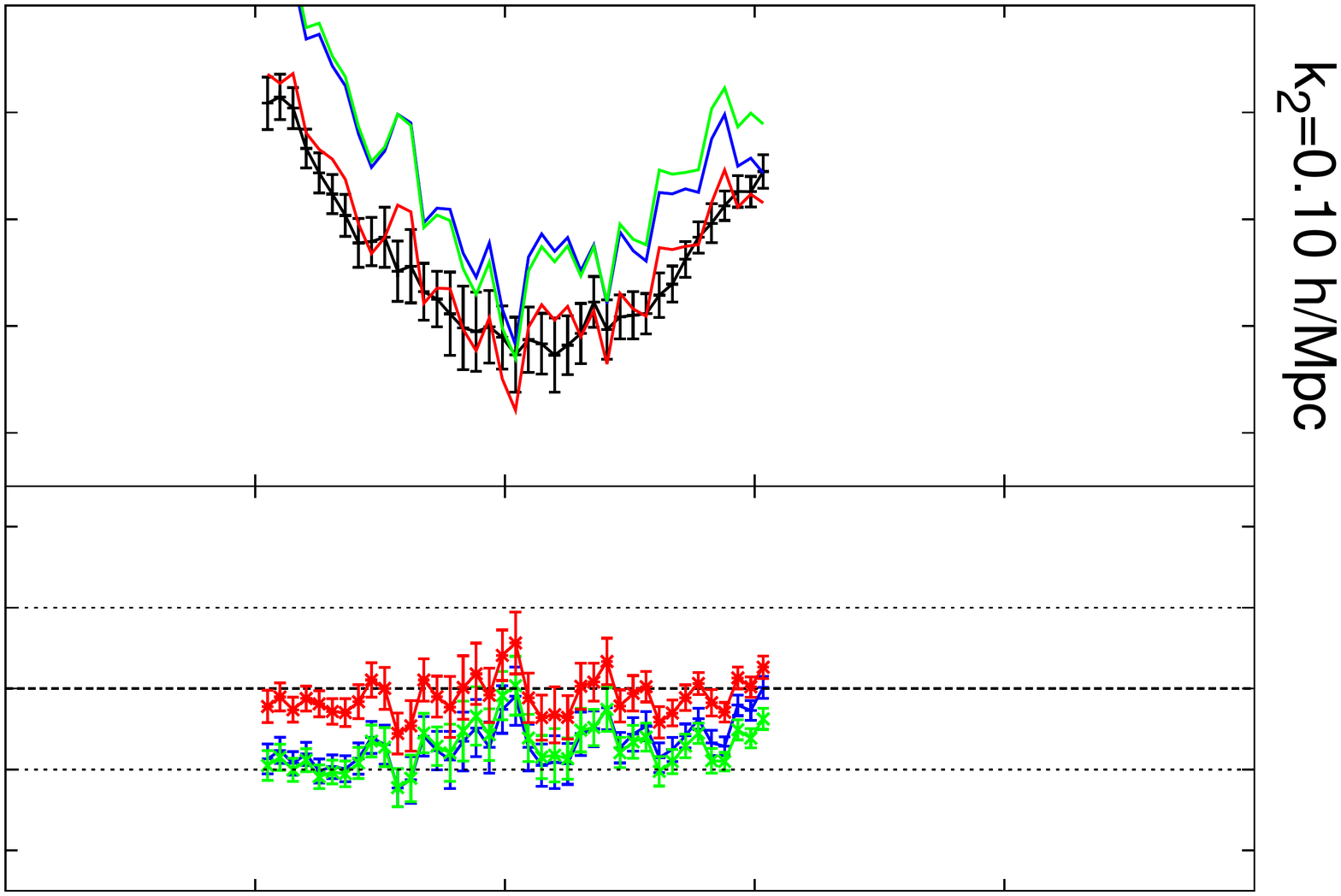}

\includegraphics[clip=false, trim= 80mm 10mm 22mm 35mm,scale=0.25]{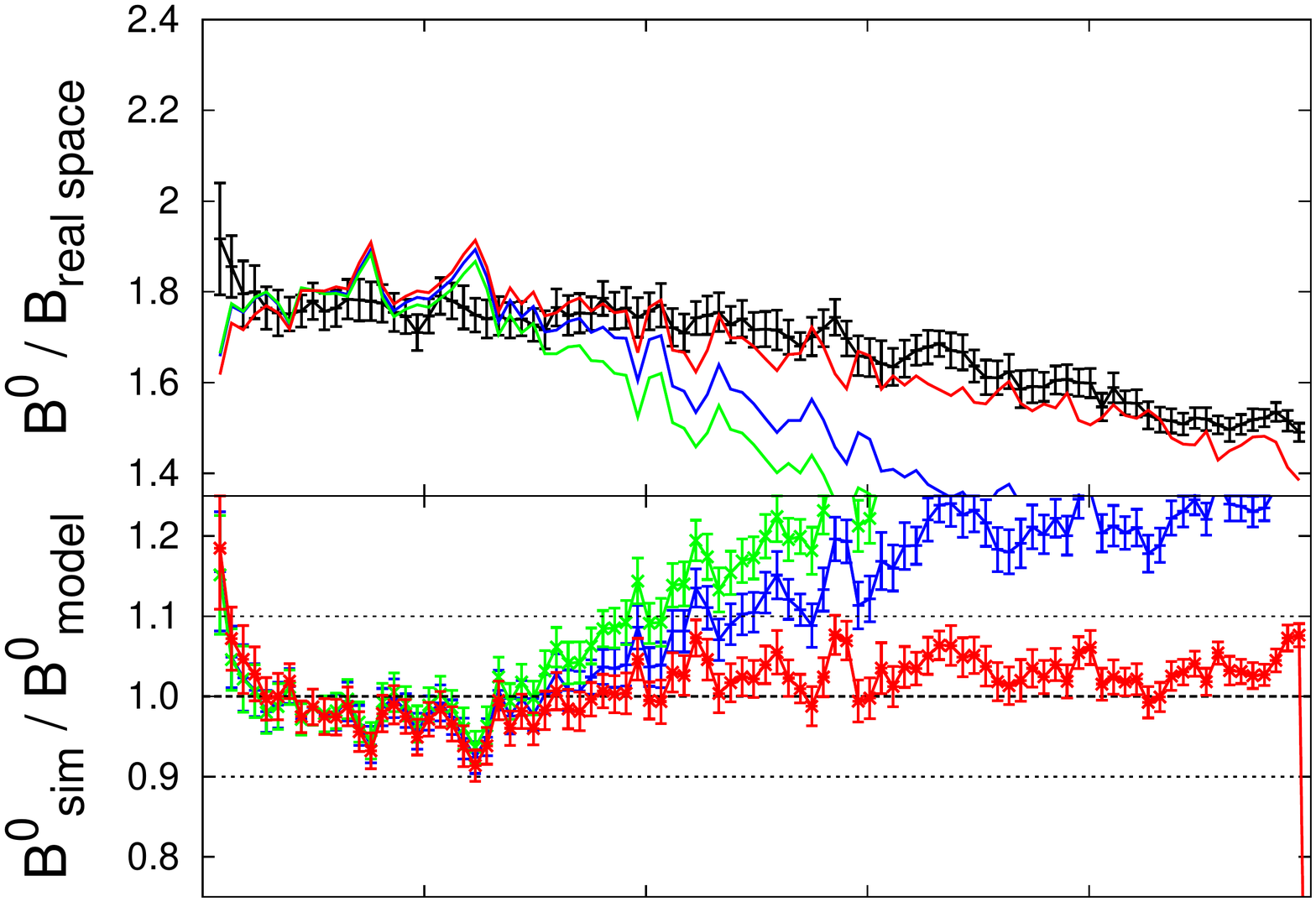}
\includegraphics[clip=false,trim= 25mm 10mm 22mm 35mm, scale=0.25]{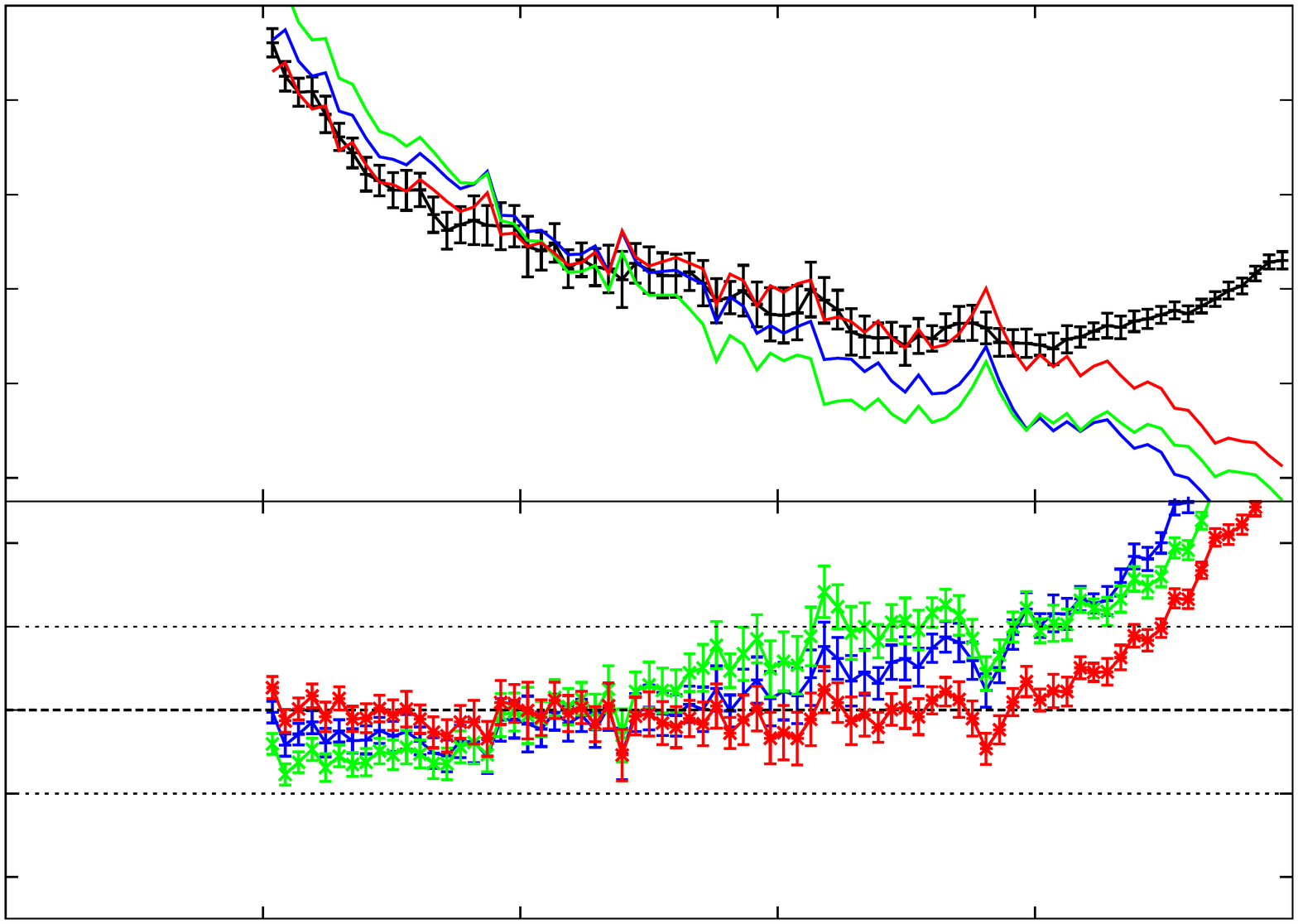}
\includegraphics[clip=false,trim= 25mm 10mm 80mm 35mm, scale=0.25]{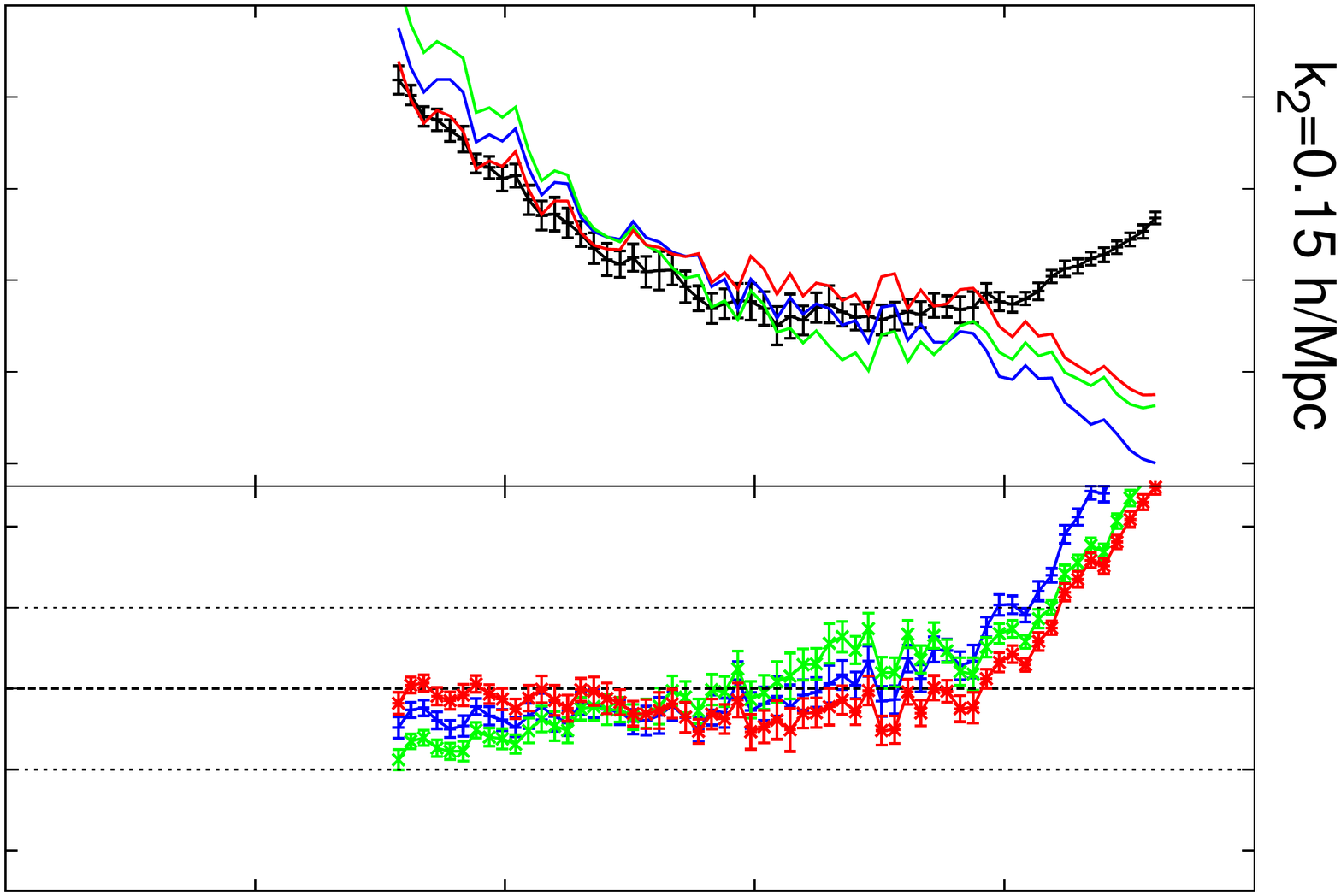}

\includegraphics[clip=false, trim= 80mm 10mm 22mm 35mm,scale=0.25]{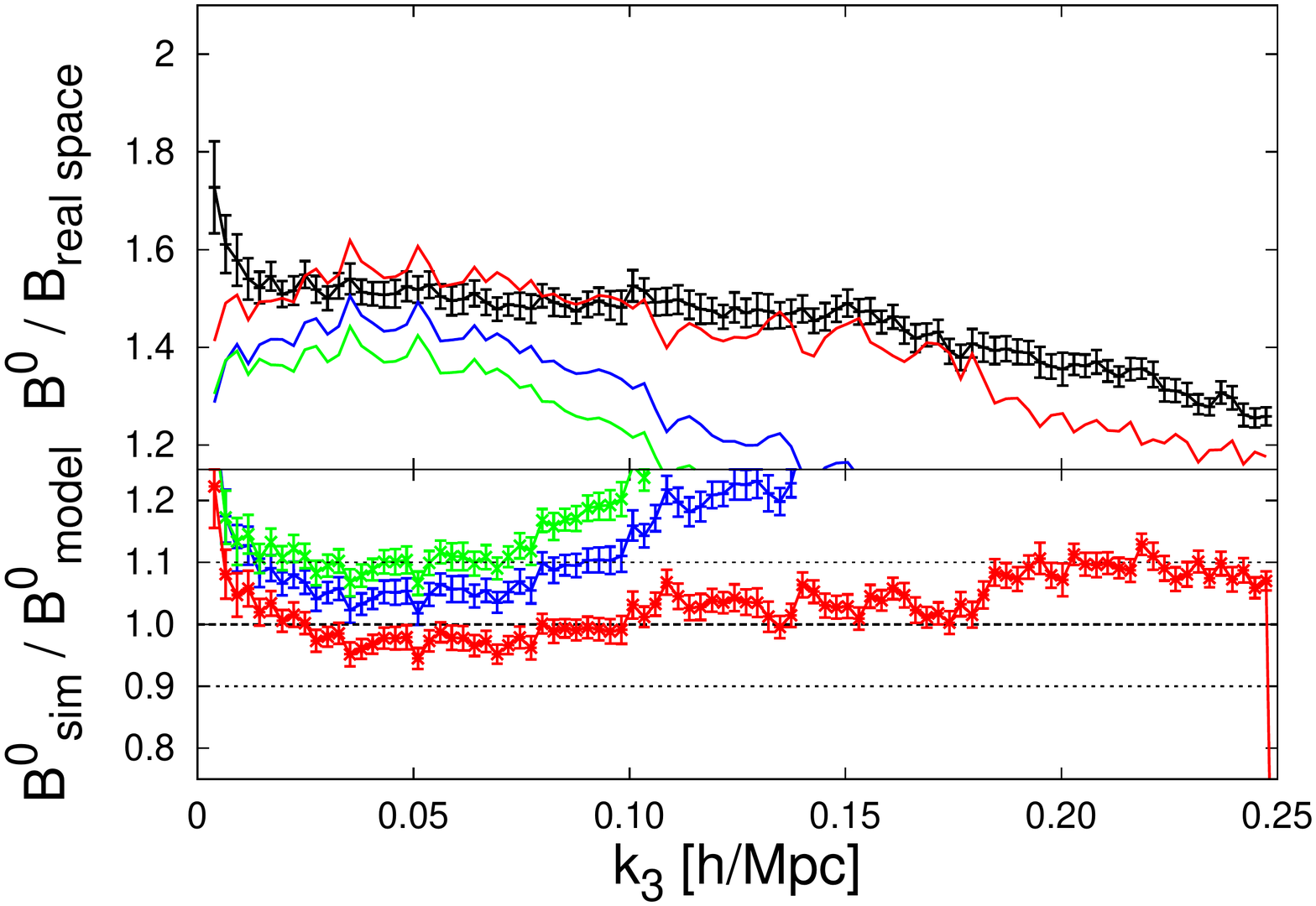}
\includegraphics[clip=false,trim= 25mm 10mm 22mm 35mm, scale=0.25]{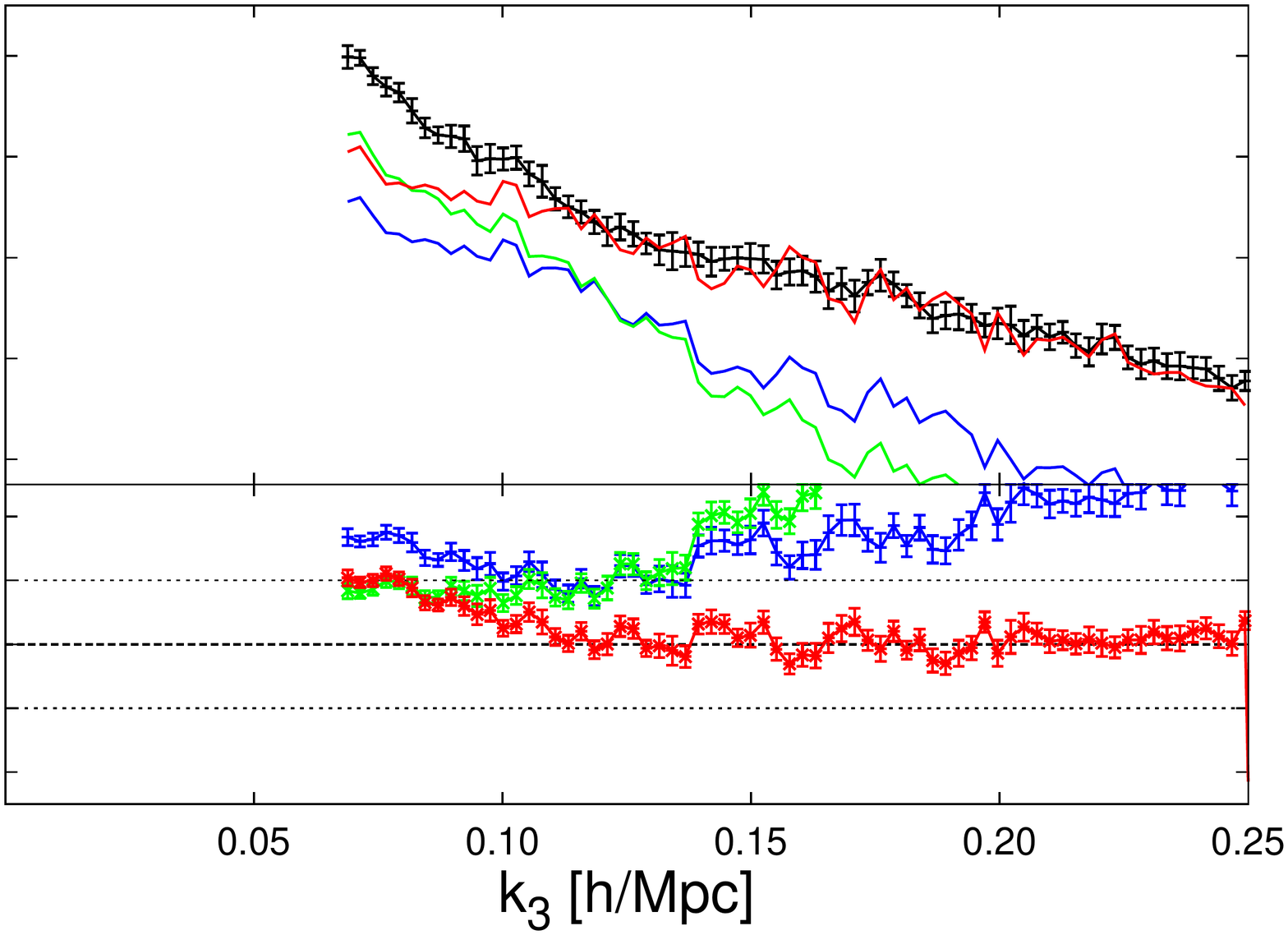}
\includegraphics[clip=false,trim= 25mm 10mm 80mm 35mm, scale=0.25]{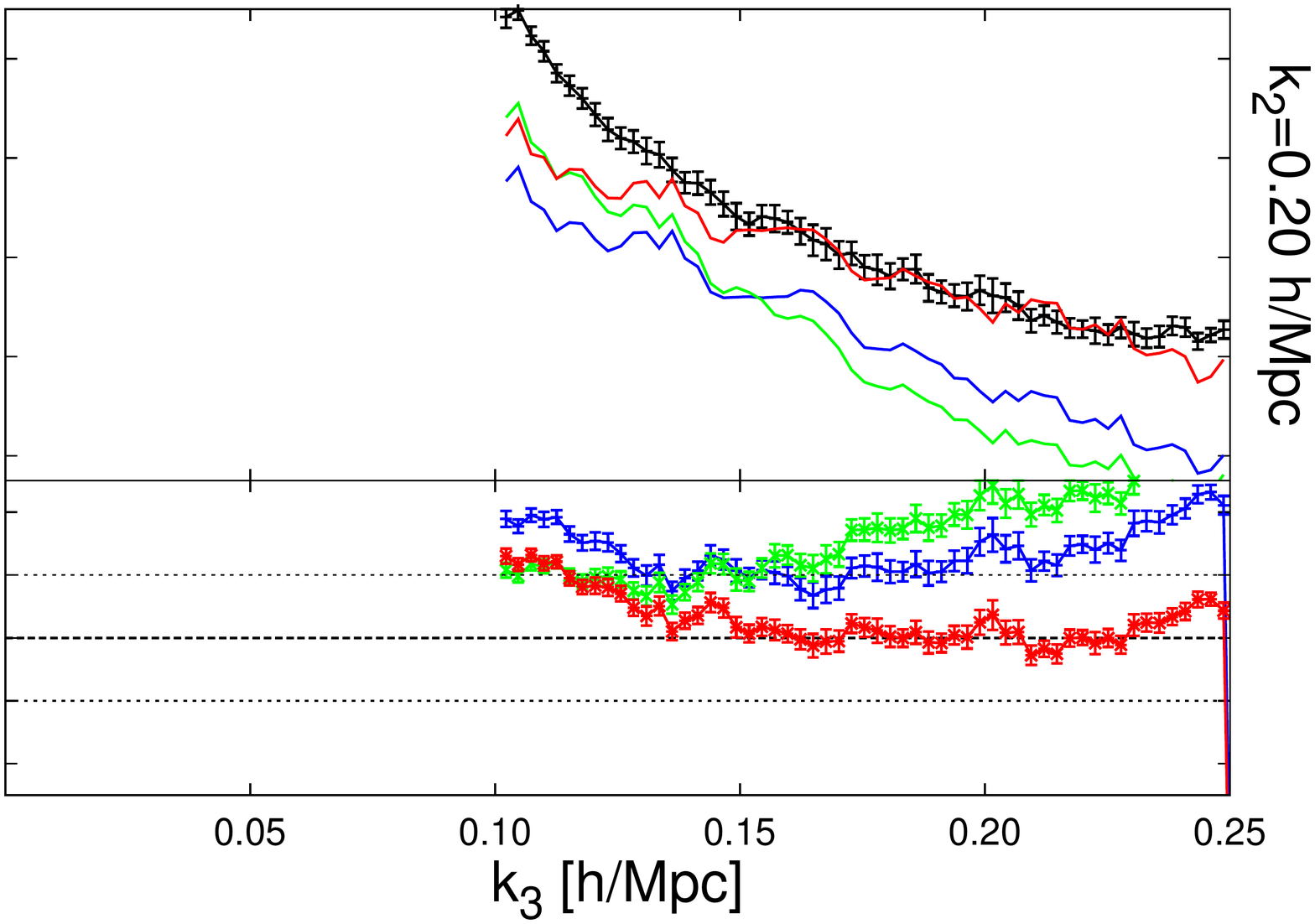}

\caption{Same notation as in Fig.~\ref{bis1}. All panels at $z=0.5$.}
\label{bis2}
\end{figure}

\begin{figure}
\centering
\includegraphics[clip=false, trim= 80mm 10mm 22mm 35mm,scale=0.25]{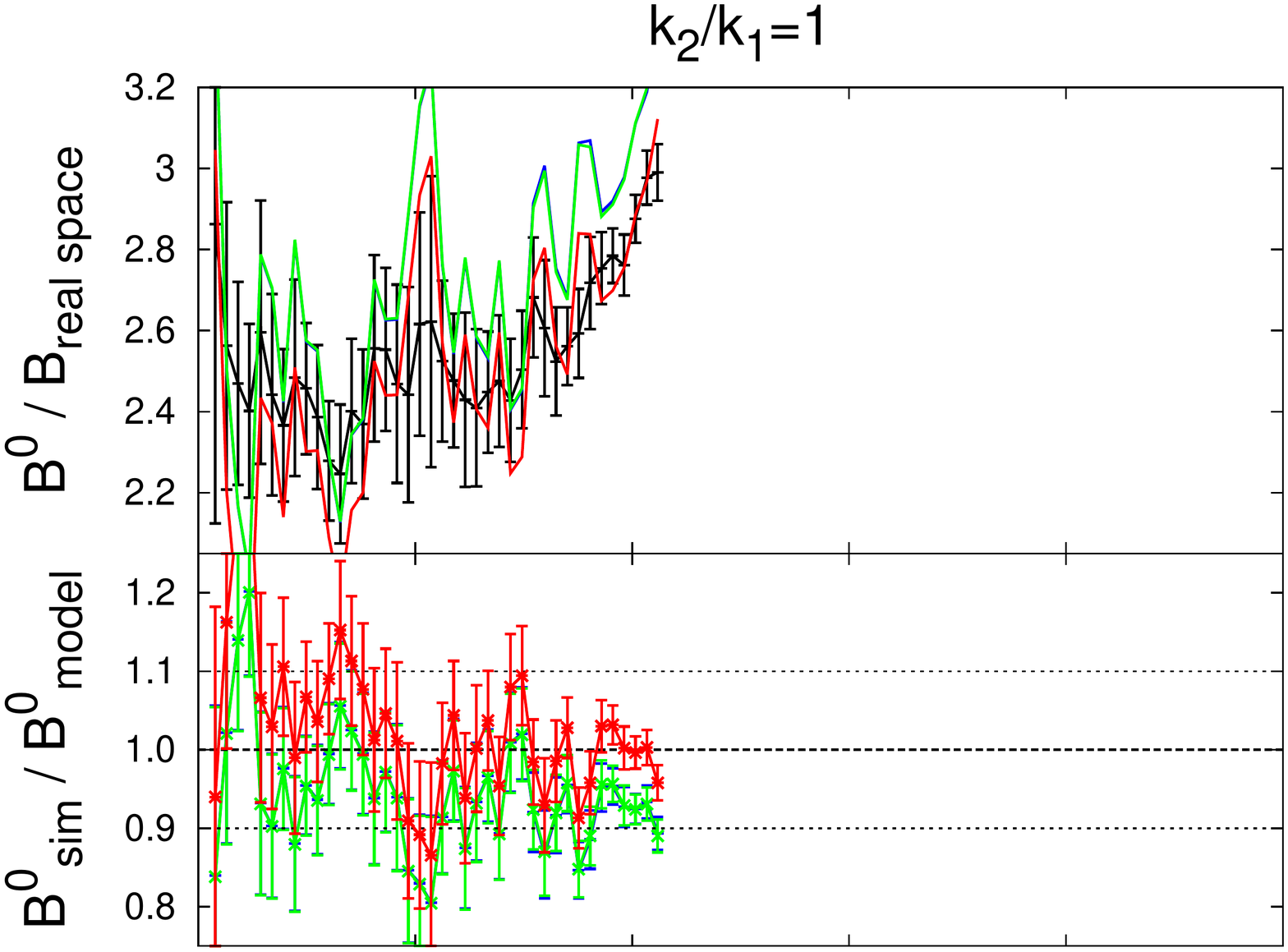}
\includegraphics[clip=false,trim= 25mm 10mm 22mm 35mm, scale=0.25]{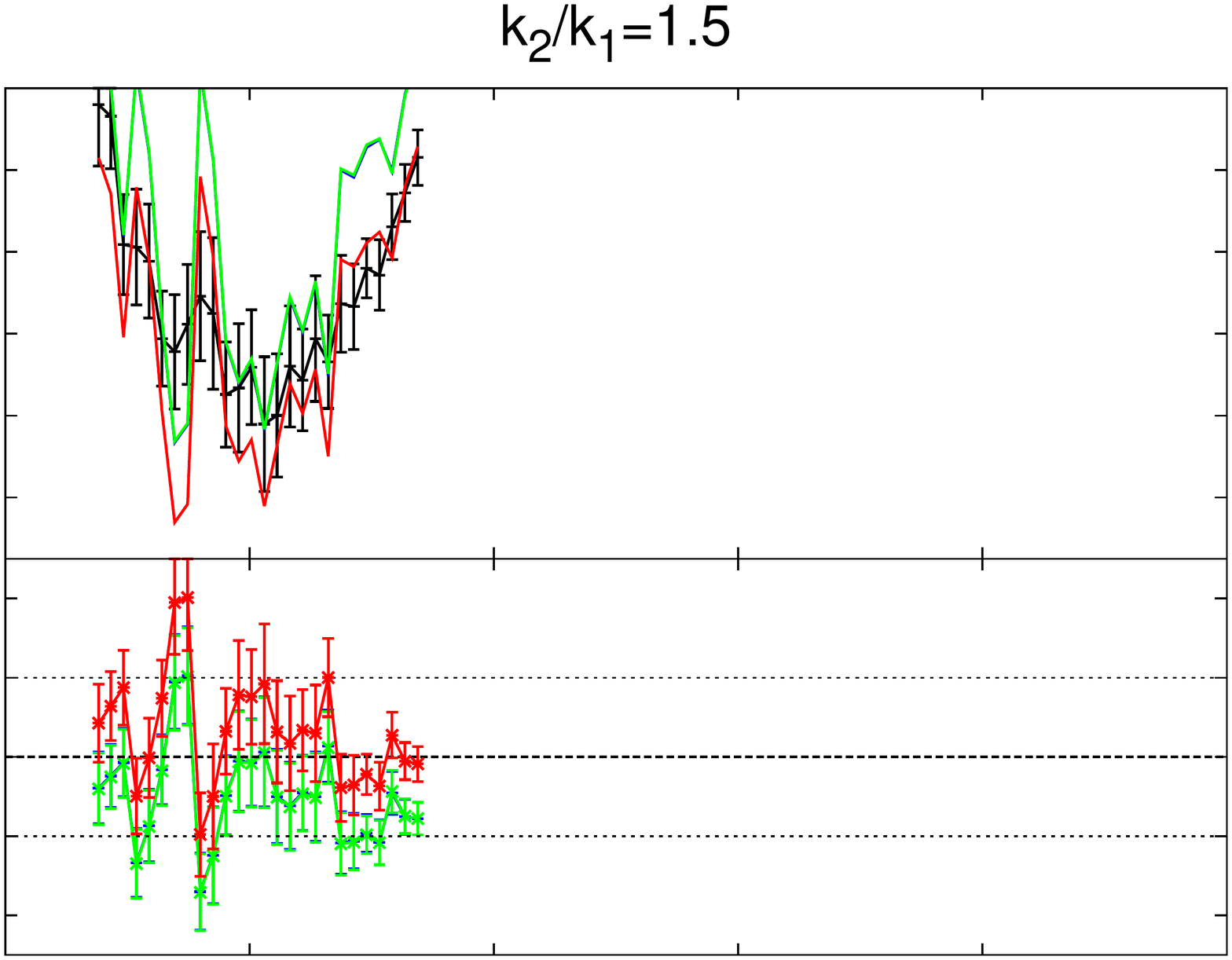}
\includegraphics[clip=false,trim= 25mm 10mm 80mm 35mm, scale=0.25]{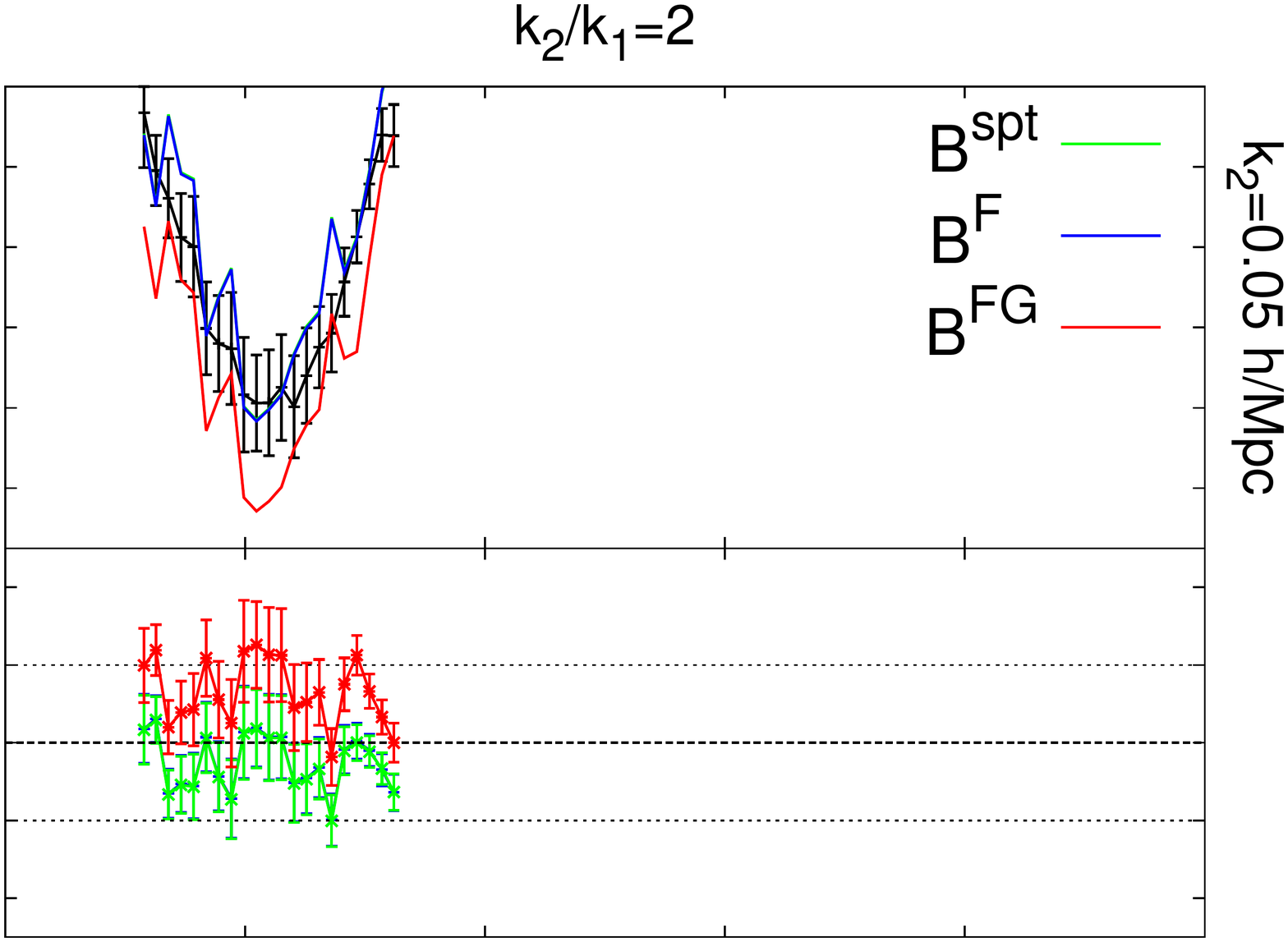}

\includegraphics[clip=false, trim= 80mm 10mm 22mm 35mm,scale=0.25]{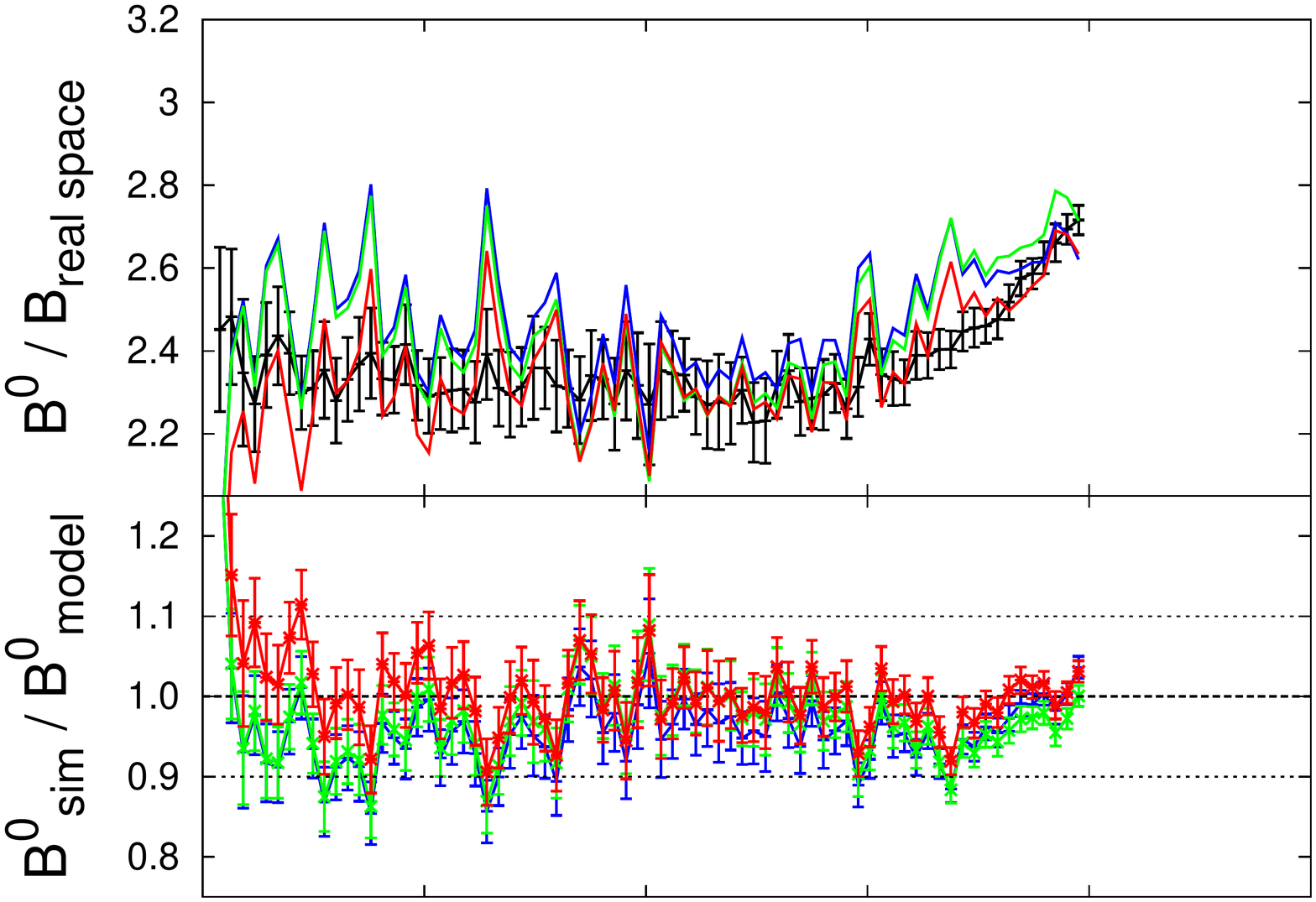}
\includegraphics[clip=false,trim= 25mm 10mm 22mm 35mm, scale=0.25]{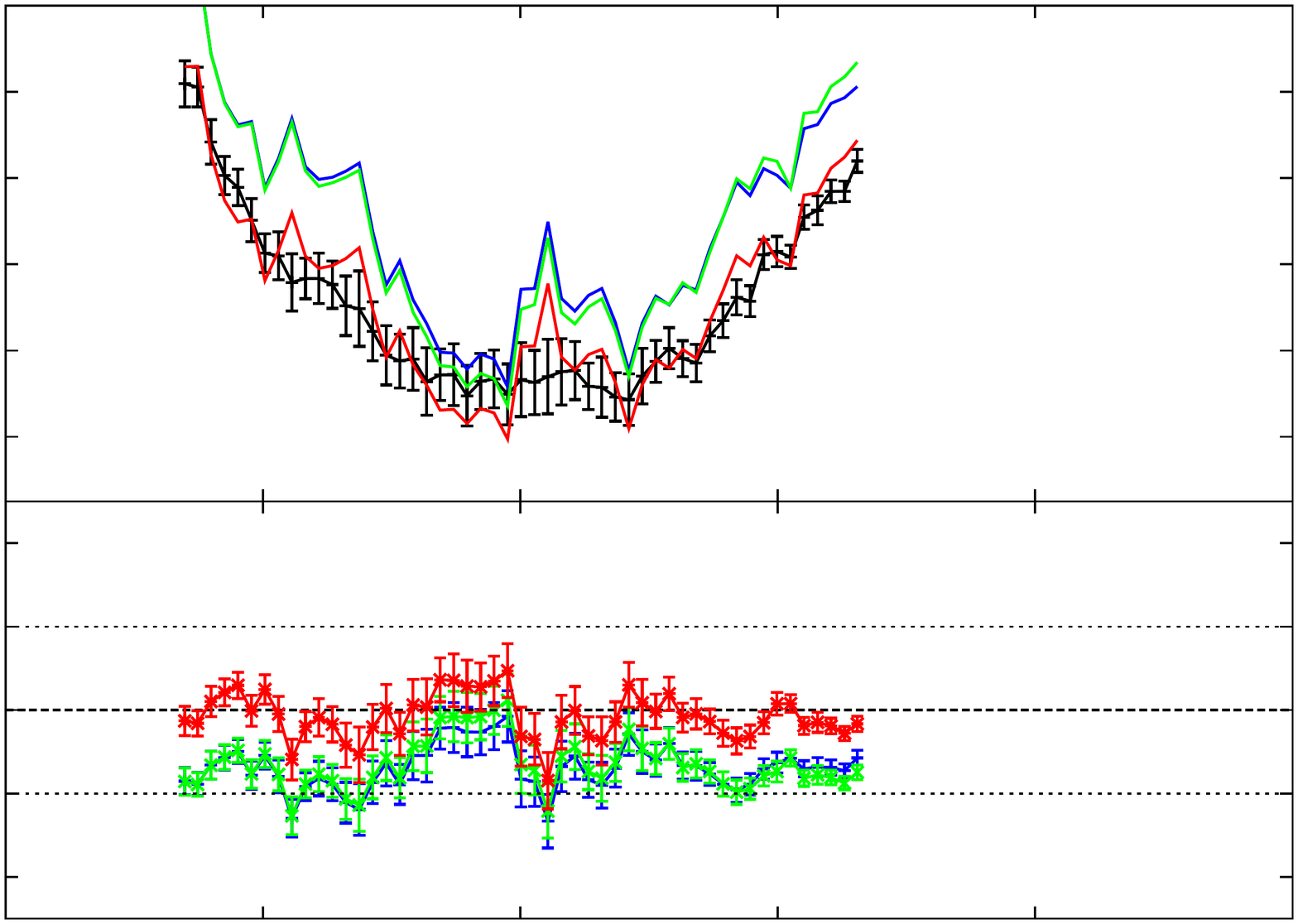}
\includegraphics[clip=false,trim= 25mm 10mm 80mm 35mm, scale=0.25]{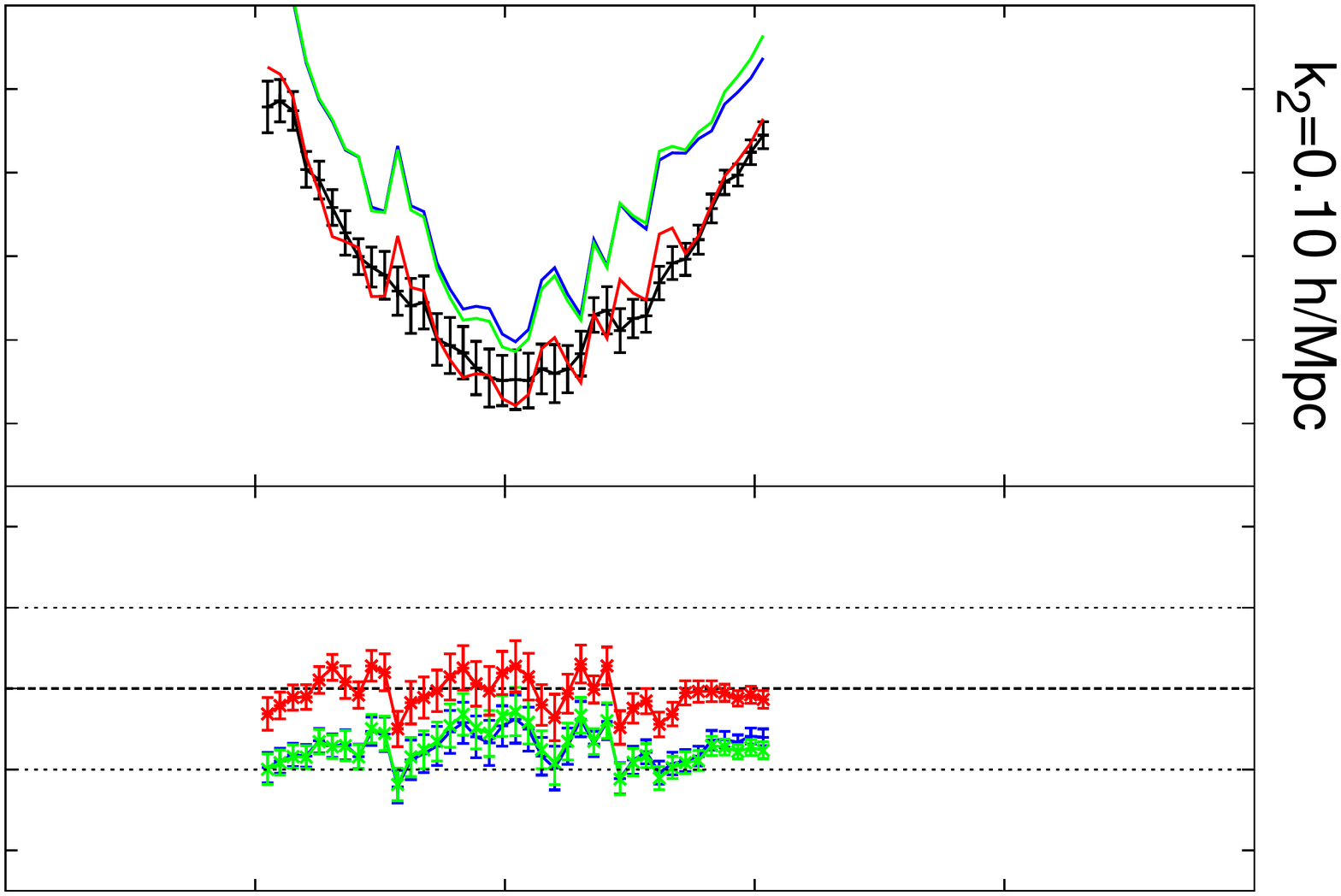}

\includegraphics[clip=false, trim= 80mm 10mm 22mm 35mm,scale=0.25]{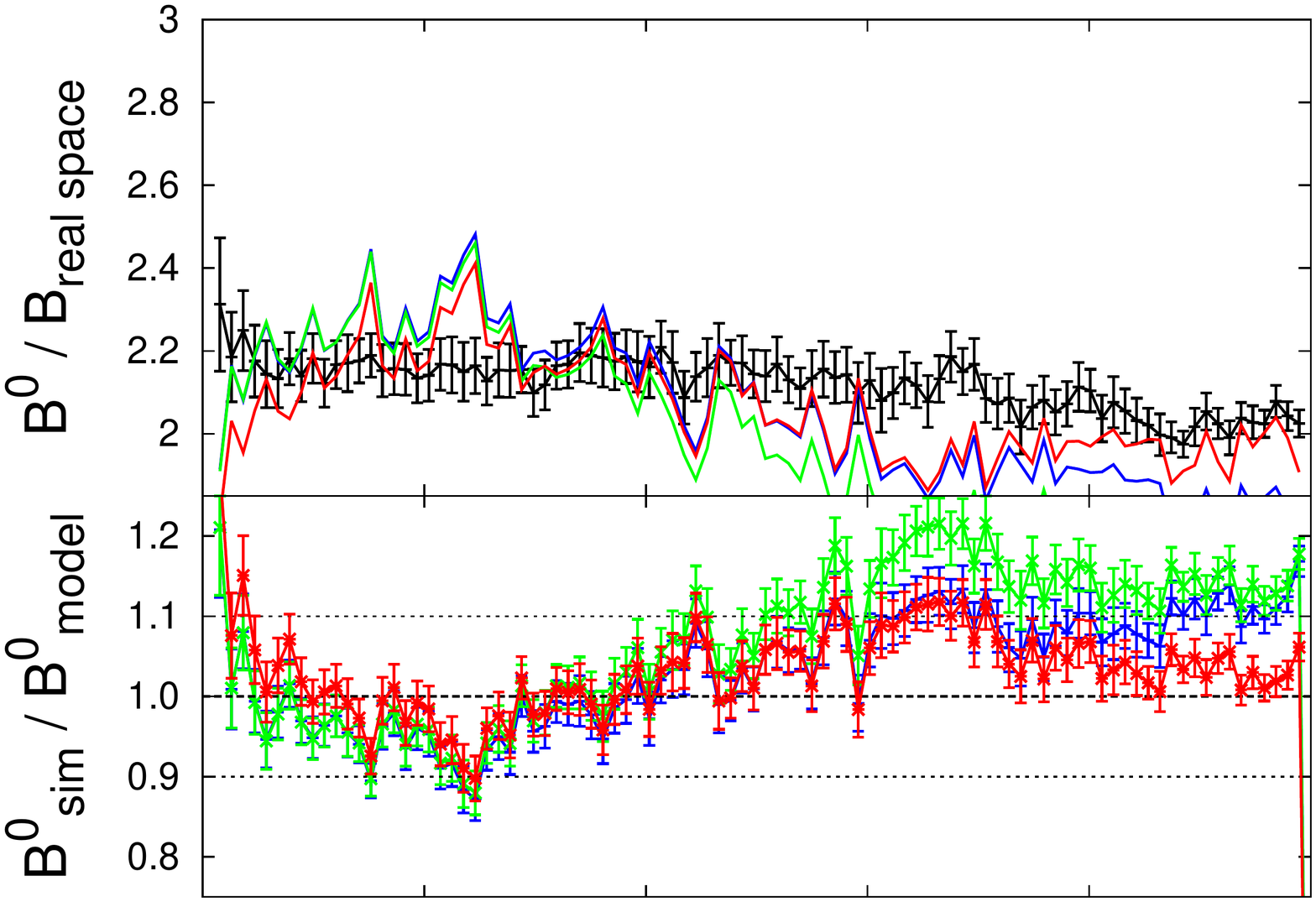}
\includegraphics[clip=false,trim= 25mm 10mm 22mm 35mm, scale=0.25]{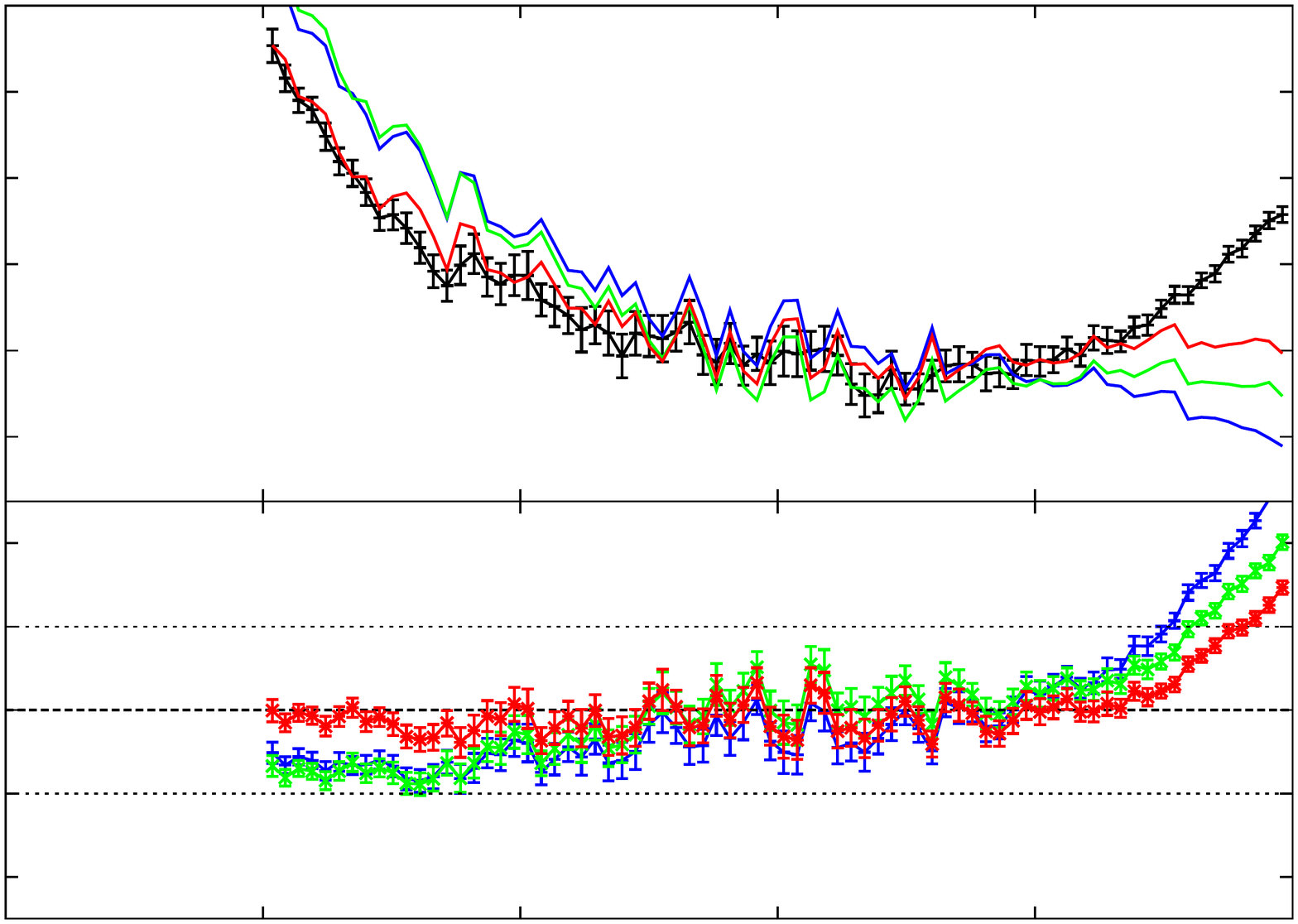}
\includegraphics[clip=false,trim= 25mm 10mm 80mm 35mm, scale=0.25]{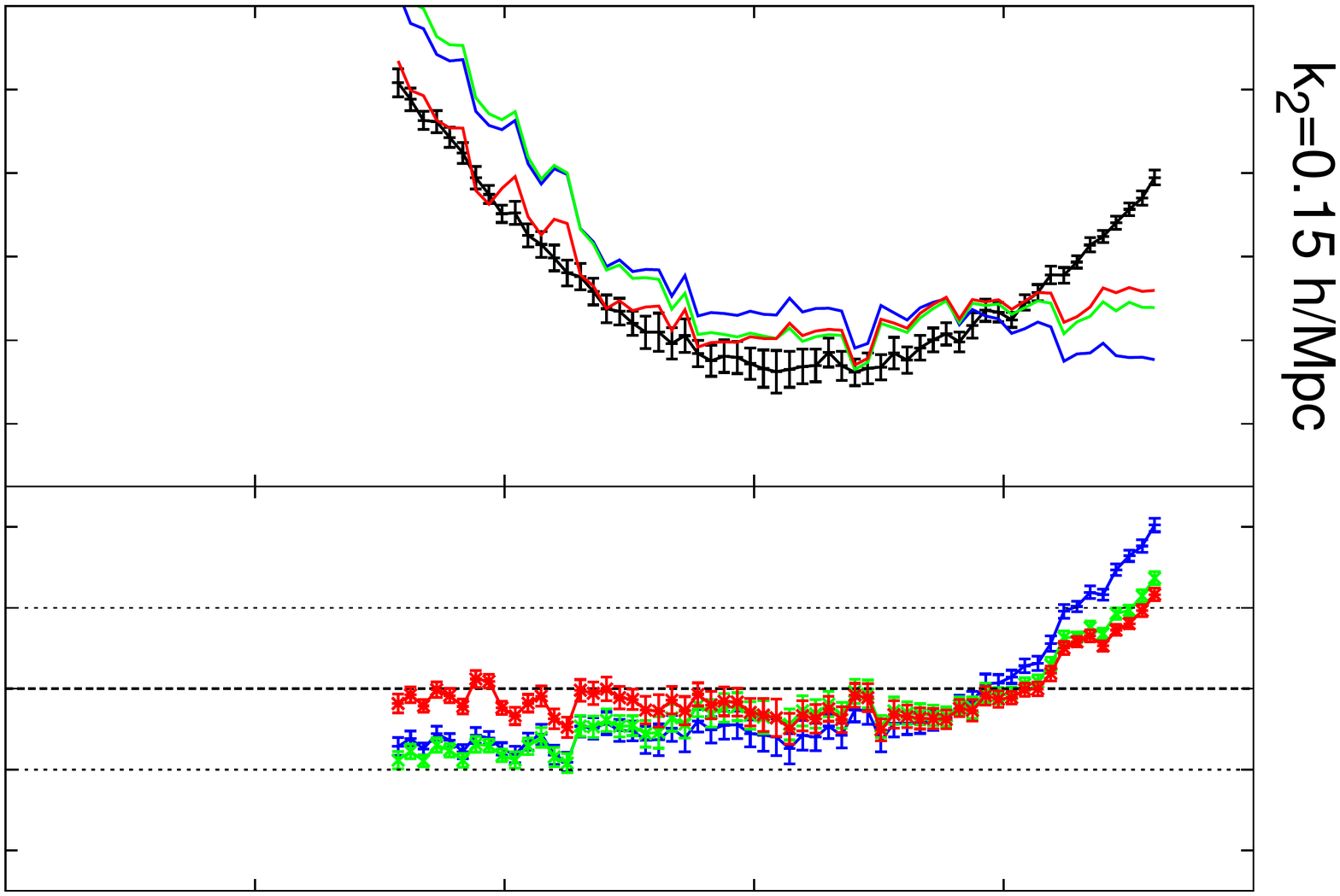}

\includegraphics[clip=false, trim= 80mm 10mm 22mm 35mm,scale=0.25]{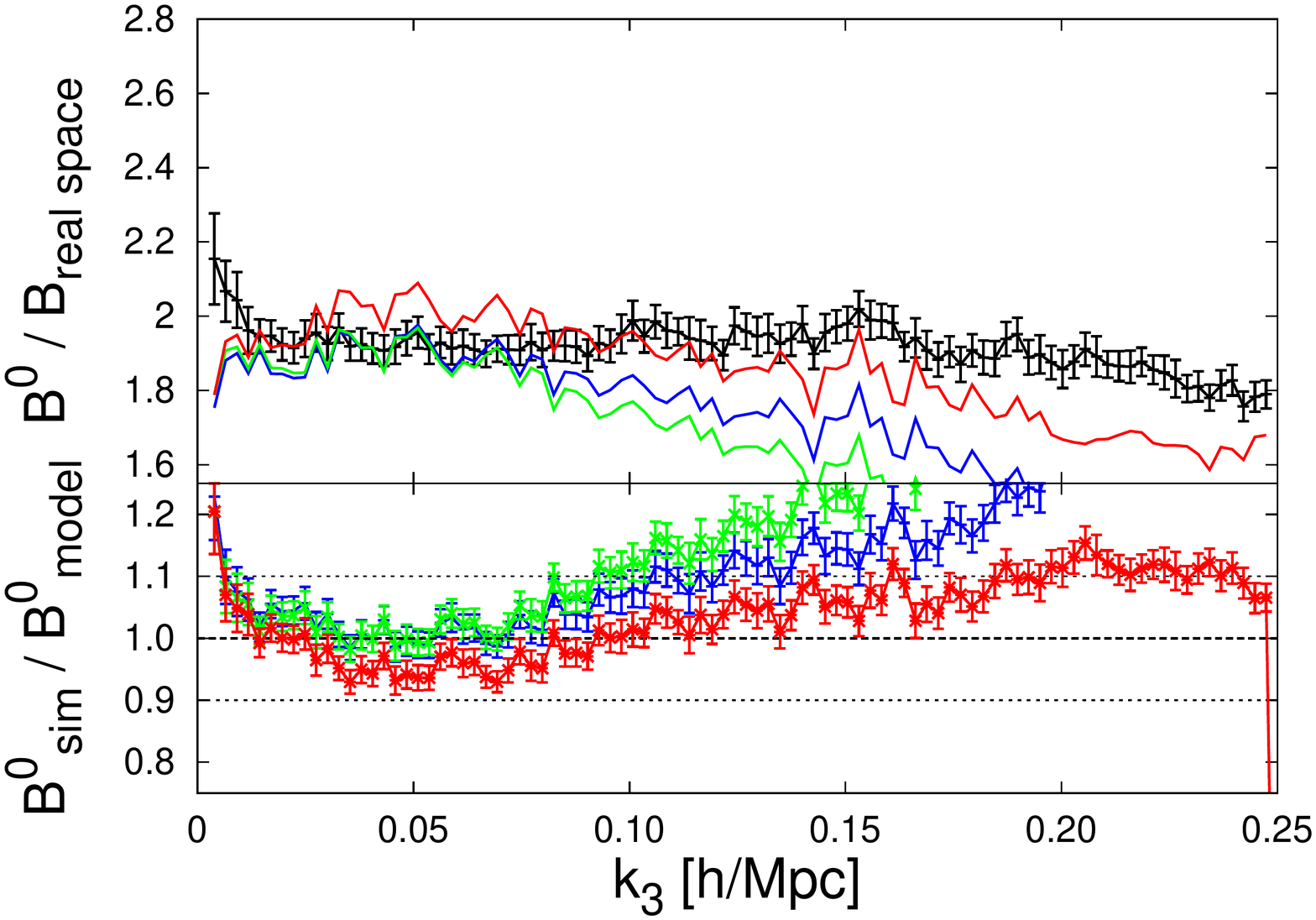}
\includegraphics[clip=false,trim= 25mm 10mm 22mm 35mm, scale=0.25]{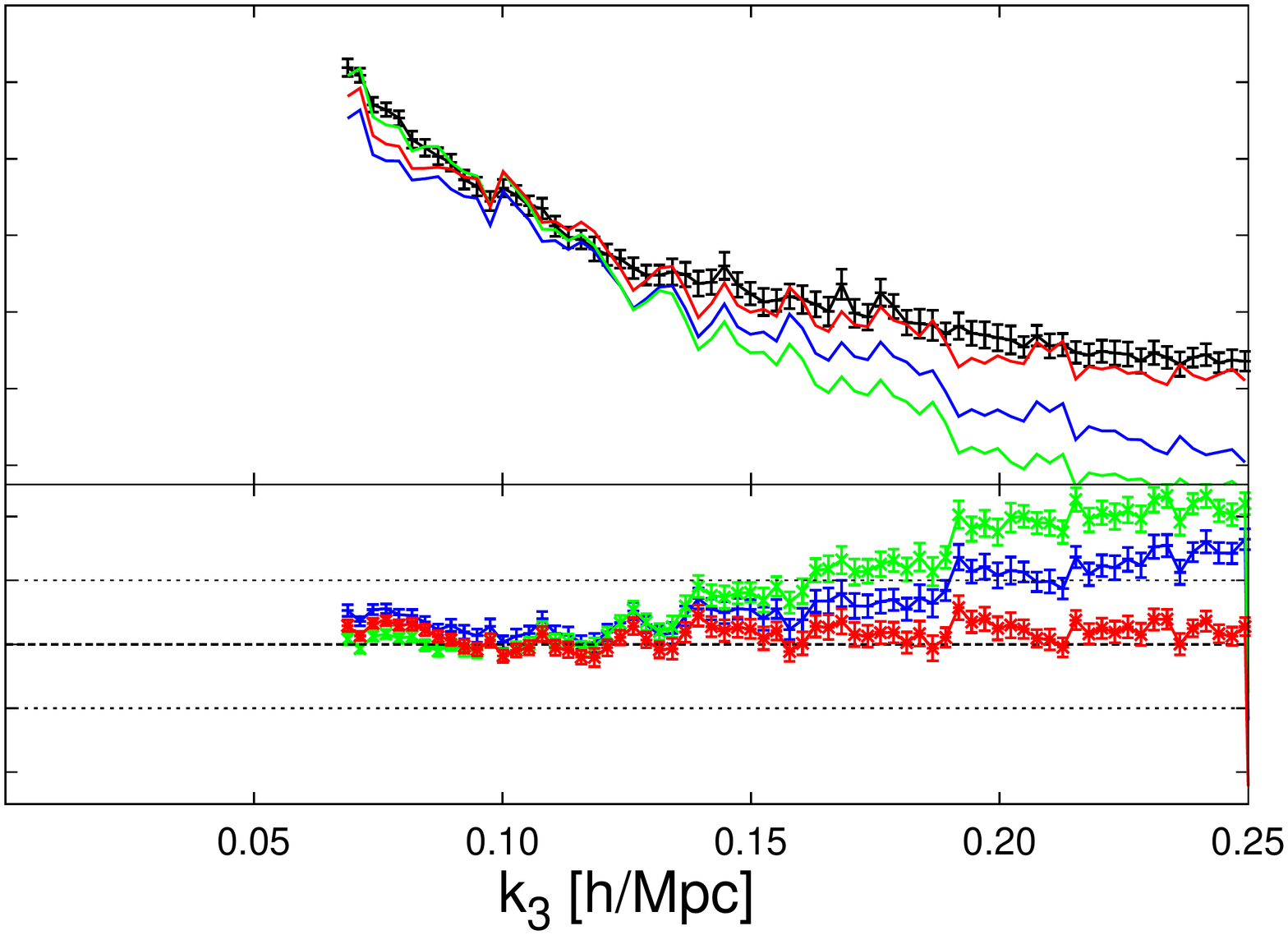}
\includegraphics[clip=false,trim= 25mm 10mm 80mm 35mm, scale=0.25]{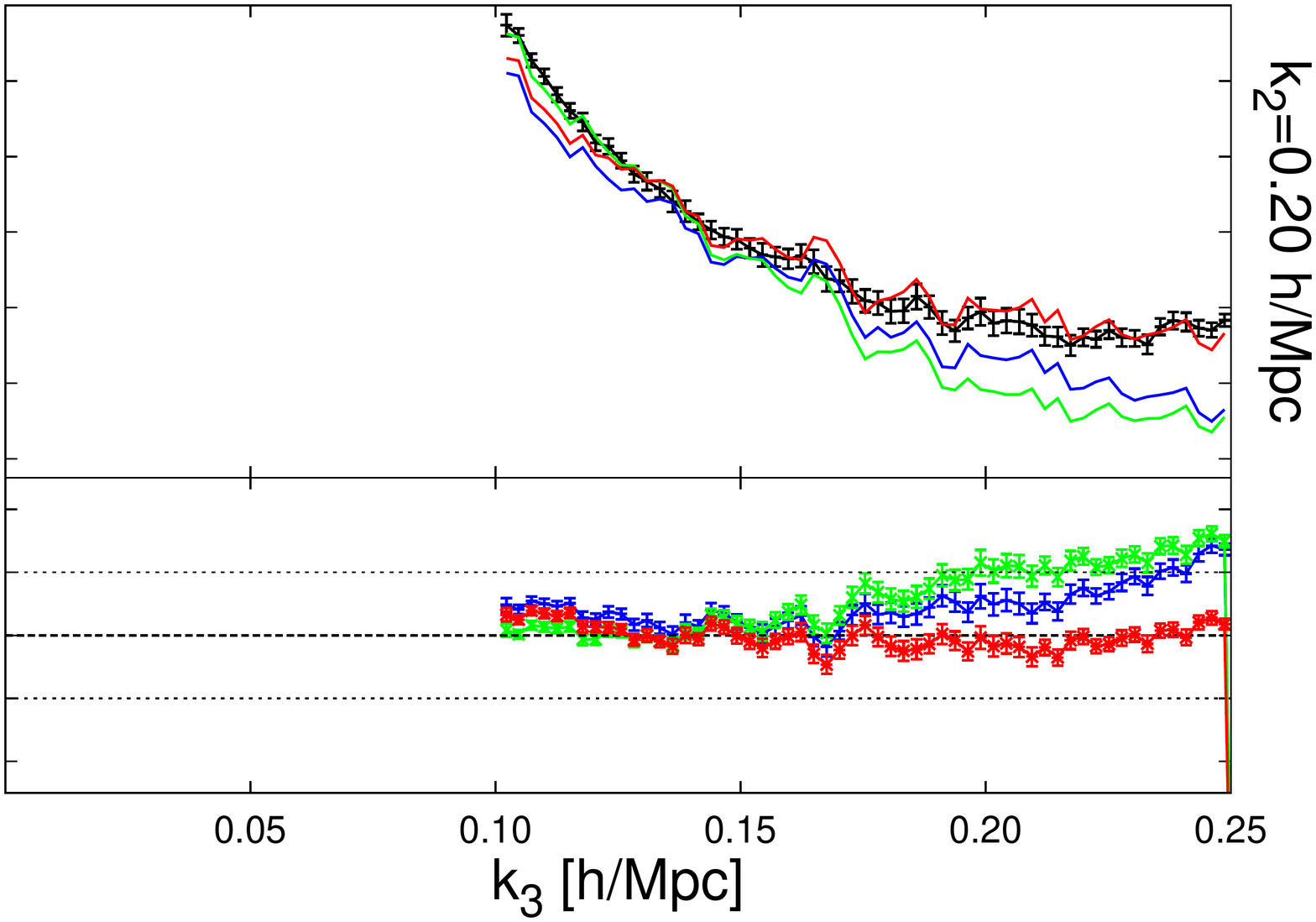}

\caption{Same notation that in Fig.~\ref{bis1}. All panels at $z=1$.}
\label{bis3}
\end{figure}

\begin{figure}
\centering
\includegraphics[clip=false, trim= 80mm 10mm 22mm 35mm,scale=0.25]{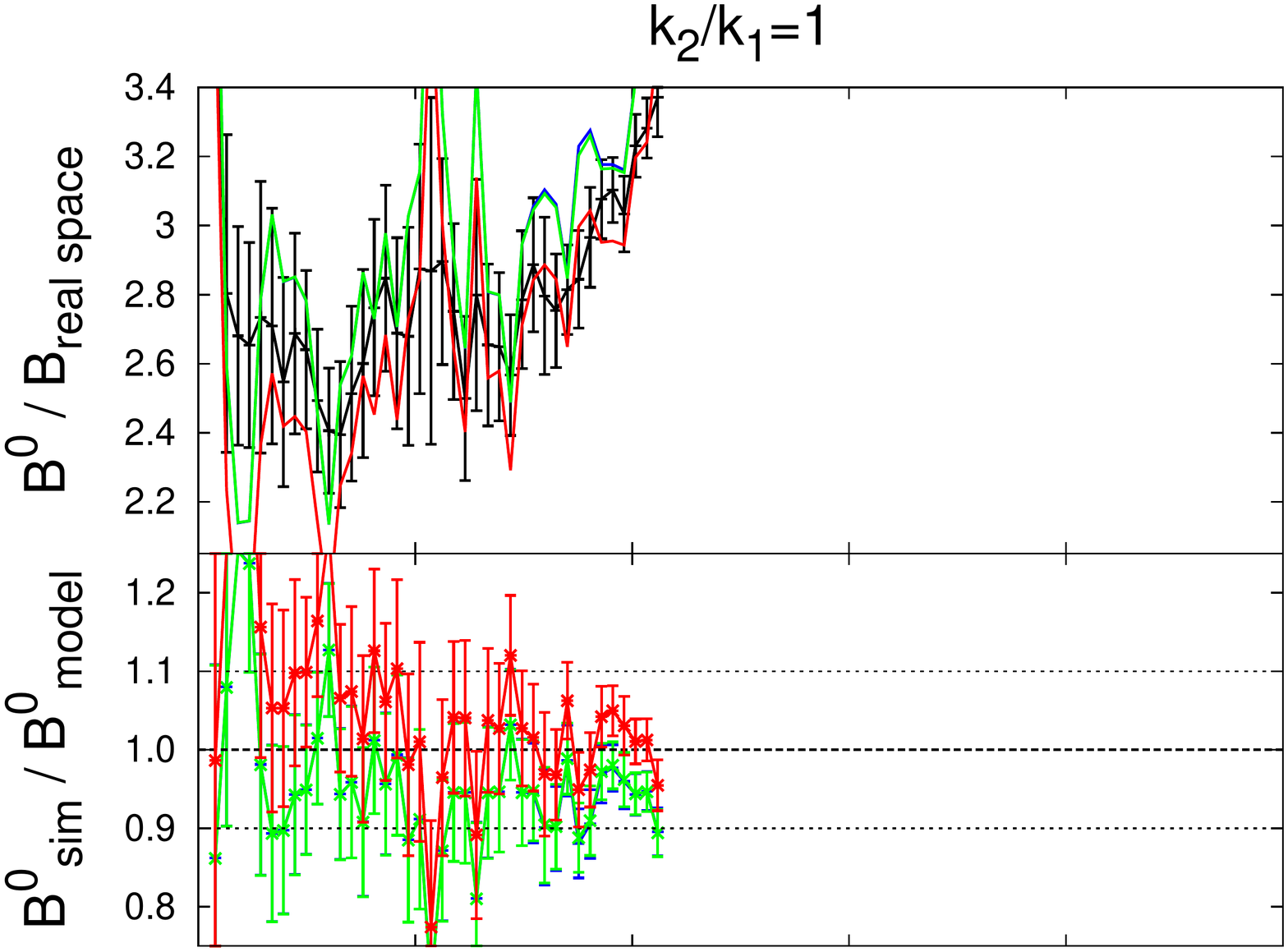}
\includegraphics[clip=false,trim= 25mm 10mm 22mm 35mm, scale=0.25]{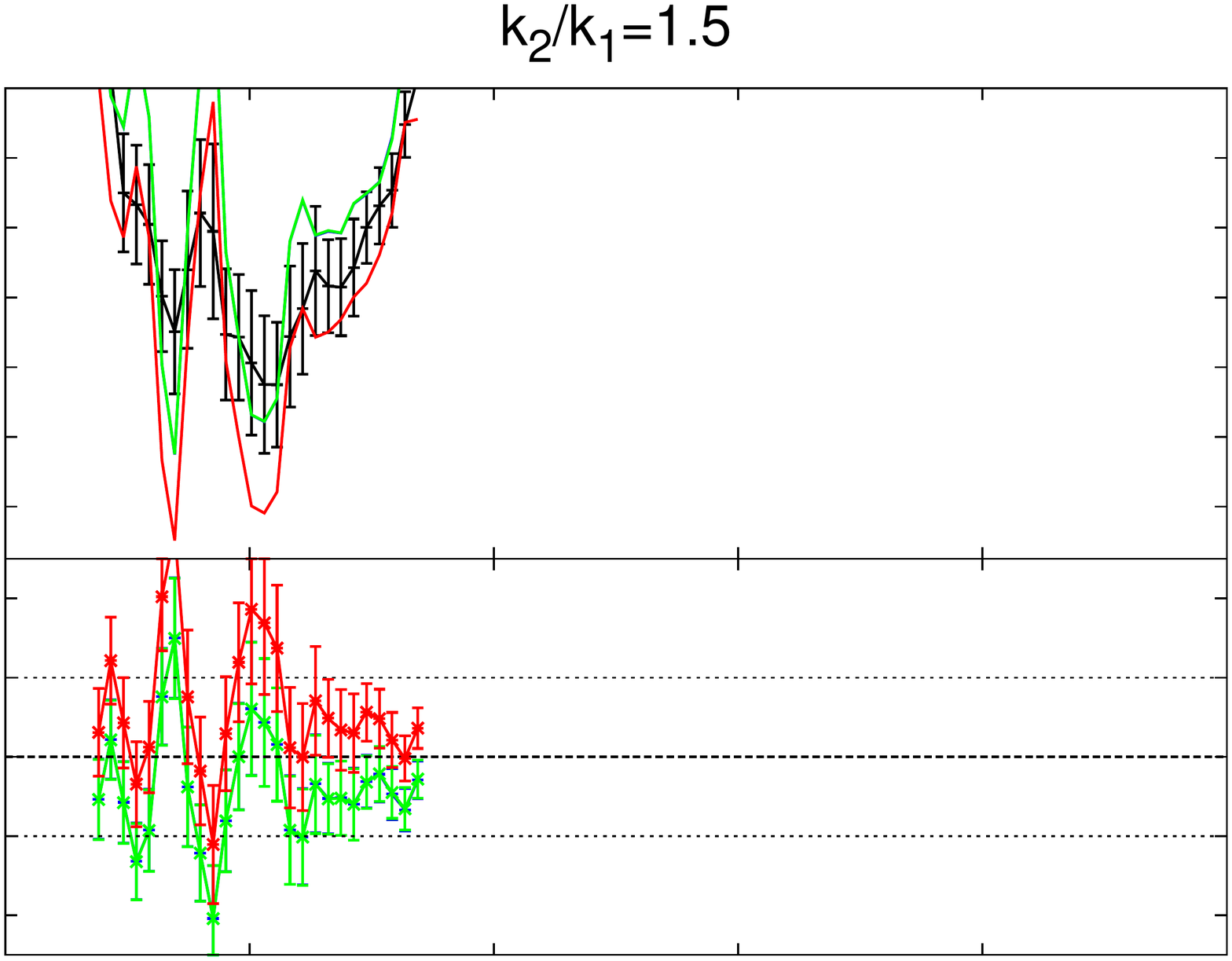}
\includegraphics[clip=false,trim= 25mm 10mm 80mm 35mm, scale=0.25]{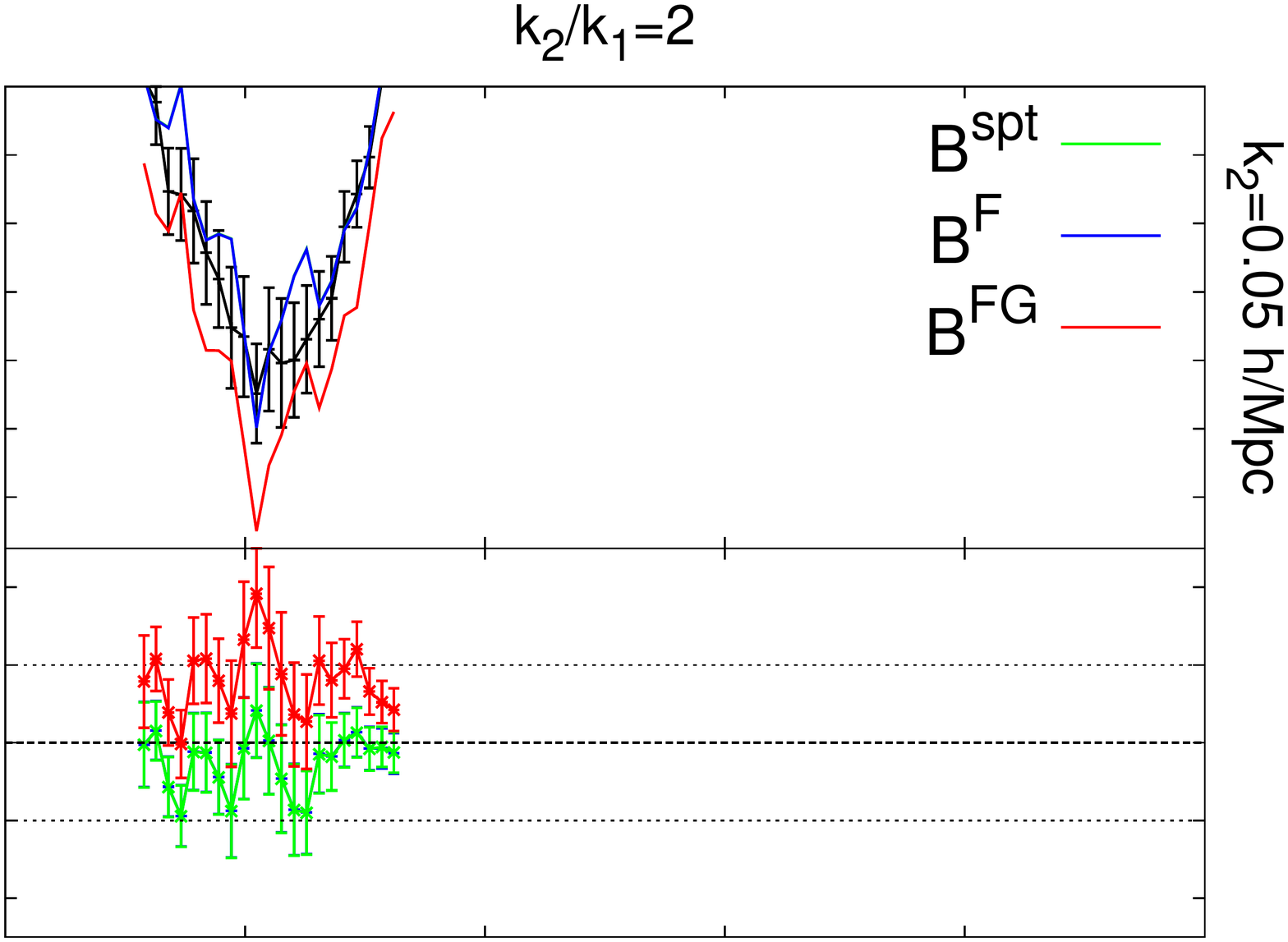}

\includegraphics[clip=false, trim= 80mm 10mm 22mm 35mm,scale=0.25]{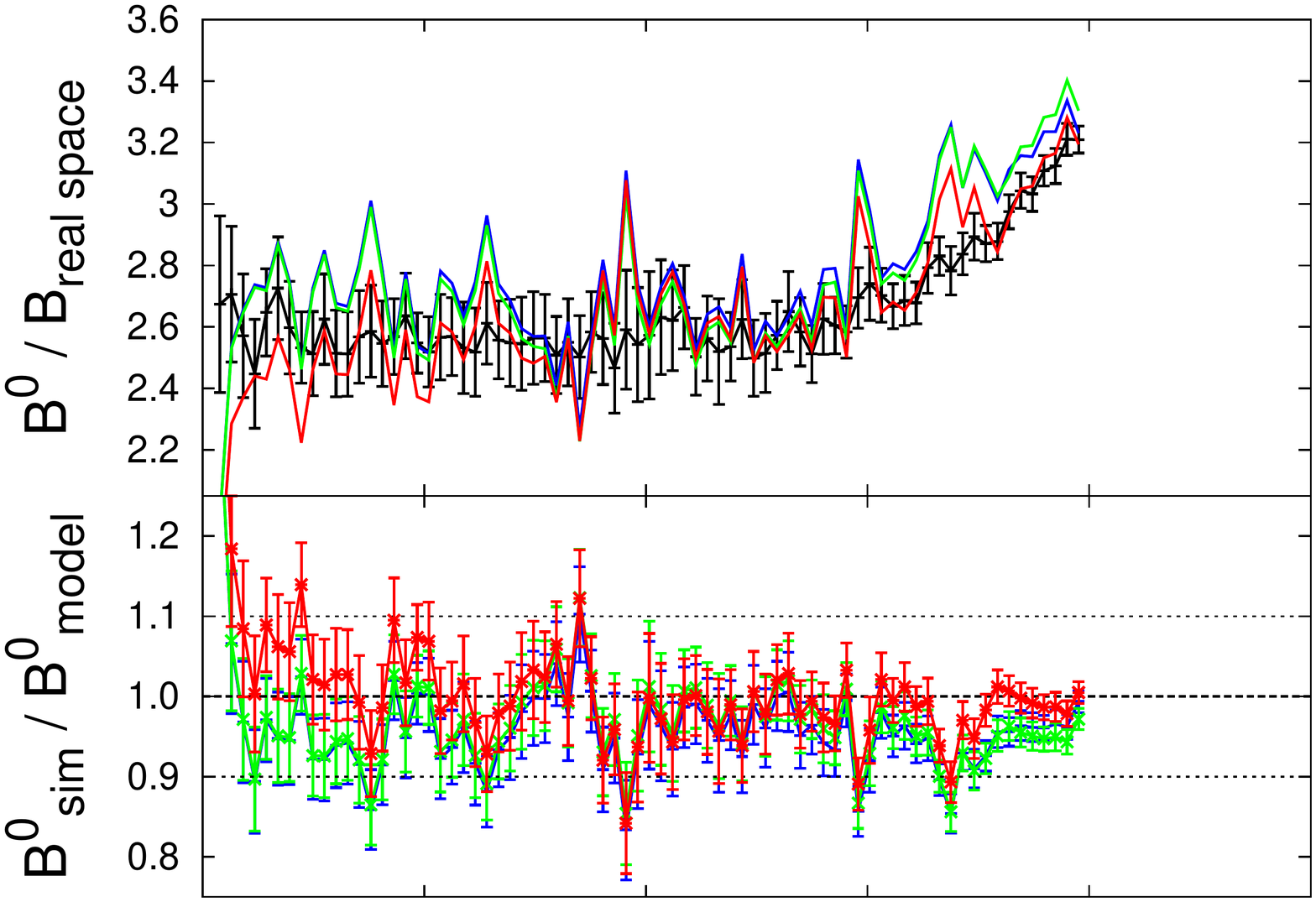}
\includegraphics[clip=false,trim= 25mm 10mm 22mm 35mm, scale=0.25]{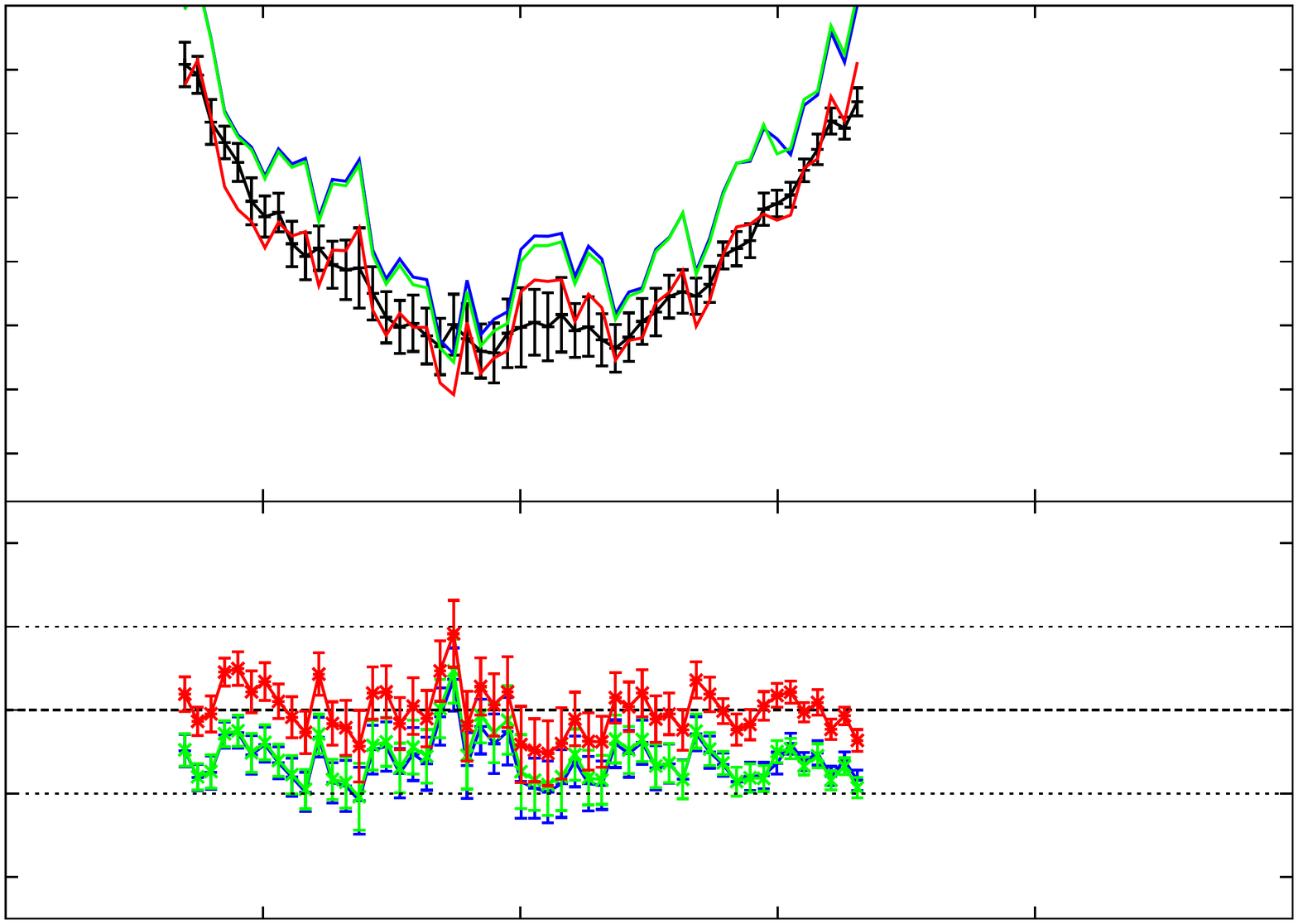}
\includegraphics[clip=false,trim= 25mm 10mm 80mm 35mm, scale=0.25]{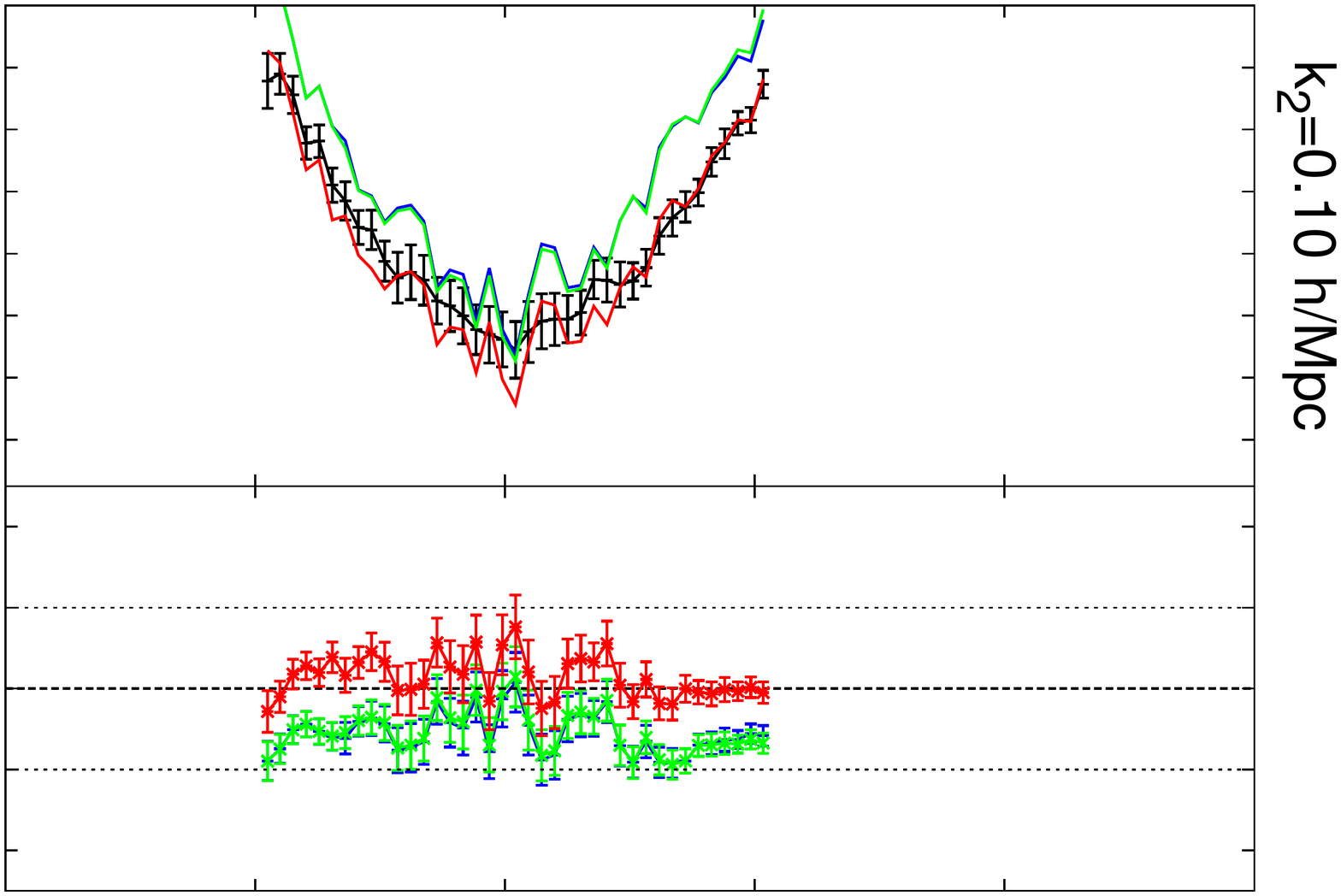}

\includegraphics[clip=false, trim= 80mm 10mm 22mm 35mm,scale=0.25]{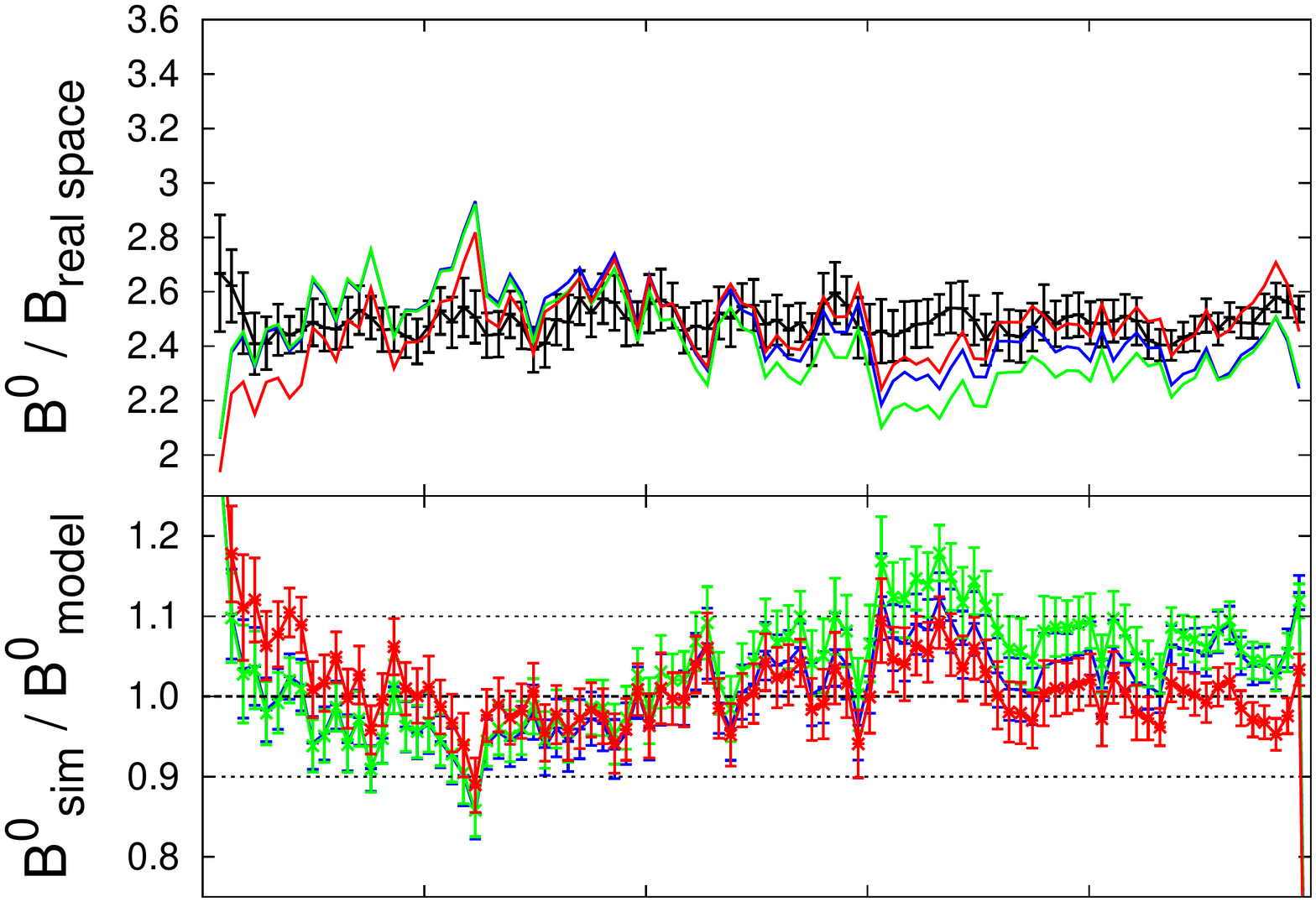}
\includegraphics[clip=false,trim= 25mm 10mm 22mm 35mm, scale=0.25]{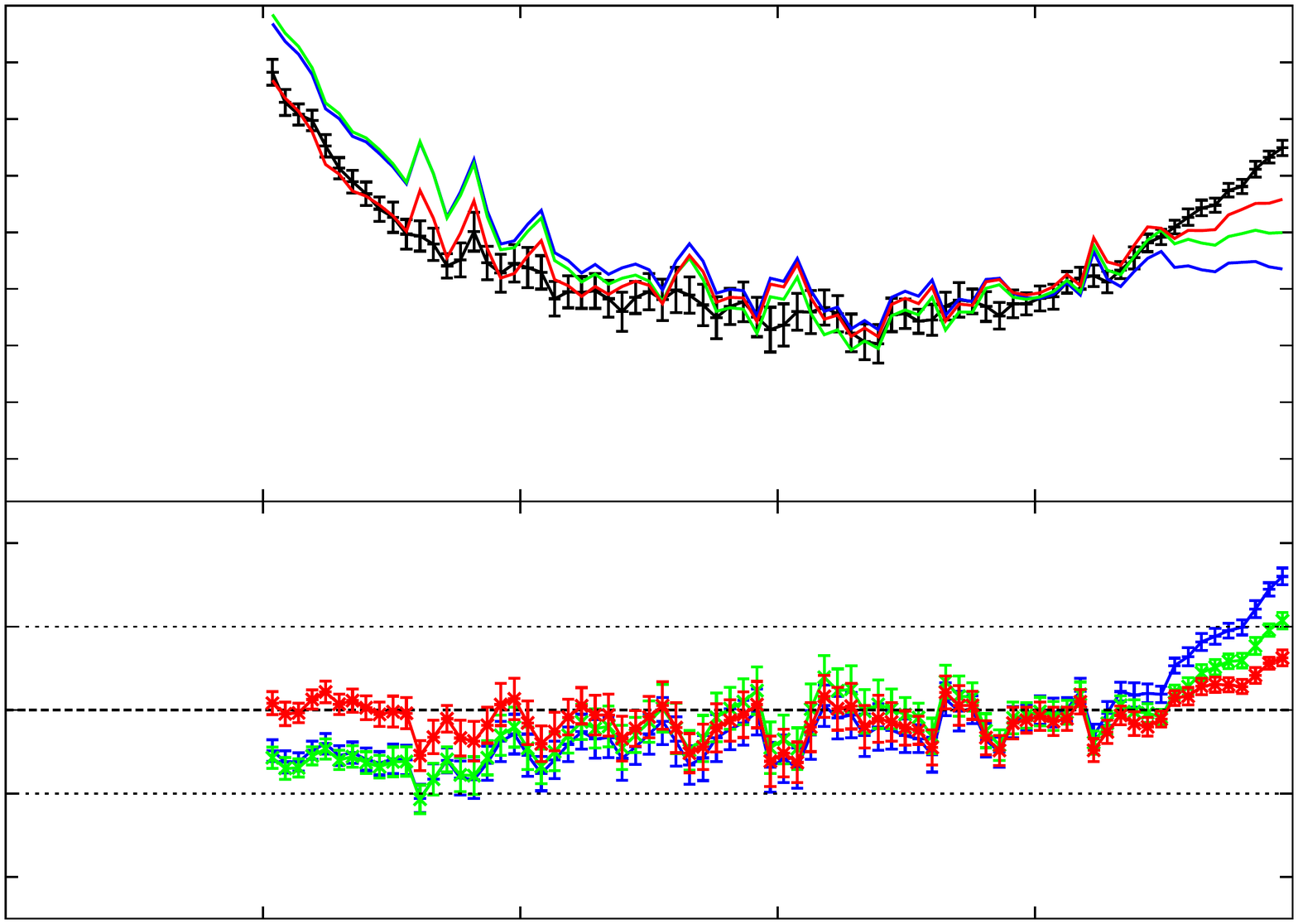}
\includegraphics[clip=false,trim= 25mm 10mm 80mm 35mm, scale=0.25]{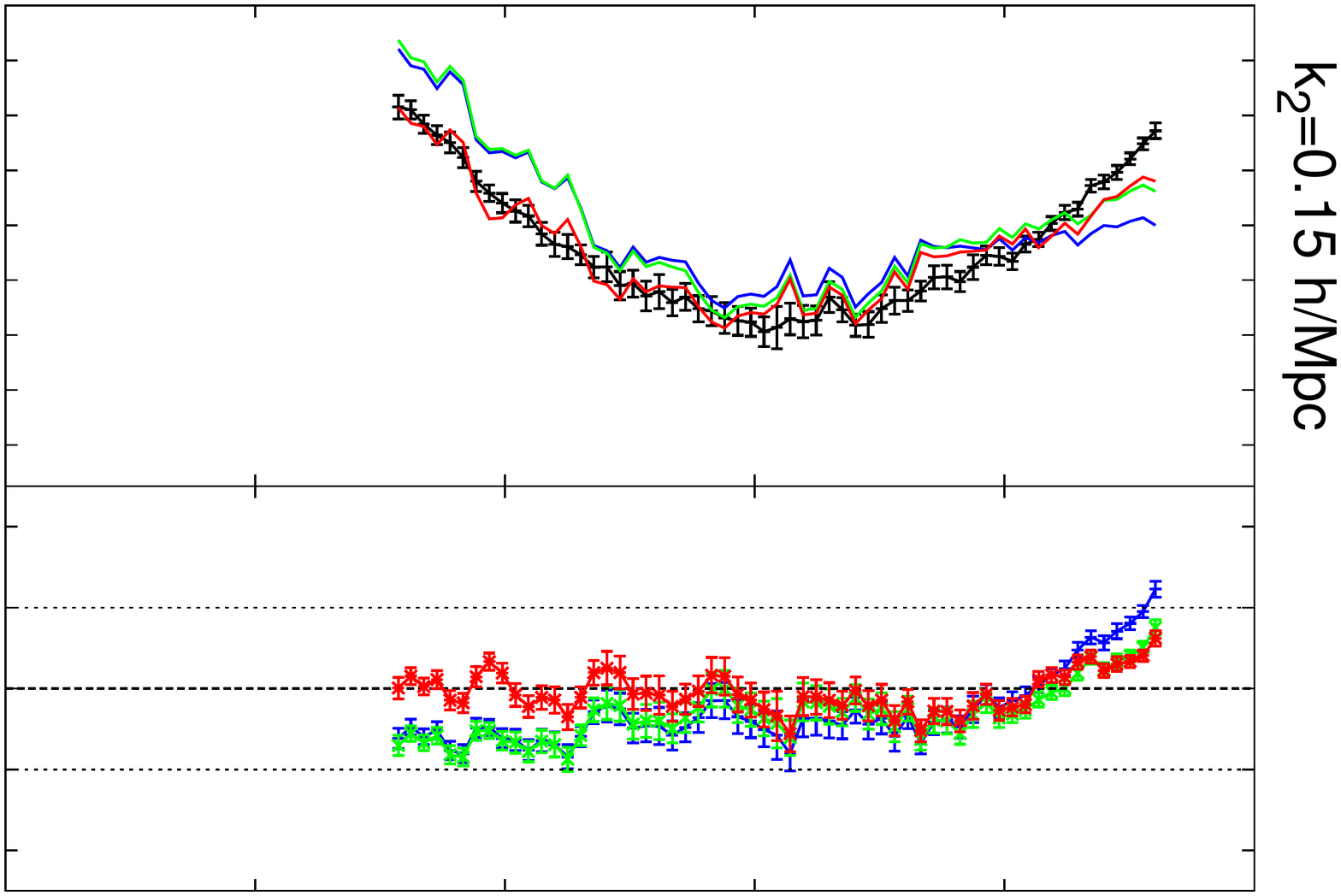}

\includegraphics[clip=false, trim= 80mm 10mm 22mm 35mm,scale=0.25]{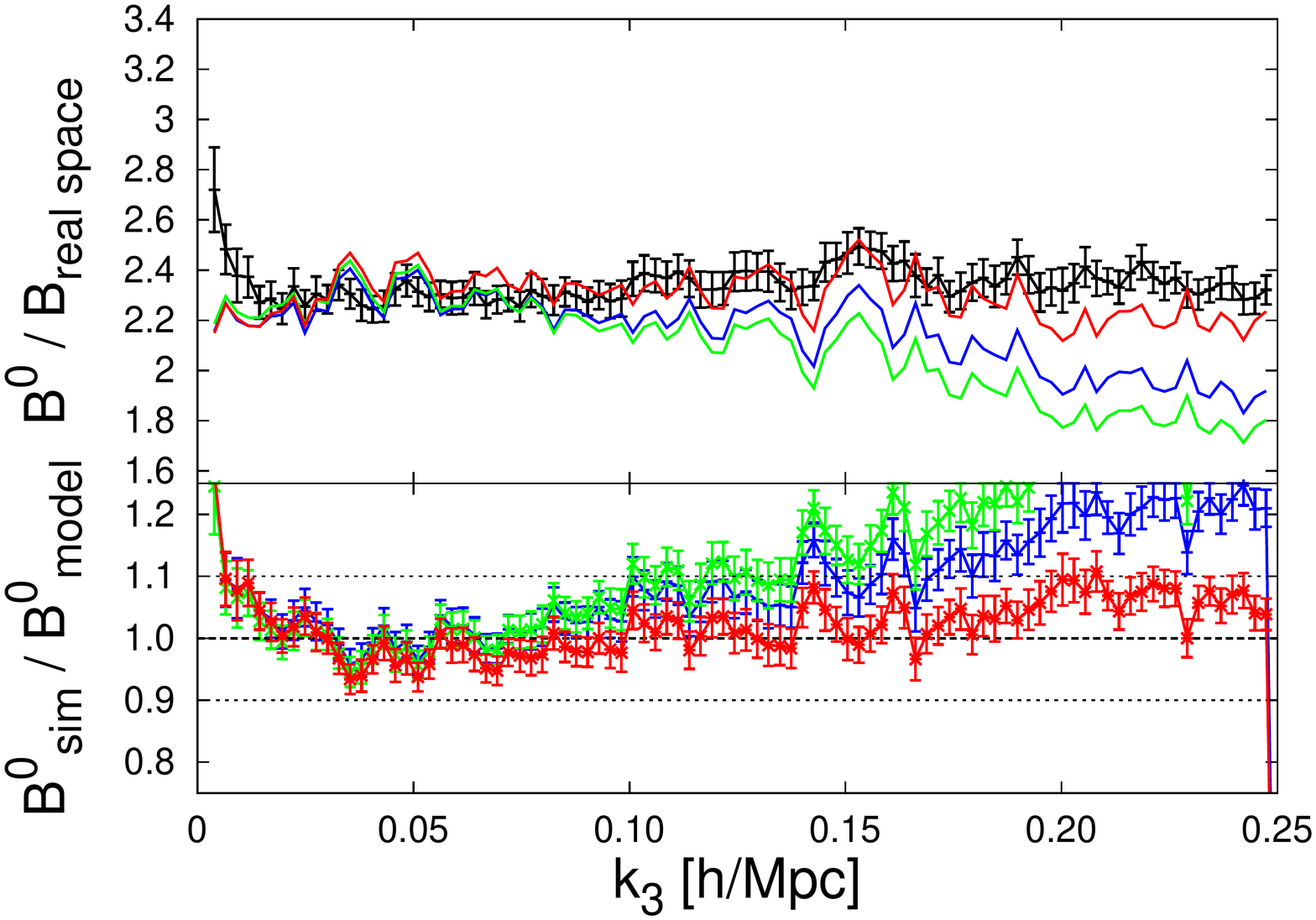}
\includegraphics[clip=false,trim= 25mm 10mm 22mm 35mm, scale=0.25]{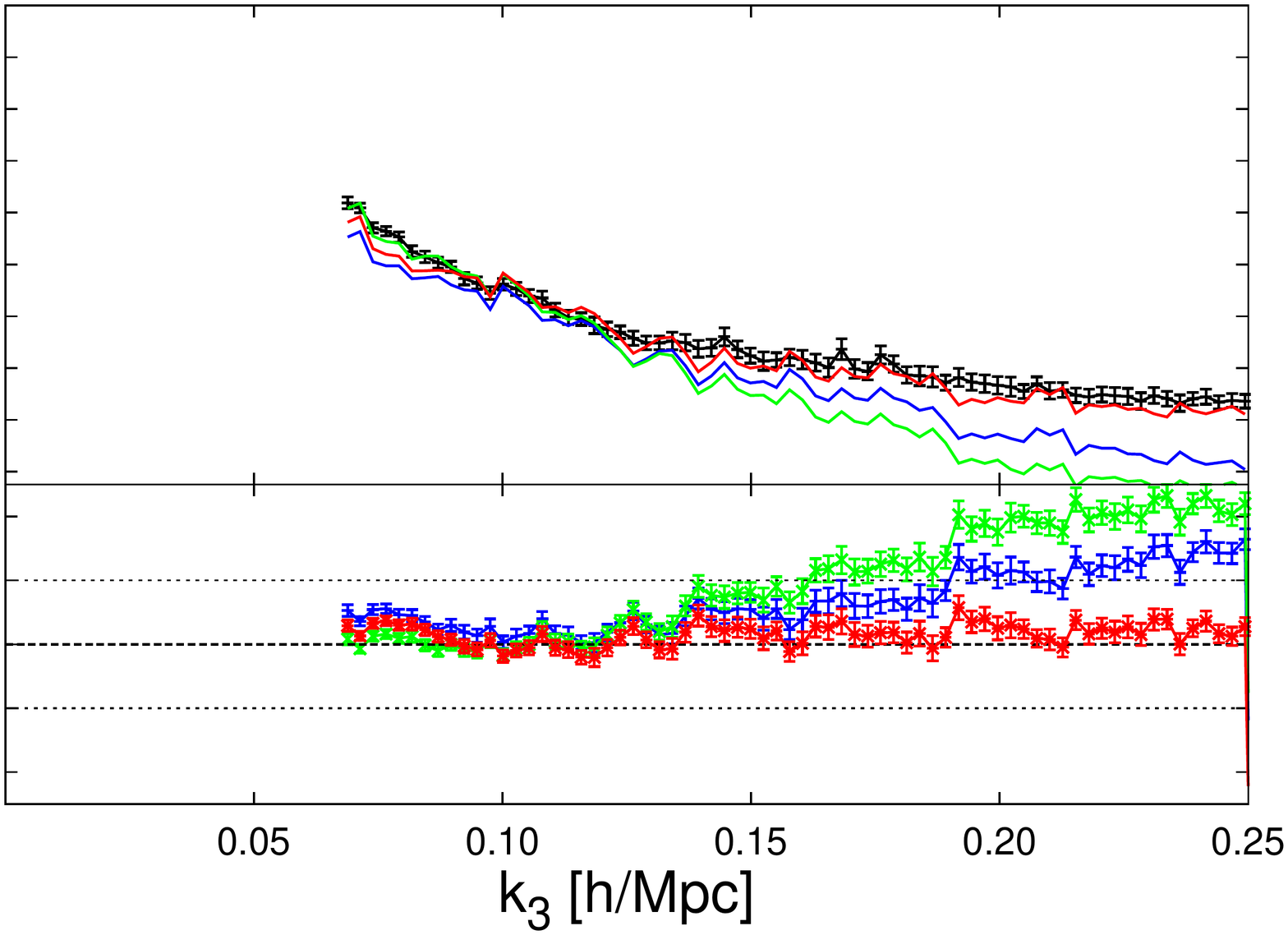}
\includegraphics[clip=false,trim= 25mm 10mm 80mm 35mm, scale=0.25]{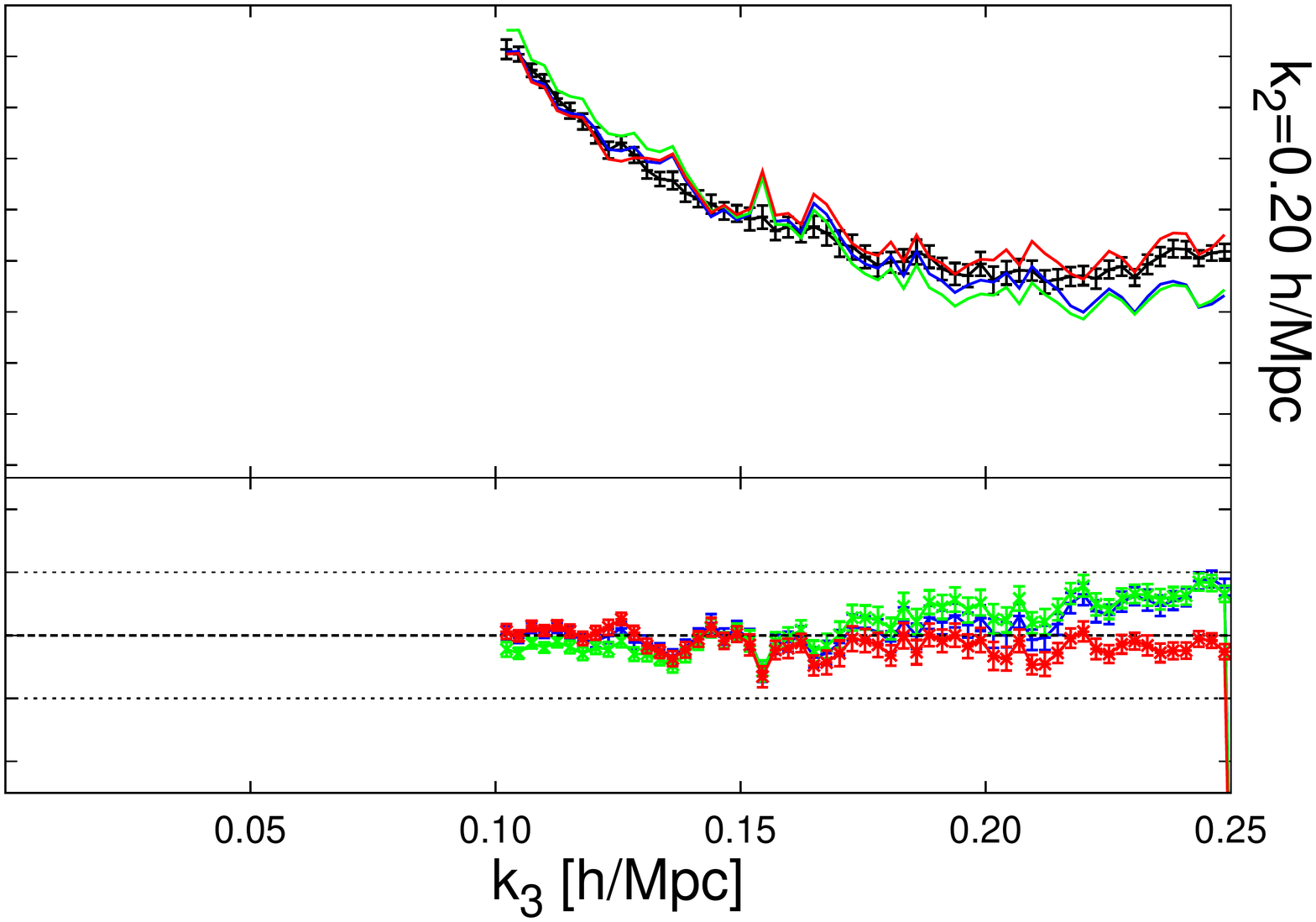}

\caption{Same notation that in Fig.~\ref{bis1}. All panels at $z=1.5$.}
\label{bis4}
\end{figure}

In Figs.~\ref{bis1}-\ref{bis4} we compare the three approaches, $B^{\rm spt}$ (green lines), $B^{\rm F}$ (blue lines), $B^{\rm FG}$  (red lines), with the N-body dark matter monopole bispectrum  (black symbols) for $z=0,\, 0.5,\,1.0,\,1.5$ respectively. The error-bars correspond to the error on the mean of 60 realisations, with a total volume of $V\simeq829\,[{\rm Gpc}/h]^3$. In the top sub-panels the redshift-space bispectrum monopole is shown. For visualisation reasons this quantity  has been normalised by the measurement of the bispectrum in real space. In the lower sub-panel we plot ratio between the measured bispectrum and the model. The oscillatory features in some of the panels are due to the high correlations between similar triangles.

The accuracy of the different models depends noticeably on the $k$-range, redshift and triangular shapes we are considering. As a general trend, we observe that at large scales and high redshifts  the three models studied here do not show large differences and describe well the N-body measurements. This makes sense, since in the large scale limit, $F_2^{\rm eff}\rightarrow F_2$, and $G_2^{\rm eff}\rightarrow G_2$. On the other hand, at low redshift and small scales the three models present different predictions. Typically, $B^{FG}$ best describes the N-body data, followed first by $B^F$ and finally by $B^{\rm spt}$. This is the expected behaviour, given the number of free parameters and complexity of each model. 

For $B^{FG}$, at $z=0$ we see that the differences between the model and the N-body predictions are $\leq10\%$ when $k_i\leq0.15\,h/{\rm Mpc}$, and $\lesssim5\%$ when $k_i\leq0.10\,h/{\rm Mpc}$. 
For $z=0.5$, the agreement between $B^{FG}$ and N-body simulations is $\leq10\%$ when $k_i\leq0.20\,h/{\rm Mpc}$ and $\lesssim5\%$ when $k_i\leq0.15\,h/{\rm Mpc}$. For $z=1.0$  we observe that for the whole range studied here, $k\leq0.25\,h/{\rm Mpc}$, we always have an accuracy of $\leq10\%$, whereas when we restrict it to $k_i\leq0.17\,h/{\rm Mpc}$, the accuracy increases to  $\lesssim5\%$. Finally for $z=1.5$ we observe that the accuracy is  $\lesssim5\%$ for $k\leq0.20\,h/{\rm Mpc}$.  Note that since the different bispectrum shapes of each panel are very correlated a ``$\chi^2$-by-eye" is not suitable in order to test the goodness of the fit. However, here we are trying to identify the range of validity of the model, knowing that the model will work at large scales and will break down at some smaller scale. 
This will just be used for guidance,  a more rigorous and quantitative analysis will be presented in \S~\ref{applications_section}. 
In particular in Fig. \ref{sigma8_f_kmax} we will recover the bias parameters as a function of $k_{\rm max}$  using the $B^{FG}$ model and it will be shown that the results are unbiased respect to the true values, confirming the validity of the model in the proposed range.

 We also observe that for the squeezed triangles, those with $k_1\sim k_2$ and $k_3\leq0.02\,h{\rm Mpc}^{-1}$, all the theoretical models under-predict the measured bispectrum at all $z$, especially when $k_1$ and $k_2$ are $\geq0.10\,h{\rm Mpc}^{-1}$. 
However, this mismatch between theory and simulations does not have an important impact in the the best-fitting values of ${\bf a}^F$ and ${\bf a}^G$ due to the low signal-to-noise contained in these shapes.

\begin{table}
\begin{center}
\begin{tabular}{|c|c|c|c|c|}
\hline
 $z$ & 0 & 0.5 & 1.0 & 1.5 \\
 \hline
 \hline
$\sigma_0^B (B^{\rm spt})$ & 43.  & 33.7 & 25.0 & 20.3\\
\hline
$\sigma_0^B (\,B^{F}\,)$ & 51.93 & 38.5 & 28.28 & 23.56 \\
  \hline
  $\sigma_0^B (B^{FG})$ & 41.67 & 26.9 & 17.4 & 11.67\\
\hline
$\sigma_0^P$ & 7.8 & 7.0 & 6.2 & 5.6\\
\hline
\end{tabular}
\end{center}
\caption{Best-fitting values for $\sigma_0^B(z)$ (in Mpc/$h$) for different bispectrum models for dark matter (see text for details). These values correspond to the different models shown in Figs.~\ref{bis1} -~\ref{bis4}. In the last row for comparison we also show the values for $\sigma_{\rm FoG}^P$ from the monopole-to-quadrupole ratio.}
\label{sigmab_table}
\end{table}

In Table~\ref{sigmab_table} we report the best-fitting values of $\sigma_0^B$, for the different models used, and for the different redshifts.
We note that, for each redshift, the value of $\sigma_0^B$ depends strongly on the model. In particular, $B^{FG}$ requires a smaller $\sigma_0^B$ for describing the bispectrum than the other two models. A possible explanation for this behaviour, is that  for $B^{\rm spt}$ and $B^{F}$ this parameter absorbs higher levels of systematic imperfections of the modelling than  for $B^{FG}$. 
We assume that when we use $B^{FG}$ to model the halo bispectrum, setting $D_{\rm FoG}^B$ to 1 will produce a good estimate for the halo bispectrum, since $D_{\rm FoG}^B$ should correct only for the FoG and not for any other systematic effects. We study this in detail in \S~\ref{haloes_section}. We also note that the ratio between $\sigma_0^P$ and $\sigma_0^B$ for any of the models is not constant as a function of $z$.

In this section, the models for ${\bf a}^F$ and ${\bf a}^G$ are shown to describe well the redshift space dark matter bispectrum for a unique set of cosmological parameters (the one listed as Sim DM in Table \ref{table_sims}). For any different set of cosmological parameters, especially for $\Omega_m$ that regulates the  distortions in the redshift space through $f\approx \Omega_m^{0.55}$, and for $\sigma_8$ that regulates  the amplitude of the clustering,  the best-fitting values for  ${\bf a}^F$ and ${\bf a}^G$ might be different, which would limit the applicability of the fitting formula. However, since ${\bf a}^F$ and ${\bf a}^G$ have been fitted simultaneously to different epochs within the range $0\leq z \leq 1.5$, they already contain information of different fluctuation amplitudes, i.e. on different values of $\sigma_8$. The same thing occurs with $f$: since $f$ changes with $z$, ${\bf a}^F$ and ${\bf a}^G$ were constrained using bispectra with different values of $f$. However, the variation that redshift evolution induces on the parameters $f$ and $\sigma_8$ is not independent, since both parameters vary at the same time. In order to isolate the effect of varying $f$ from the effect of varying $\sigma_8$,  we would need to change $\Omega_m$  while keeping the amplitude of the power spectrum in real space constant. We do so in Appendix \ref{cosmo_appendix}, where we show how the fitting formula (with the same values for the ${\bf a}^F$ and ${\bf a}^G$  parameters) is able to describe the dark matter bispectrum of two extra cosmologies: a low-$\Omega_m$ cosmology where $\Omega_m$ has been lowered to 0.2 and a high-$\Omega_m$ cosmology where $\Omega_m$ is 0.4. In addition to the Sim DM cosmology (or fiducial cosmology) used in this section, these two extra cosmologies provide a test for the logarithmic growth rate that covers the range $0.4\lesssim f(z=0) \lesssim 0.6$. 
From Fig. \ref{bis5} in Appendix \ref{cosmo_appendix}, we see that the fitting formula is able to describe very well the redshift space bispectrum of the low- and high-$\Omega_m$ cosmologies.  On the scales on which the fitting formula described well the measurements for the fiducial cosmology the agreement between predictions and measurements are always below 5\% for all cosmologies considered.   Therefore, we conclude that the  fitting formula we present here is accurate enough in describing the bispectrum over the range of interest for the relevant cosmological parameters.

 In the next section, we apply model $B^{FG}$ with the fitted ${\bf a}^G$ parameters to describe the monopole bispectrum of haloes with a similar cosmology to the one used in this section.

\section{Extension to biased tracers}\label{haloes_section}

In this section we aim to show how the $B^{FG}$ model (with the values of the ${\bf a}^G$ parameters extracted in \S  \ref{bis_section}) can be used to describe the bispectrum of N-body haloes. These N-body haloes have  the Sim HC $\Lambda$CDM cosmology of Table \ref{table_sims}, which is very similar to the cosmology used to extract the ${\bf a}^G$ values. The main purpose of this section is to show that the ${\bf a}^G$ set found in \S\ref{bis_section} does not depend on the FoG feature, and that $B^{FG}$ is suitable to be applied to  dark matter tracers and therefore suitable to be applied to galaxy surveys. We also compare $B^{FG}$ with the predictions of model $B^{\rm spt}$ to see the improvement.

In order to describe the halo biasing, we use the non-local and non-linear bias model presented in Eq.~\ref{delta_galaxy_k}.
Since we are dealing with haloes, we cannot ignore the contribution of shot noise. Due to halo exclusion and clustering we expect some deviations from the Poisson noise prediction. In order to account for that, we use the prescription described in \S~\ref{sec:shot_noise}.

We start by determining  the bias parameters and $A_{\rm noise}$ from the real space power spectrum and bispectrum assuming fixed true values for $f$, $D(z)$, $\sigma_8$ and the shape of the linear power spectrum. In order to do so we apply the methodology described in \S~\ref{sec:estimation}, using 20 realisations of N-body haloes. We use the model for real and redshift space power spectra described in \S~\ref{theory_section}. We refer to these models as 2LRPT for real space and TNS-2LRPT for redshift space power spectrum.

For $k_{\rm max}=0.15\,h/{\rm Mpc}$,  when combining  the  real space power spectrum and bispectrum, we find that $b_1=2.050\pm0.014$, $b_2=0.31\pm0.05$, $A_{\rm noise}=0.13\pm0.06$, where the error-bars correspond to the volume of one realisation, $V=3.375\,[{\rm Gpc}/h]^3$. We use these values as reference to test the accuracy of the description of the halo power spectrum and bispectrum in redshift space\footnote{We have checked that $b_1$ and $b_2$ do not change significantly with $k_{\rm max}$ as it can be seen in black dashed lines of the right panels of Fig. \ref{sigma8_f_kmax_halo}.}.

\begin{figure}
\centering
\includegraphics[clip=false, trim= 13mm 0mm 13mm 10mm,scale=0.30]{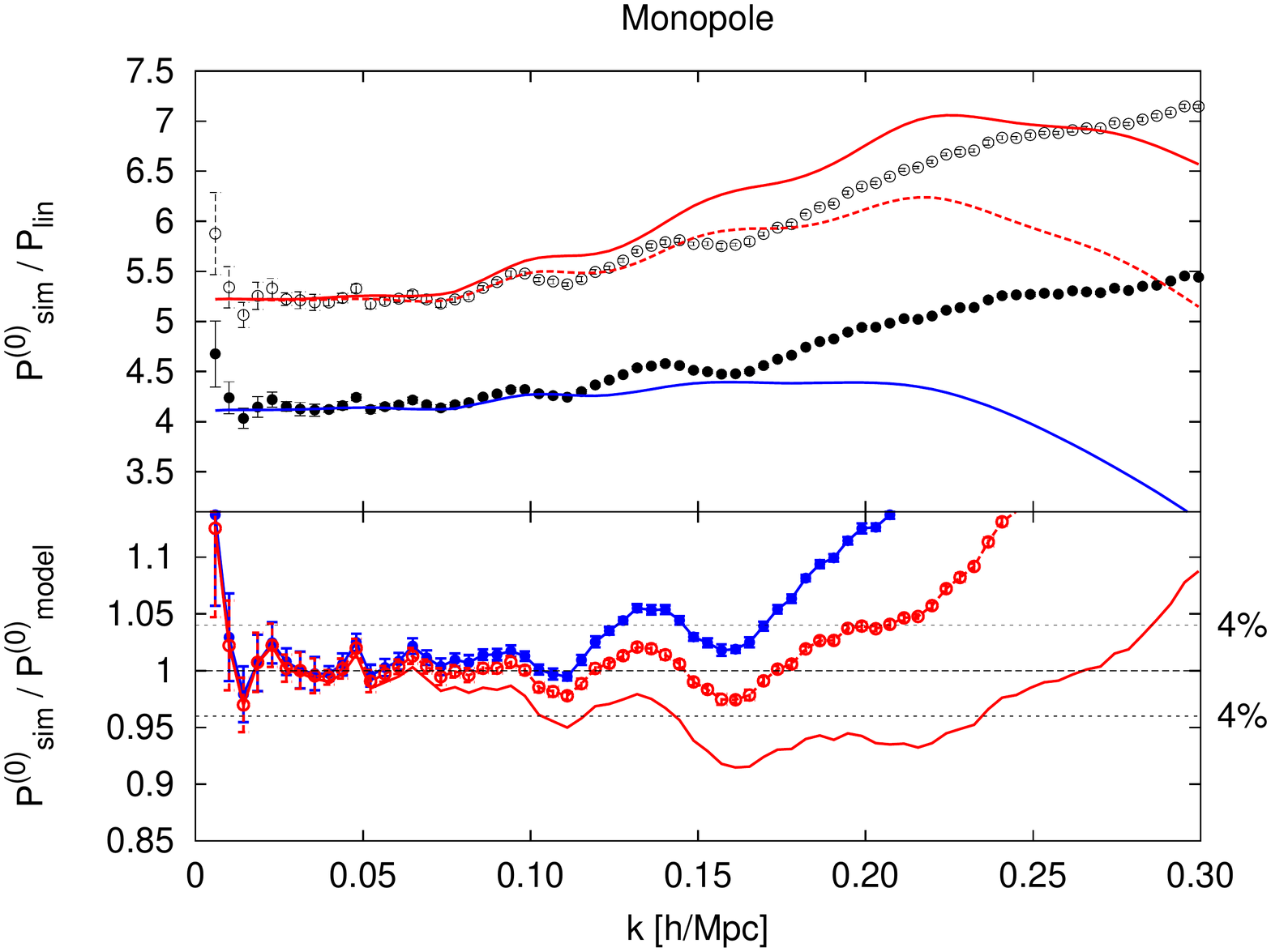}
\includegraphics[clip=false, trim= 13mm 0mm 13mm 10mm,scale=0.30]{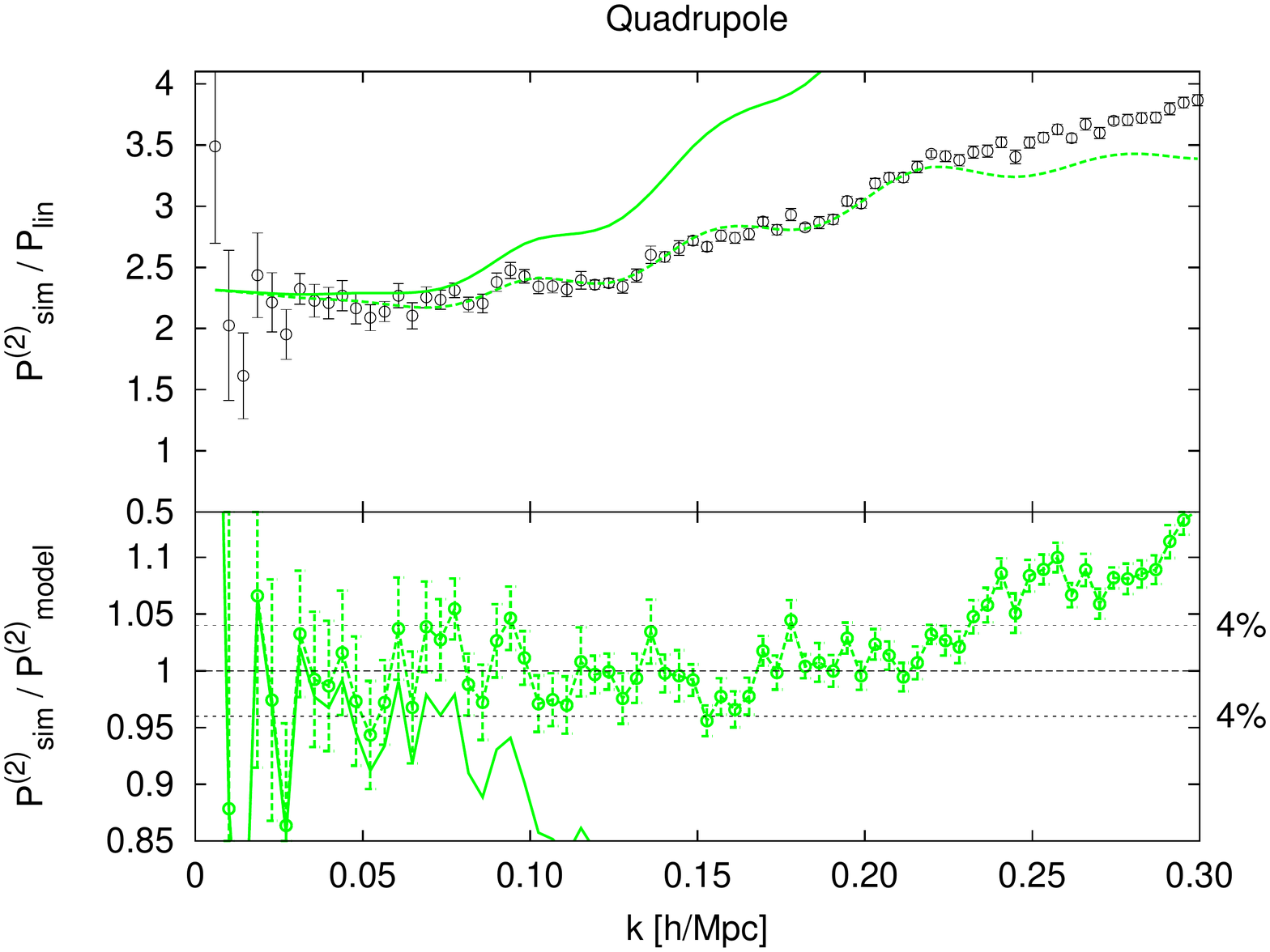}
\caption{Top left: Halo power spectrum in real space (filled black circles) and power spectrum monopole (empty black circles) normalised to the linear model. The real space power spectrum according to the 2LRPT model is shown as blue line. The red lines depict the TNS-2LRPT prediction for the redshift-space monopole.  Top right: redshift space halo power spectrum quadrupole (open circles) and TNS-2LRPT (green lines). In both cases solid line corresponds to  $\sigma_{\rm FoG}^P$ is set to 0 and dashed line when $\sigma_{\rm FoG}^P$ is set to 2.44\, Mpc/$h$.  In all the cases the bias parameters has been set to $b_1=2.05$, $b_2=0.31$ and the  noise  parameter $A_{\rm noise}=0.13$. Bottom panels: Relative deviation between each model from top sub-panel and the measurement from N-body simulations.}
\label{halo_ps}
\end{figure}

Fig.~\ref{halo_ps} presents a comparison between the measured power spectrum and the prediction of the model for $z=0.55$. The real space power spectrum is displayed  in the left panel (filled black circles). The redshift space power spectrum monopole and quadrupole are presented as empty black circles in the left and right panels respectively. These data are compared to the model for the real space power spectrum (blue line), monopole (red lines) and quadrupole (green lines). For the redshift space multipoles, the solid lines correspond to the assumption of  $\sigma_{\rm FoG}^P=0$, whereas the dashed lines have $\sigma_{\rm FoG}^P$ as free parameter. In this case, we find that $\sigma_{\rm FoG}^P=2.44\, {\rm Mpc}/h$  is the best-fitting value obtained from a joined fit to the power spectrum monopole and quadrupole,  using the technique described in \S\ref{sec:estimation}. Therefore, we  see the necessity of including a FoG-like damping term in the power spectrum even for describing the clustering of massive haloes. This feature was also reported by the authors of the model \cite{TNS_halo}.
When $\sigma_{\rm FoG}^P$ is treated as a free parameter, the TNS-2LRPT model is able to reproduce the halo power spectrum monopole and quadrupole with a $\sim4\%$ accuracy for   $k\lesssim0.22\,h/{\rm Mpc}$ at $z=0.55$. 

\begin{figure}
\centering
\includegraphics[scale=0.25]{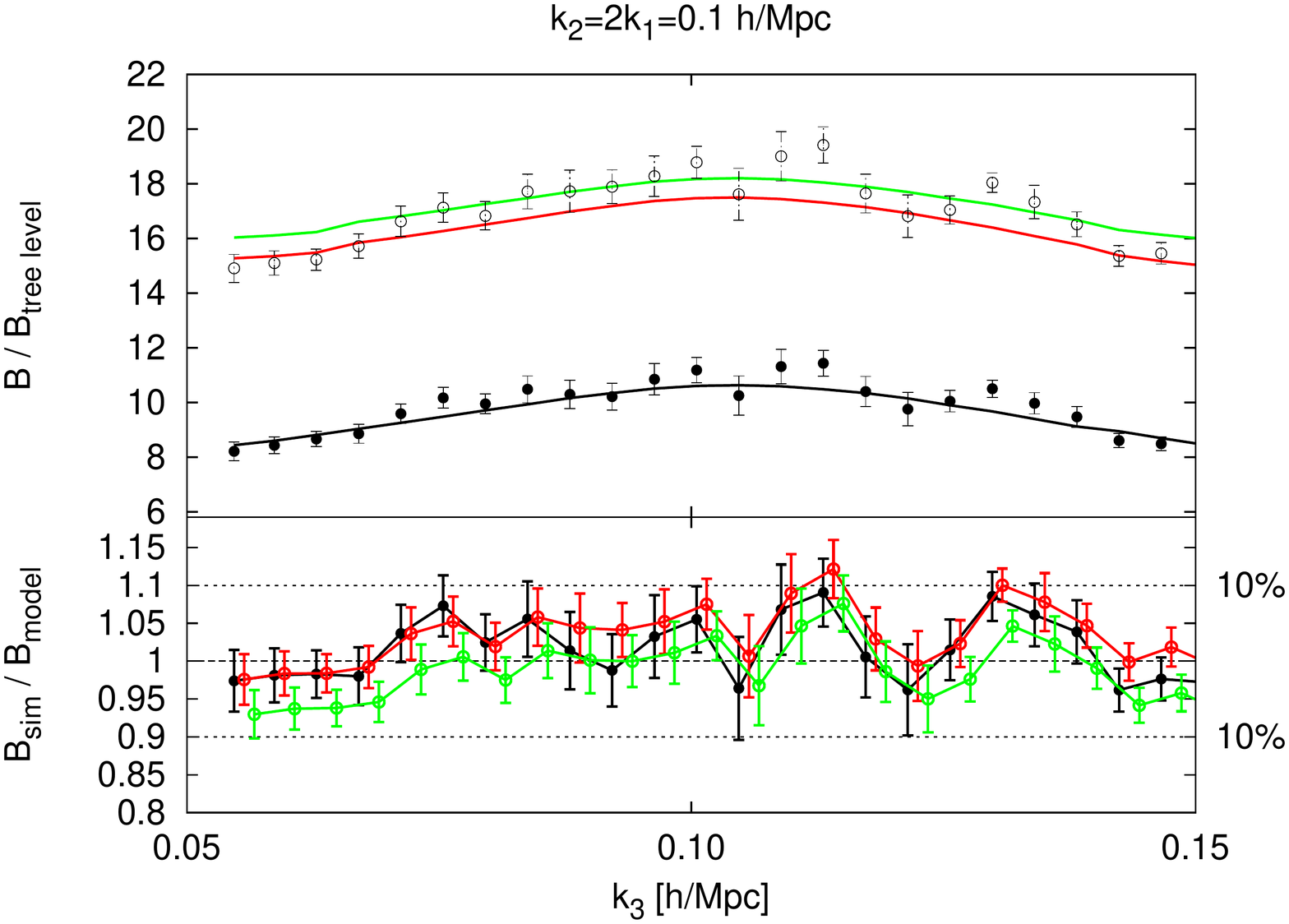}
\includegraphics[scale=0.25]{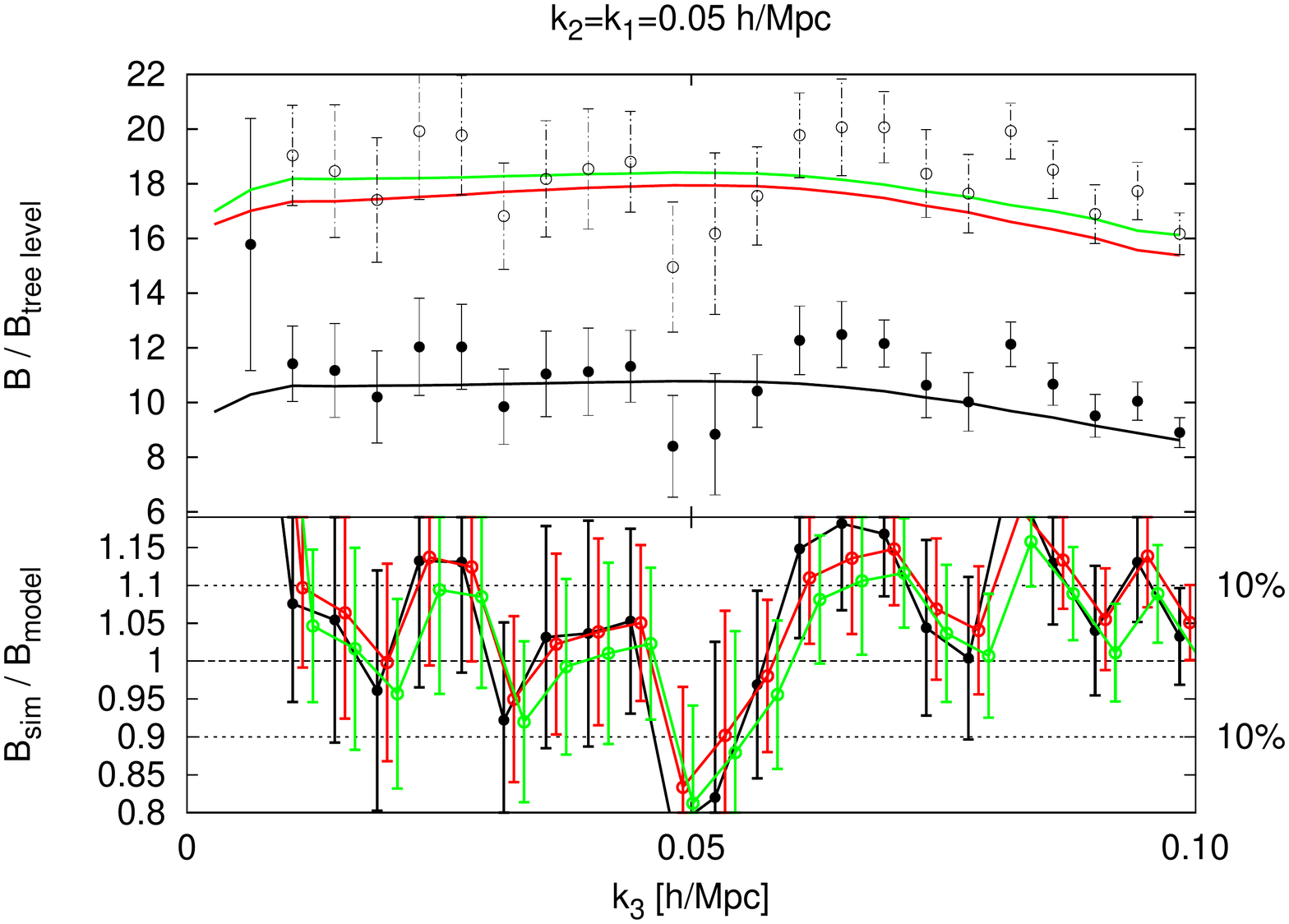}

\includegraphics[scale=0.25]{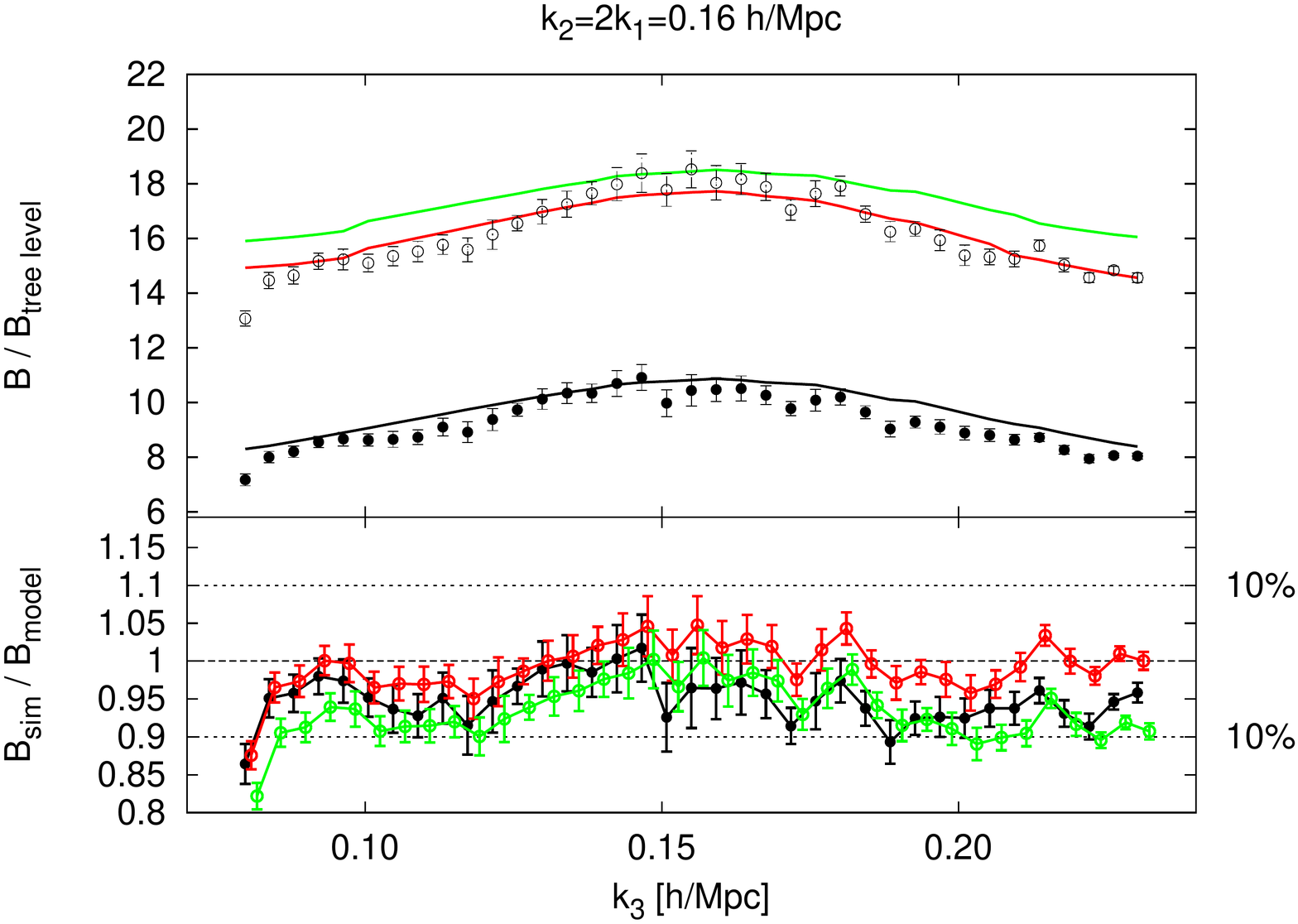}
\includegraphics[scale=0.25]{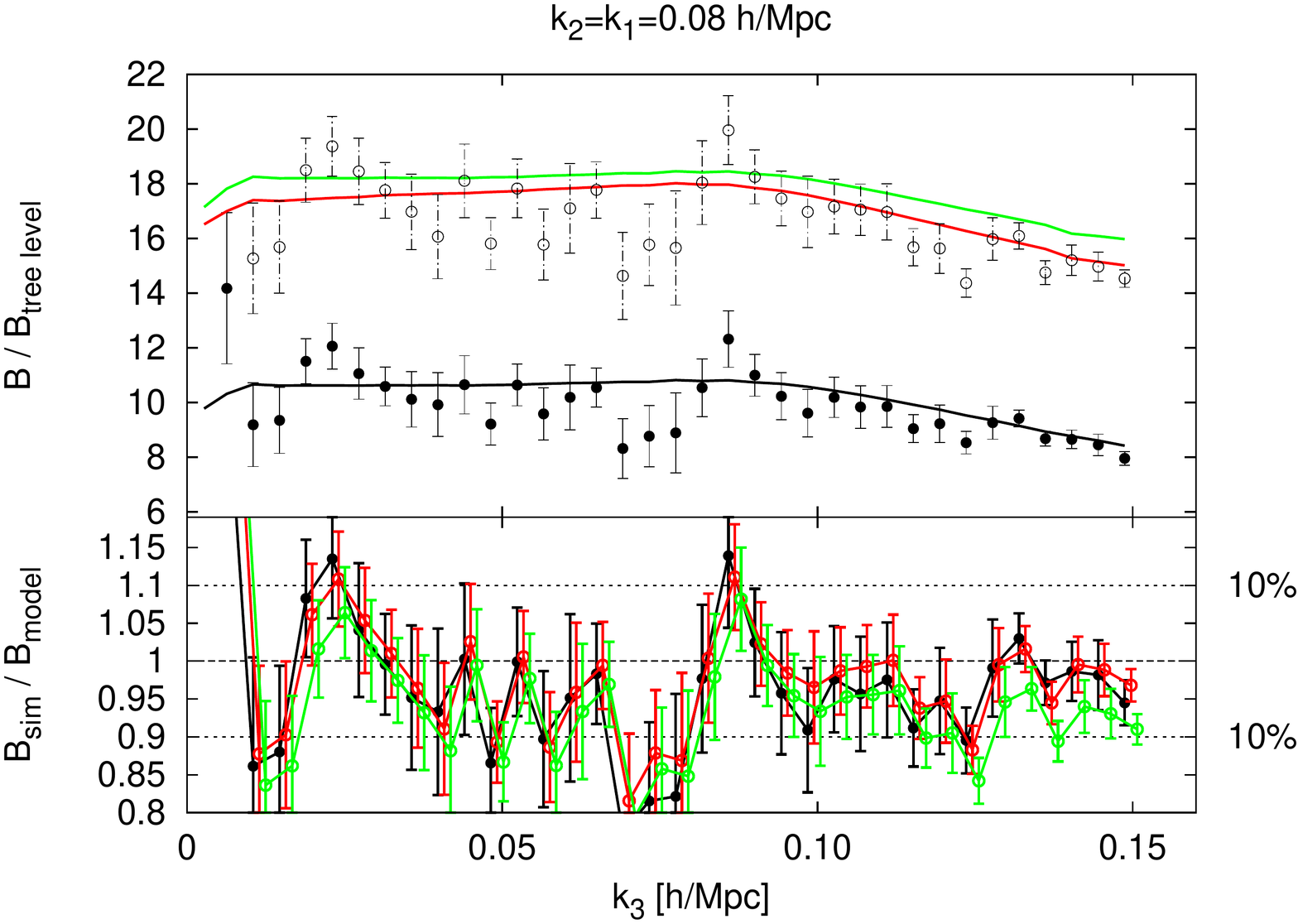}

\includegraphics[scale=0.25]{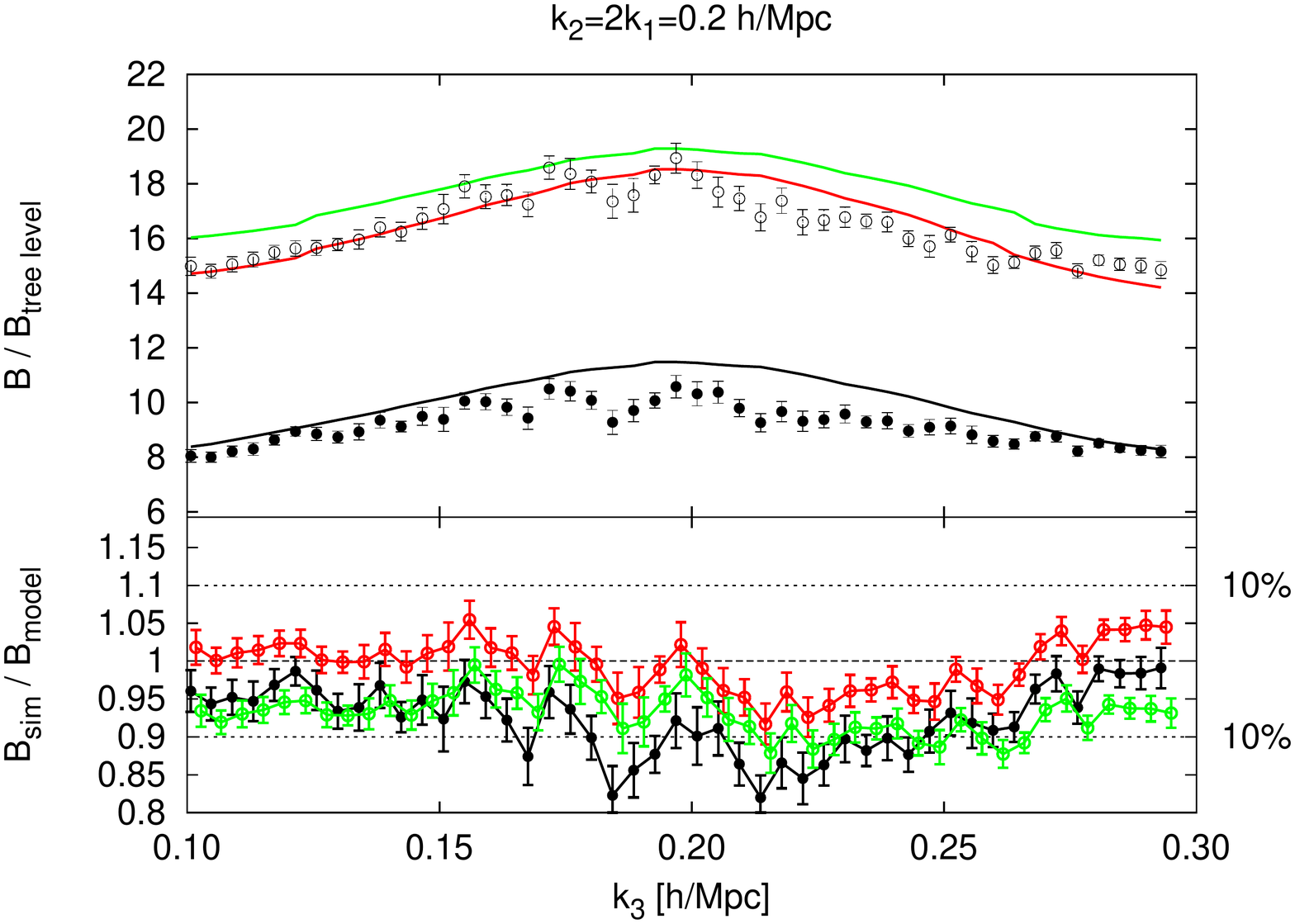}
\includegraphics[scale=0.25]{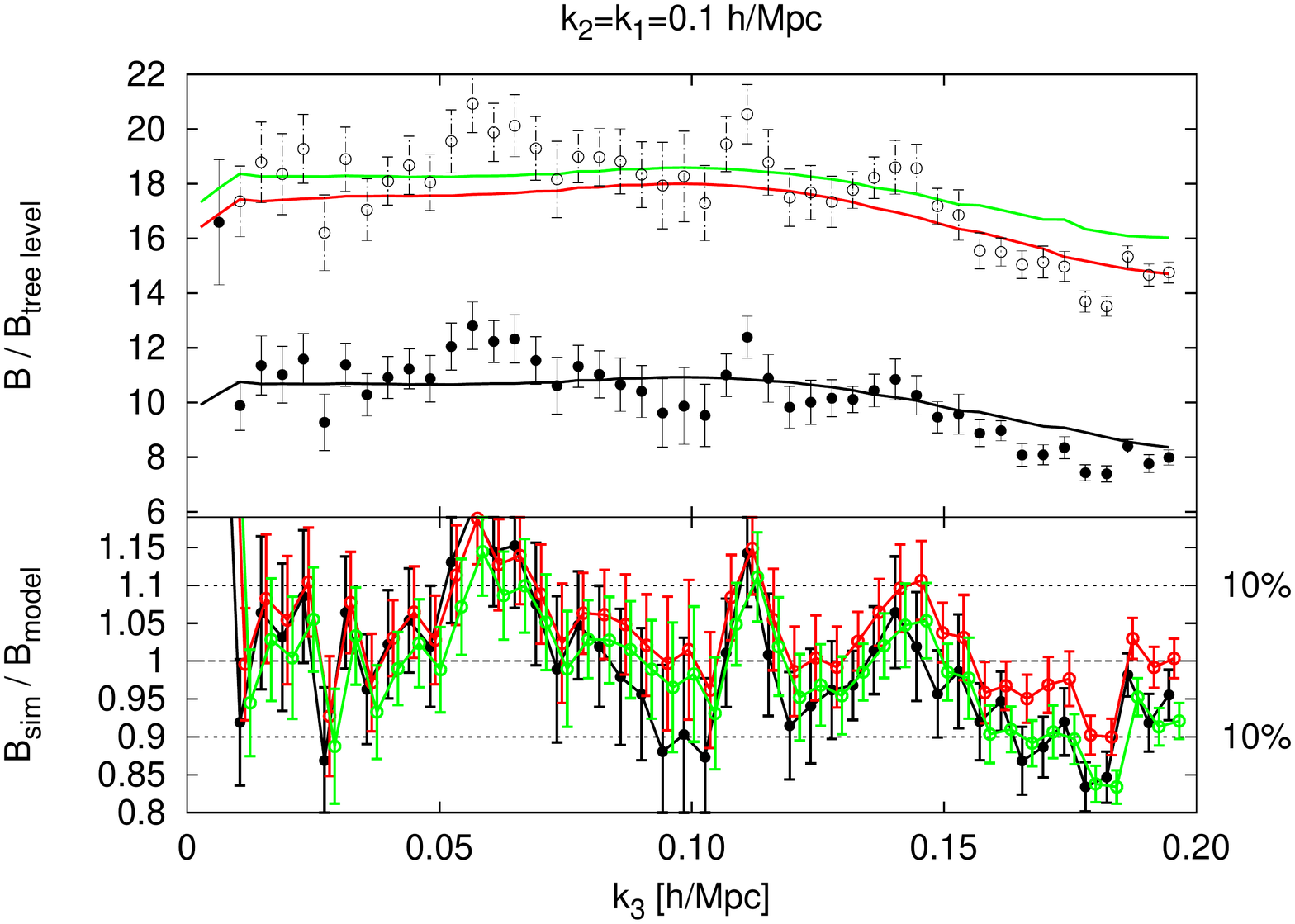}
\caption{Top sub-panels: Halo bispectrum in real space (filled black circles) and bispectrum monopole (empty black circles) normalised to the tree-level prediction of the matter bispectrum in real space. Black solid line is the prediction for real space halo bispectrum according to the model of  Eq.~\ref{B_ggg2} with the $F_2^{\rm eff}$ kernel of Eq.~\ref{Fkernel_hgm}. Green and red line are the predictions for the monopole halo bispectrum according to $B^{\rm spt}$ and $B^{FG}$ models respectively where the $\sigma_{\rm FoG}^B$ parameter has been set to 0. Different panels show different triangular configuration as indicated at the top. Bottom sub-panels show for each model the ratio of the measurements from N-body haloes and the model prediction. The bias parameters and noise factor are the same as in Fig.~\ref{halo_ps}. All panels at $z=0.55$.}
\label{halo_bs}
\end{figure}

Fig.~\ref{halo_bs} presents the real-space halo bispectrum (black filled circles) and the redshift-space monopole halo bispectrum (black empty circles) for different scales and triangle shapes, as indicated in the different panels. We  also show  the prediction of the different models. The black solid line shows the real space prediction of Eq.~\ref{B_ggg2} with the $F_2^{\rm eff}$ kernel of Eq.~\ref{Fkernel_hgm}. The coloured lines show the bispectrum model predictions, $B^{\rm spt}$ (green lines) and $B^{\rm FG}$ (red lines), when $\sigma_{\rm FoG}^B=0$. Note that the bias parameters and $A_{\rm noise}$ take the same value as for the power spectrum shown in Fig.~\ref{halo_ps}.
In general we see a moderate improvement for $B^{FG}$ over $B^{\rm spt}$ especially for folded triangles of the form $k_1+k_2\simeq k_3$ and $\left|k_1-k_2\right|\simeq k_3$. Thus the set of ${\bf a}^G$ derived from dark matter is able to predict the halo bispectrum when $\sigma_{\rm FoG}^P$ is set to 0. This suggests that, as the friends-of-friends haloes do not have a FoG component, ${\bf a}^G$ does not contain any significant FoG feature.
 We also observe that for the $k_2=2k_1=0.16\,h{\rm Mpc}^{-1}$ the data point corresponding to $k_3\simeq0.08\,h{\rm Mpc}^{-1}$ is not as good a match compared with the neighboring  points. However, being just one single data point, this deviation is not statistically important.
 
We see that on the scales considered $B^{FG}$ is in general able to describe with $5$ to $10\%$ percent error the halo bispectrum with the bias parameters derived from real spaces quantities. However, we are also interested in seeing whether the bias parameters estimated from redshift space statistics  are similar to those obtained in real space.

\begin{figure}
\centering
\includegraphics[clip=false, trim= 10mm 0mm 12mm 0mm,scale=0.28]{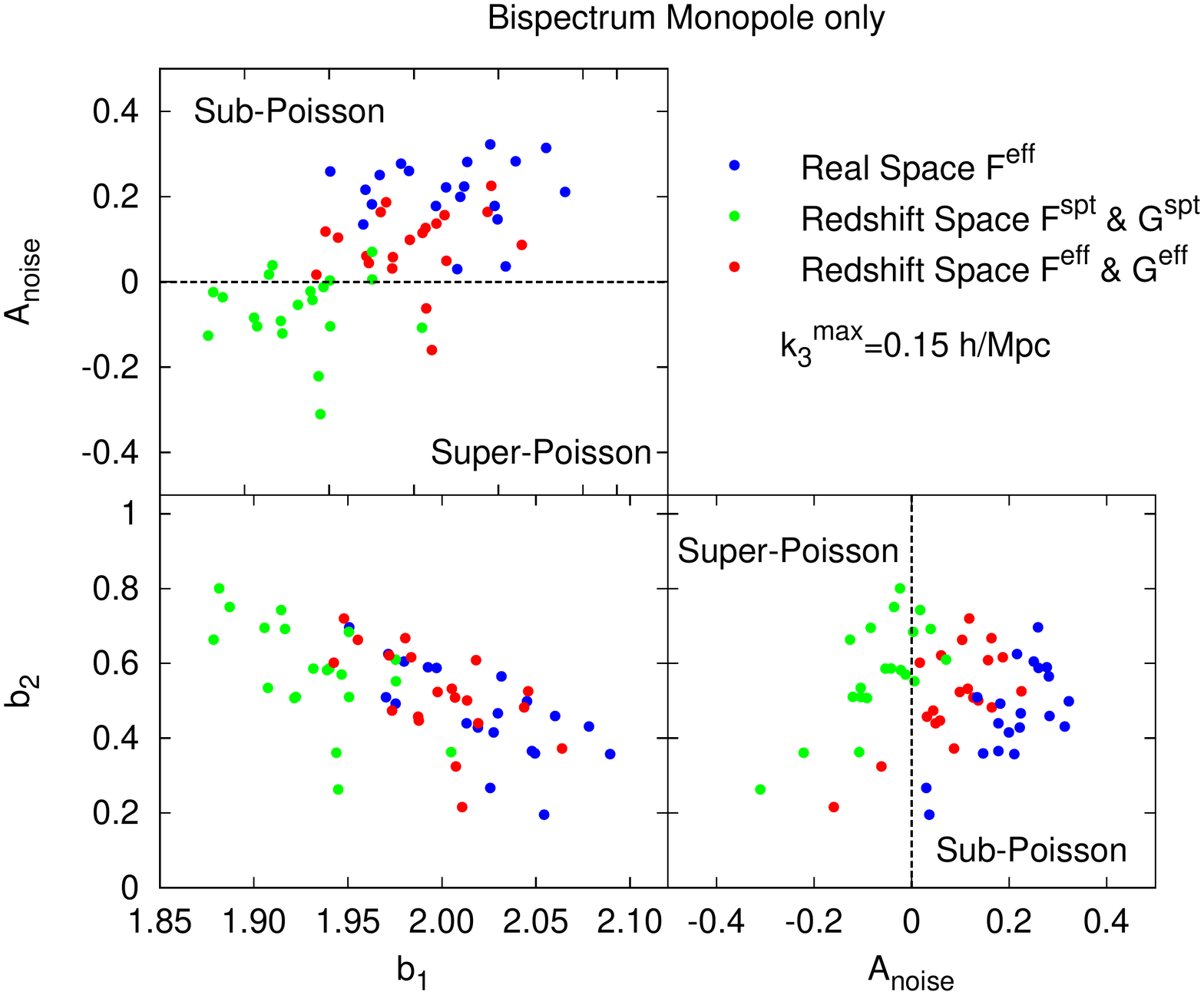}
\includegraphics[clip=false, trim= 10mm 0mm 12mm 0mm,scale=0.28]{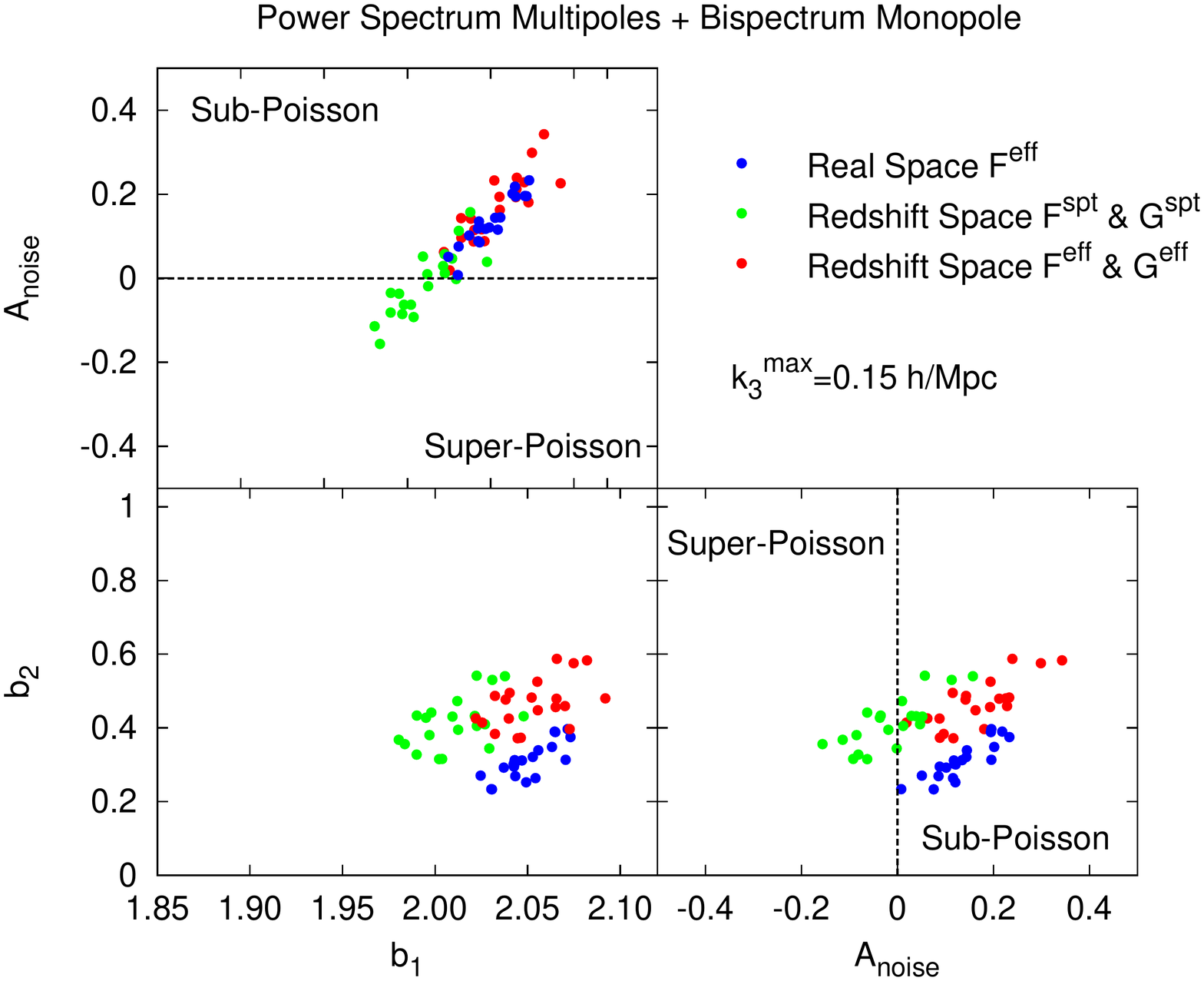}
\caption{Best-fitting parameters from halo bispectrum only measurements (left panel) and from the combination of the power spectrum multipoles and bispectrum (right panel). Only $b_1$, $b_2$, $A_{\rm noise}$ and $\sigma_{\rm FoG}^P$ are allowed to vary. Blue dots mark the best-fitting parameters from real space fits. Green and red points are the best-fitting parameters from redshift space using $B^{\rm spt}$ (green points) and $B^{\rm FG}$ (red points) for describing the bispectrum. Note that for the fits shown in the right panel, $\sigma_{\rm FoG}^P$ was allowed to vary,  although it is not shown for clarity. $f$ and $\sigma_8$ have been set to their true values.}
\label{scatter_halo}
\end{figure}

  In Fig~\ref{scatter_halo} we apply the method from \S~\ref{sec:estimation} to estimate the best-fitting values of $b_1$, $b_2$ and $A_{\rm noise}$. Each dot corresponds to the set of parameters that minimise $\chi_{\rm diag.}^2$ for each of the 20 realisations. The mean value and its dispersion corresponds to the estimator of the parameter set. These quantities are estimated from real space statistics (blue points) and redshift space statistics using $B^{\rm spt}$ (green points) and $B^{\rm FG}$ (red points). Left panels display the results using bispectrum information only, whereas  the right panel combines power spectrum and bispectrum measurements.
From the bispectrum only panels we see that the $B^{\rm FG}$ model (red points)  agrees very well with the real space model (blue points) for the bias parameters $b_1$ and $b_2$, but $B^{\rm spt}$ (green points) tends to underestimate the value of $b_1$ and overestimate the value for $b_2$. Regarding the $A_{\rm noise}$ parameter, we see a small disagreement between the real and the redshift prediction even for the model $B^{\rm FG}$. This is not necessarily a problem since we expect an extra clustering in redshift space due to the redshift space distortions. This could mean that the noise is actually larger than it is in real space, matching the trend observed.
From the right panel of Fig.~\ref{scatter_halo} we observe a similar behaviour between the two models: $B^{FG}$ is able to recover a $b_1$  consistent with the fit in real space. We see that there is a moderate discrepancy between the  real and redshift-space predictions of $b_2$, where the prediction in redshift space tends to overestimate $b_2$ with respect to the real space prediction. Note that in this case, we allow $\sigma_{\rm FoG}^P$ as a free parameter, although is not shown and $\sigma_{\rm FoG}^B$ is always set to 0.

\begin{table}
\begin{center}
\begin{tabular}{|c|c|c|c|}
\hline
  & Real Space & Redshift Space $B^{\rm spt}$ & Redshift Space $B^{\rm FG}$  \\
  \hline
  \hline
  $b_1$ & $2.050\pm0.014$ & $2.011\pm0.018$ & $2.053\pm0.019$ \\
\hline
  $b_2$ & $0.31\pm0.05$ & $0.41\pm0.07$ & $0.47\pm0.06$ \\
\hline
$A_{\rm noise}$ & $0.13\pm0.06$ & $-0.01\pm0.07$ & $0.17\pm0.08$ \\
\hline
$\sigma_{\rm FoG}^P$ [Mpc/$h$] & 0 & $2.31\pm0.13$ & $2.44\pm0.12$\\
\hline
\end{tabular}
\end{center}
\caption{Recovered parameters, $b_1$, $b_2$, $A_{\rm noise}$ and $\sigma_{\rm FoG}^P$ for haloes, when the power spectrum and bispectrum are used. The different columns are measurements in real space (left column) and redshift space when $B^{\rm spt}$ and $B^{\rm FG}$ are used to describe the bispectrum (central and right column respectively). The maximum scale is set to $k_{\rm max}=0.15\,h/{\rm Mpc}$. These values are obtained from the data shown in the right panel of Fig.~\ref{scatter_halo}.}
\label{bias_table}
\end{table}
In Table~\ref{bias_table}, we provide the estimated values of the bias parameters, as well as $A_{\rm noise}$ and $\sigma_{\rm FoG}^P$ corresponding to the   right panel of Fig.~\ref{scatter_halo}. The left column corresponds to the (mean of the) blue distribution of points, the central column to the green, and the right column to the red in Fig.~\ref{scatter_halo}. The error-bars correspond to $1\sigma$ of a volume of $V=3.375\,[{\rm Gpc}/h]^3$. This enable us to quantify how the measurements of $b_1$ obtained with  the $B^{\rm FG}$ model, compares with the real space predictions. 

We conclude that the $B^{FG}$ model, with the set of ${\bf a}^G$ parameters presented in \S~\ref{bis_section}, in combination with the TNS-2LRPT model for the redshift space  power spectrum,  is able to consistently recover  the values of $b_1$ in real and redshift space when the power spectrum and bispectrum statistics are analysed. On the other hand, we find that if $B^{\rm spt}$ model is used instead, in redshift space $b_1$ is  underestimated. For both models the obtained  value of $b_2$ is not consistent in real and redshift space. However, the cosmological constraints that galaxy surveys provide on $f$ and $\sigma_8$, comes with a degeneration with the $b_1$ parameter, whereas $b_2$ is often unused. Because of this, is more important to have a model that predicts unbiased estimates on $b_1$, which we will use to determine for example $b_1\sigma_8$ when monopole and quadrupole power spectra are measured, than to predict correctly $b_2$ that has not been used to constraint any cosmological parameter, such as $\sigma_8$ or $f$.

\section{Applications to cosmology}\label{applications_section}

In this section we show how the $B^{\rm FG}$ model can be used to constrain $f$ and extract the bias parameters as well as $\sigma_8$ from power spectrum and bispectrum measurements. Combining the power spectrum multipoles and the bispectrum monopole allows us to disentangle the large scale degeneracy that typically ties $b_1$, $\sigma_8$ and $f$ together. In order to study these degeneracies, we start by recovering these parameters from dark matter fields. However, we are also interested in applying this technique to N-body haloes, which may suffer from different, and more realistic systematics errors.

\subsection{Dark matter field}\label{applications_dm}
\begin{figure}
\centering
\includegraphics[scale=0.4]{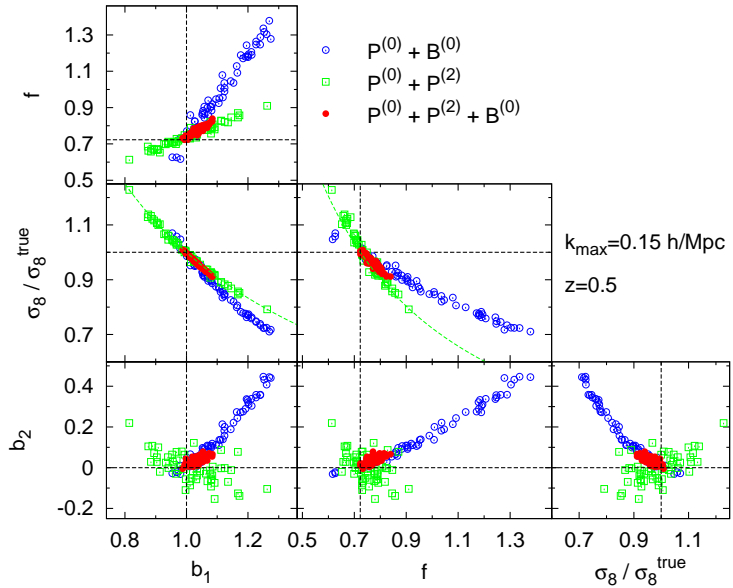}
\caption{Best-fitting parameters for dark matter simulations in redshift space at $z=0.5$ for $k_{\rm max}=0.15$ when different statistics are used: blue points correspond to $P^{(0)}+B^{(0)}$, green points to $P^{(0)}+P^{(2)}$ and red points to $P^{(0)}+P^{(2)}+B^{(0)}$ as indicated. The dashed black lines mark the  true values. The green dashed lines mark the $b_1\propto\sigma_8^{-1}$ and the $f\propto\sigma_8^{-1}$ relations.  Note that $b_1$, $b_2$, $f$, $\sigma_8$, $\sigma_0^P$, $\sigma_0^B$ are varied as free parameters, although only $b_1$, $b_2$, $f$ and $\sigma_8$ are shown for clarity.}
\label{sigma8_f_scatter}
\end{figure}

The different panels of Fig.~\ref{sigma8_f_scatter} display the  distributions of the best-fitting parameters values  obtained from the  dark matter N-body simulations for $z=0.5$ and $k_{\rm max}=0.15\,h/{\rm Mpc}$ when different statistics are used: power spectrum monopole and quadrupole (green symbols), power spectrum and bispectrum monopole (blue symbols) and power spectrum monopole, quadrupole and bispectrum monopole (red symbols). As in Fig.~\ref{scatter_halo}, we have applied the method of \S~\ref{sec:estimation}, where each point corresponds to the set of parameters that minimises $\chi_{\rm diag.}^2$ for a single realisation. We consider as free parameters $\{ b_1, b_2, \sigma_{\rm FoG}^P, \sigma_{\rm FoG}^B, f,\sigma_8\}$. Note that we assume that the shot noise is given by Poisson statistics, hence $A_{\rm noise}=0$. We have checked that for dark matter particles the role that $A_{\rm noise}$ plays is negligible, since the number density of particles is very high. For clarity, in Fig.~\ref{sigma8_f_scatter} we only display the parameter-space projection for $\{b_1,b_2,\sigma_8,f\}$, which are the parameters we are interested in. The black dashed lines show the  reference (true) values  of the parameters. We see that when the power spectrum monopole and quadrupole are used, $f$, $\sigma_8$ and $b_1$ are only constrained in the following combinations: $f\sim\sigma_8^{-1}$ and $b_1\sim\sigma_8^{-1}$. These relations can be analytically extracted by a simple inspection of the large scale limits of the model. We also notice that  this is not the case for $b_2$, because it is a second-order parameter in the power spectrum. We also see that when the power spectrum and bispectrum monopole are used, the parameters are again constrained in combination, but the combination is different from the monopole-to-quadrupole case. Furthermore, we see that in this case, $b_2$ is also constrained only in combination with $\sigma_8$ and $b_1$. This is because in the bispectrum $b_2$ appears at leading order. Since these parameter combinations are different for $P^{(0)} + P^{(2)}$ (green symbols) and $P^{(0)} + B^{(0)}$ (blue symbols), we  break the degeneracies between $b_1$, $b_2$, $\sigma_8$ and $f$ when we combine them all: $P^{(0)}+P^{(2)}+B^{(0)}$. In this case, we observe that the estimated parameters are close to their reference values (marked as black dashed lines), although there are some differences. We see that the mean of the scatter in the $f$-$\sigma_8$ plane obtained from $P^{(0)}+B^{(0)}$ is biased. However, the scatter does include the true values of these parameters, and indicates that there is well defined degeneracy direction which crosses the true values. When we include the constraint from $P^{(2)}$, we see that the degeneracy is broken, given $f$ and $\sigma_8$ within a few percent of the true values.  We are interested in quantifying these deviations for different redshifts, and also as a function of the maximum scale used for the analysis. 

\begin{figure}
\centering
\includegraphics[scale=0.3]{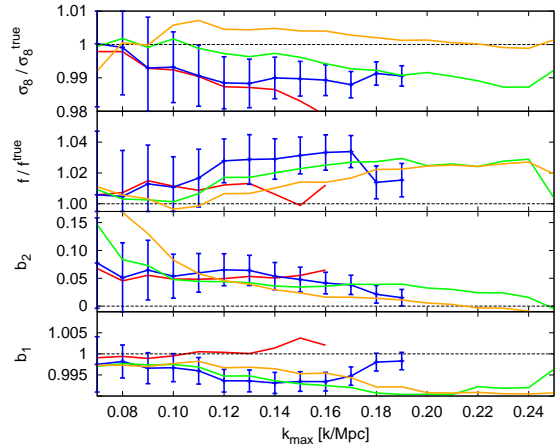}
\caption{Parameters as a function of $k_{\rm max}$ obtained  from  combining the dark matter $P^{(0)}$, $P^{(2)}$ and $B^{(0)}$. 
Although only $b_1$, $b_2$, $\sigma_8$ and $f$ are shown,  $\sigma_0^P$ and $\sigma_0^B$ are also varied in the fit. The colour indicate the redshift: $z=0$ (red), $z=0.5$ (blue),  $z=1$ (green) and  $z=1.5$ (orange). Values are the mean of the best-fittings of the simulations and errors correspond to $1\sigma$  dispersion among simulations for a volume of $13.8\,[{\rm Gpc}/h]^3$. Error-bars are only shown for $z=0.5$ for clarity;  the relative errors are similar for the other redshifts. The black dashed lines are the reference values.}
\label{sigma8_f_kmax}
\end{figure}

In Fig.~\ref{sigma8_f_kmax} we present the estimates of $b_1$, $b_2$, $f$ and $\sigma_8$ as a function of $k_{\rm max}$ from 60 realisations of dark matter N-body simulations for $z=0$ (red lines), $z=0.5$ (blue lines), $z=1$ (green lines) and $z=1.5$ (orange lines), when the power spectrum monopole, quadrupole and bispectrum monopole are used. The error-bars correspond to $1\sigma$ dispersion of a realisation of volume $V=13.8\,[{\rm Gpc}/h]^3$. We only display error-bars for $z=0.5$ for clarity, since the relative errors are similar for the other redshifts. The black dashed lines indicate  the reference values of the parameters.

We observe that for the $z>0$ redshift snapshots, the $b_1$ parameter is underestimated by $\sim0.5\%$. For all redshifts,  $b_2$ is overestimated by 0.05; $f$ is typically overestimated by  $3-4\%$, whereas $\sigma_8$ is underestimated at some  redshifts and overestimated at others, but typically  by $\leq1\%$. 

To summarise,  for dark matter particles, we are able to recover the correct bias parameters as well as $\sigma_8$ and the logarithmic growth rate with  few percent accuracy (3-4\%), when the power spectrum monopole and quadrupole are used in combination with the bispectrum monopole. 

\subsection{Dark matter haloes}\label{applications_haloes}
\begin{figure}
\centering
\includegraphics[scale=0.4]{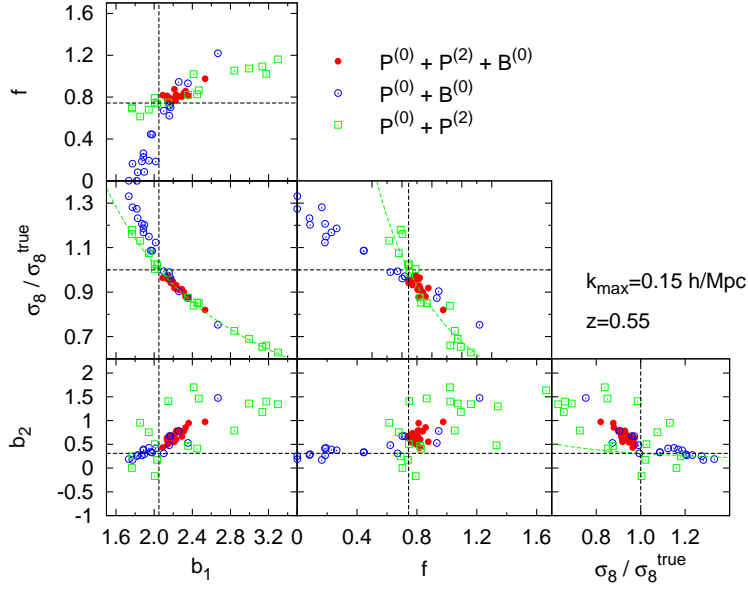}
\caption{Same as Fig.~\ref{sigma8_f_scatter} but for dark matter haloes at $z=0.55$. In black dashed lines the true values of $f$ and $\sigma_8$ are plotted. Also in black dashed lines, the real space best-fitting values for $b_1$ and $b_2$ are plotted.}
\label{sigma8_f_scatter_halo}
\end{figure}

In this section we aim to repeat the above analysis, but now for dark matter haloes. In this case we fix $\sigma^B_{\rm FoG}=0$, as we do not have FoG features for haloes, and we allow $A_{\rm noise}$ to be free. Therefore, the $\bf\Psi$ set of free parameters corresponds to $\{b_1,b_2,\sigma_{\rm 0}^P,A_{\rm noise}, f,\sigma_8\}$. Fig.~\ref{sigma8_f_scatter_halo} is similar to  Fig.~\ref{sigma8_f_scatter}, but using the halo catalogue instead of dark matter particles. In this case, the redshift is $z=0.55$ and, as before, the maximum scale used for the fit is $k_{\rm max}=0.15\,h/{\rm Mpc}$. 

From Fig.~\ref{sigma8_f_scatter_halo} we see that when the  $P^{(0)}+P^{(2)}$ and $P^{(0)}+B^{(0)}$ statistics are used, degeneracies  appear among the parameters $f$, $\sigma_8$ and $b_1$, in a similar way to the dark matter case. Adding a third statistic breaks the degeneracies.  Fig.~\ref{sigma8_f_scatter_halo} shows that the best-fitting values for $f$ and $\sigma_8$ are $\sim10\%$ biased from the reference values, which are marked by black dashed lines.  The black dashed lines show the real space best-fitting values for $b_1$ and $b_2$ (see Table \ref{bias_table}).  

\begin{figure}
\centering
\includegraphics[scale=0.25]{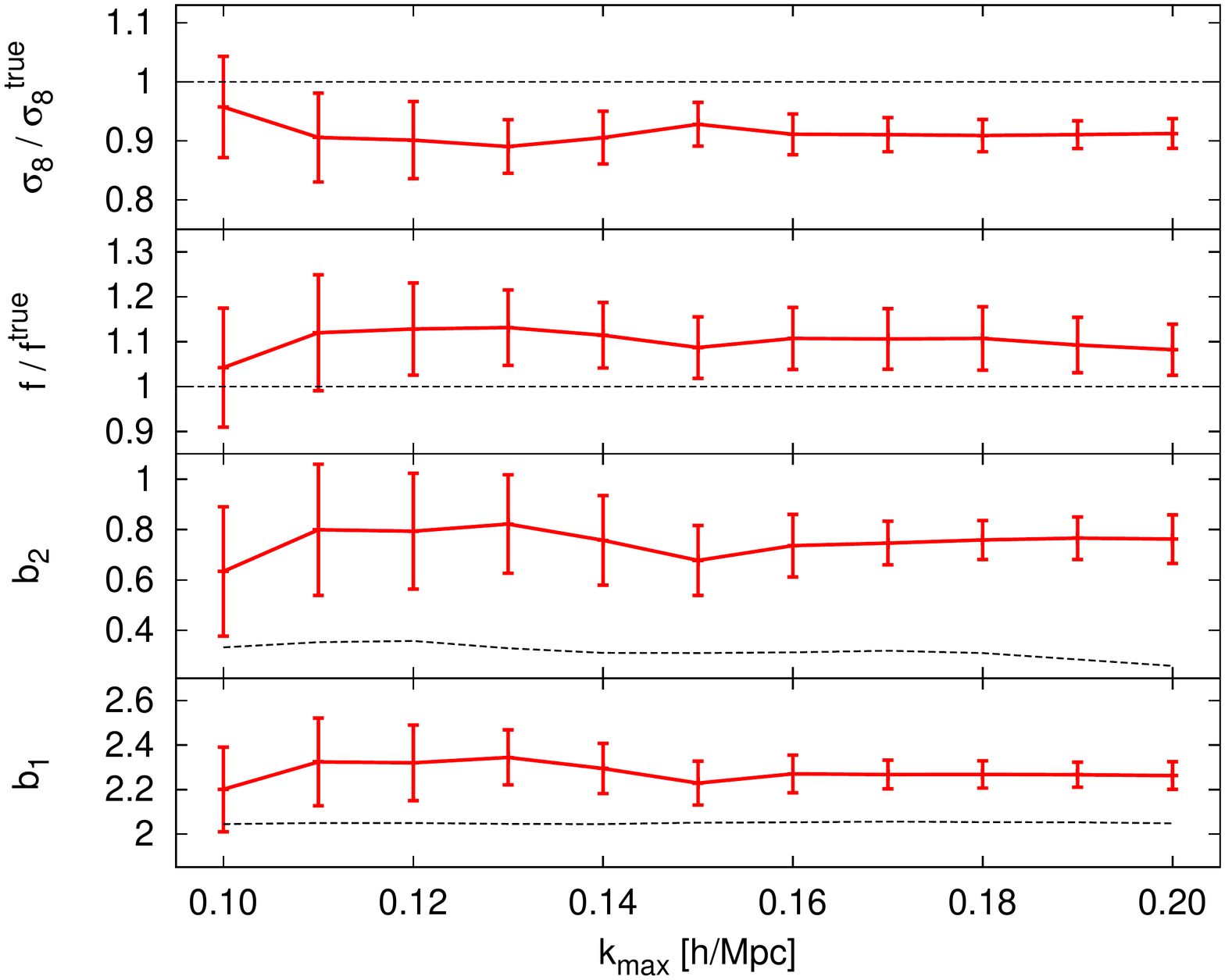}
\includegraphics[scale=0.25]{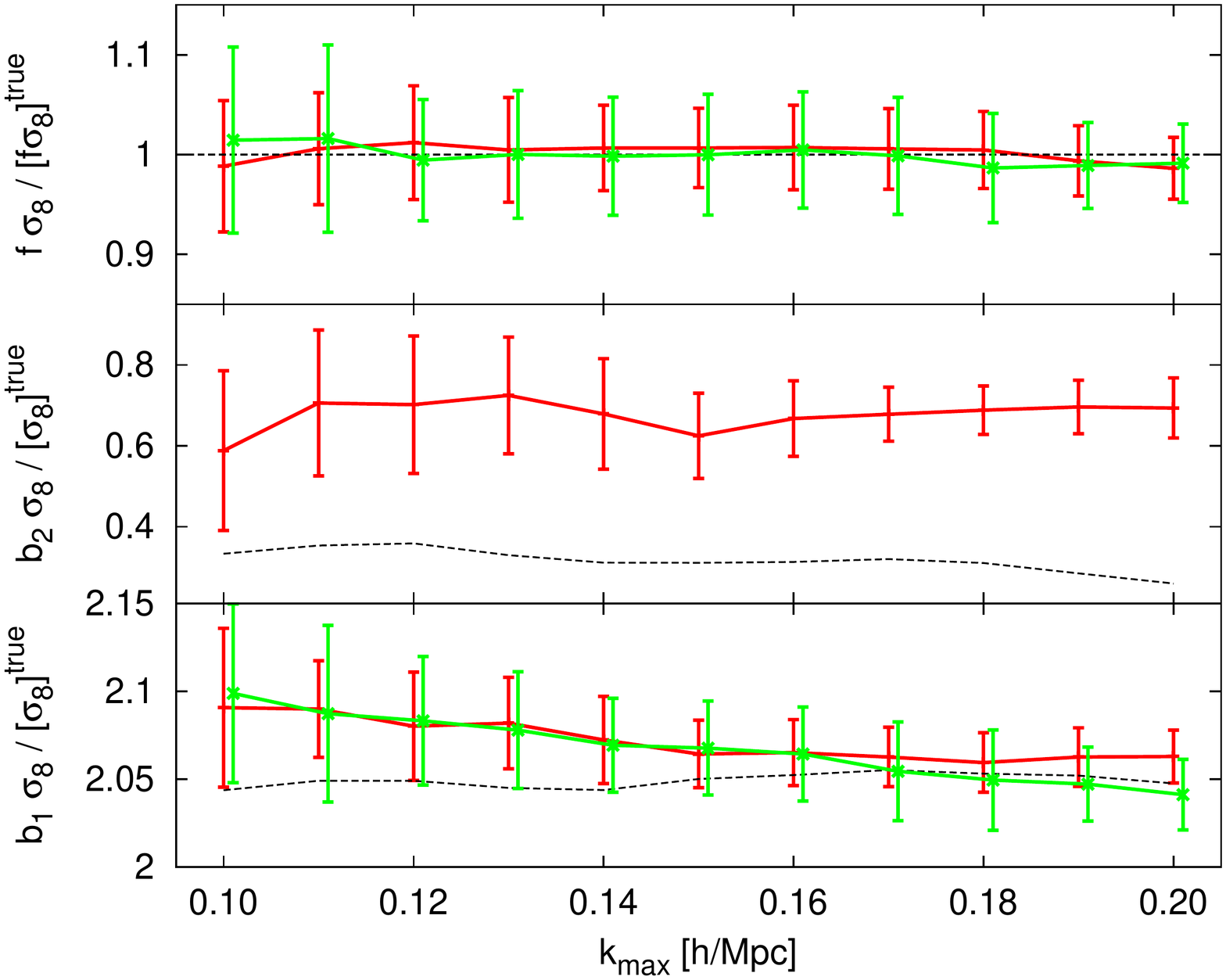}
\caption{Left panel shows the best-fitting parameters for dark matter haloes as a function of $k_{\rm max}$ when $P^{(0)}$, $P^{(2)}$ and $B^{(0)}$ are jointly fitted. Although only $b_1$, $b_2$, $\sigma_8$ and $f$ are shown, $\sigma_0^P$ and $A_{\rm noise}$ are also varied in the fit. The right panel indicates the error of the combo variables $f\sigma_8$, $b_1\sigma_8$ and $b_2\sigma_8$ when $P^{(0)}$, $P^{(2)}$ and $B^{(0)}$ are used (red lines) and when $P^{(0)}$ and $P^{(2)}$ are used (green lines). Errors correspond to $1\sigma$ with a volume of $3.375\,[{\rm Gpc}/h]^3$. Black dotted lines show the true values for $\sigma_8$ and $f$, as well as the real-space values for $b_1$ and $b_2$ when $\sigma_8$ and $f$ are set to their true values.}
\label{sigma8_f_kmax_halo}
\end{figure}

In the left panel of Fig.~\ref{sigma8_f_kmax_halo} we show the mean values of  $f$, $\sigma_8$, $b_1$ and $b_2$ as a function of $k_{\rm max}$ when all $P^{(0)}$, $P^{(2)}$ and $B^{(0)}$ are used. The black dashed lines, show the reference values:  true values for $f$ and $\sigma_8$ and  values of $b_1$ and $b_2$  estimated from the real space power spectrum and bispectrum when $f$ and $\sigma_8$ were fixed (which we refer to as reference values), as in \S~\ref{haloes_section}. In the right panel of Fig.~\ref{sigma8_f_kmax_halo}, we have combined the variables into $f\sigma_8$, $b_1\sigma_8$ and $b_2\sigma_8$, using $P^{(0)}+P^{(2)}+B^{(0)}$   (red lines) as well as $P^{(0)}+P^{(2)}$ (green lines).  In both panels, the error-bars correspond to $1\sigma$ with a volume of $3.375\,[{\rm Gpc}/h]^3$ and are estimated using the method described in $\S~\ref{sec:estimation}$.

From the left panel of Fig.~\ref{sigma8_f_kmax_halo}, we see that, for all $k_{\rm max}$ explored, $\sigma_8$ is underestimated by $\sim10\%$, whereas $f$ and $b_1$ are overestimated by $\sim10\%$.  Also $b_2$ is  overestimated by about $\Delta b_2\sim0.4$ with respect to the real space findings. We note a significant deviation of these parameters from their reference values compared to those obtained from dark matter particles in Figs.~\ref{sigma8_f_scatter} -~\ref{sigma8_f_kmax}.  These offsets may result from the breakdown of the local Lagrangian  halo bias model that relates the dark matter field with the halo field, or the mapping between real and redshift space of biased tracers, where we have assumed no velocity bias.

Table~\ref{table5}  (second column) shows the results corresponding to the left panel of Fig.~\ref{sigma8_f_kmax_halo} for  $k_{\rm max}=0.15\,h/{\rm Mpc}$. Also shown (third column) are the  results from real space (where $f$ and $\sigma_8$ have been fixed to their true values) for comparison, which are the dashed black lines in Fig.~\ref{sigma8_f_kmax_halo}.

\begin{table}[htdp]
\begin{center}
\begin{tabular}{|c|c|c|}
\hline
$k_{\rm max}=0.15\,h/{\rm Mpc}$ & $P^{(0)}+P^{(2)}+B^{(0)}$  & $P+B$ \\
 \hline
 \hline
 $b_1$ & $2.23\pm0.10$ & $2.050\pm0.014$  \\
\hline
 $b_2$ & $0.68\pm0.14$ & $0.31\pm0.05$  \\
\hline
 $\sigma_8$ & $0.742\pm0.030$ & 0.80  \\
\hline
 $f$ & $0.809\pm0.051$ & $0.744$  \\
\hline
\end{tabular}
\end{center}
\caption{Recovered parameters, $b_1$, $b_2$, $f$ and $\sigma_8$ for haloes, when the power spectrum and bispectrum are used. The different columns are measurements in real space (``P+B" column) and redshift space when $P^{(0)}$, $P^{(2)}$ and $B^{(0)}$ are used. The maximum scale is set to $k_{\rm max}=0.15\,h/{\rm Mpc}$. The measurements without error-bars are set to their true values. These values corresponds to the left panel of Fig.~\ref{sigma8_f_kmax_halo}.}
\label{table5}

\end{table}

In the right panel of Fig.~\ref{sigma8_f_kmax_halo} we show the the predictions for the parameter combinations, $f\sigma_8$, $b_1\sigma_8$ and $b_2\sigma_8$ as a function of $k_{\rm max}$ estimated from $P^{(0)}+P^{(2)}$ (green lines) and $P^{(0)}+P^{(2)}+B^{(0)}$ (red lines). Since the combination $P^{(0)}+P^{(2)}$ is not able to estimate efficiently $b_2\sigma_8$, we do not show it in this case. 

We note that neither $P^{(0)}+P^{(2)}$ nor $P^{(0)}+P^{(2)}+B^{(0)}$ present any significant offset on $f\sigma_8$ at any scale. On the other hand, a small systematic offset on $b_1\sigma_8$ is observed with respect to the real space predictions, of order $1\sigma$ at small $k_{\rm max}$. 
 Finally, $b_2\sigma_8$ presents a similar systematic to that  observed for $b_2$ alone in the left panel. Therefore, we see that the systematic offsets reported for $b_1$, $\sigma_8$ and $f$ cancel almost perfectly when we work with $f\sigma_8$ and $b_1\sigma_8$. In this case, we see that the predictions from $P^{(0)}+P^{(2)}$ and $P^{(0)}+P^{(2)}+B^{(0)}$ are very similar and we quantify that, by adding the bispectrum monopole to the power spectrum monopole and quadrupole, the error on $f\sigma_8$ and $b_1\sigma_8$ reduces by $\sim30-40\%$ at all scales. These cancellations may only hold for the fiducial cosmology and measured halo population. For different populations of galaxies or if the cosmological model were different this may not hold, and we would require further mocks to test them.

 Table~\ref{table6} shows the results corresponding to the right panel of Fig.~\ref{sigma8_f_kmax_halo} when different parameters of interest, $b_1\sigma_8$, $b_2\sigma_8$ and $f\sigma_8$ are estimated from $P^{(0)}+P^{(2)}$ and $P^{(0)}+P^{(2)}+B^{(0)}$ for $k_{\rm max}=0.15\,h/{\rm Mpc}$. We have normalised these quantities by the true  values of $\sigma_8$ and $f$ for clarity.

\begin{table}[htdp]
\begin{center}
\begin{tabular}{|c|c|c|c|}
\hline
 & $P^{(0)}+P^{(2)}+B^{(0)}$ & $P^{(0)}+P^{(2)}$  &  $P+B$ \\
 \hline
 \hline
 $b_1\sigma_8/[\sigma_8]^{\rm true}$ & $ 2.064\pm0.019$ & $2.068\pm0.027$ & $2.050\pm0.014$  \\
\hline
 $b_2\sigma_8/[\sigma_8]^{\rm true}$ & $ 0.62\pm0.11$ & $0.66\pm0.43$ & $0.31\pm0.05$  \\
\hline
 $f\sigma_8/[f\sigma_8]^{\rm true}$ & $1.007\pm0.040$ & $1.000\pm0.055$ & 1  \\
\hline
\end{tabular}

\end{center}
\caption{Recovered parameters, $b_1\sigma_8$, $b_2\sigma$ and  $f\sigma_8$ for haloes, when different statistics are used: first column $P^{(0)}+P^{(2)}+B^{(0)}$, second column $P^{(0)}+P^{(2)}$ and third column $P+B$ (real space quantities with $f$ and $\sigma_8$ set to true values). The maximum scale is set to $k_{\rm max}=0.15\,h/{\rm Mpc}$. The numbers without error-bars are set to their true values. Numbers in this table correspond to  the right panel of Fig.~\ref{sigma8_f_kmax_halo}.}
\label{table6}

\end{table}

\section{Conclusions}\label{conclusions_section}

The main goal of this paper is to provide an empirical formula for the redshift space bispectrum monopole for the dark matter field and  for  biased tracers such as galaxies or haloes. The statistical power of present and forthcoming surveys imply that the accuracy of existing  analytic descriptions is not sufficient  considering the statistical power of current surveys. The bispectrum statistic offers additional complementary information to that contained in the power spectrum multipoles, which, in principle, helps  reduce error-bars and break degeneracies among cosmological parameters.

In \S~\ref{bis_section} we have extended the real space dark matter bispectrum formula presented in \cite{bispectrum_fitting}  to account for the redshift-space distortions at the level of the bispectrum monopole.  We refer to this new formula as $B^{FG}$. We have proceeded by modifying the standard perturbation theory  velocity kernel $G_2$ to an effective kernel $G_2^{\rm eff}$ with nine free parameters,  ${\bf a}^G$. We have constrained the values of these parameters using measurements of the redshift space bispectrum monopole from dark matter N-body simulations (for a total  volume of $\sim 829$[Gpc/$h$]$^3$) at four different redshift, $z=0\,, 0.5\,, 1.0\,, 1.5$ for the $k_2/k_1=1.0,\, 1.5,\, 2.0,\, 2.5$. With this,  $B^{FG}$ is able to describe the dark matter bispectrum monopole in redshift space with a precision of $\lesssim5\%$ for  $k\leq0.10\,h/{\rm Mpc}$ at $z=0$; for $k\leq0.15\,h/{\rm Mpc}$ at $z=0.5$; for $k\leq0.17\,h/{\rm Mpc}$ at $z=1.0$ and for $k\leq0.20\,h/{\rm Mpc}$ at $z=1.5$.
 For squeezed triangles, those with $k_1\sim k_2$ and $k_3\leq0.02\,h{\rm Mpc}^{-1}$, $B^{FG}$ under-predicts the measured bispectrum at all redshifts, especially when $k_1$ and $k_2$ are $\geq0.10\,h{\rm Mpc}^{-1}$.

In Appendix \ref{appendixb} we have shown that the $B^{FG}$ model can also be applied to those triangles that were not used for calibrating the  ${\bf a}^F$ and ${\bf a}^G$ parameters. In particular, we have checked that $B^{FG}$ provides a description of the dark matter redshift space bispectrum for the shapes $k_2/k_1=1.25,\, 1.75,\, 2.25$ with the same accuracy of the set of shapes used for the fitting.

In Appendix \ref{cosmo_appendix} we have shown how $B^{FG}$ is able to describe the bispectrum in redshift space for two cosmologies with different values of $\Omega_m$, and therefore a different strength of redshift space distortions. We have found that the effective kernels $F^{\rm eff}$ and $G^{\rm eff}$, with the values of the  ${\bf a}^F$ and ${\bf a}^G$ obtained from fitting to a cosmology with $\Omega_m=0.27$, are able to describe the bispectrum of cosmologies with $0.2 \lesssim \Omega_m \lesssim 0.4$ (and therefore with $0.4\lesssim f(z=0) \lesssim 0.6$), with similar accuracy to the one measured for the fiducial cosmology.

In \S~\ref{haloes_section} we have proceeded to  combine the predictions of $B^{FG}$, with the non-local and non-linear bias model \citep{McDonald_Roy}, in order to provide a theoretical description of the bispectrum in redshift space for dark matter haloes. We find that $B^{FG}$ provides a better description of the halo bispectrum in redshift space than the standard perturbation theory leading order prediction. In this case $B^{FG}$ predicts with a $\lesssim5\%$ accuracy the halo bispectrum in redshift space for $k\lesssim0.15\,h/{\rm Mpc}$ at $z=0.55$.   For comparison, the SPT approach would perform similarly only up to $k\lesssim 0.06$: in other words, the extension reduces the statistical error-bars  as much as increasing the survey volume by a factor of four would.

To demonstrate the power of adding  the bispectrum information  to the power spectrum, we have combined the bispectrum model $B^{FG}$ with the power spectrum monopole and quadrupole model of  \citep{TNS_halo} and \cite{ps_model}. First we have  extracted the bias parameters $b_1$ and $b_2$ from simulations when other cosmological parameters such as $f$ and $\sigma_8$ were fixed to their true values. We have found that $B^{FG}$ is able to predict the same large scale bias parameter,  $b_1$ in real and redshift space, whereas standard perturbation theory approach for the  redshift space  kernel underestimates $b_1$ in redshift space with respect to real space by 2\% (this is large enough to be statistically significant given the size of the simulations).  
In \S~\ref{applications_section}  we have further explored  the performance of the modelling proposed here in extracting $f$, $\sigma_8$ as well as the bias parameters $b_1$ and $b_2$ from the power spectrum monopole, $P^{(0)}$, quadrupole, $P^{(2)}$  and the bispectrum monopole, $B^{(0)}$. Our main findings are as follows:
\begin{enumerate}

\item For the dark matter field no systematics offsets larger than few percent are found for $b_1$, $b_2$ and $f$ when $P^{(0)}$, $P^{(2)}$ and $B^{(0)}$ are used. 

\item For the dark matter halo catalogue, when the parameters $\{b_1,b_2,f,\sigma_8\}$ are estimated from $P^{(0)}+P^{(2)}$, no systematic offsets appear for $f\sigma_8$ and a $\sim1\%$ systematic error is found for $b_1\sigma_8$ with respect to the real space prediction.  If we add $B^{(0)}$ to these two statistics, the errors on $f\sigma_8$ and $b_1\sigma_8$ combinations are reduced by about $\sim30-40\%$, regardless of the value of $k_{\rm max}$ and no additional systematic errors are evident. Adding $B^{(0)}$ allows us to measure also  $b_2\sigma_8$. In this case we do find a systematic error of $\sim50\%$ compared  to the real space prediction. 
 
\item Combining $P^{(0)}$, $P^{(2)}$ and $B^{(0)}$ allows us also to estimate the variables $b_1$, $f$ and $\sigma_8$ separately. In this case we find that $b_1$ and $f$ are underestimated by $\sim10\%$ and $\sigma_8$ is overestimated by a similar amount for $0.10\leq k_{\rm max}\,[h/{\rm Mpc}]\leq0.20$. Note that for the dark matter case these systematics were smaller than $3-4\%$.

\item It is likely that the systematics found for $f$, $b_1$ and $\sigma_8$ are due to a limitation of the halo bias modelling when describing the power spectrum and bispectrum in redshift space.  For this particular halo population, these systematics can be tamed if we work with the combinations $f\sigma_8$ and $b_1\sigma_8$. 
 
\end{enumerate}

The bispectrum fitting formula presented in this paper  may be  useful for  and directly  applicable to any galaxy survey  when  redshift space distortions in the bispectrum must be accounted for. 
 While for this particular halo population and cosmology there is no evidence for important systematic offsets when measuring $f\sigma_8$, the combination of power spectrum monopole, quadrupole and bispectrum  monopole allow us to break the degeneracy between $f$ and $\sigma_8$. In this case systematic shifts of $\sim10\%$ appear,  which are of the order of the statistical errors for current  state-of-the-art surveys.

Clearly more work, especially in understanding the interplay between biasing and redshift-space distortions and their combined effects on clustering, is needed in order to reduce these systematic shifts and bring them below the statistical errors of future surveys.

\section*{Acknowledgments}
 We thank Beth Reid for providing the N-body halo catalogues used in this paper.
 
HGM is grateful for support from the UK Science and Technology Facilities Council through the grant ST/I001204/1. LV is supported by European Research Council under the European Communities Seventh Framework Programme grant FP7-IDEAS-Phys.LSS and acknowledges Mineco grant FPA2011-29678- C02-02.
WJP is grateful for support from the UK Science and Technology Facilities Research Council through the grant ST/I001204/1, and the European Research Council through the grant ÒDarksurveyÓ.

Numerical computations were done on Hipatia ICC-UB BULLx High Performance Computing Cluster at the University of Barcelona. The simulations for N-body haloes used in \S~\ref{haloes_section} of this  paper were analysed at the National Energy Research Scientific Computing Center, the Shared Research Computing Services Pilot of the University of California and the Laboratory Research Computing project at Lawrence Berkeley National Laboratory. 

\appendix

\appendix
\section{Explicit expressions for the $a$, $b$, $c$ functions}
\label{fit_kernels}

The functions $a$, $b$ and $c$ are defined as,
\begin{eqnarray}
 \nonumber \label{abc_new} {a}(n,k,{\bf a})&=&\frac{1+\sigma_8^{a_6}(z)[0.7Q_3(n)]^{1/2}(q a_1)^{n+a_2}}{1+(q a_1)^{n+a_2}}, \\
 {b}(n,k,{\bf a})&=&\frac{1+0.2a_3(n+3)(q a_7)^{n+3+a_8}}{1+(q a_7)^{n+3.5+a_8}}, \\
\nonumber {c}(n,k,{\bf a})&=&\frac{1+4.5a_4/[1.5+(n+3)^4](q a_5)^{n+3+a_9}}{1+(q a_5)^{n+3.5+a_9}}.
\end{eqnarray}
where $q\equiv k/k_{\rm nl}$ with $k_{\rm nl}(z)$ a characteristic scale defined as,
\begin{eqnarray}
 \frac{k_{\rm nl}(z)^3P^{\rm lin}(k_{\rm nl},z)}{2\pi^2}\equiv1;
\end{eqnarray}
 $n$ is the slope of the smoothed linear power spectrum,
\begin{eqnarray}
 n(k)\equiv\frac{d\log P_{\rm nw}^{\rm lin}(k)}{d\log k},
 \end{eqnarray}
and ${\bf a}=\{a_1,\ldots,a_9\}$, is a set of 9 free parameters to be fitted by comparison to N-body simulations. The function $Q_3(n)$ is defined as,
\begin{equation}
Q_3(n)\equiv\frac{4-2^n}{1+2^{n+1}}.
\end{equation}

\section{Bispectrum fitting formula for other triangular shapes}
\label{appendixb}
\begin{figure}
\centering
\includegraphics[clip=false, trim= 80mm 10mm 22mm 35mm,scale=0.25]{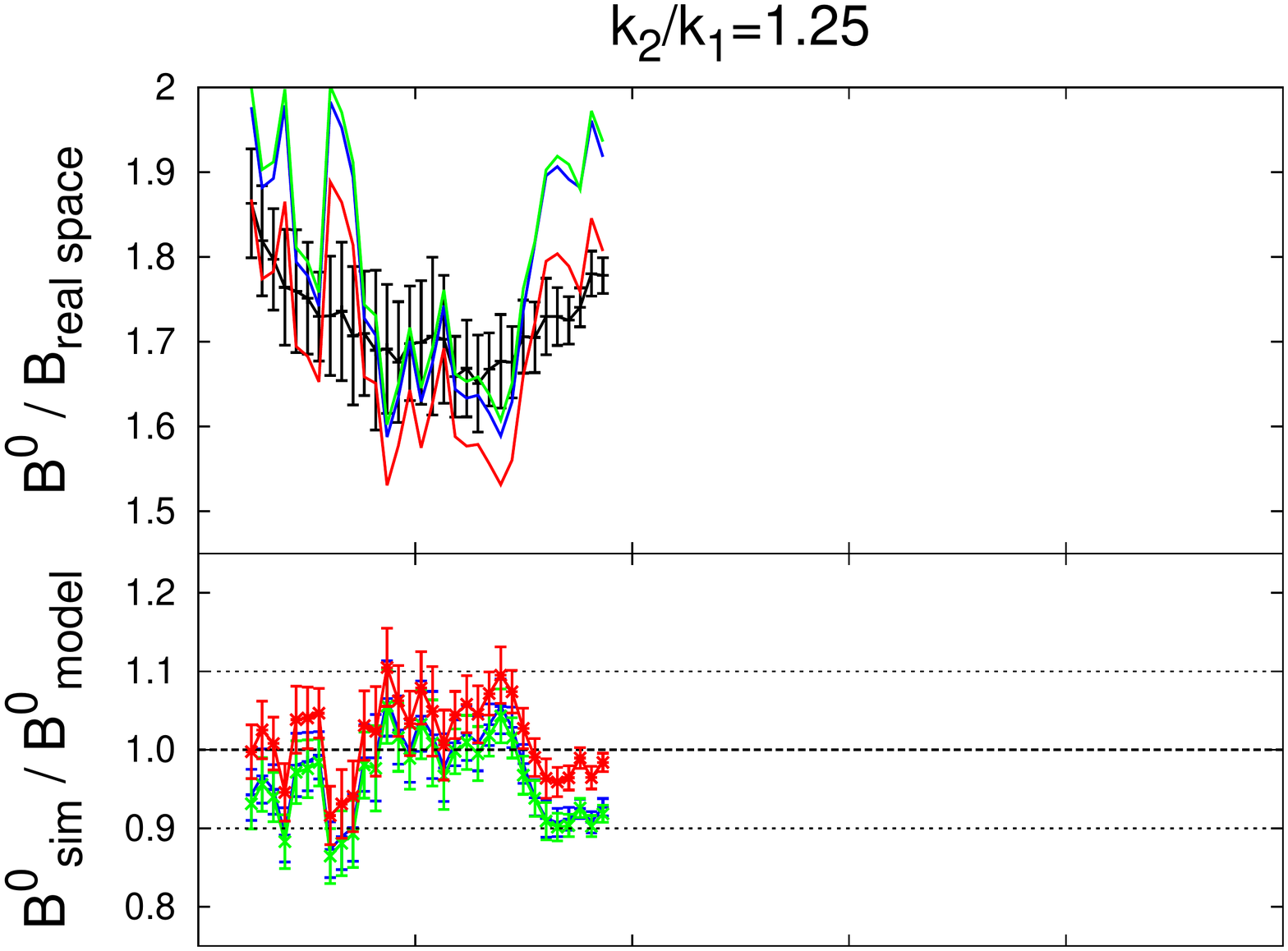}
\includegraphics[clip=false,trim= 25mm 10mm 22mm 35mm, scale=0.25]{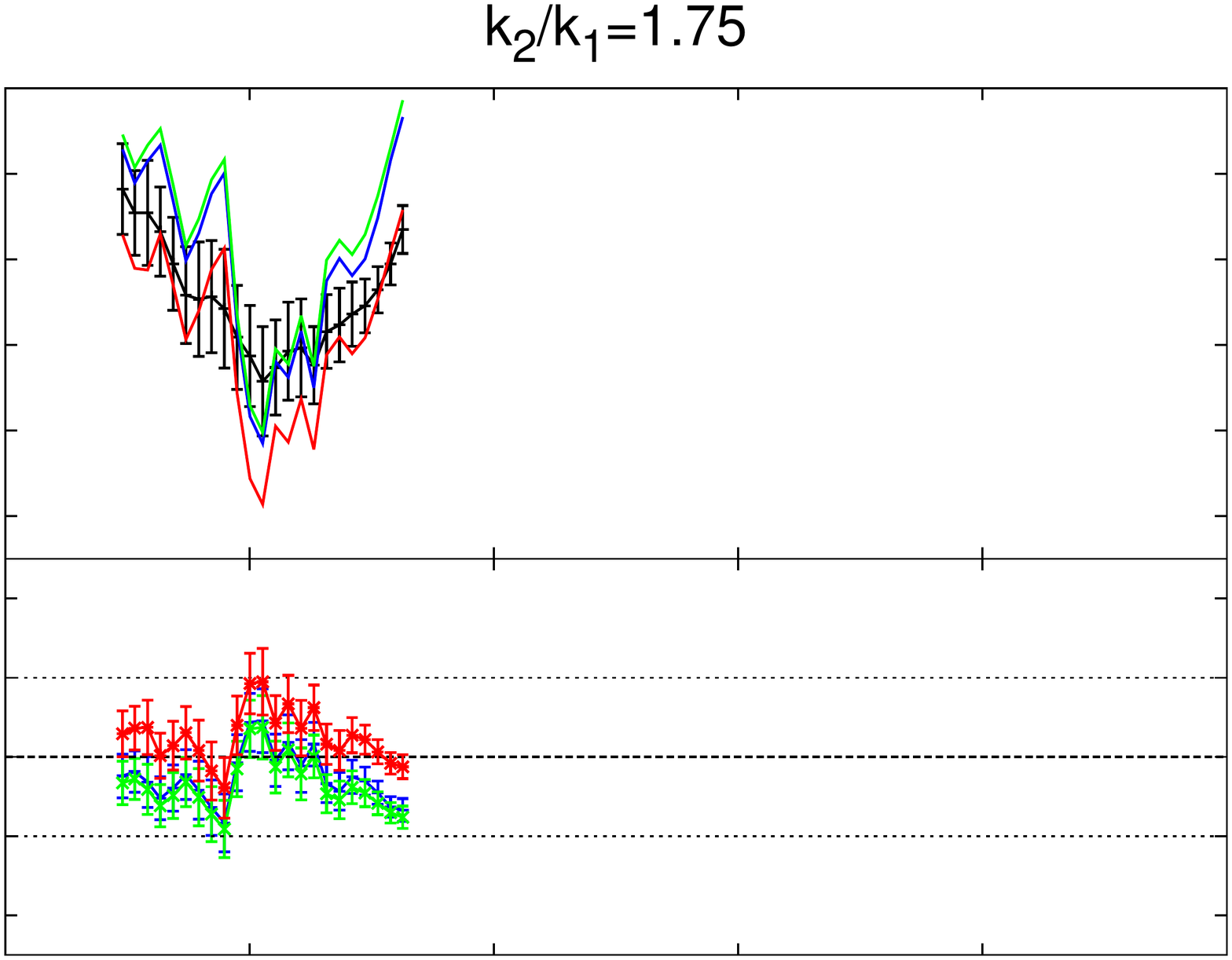}
\includegraphics[clip=false,trim= 25mm 10mm 80mm 35mm, scale=0.25]{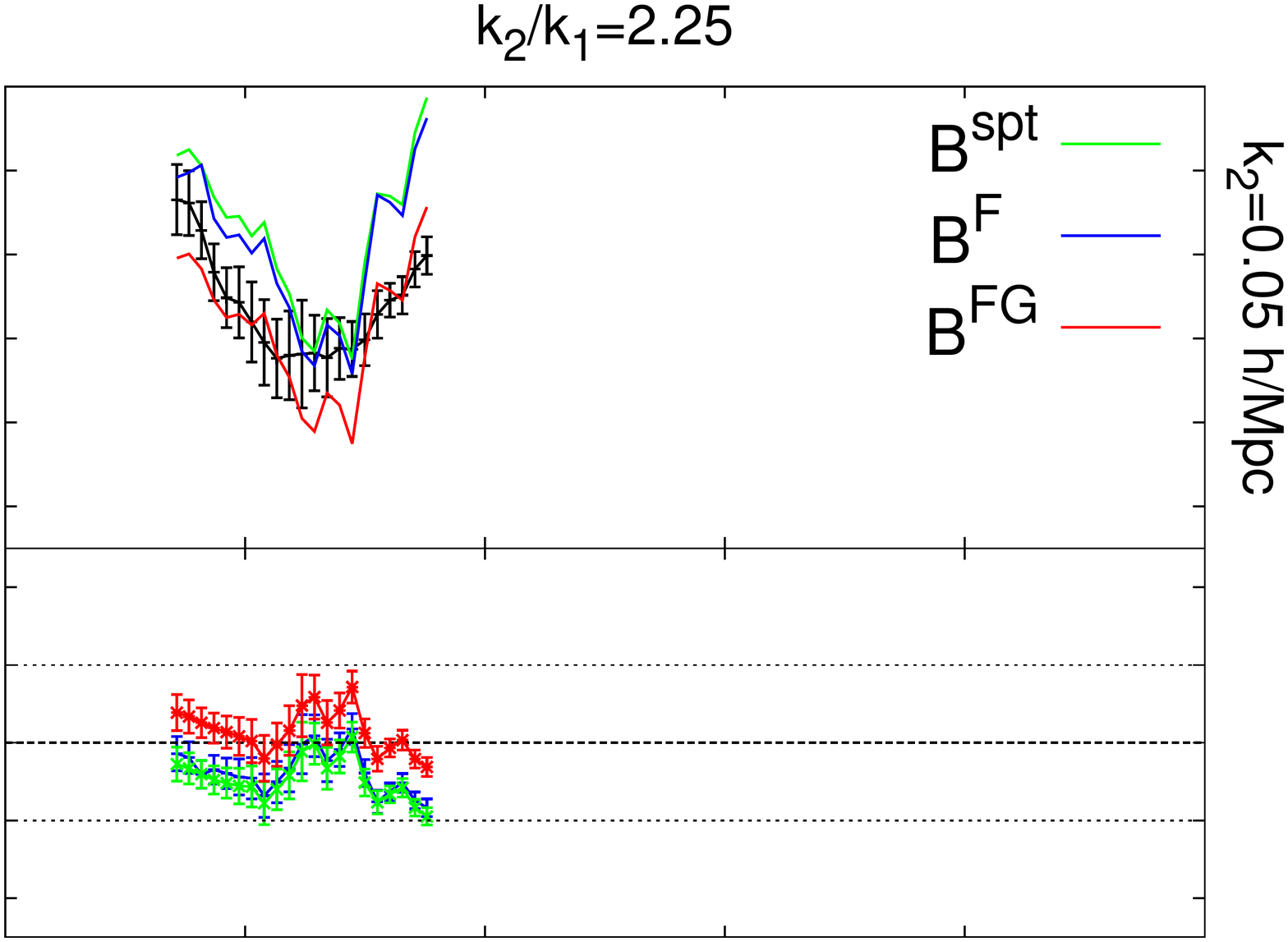}

\includegraphics[clip=false, trim= 80mm 10mm 22mm 35mm,scale=0.25]{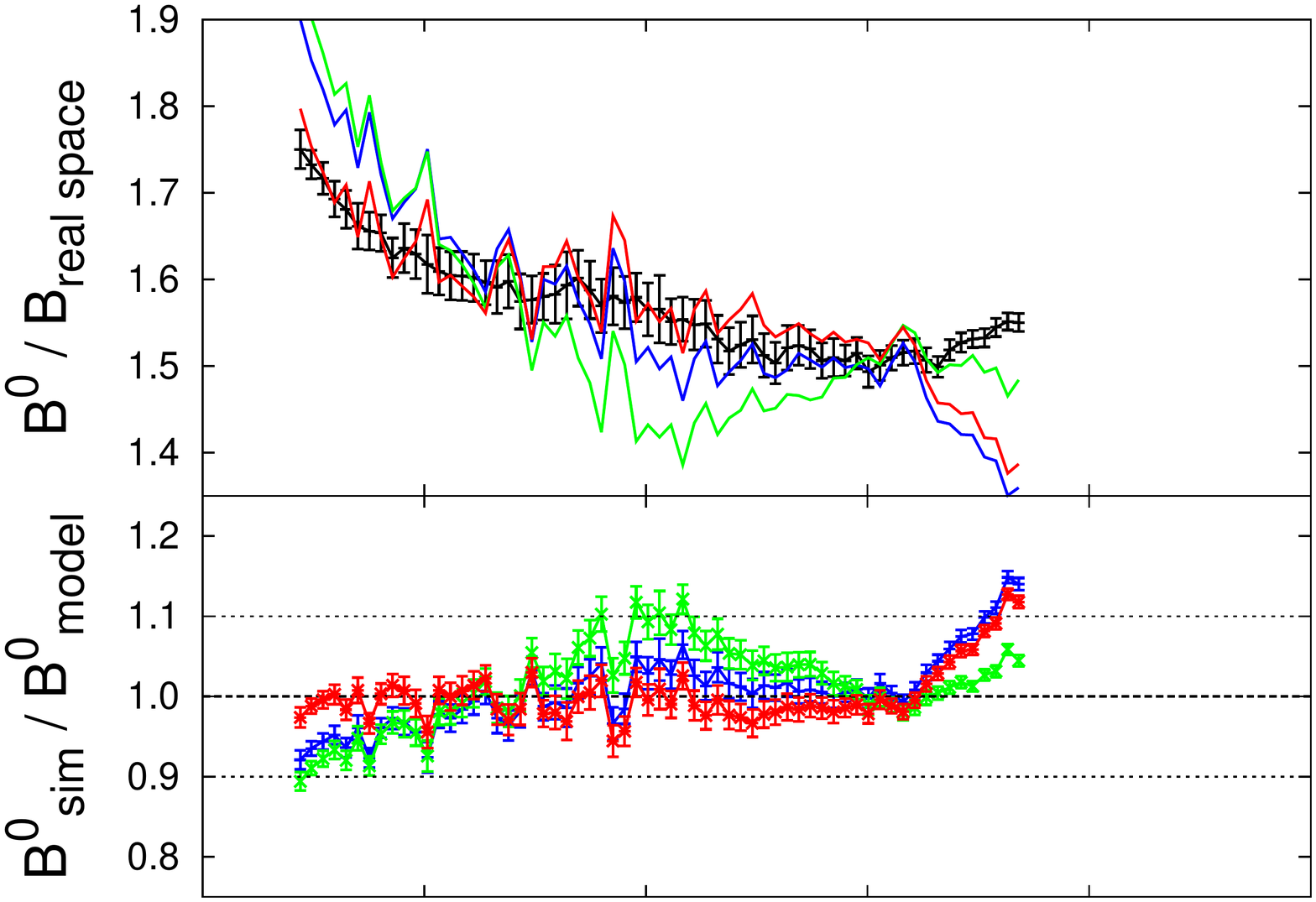}
\includegraphics[clip=false,trim= 25mm 10mm 22mm 35mm, scale=0.25]{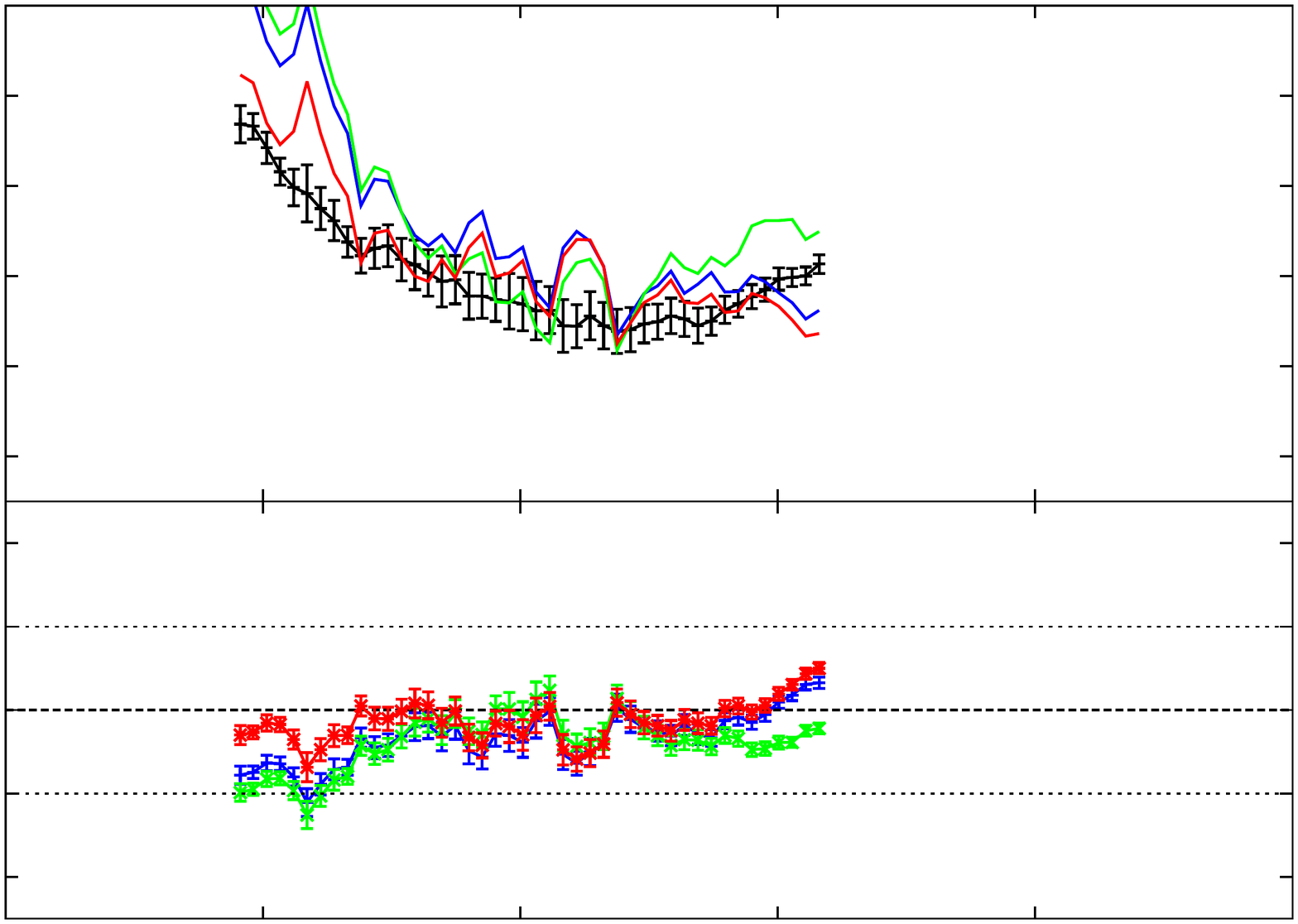}
\includegraphics[clip=false,trim= 25mm 10mm 80mm 35mm, scale=0.25]{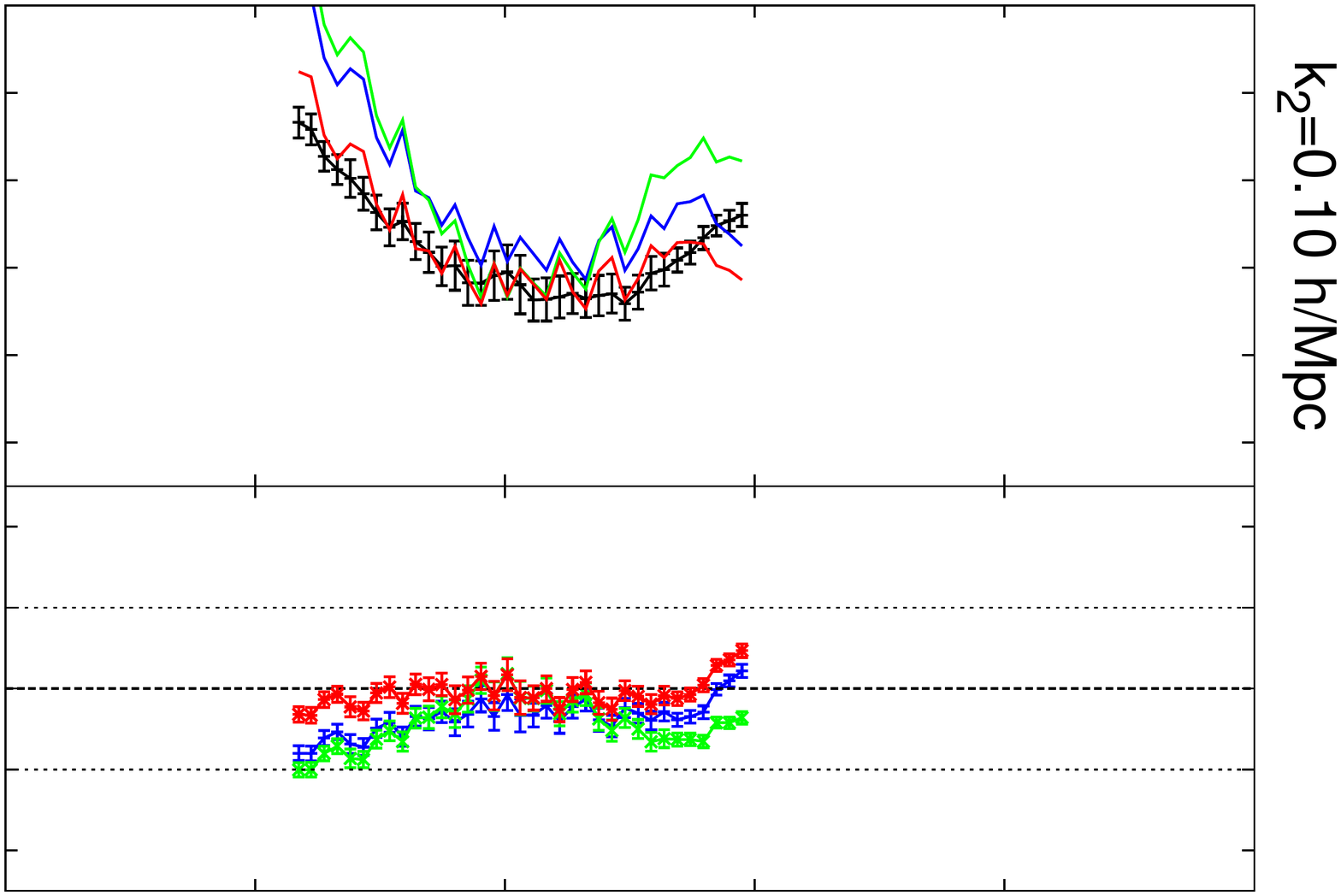}

\includegraphics[clip=false, trim= 80mm 10mm 22mm 35mm,scale=0.25]{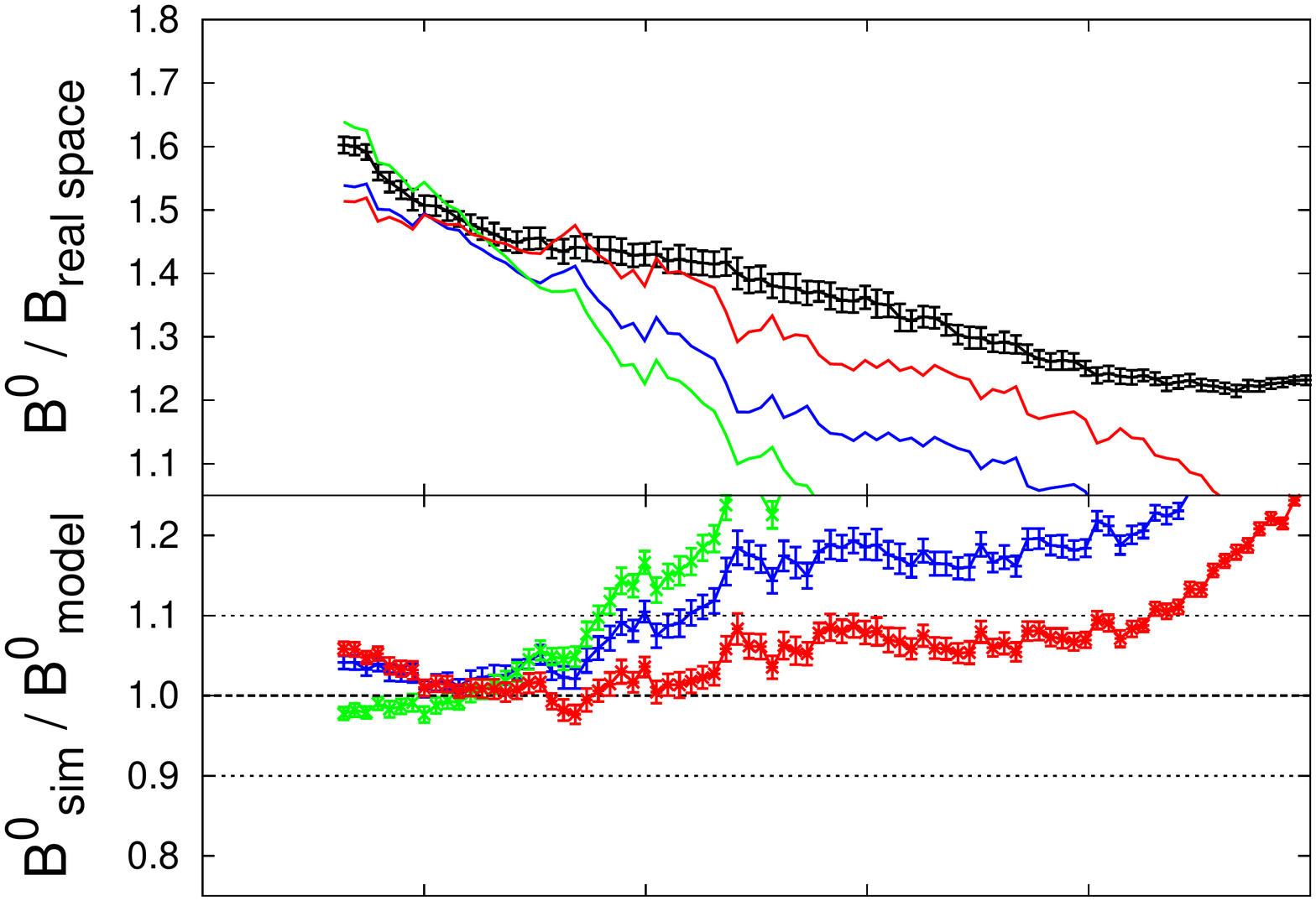}
\includegraphics[clip=false,trim= 25mm 10mm 22mm 35mm, scale=0.25]{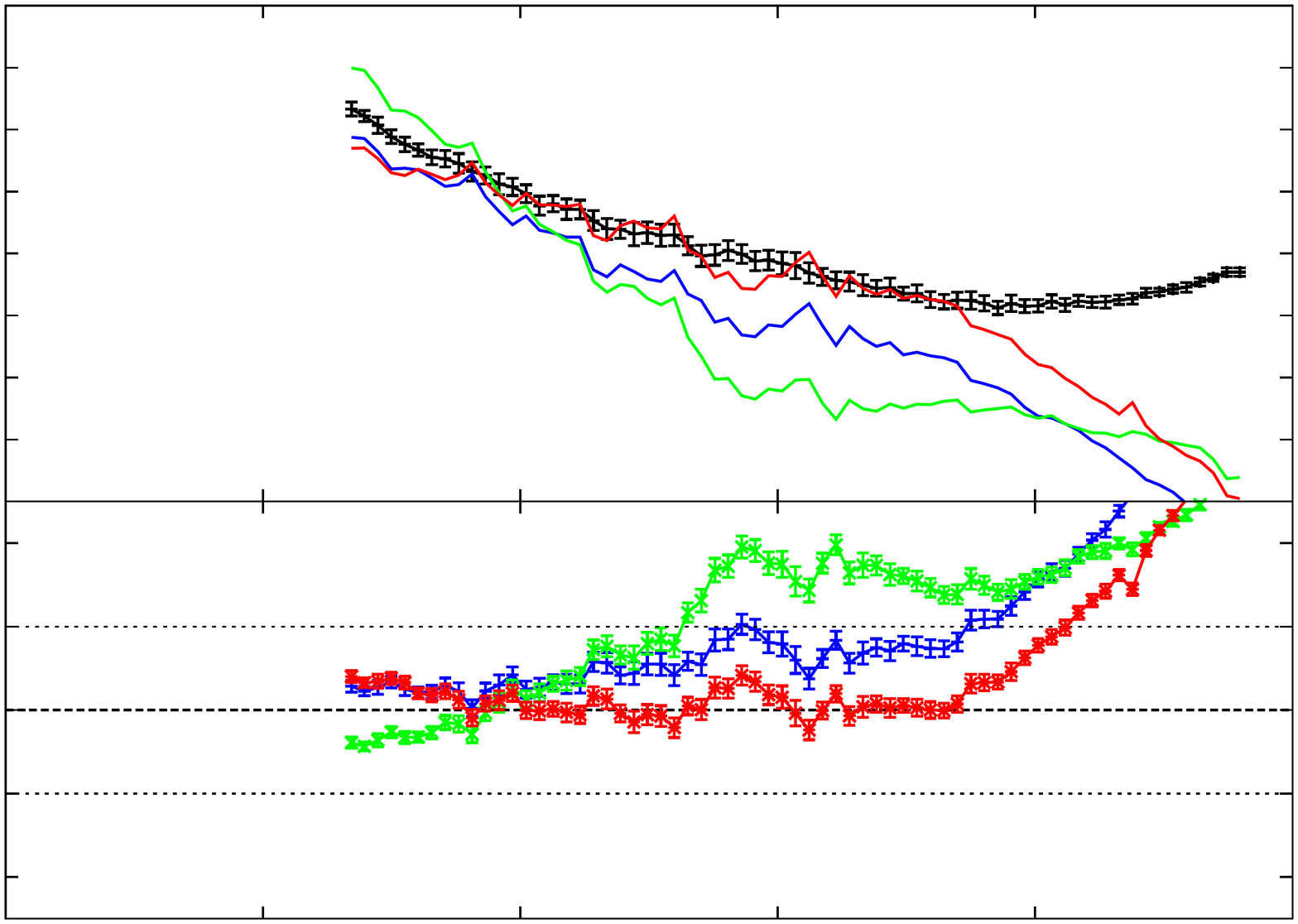}
\includegraphics[clip=false,trim= 25mm 10mm 80mm 35mm, scale=0.25]{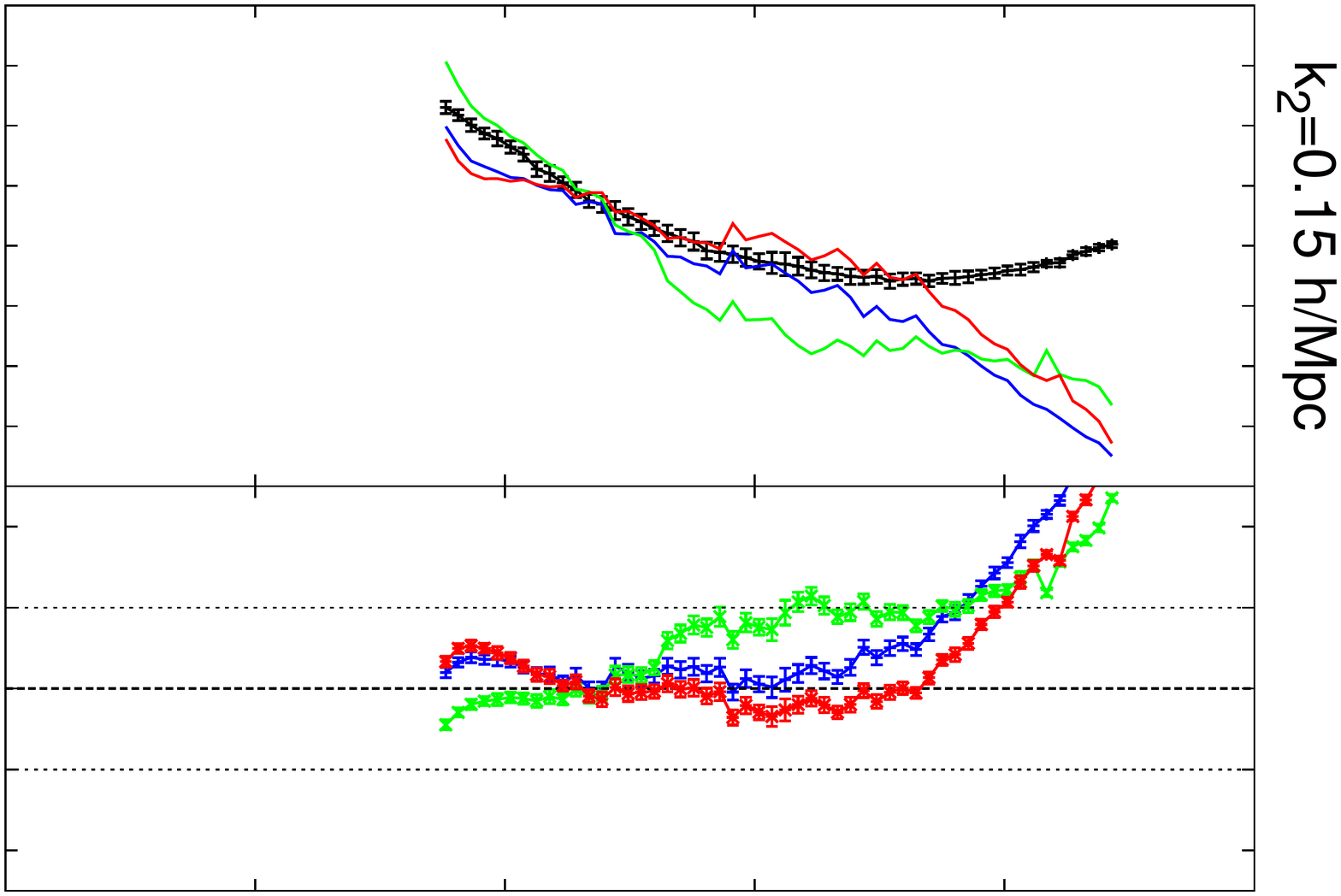}

\includegraphics[clip=false, trim= 80mm 10mm 22mm 35mm,scale=0.25]{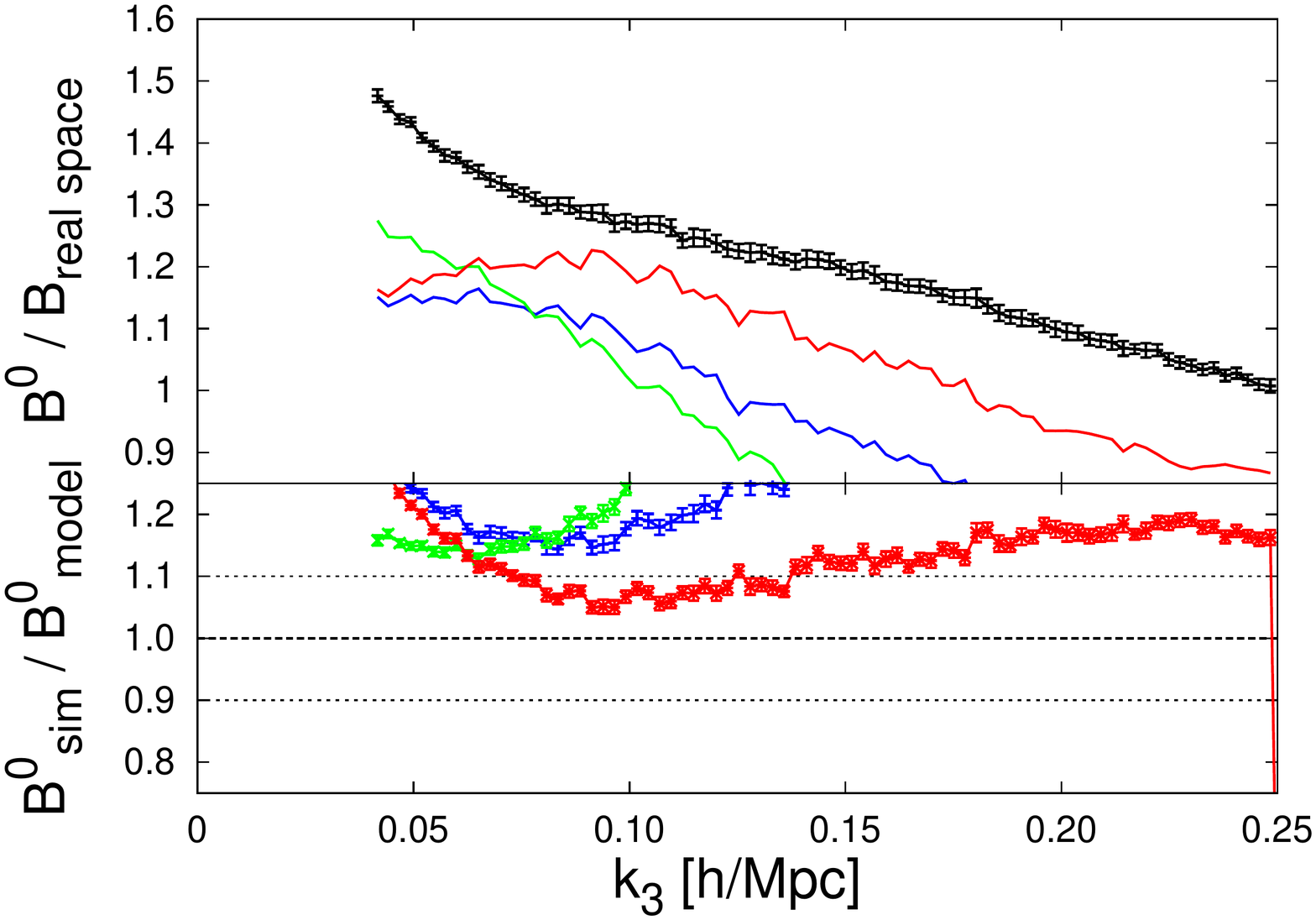}
\includegraphics[clip=false,trim= 25mm 10mm 22mm 35mm, scale=0.25]{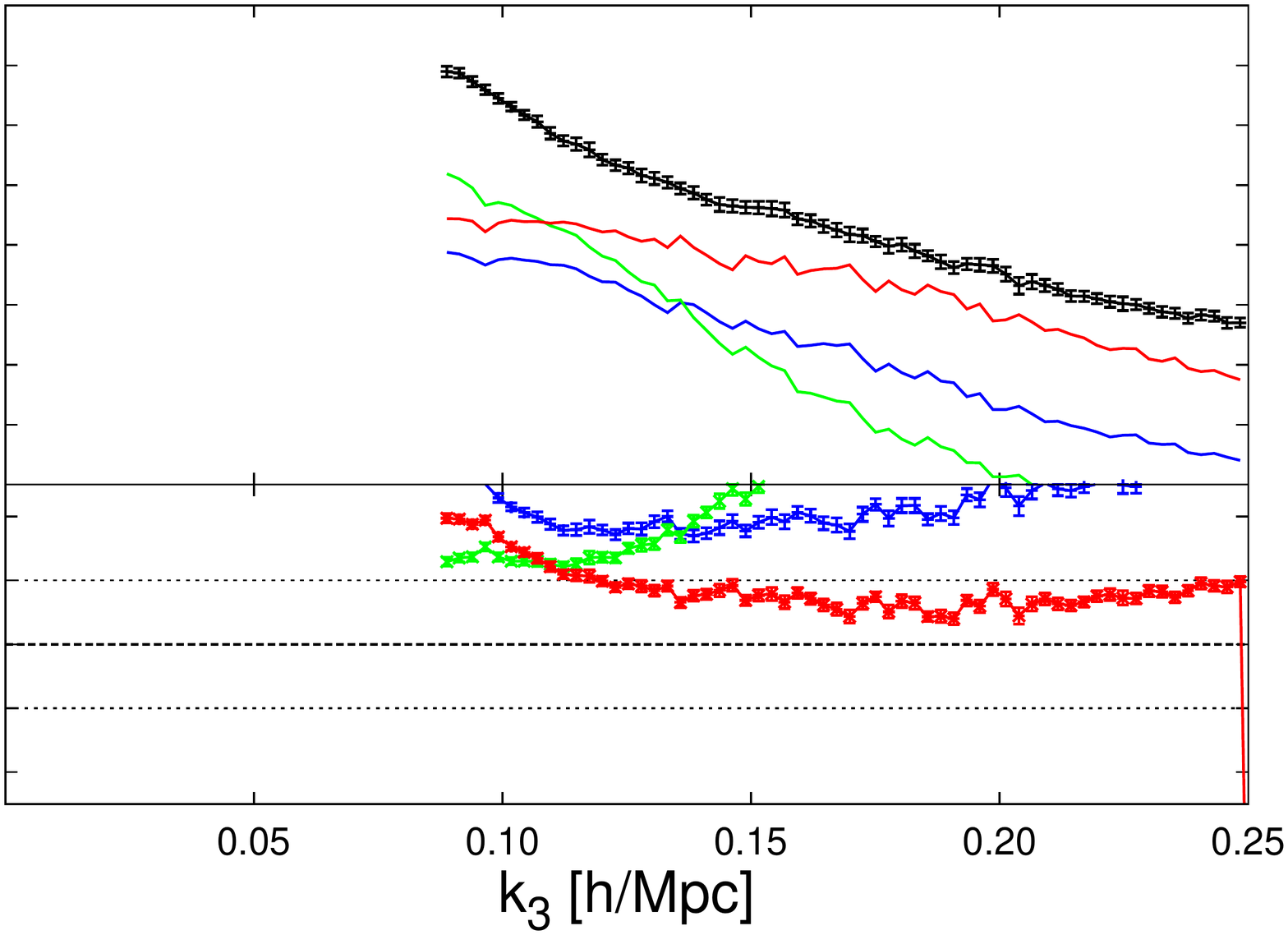}
\includegraphics[clip=false,trim= 25mm 10mm 80mm 35mm, scale=0.25]{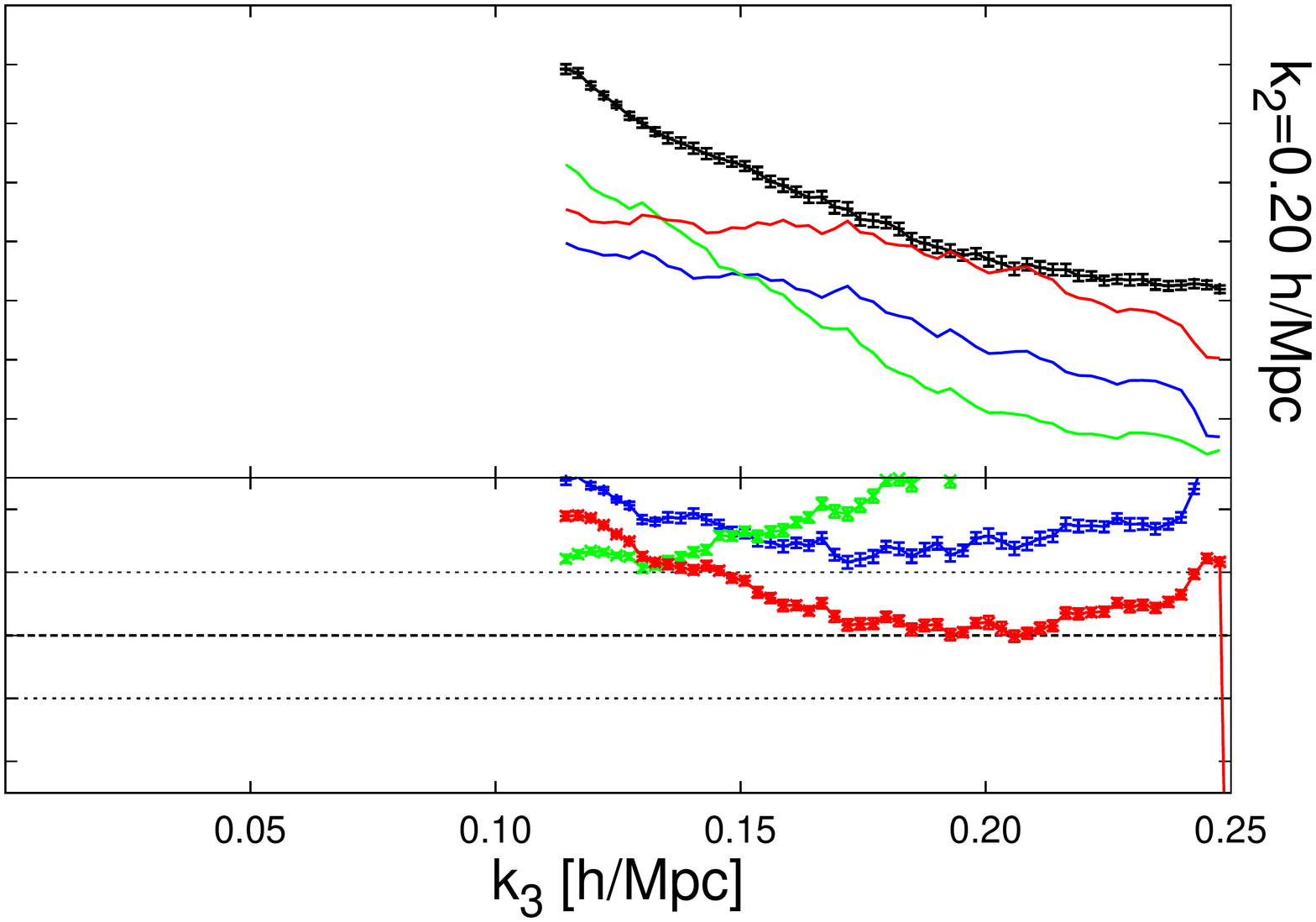}

\caption{Top sub-panels: dark matter monopole bispectrum normalised to the real space matter bispectrum for different triangle configurations. First column, second and third column panels are triangles with $k_2/k_1=1.25$, 1.75 and 2.25 respectively. Different rows show different scales: first, second, third and forth rows correspond to $k_2=0.05$, 0.10, 0.15 and 0.20\, $h$/Mpc as indicated. Black symbols correspond to N-body simulations whereas colour lines to the different models based on Eq.~\ref{Bspt}: $B^{\rm spt}$ (green lines), $B^{ \rm F}$ (blue line) and $B^{\rm FG}$ (red) (see text for description). Bottom sub-panels: relative deviation of the dark matter measurement to each of the  models. All panels  are at $z=0$.}
\label{bis_different_triangles}
\end{figure}

In Fig.~\ref{bis_different_triangles} we show the fitting formula for a set of triangles, $k_2/k_1=1.25$, 1.75, 2.25,  which have not been used for fitting the ${\bf a}^F$ / ${\bf a}^G$ parameters, at $z=0$. The colour notation is the same used in Fig.~\ref{bis1}-\ref{bis4}.  We see that the agreement level between the measured bispectrum and the $B^{FG}$ model is comparable to that observed in Fig.~\ref{bis1}. Therefore, we conclude that the fitting formula is able to describe the bispectrum of other triangular configurations, beyond those used for constraining the ${\bf a}^F$ and ${\bf a}^G$ parameters. 

\section{Information contained in the used shapes}
\label{variance_appendix}
In this appendix we estimate the amount of information contained in the triangular shapes used for constraining ${\bf a}^F$ and ${\bf a}^G$: $k_2/k_1=1.0,\,1.5,\, 2.0,\, 2.5$, which we refer as the {\it used}-set; relative to the information contained in all the shapes, which we refer as the {\it all}-set. In order to do this, we have measured all the shapes of the dark matter bispectrum of 60 different realisations in real space, and we have estimated the $b_1$ and $b_2$ parameters, setting all the remaining cosmological parameters to their true values. We have used the $F_2^{\rm eff}$ kernel with the fitting formula from \cite{bispectrum_fitting}.

\begin{figure}
\centering
\includegraphics[scale=.4]{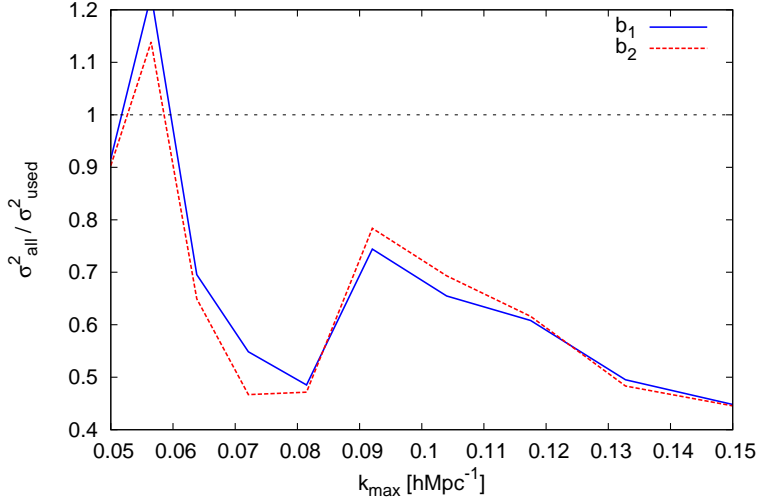}
\caption{Ratio of variances as a function of the maximum $k$ used of the bias parameters estimated from the measured bispectrum. We have plotted the variance of the  of the  {\it all}-set, which incorporates all possible triangular configurations, relative to the variance of the  {\it used}-set of triangles: $k_2/k_1=1.0,\,1.5,\, 2.0,\, 2.5$.  Blue solid is for $b_1$ and red dashed line for $b_2$, where the remaining parameters have been set to their true values.  }
\label{variances}
\end{figure}

 In Fig. \ref{variances} we show the ratios of variances of the set $k_2/k_1=1.0,\,1.5,\, 2.0,\, 2.5$, $\sigma^2_{\rm used}$, and the whole set of triangles, $\sigma^2_{\rm all}$,   for the bias parameters, $b_1$ (blue solid lines) and $b_2$ (red dashed lines), as a function of $k_{\rm max}$, where the condition $k_1\,\, \&\,\, k_2\,\, \&\,\, k_3 <k_{\rm max}$ has been imposed. For estimating the variances of $b_1$ and $b_2$, the methodology of \S\ref{sec:estimation} has been applied.  

We see that when $k_{\rm max}$ is small, both sets of triangles contain the similar information (the ratio of variances is close to 1), indicating that at these scales the set of triangles $k_2/k_1=1.0,\,1.5,\, 2.0,\, 2.5$ contains  very similar information that the whole set\footnote{Note that at large scales we obtain a ratio which is higher than 1. We interpret this as noise, since by definition we should always obtain $\sigma_{\rm all}\leq\sigma_{\rm used}$.}.  
As we extend the analysis to smaller scales the information content of the two sets of triangles becomes different. 
This behaviour seems contradictory, since one expects the correlation between triangles to be high at small scales and low at large scales (due to the mode coupling). On the other hand, we expect that when the values of $|{\bf k}_i|$ are close to the $k$-bin used (in our case close to the fundamental-$k$),  the number of fundamental triangles that fulfill the condition $k_2/k_1=1.0,\,1.5,\, 2.0,\, 2.5$ within the $k$-bin, is very close to the total number of fundamental triangles, just because the $k$-bin is of the order of magnitude of the size of the triangle. However, as we go to smaller scales, the values of $|{\bf k}_i|$ are much larger than the value of $k$-bin used, and consequently the total number of fundamental triangles that now fulfill the  condition $k_2/k_1=1.0,\,1.5,\, 2.0,\, 2.5$, is very different that the total number of fundamental triangles, and we expect that the information content of these to sets is more significant.  These arguments can be applied to  any dark matter tracer, since they are based in geometrical aspects of the estimation of the bispectrum and not on the actual clustering and distribution of particles. 

In Fig. \ref{variances} we see that when $k_{\rm max}\simeq0.1\,h/{\rm Mpc}$, the ratio of the variances is about 0.7. If we assume that the information is proportional to the variance, then we can say that the set of $k_2/k_1=1.0,\,1.5,\, 2.0,\, 2.5$ contains about $\sim70\%$ of the information of the whole set of triangles at $k\leq0.1\,h{\rm Mpc}^{-1}$. This number reduces as we go to smaller scales, and reaches $\sim45\%$ for $k\leq0.15\,h{\rm Mpc}^{-1}$. We have not explored smaller scales due to computational time limitations: as we go to smaller scales, the number of total triangles scales with $\sim k^6$, and therefore the $k$-space to explore increases with this power law. Thus, Fig. \ref{variances} provides useful information about the order of magnitude of information lost by just selecting the $k_2/k_1=1.0,\,1.5,\, 2.0,\, 2.5$ set of shapes.

\section{Bispectrum fitting formula for cosmologies with different $\Omega_m$-values}\label{cosmo_appendix}

In this section we show how the $B^{FG}$ fitting formula presented in \S\ref{bis_section} is able to describe the redshift space dark matter bispectrum corresponding to cosmologies with different values of $f$. Since ${\bf a}^F$ and ${\bf a}^G$ have been fitted simultaneously to different epochs within the range $0\leq z \leq 1.5$, $B^{FG}$ is already designed to fit model with different fluctuation amplitudes, i.e. $\sigma_8(z)$\footnote{Also the fitting formula in real space \citep{bispectrum_fitting} was proved to describe with $\lesssim10\%$ accuracy the dark matter bispectrum of different cosmologies with $\sigma_8=0.7913\,, 0.834\,, 0.878\,,0.944$.}. The same thing occurs with $f$: since $f$ changes with $z$, ${\bf a}^G$ was calibrated using bispectra with different values of $f$. However,the variation that redshift evolution induces on the parameters $f$ and $\sigma_8$ is not independent, since both parameters vary at the same time. In order to isolate the effect of varying $f$ from the effect of varying the power spectrum amplitude,  we need to change $\Omega_m$ while keeping the power spectrum in real space constant. As the shape of the power spectrum is determined by $n_s$, $\Omega_b h^2$, and $\Omega_m h^2$, we keep these parameters constant when varying $\Omega_m$, whereas $\sigma_8$ is chosen such that  the amplitude of the linear power spectrum in real space at $z=0$ is the same when distances are expressed in Mpc instead of ${\rm Mpc}/h$. With this normalisation of the power spectrum, the dimensionless quantities $q=k/k_{nl}$ and the logarithmic slope of the power spectrum $n(k)$, which enter the functions $a$, $b$, and $c$ (see Eq.~\ref{abc_new}), remain unchanged.

In particular, we choose two values of $\Omega_m$: a lower value than the fiducial cosmology (Sim DM cosmology in Table \ref{table_sims}), $\Omega_m=0.2$; and a higher value, $\Omega_m=0.4$; We consider that these two extra cases cover a sufficiently broad range of $f$-values for any realistic analysis. The details about these two extra cosmological models and the fiducial model are listed in Table~\ref{table_sims_cosmo}. The simulations using the low- and high-$\Omega_m$ cosmology have the same box size in units of Mpc and the same mass resolution  as that of the fiducial cosmology described in Table \ref{table_sims}. 
In order to reduce the computational time, we run only a single realisation for each of these new cosmologies, but keep the random realization of the initial conditions the same as one particular realisation of the fiducial cosmology. Thus, by considering the relative difference in the statistical moments of these different simulations the uncertainties due to sample variance are mostly cancelled, and hence the statistical errors are significantly reduced. 
\begin{table}[htdp]
\begin{center}
\begin{tabular}{|c|c|c|c|c|c|c|c|c|c|c|}
\hline
 & $\Omega_\Lambda$ & $\Omega_m$ & $h$ & $\Omega_m h^2$   & $\sigma_8$ & $f$  \\
 \hline
 \hline
 low-$\Omega_m$ & 0.80 & 0.20 & 0.8133 & 0.1323  & 0.872 & 0.413  \\
 \hline
 high-$\Omega_m$ & 0.60 & 0.40 & 0.5751 & 0.1323  & 0.693 & 0.604  \\
\hline
 fiducial & 0.73 & 0.27 & 0.7000 & 0.1323  & 0.791 & 0.483\\
 \hline
\end{tabular}
\end{center}
\caption{Parameter of the different cosmology models studied. All of them shares the same value of  $\Omega_m h^2$, and $\sigma_8$ has been chosen such that the linear power spectrum in real space is the same for all the cosmologies. However, the different value of $\Omega_m$, induces different clustering in redshift space. The rest of cosmological parameters are kept to the values of Table~\ref{table_sims}.}
\label{table_sims_cosmo}
\end{table}

\begin{figure}
\centering
\includegraphics[scale=0.4]{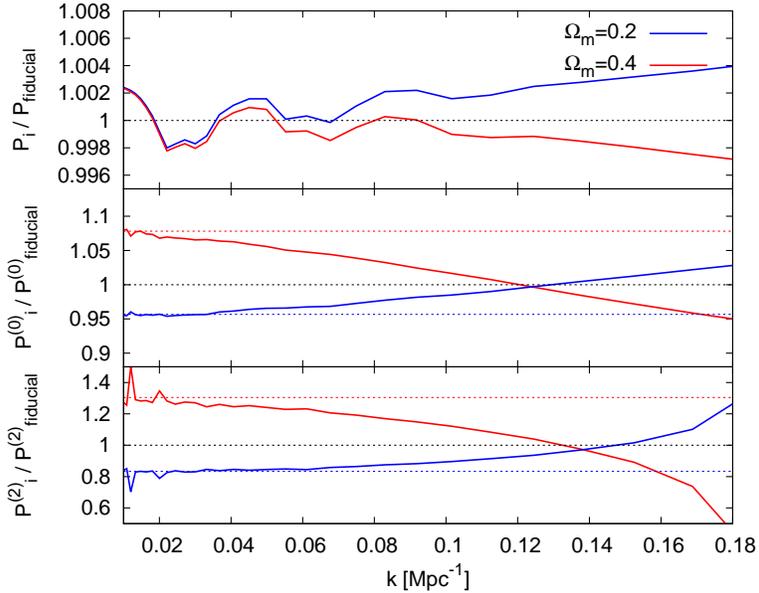}

\caption{Ratio of the power spectra of the cosmologies listed in Table~\ref{table_sims_cosmo}: low-$\Omega_m$ (blue lines), high-$\Omega_m$ (red lines),  respect to the power spectrum of the fiducial cosmology. Top panel shows the real space power spectra, middle panel the redshift space monopole, bottom panel the redshift space quadrupole. In middle and bottom panels the Kaiser predictions have been added in colour lines, following the same colour notation. We observe that at large scales, the Kaiser prediction matches with the results from simulations.   Distances expressed in Mpc. All panels at $z=0$.
}
\label{PS_P0_P2}
\end{figure}

In Fig. \ref{PS_P0_P2} we show the power spectra for the cosmologies listed in the two first rows of Table~\ref{table_sims_cosmo} divided by the power spectrum of the fiducial cosmology (third row), in real space (top panel), and the corresponding redshift space multipoles: the monopole (middle panel) and quadrupole (bottom panel). The different colour lines denote the different cosmologies: low-$\Omega_m$ (blue lines) and high-$\Omega_m$ (red lines), as labelled. All panels are at $z=0$. Note that the units in which distances are expressed are Mpc (and not ${\rm Mpc}/h$). From the top panel of Fig. \ref{PS_P0_P2} we can see that the power spectra of these three cosmologies present differences that are very small, especially at large scales where they are almost indistinguishable ($\leq0.2\%$). At mildly non-linear scales some differences arise due to the different growth histories. However, the differences between them are always $\leq 0.5\%$ for $k\leq0.18\,{\rm Mpc}^{-1}$. This is the expected behaviour, since $\sigma_8$ was chosen for each cosmology to match the linear power spectrum  of the fiducial cosmology at $z=0$. On the other hand, in the middle and bottom panels we show how the clustering in redshift space is significantly different due to the different value of $f$. On large scales, we observe (colour dashed lines) the expected differences induced by the change in the Kaiser factors $1+2/3 f + 1/5 f^2$ and $4/3 f + 4/7 f^2$ for the monopole and quadrupole, respectively.

\begin{figure}
\centering
\includegraphics[clip=false, trim= 80mm 10mm 22mm 35mm,scale=0.252]{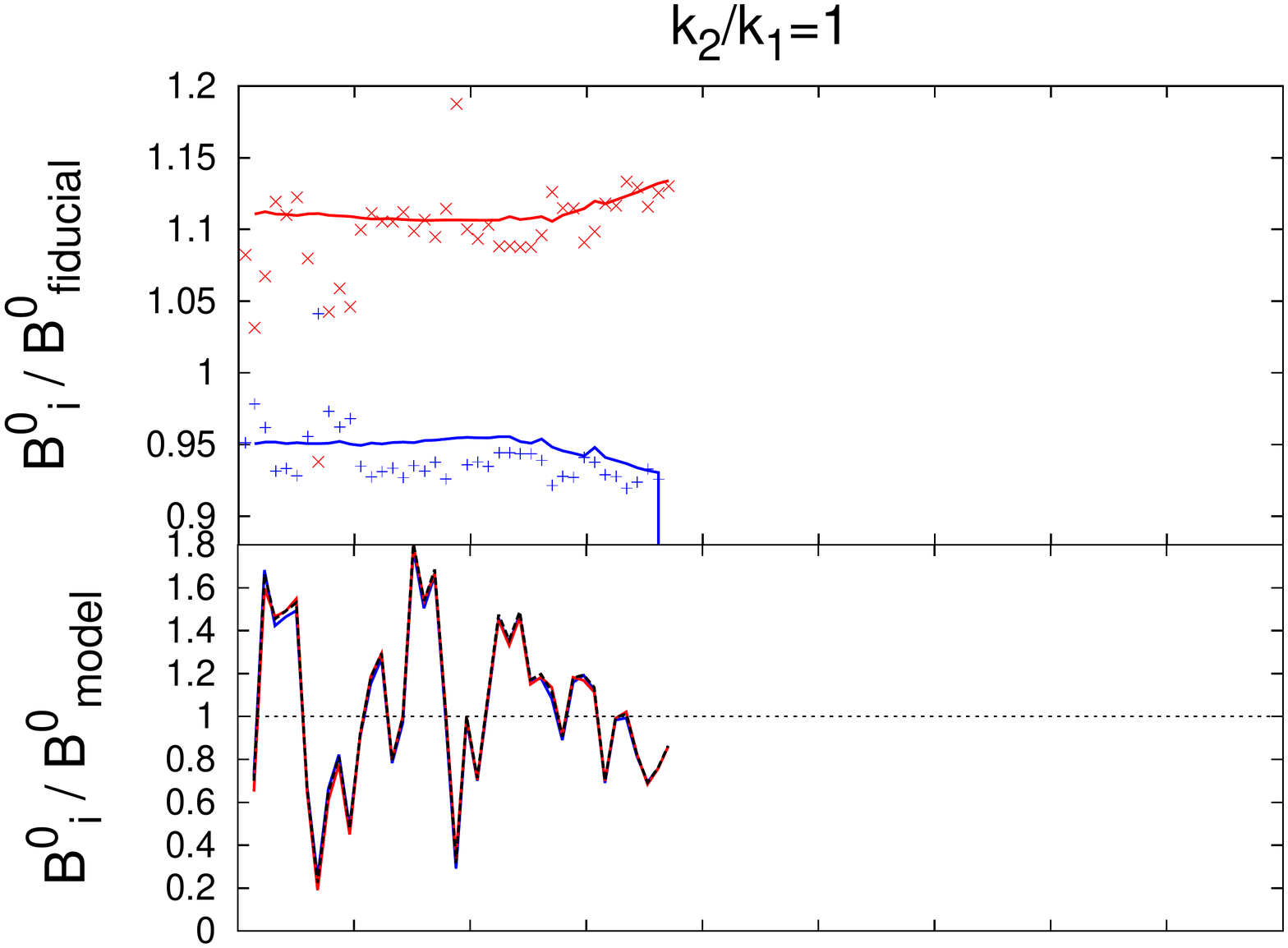}
\includegraphics[clip=false,trim= 25mm 10mm 22mm 35mm, scale=0.252]{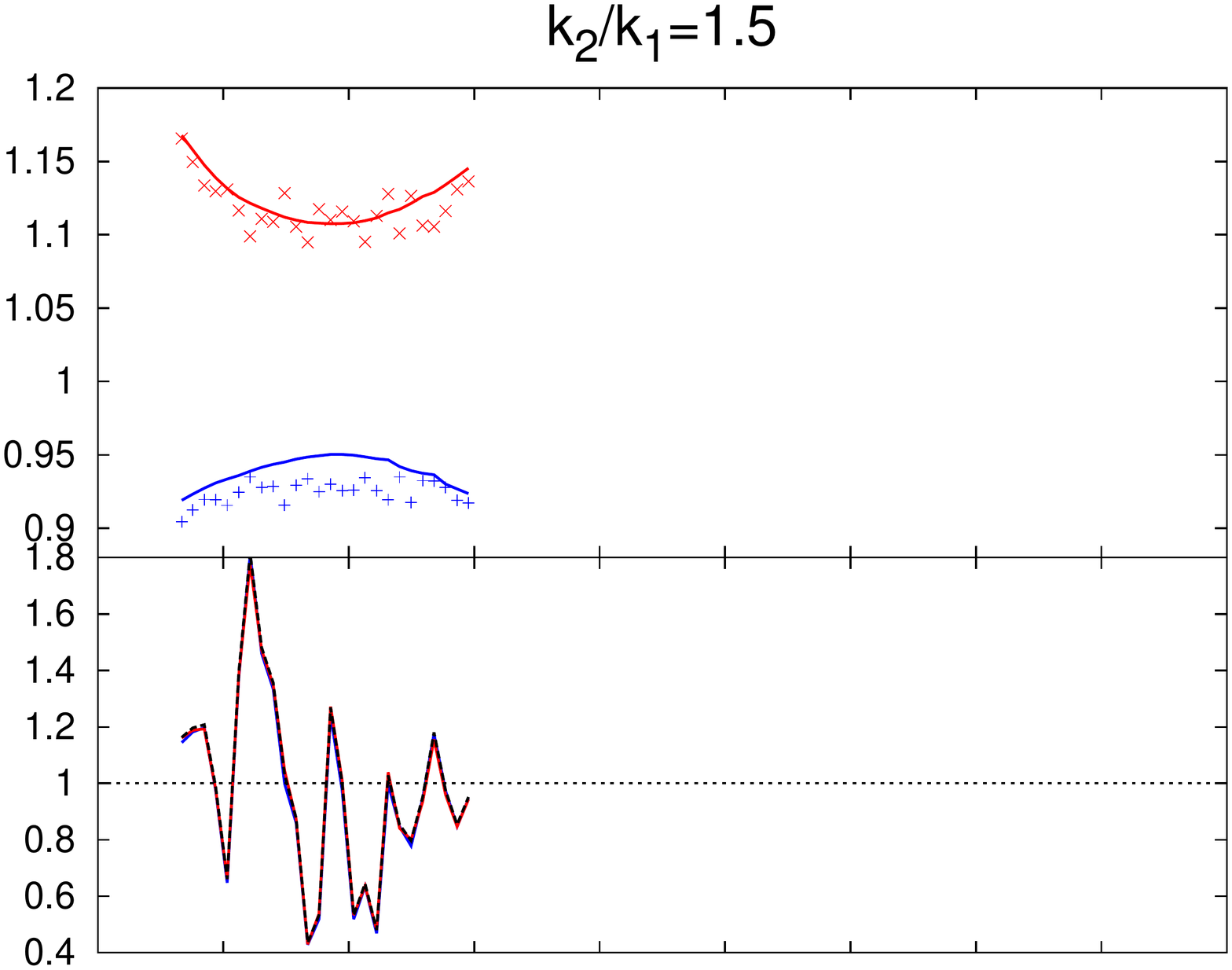}
\includegraphics[clip=false,trim= 25mm 10mm 80mm 35mm, scale=0.252]{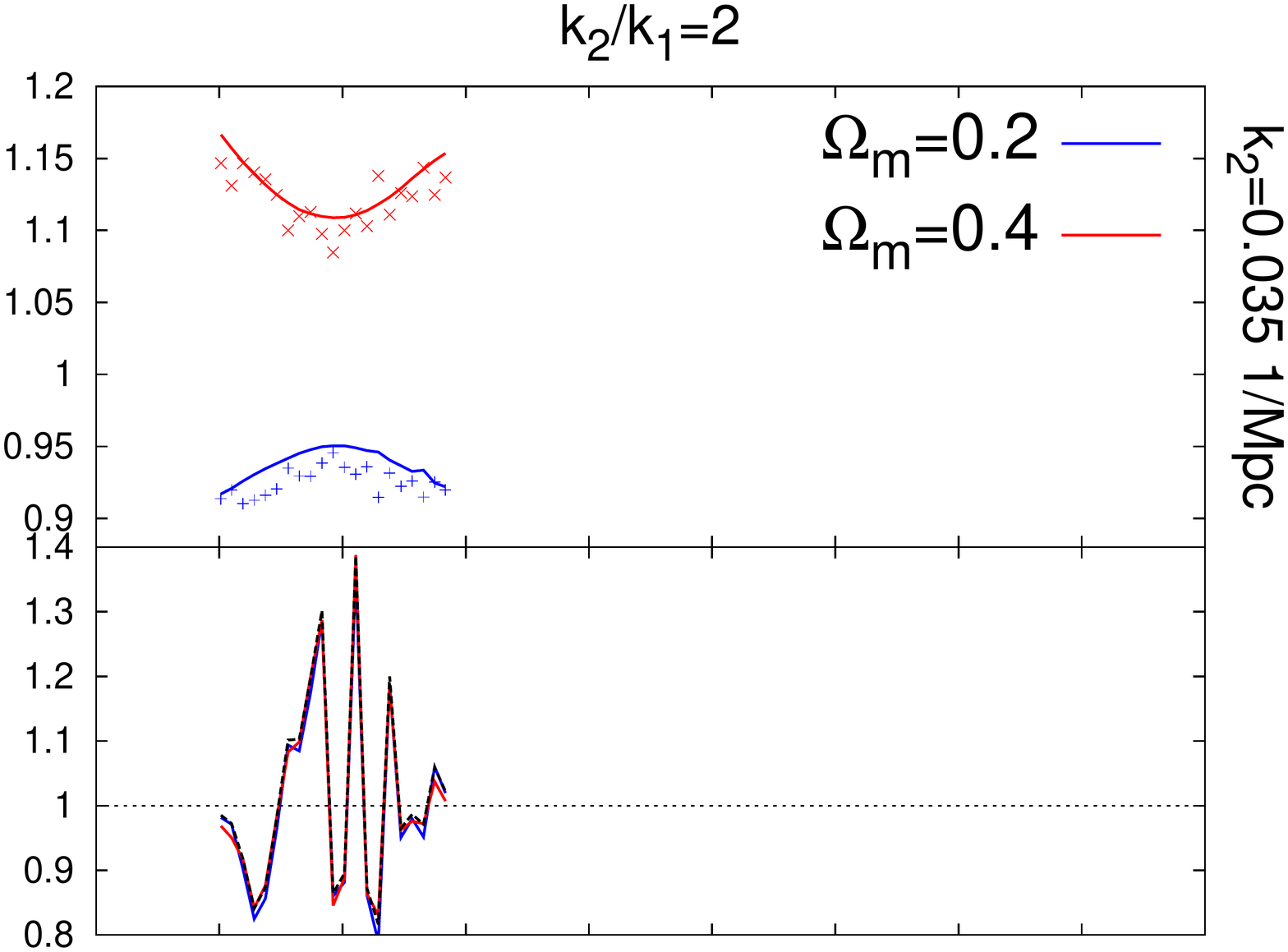}

\includegraphics[clip=false, trim= 80mm 10mm 22mm 35mm,scale=0.252]{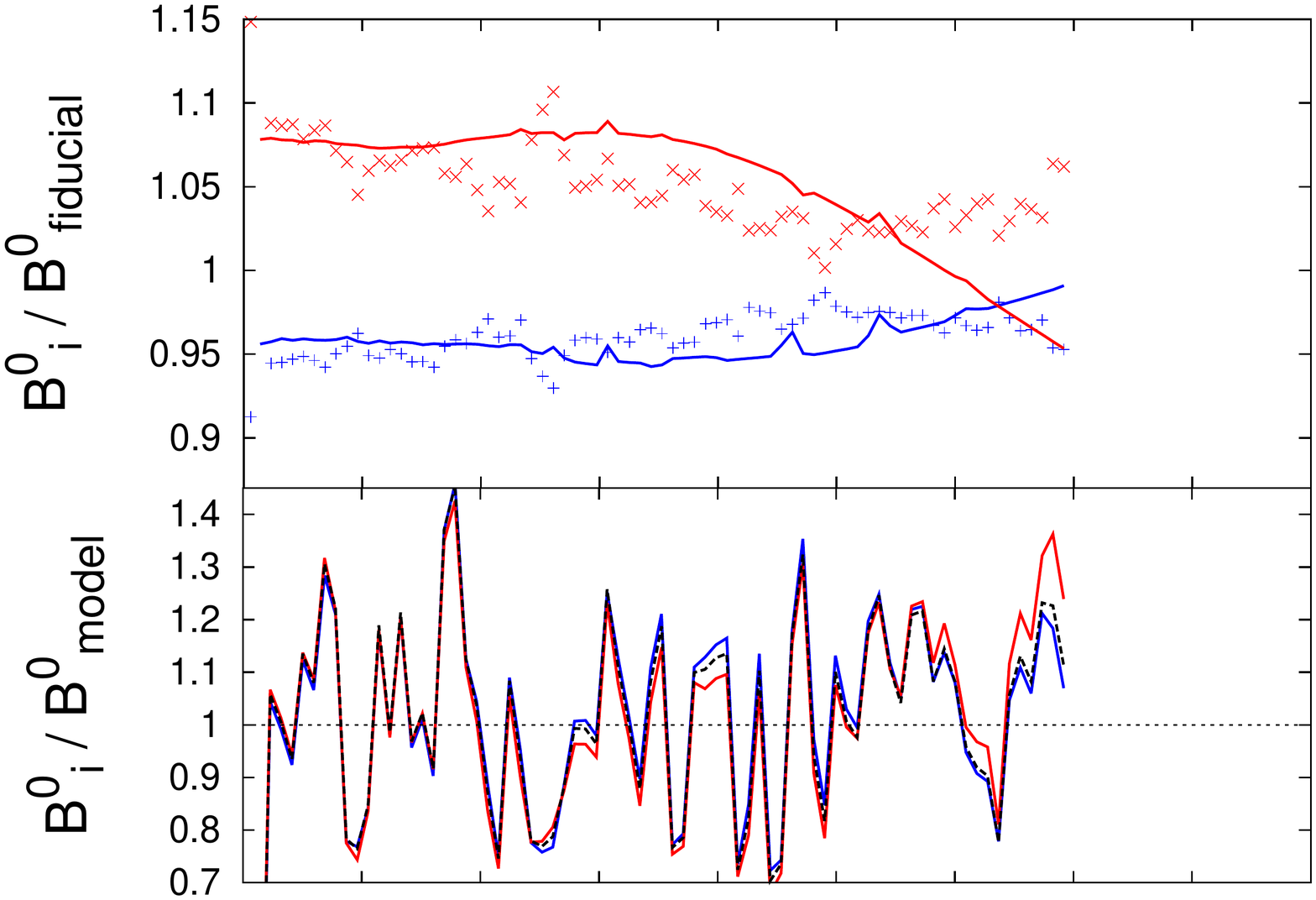}
\includegraphics[clip=false,trim= 25mm 10mm 22mm 35mm, scale=0.252]{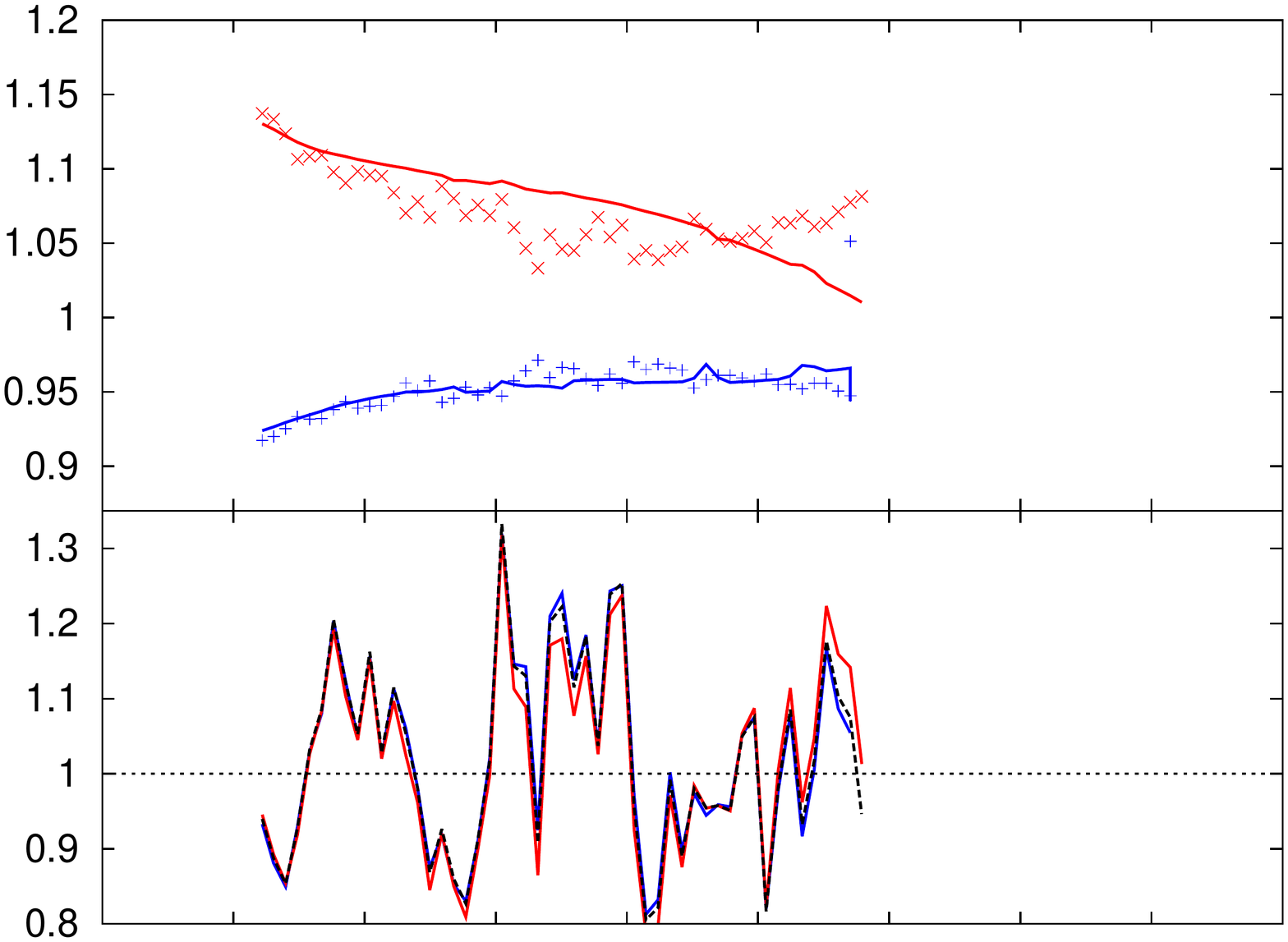}
\includegraphics[clip=false,trim= 25mm 10mm 80mm 35mm, scale=0.252]{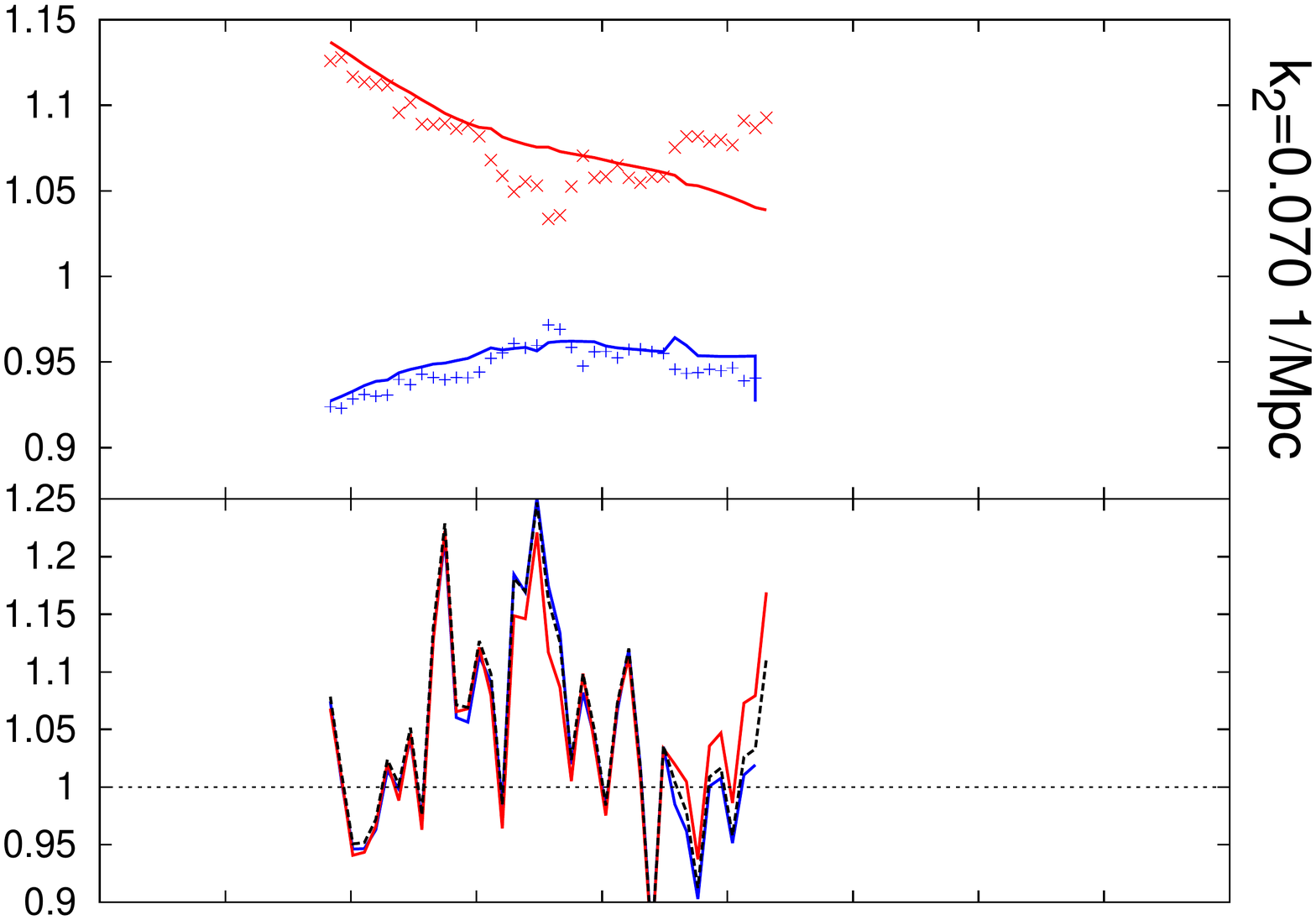}

\includegraphics[clip=false, trim= 80mm 10mm 22mm 35mm,scale=0.252]{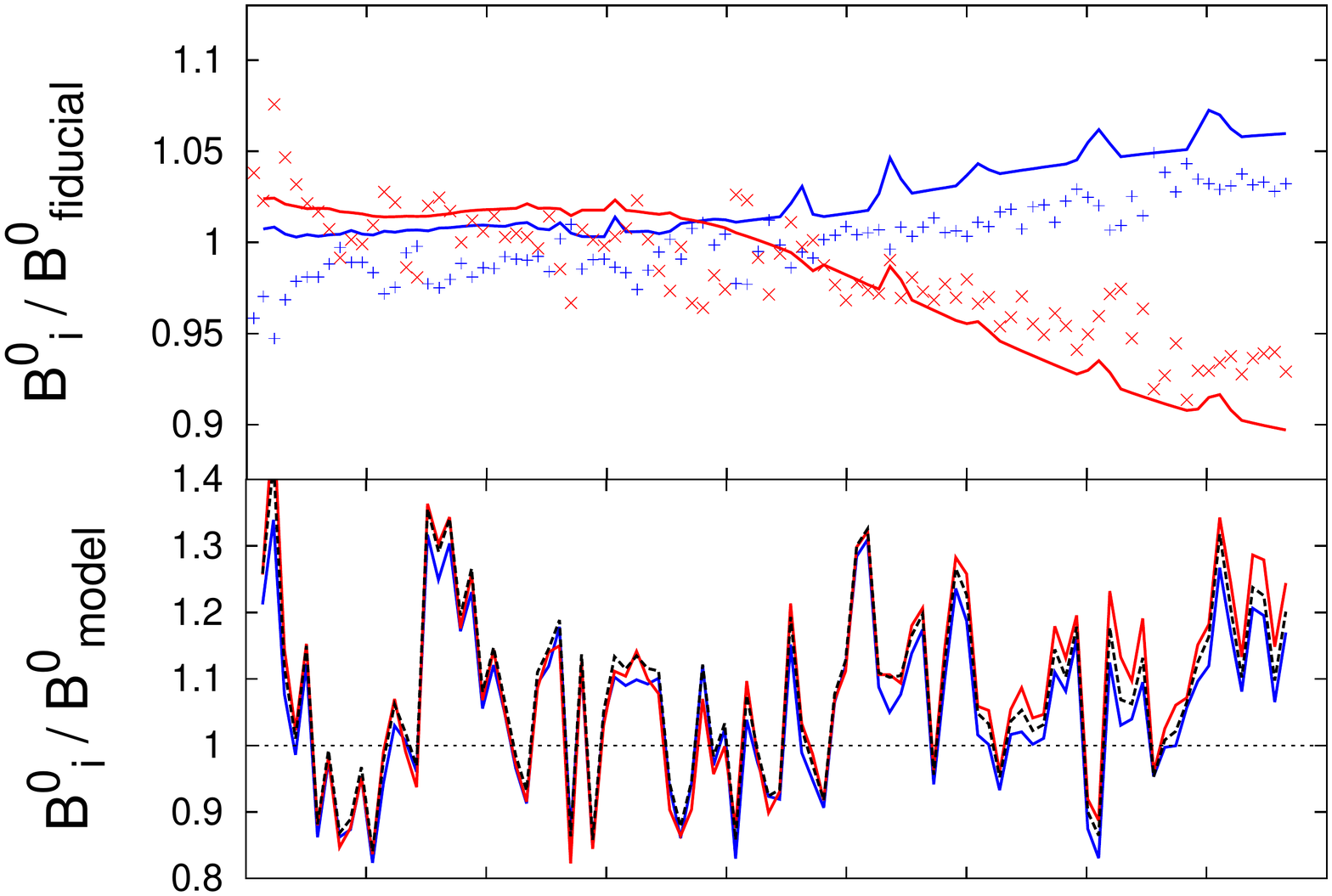}
\includegraphics[clip=false,trim= 25mm 10mm 22mm 35mm, scale=0.252]{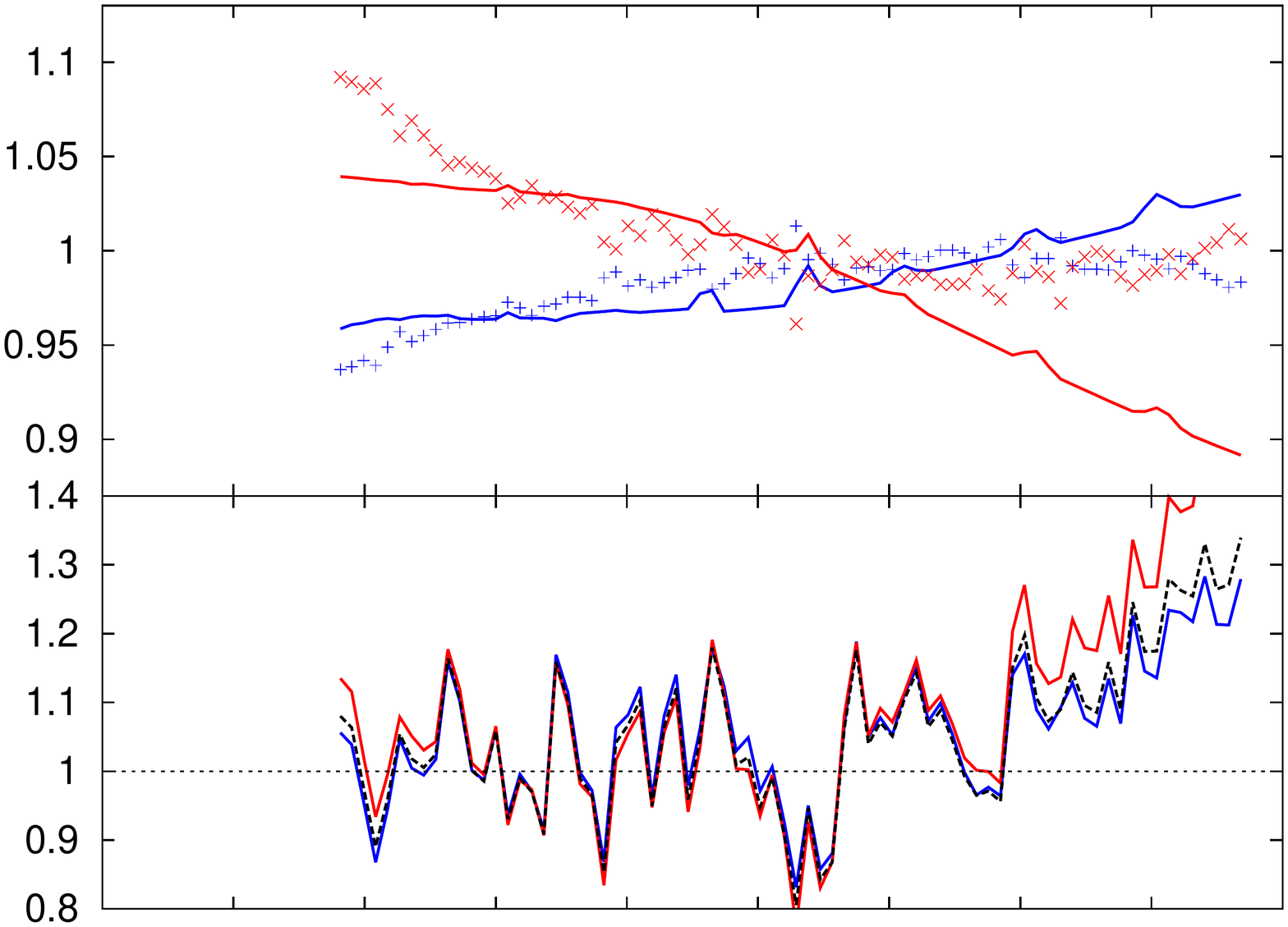}
\includegraphics[clip=false,trim= 25mm 10mm 80mm 35mm, scale=0.252]{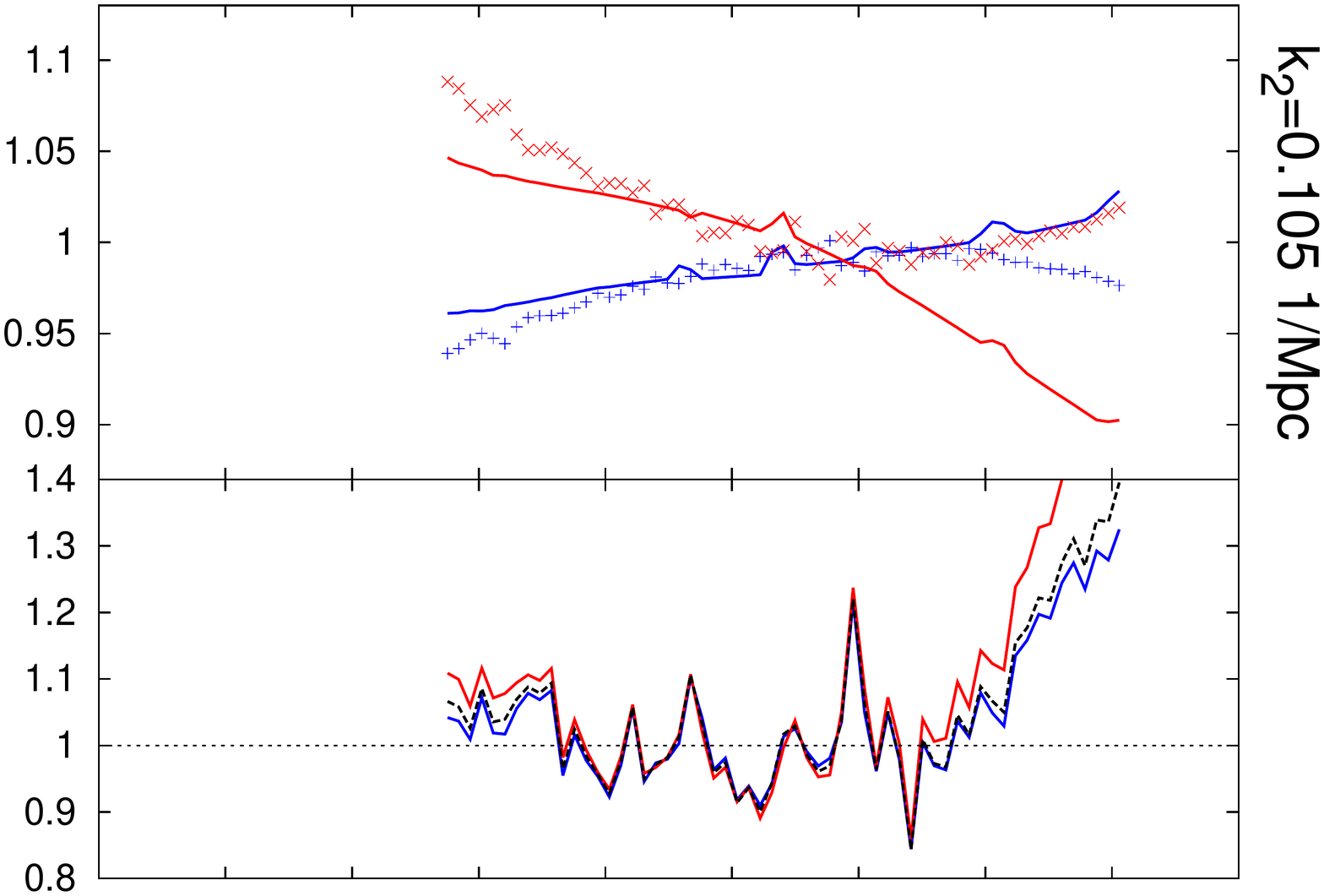}

\includegraphics[clip=false, trim= 80mm 10mm 22mm 35mm,scale=0.252]{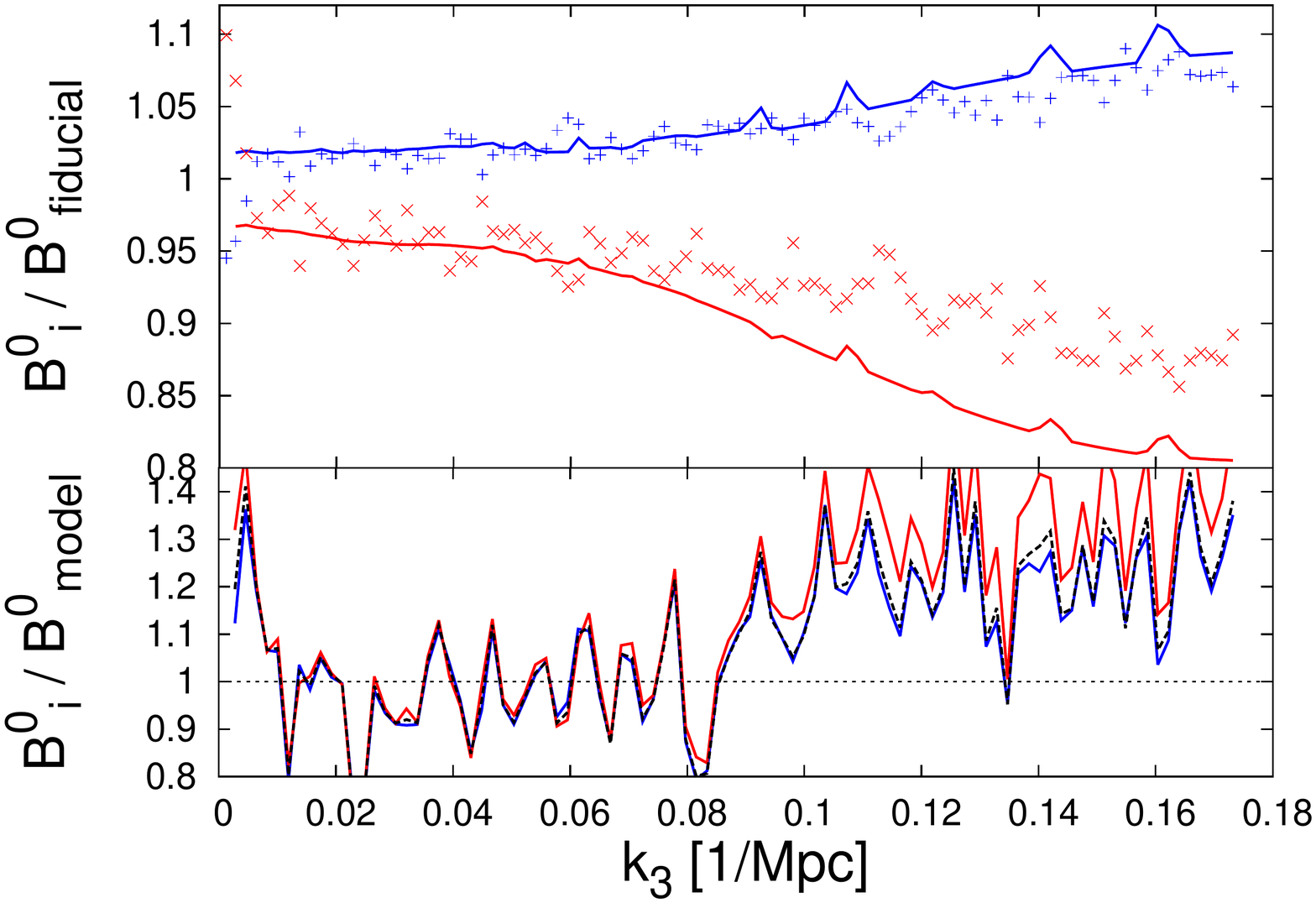}
\includegraphics[clip=false,trim= 25mm 10mm 22mm 35mm, scale=0.252]{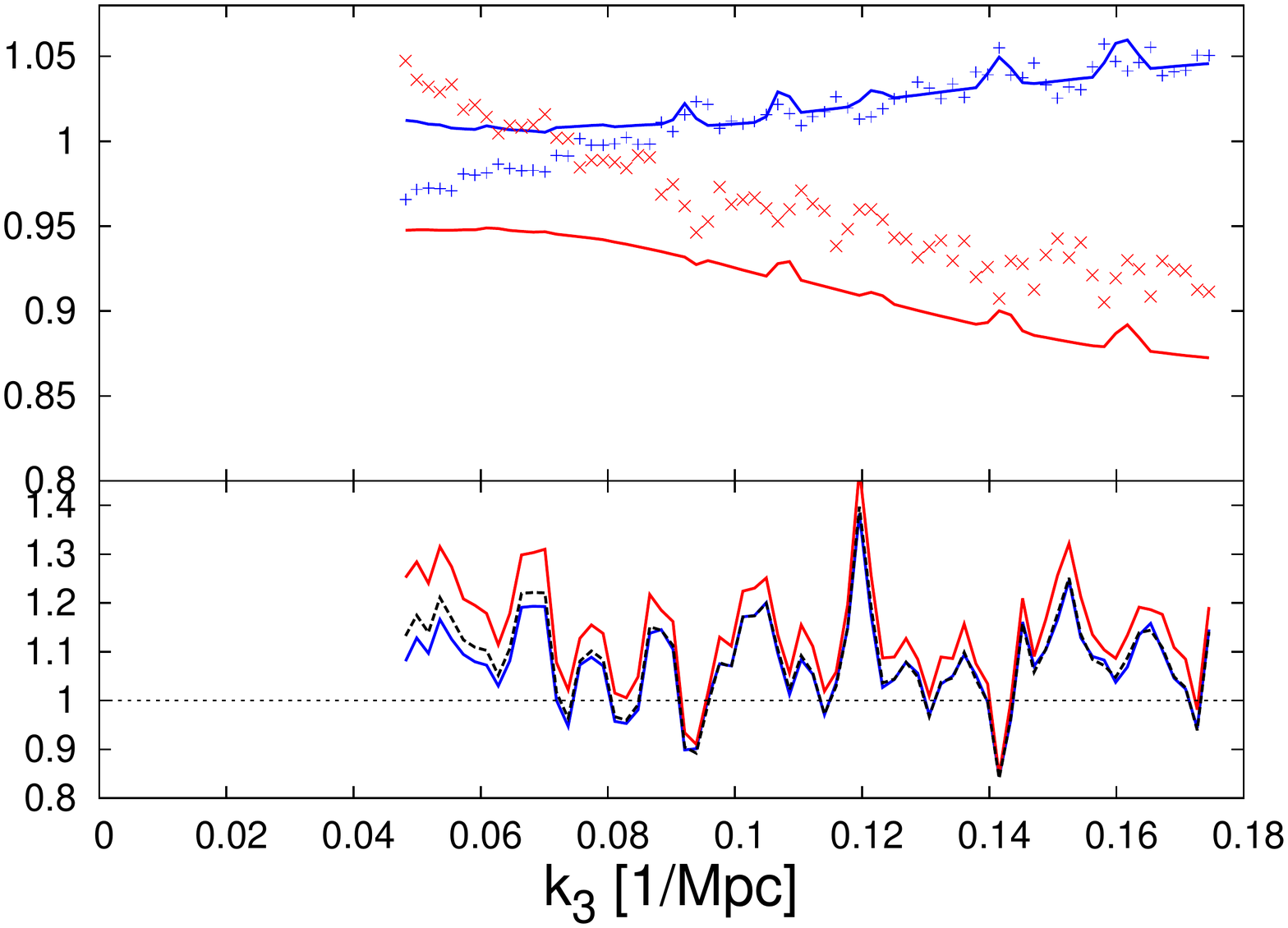}
\includegraphics[clip=false,trim= 25mm 10mm 80mm 35mm, scale=0.252]{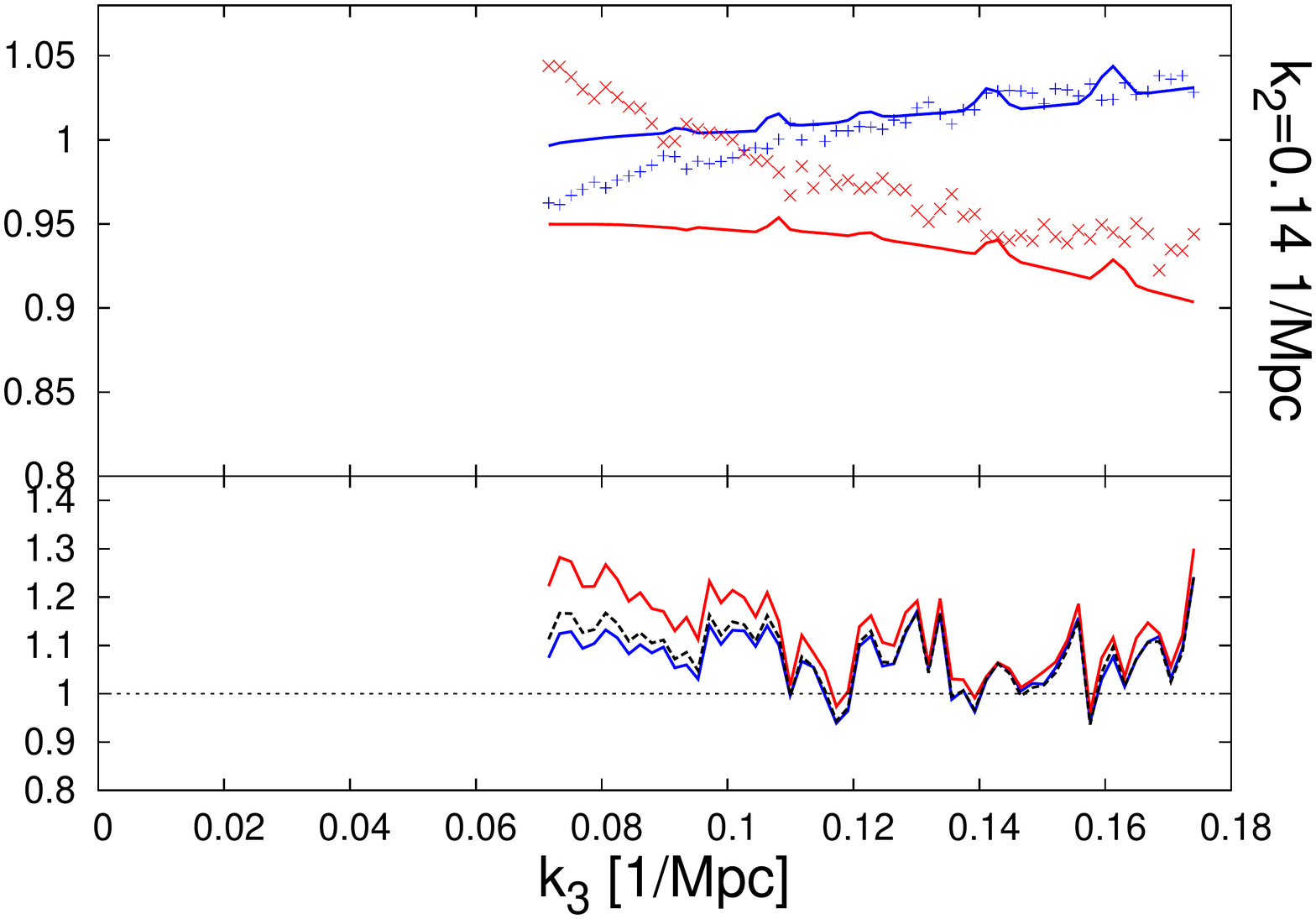}

\caption{Top sub-panels: dark matter monopole bispectrum of the low- (blue colour) and high-$\Omega_m$ (red colour) normalised to the dark matter monopole bispectrum of the fiducial cosmology for different triangle configurations. First column, second and third column panels are triangles with $k_2/k_1=1.0$, 1.5 and 2 respectively. Different rows show different scales: first, second, third and forth rows correspond to $k_2=0.05$, 0.10, 0.15 and 0.20\, $h$/Mpc as indicated. Blue and red symbols correspond to measurements from N-body simulations, whereas blue and red lines correspond to predictions of the fitting formula based on $B^{FG}$ model. Bottom sub-panels: relative deviation of dark matter bispectrum measurements to the prediction of the fitting formula for the fiducial cosmology (black dashed lines), low-$\Omega_m$ cosmology (blue solid lines) and the high-$\Omega_m$ cosmology (red solid lines). All panels  are at $z=0$.}
\label{bis5}
\end{figure}

In Fig. \ref{bis5} we show the bispectra for different scales and shapes following the same column-row notation as in Fig. \ref{bis1}-\ref{bis4}. In this case, top sub-panels show the ratio between the studied cosmology and the fiducial cosmology: $B^{(0)}_{\rm low-\Omega_m}/B^{(0)}_{\rm fiducial}$ (blue lines and symbols) and $B^{(0)}_{\rm high-\Omega_m}/B^{(0)}_{\rm fiducial}$ (red lines and symbols). Symbols show the ratio of the bispectrum measurements, whereas lines show the ratio of the theoretical predictions of $B^{FG}$. The measurements correspond to one single realisation, both for $B^{(0)}_{\rm fiducial}$, $B^{(0)}_{{\rm low}-\Omega_m}$ and $B^{(0)}_{{\rm high}-\Omega_m}$. Since the initial density realisation is the same for all these cosmologies, when dividing the bispectra,  the statistical errors due to sample variance are mostly cancelled out. Thus, computing just one realisation, we are able to detect relative changes in the bispectrum among these different cosmologies with much higher precision than the precision of the  bispectrum measurement itself.  For each cosmology, $B^{FG}$ is computed using the values of the ${\bf a}^F$ and ${\bf a}^G$ parameters presented in \S\ref{sec:halo_bispectrum} and \S\ref{bis_section}, respectively. The true values for $f$ and $\sigma_8$ are assumed, and the measured values for real space power spectrum are used. The only parameter we allow to freely vary among the different cosmologies is $\sigma_{\rm FoG}^B$, corresponding to the non-linear damping function of Eq. \ref{D_fog_sc}. The values of $\sigma_{\rm FoG}^B$ used for each cosmology are listed in Table~\ref{table_sims_cosmo22}. Note that since the differences between the real space power spectra of the cosmologies presented here are very small (see top panel of Fig. \ref{PS_P0_P2}), the variations on $B^{FG}$ from one cosmology to another are entirely due to the effective  kernels $F^{\rm eff}$ and $G^{\rm eff}$. Thus, the role that the effective kernels play when the cosmology is varied is neatly shown in the panels of Fig. \ref{bis5}.
  
For clarity, the bottom sub-panels show the relative deviation between the measured bispectrum and the prediction by $B^{FG}$. Here, since we are not dividing two measurements, the statistical errors of the bispectra do not cancel. However, we can see that the statistical fluctuations follow the same pattern  for the cosmologies studied here, as expected, since the random realisation of the initial density field is the same for all of them. We do not show the statistical errors, since we are only interested in the relative deviations among bispectra.  

From the different panels of Fig. \ref{bis5} we can infer that the differences between the measured quantities $B^{(0)}_i/B^{(0)}_{\rm fiducial}$, and the corresponding predictions by $B^{FG}$, among  the fiducial, the low-$\Omega_m$ and the high-$\Omega_m$ cosmologies are typically $\leq 5\%$, at the scales where $B^{FG}$ can be trusted for the fiducial cosmology (see Fig. \ref{bis1}). Beyond these scales, the deviations can be higher, but these cases are not of practical interest, since the fitting formula is not a good description for the fiducial cosmology either.  

To summarise,  the effective kernels $F^{\rm eff}$ and $G^{\rm eff}$, with the values of the  ${\bf a}^F$ and ${\bf a}^G$ parameters presented in previous sections, are able to describe the bispectrum of cosmologies with $0.2 \lesssim \Omega_m \lesssim 0.4$ (and therefore with $0.4\lesssim f(z=0) \lesssim 0.6$), with similar accuracy to that shown  for the fiducial cosmology in \S\ref{bis_section}. Outside this range, the fitting formula may describe less well the measured bispectrum at mildly non-linear scales. However, describing a universe with $\Omega_m>0.4$ or $\Omega_m<0.2$ may not be of practical interest, since these values of $\Omega_m$ are in $>3\sigma$ tension with the measurements of CMB \citep{WMAP,Planck}.

\begin{table}[htdp]
\begin{center}
\begin{tabular}{|c|c|c|c|c|c|c|c|c|c|c|}
\hline
 & fiducial & low-$\Omega_m$ & high-$\Omega_m$  \\
 \hline
 \hline
 $\sigma_{\rm FoG}^{B}$ [Mpc] & 58.28 & 67.42 & 52.16 \\
 \hline
\end{tabular}
\end{center}
\caption{Best-fitting values for the FoG damping parameters for the different cosmologies. These values correspond to the fitting formula, $B^{FG}$, plotted in Fig. \ref{bis5}. }
\label{table_sims_cosmo22}
\end{table}%

\end{document}